\documentclass[12pt,a4paper]{report}

\usepackage{latexsym}
\usepackage{amssymb,amsfonts,amsmath}
\usepackage{graphicx} 
\usepackage{indentfirst}
\usepackage{bbm}
\usepackage{amssymb}
\usepackage{verbatim}
\usepackage{amsmath, amsthm,amssymb}
\usepackage{mathrsfs}
\usepackage{hyperref}
\usepackage{amsfonts}
\usepackage{dsfont}
\usepackage{cite}
\usepackage{pdfpages}
\usepackage[export]{adjustbox}

\parskip=\medskipamount

\newcommand {\cA}{{\cal A}}
\newcommand {\cB}{{\cal B}}
\newcommand {\cC}{{\cal C}}
\newcommand {\cD}{{\cal D}}
\newcommand {\cE}{{\cal E}}

\newcommand {\cL}{{\cal L}}
\newcommand {\cM}{{\cal M}}
\newcommand {\cN}{{\cal N}}

\newcommand {\cR}{{\cal R}}
\newcommand {\cS}{{\cal S}}

\newcommand {\cV}{{\cal V}}
\newcommand {\cW}{{\cal W}}

\def\a{\alpha}

\def\b{\beta}
\def\c{\chi}
\def\d{\delta}
\def\e{\epsilon}
\def\f{\phi}
\def\g{\gamma}
\def\G{\Gamma}

\def\j{\psi}

\def\l{\lambda}
\def\m{\mu}

\def\o{\omega}

\def\q{\theta}
\def\r{\rho}
\def\s{\sigma}
\def\t{\tau}

\def\x{\xi}
\def\z{\zeta}

\def\F{\Phi}
\def\J{\Psi}
\def\L{\Lambda}
\def\O{\Omega}

\def\S{\Sigma}
\def\U{\Upsilon}

\def\tr{{\rm tr}}
\def\rd{{\rm d}}
\def\ri{{\rm i}}
\def\re{{\rm e}}

\newcommand{\ad}{{\dot{\alpha}}}                           
\newcommand{\bd}{{\dot{\beta}}}                            
\newcommand{\ve}{\varepsilon}                            
\newcommand{\cDB}{{\bar\cD}}                            

\newcommand{\ab}{{\a\b}}

\newcommand{\pa}{\partial}                           
\newcommand{\hf}{\frac12}

\newcommand{\vf}{\varphi}

\newcommand{\be}{\begin{equation}}
\newcommand{\ee}{\end{equation}}
\newcommand{\bea}{\begin{eqnarray}}
\newcommand{\eea}{\end{eqnarray}}
\newcommand{\non}{\nonumber}
\newcommand{\1}{{\underline{1}}}
\newcommand{\2}{{\underline{2}}}

\newcommand{\bm}[1]{\mbox{\boldmath$#1$}}

\def\double #1{#1{\hbox{\kern-2pt $#1$}}}

\newcommand{\gd}{{\dot\g}}

\newcommand{\bcD}{{\bm \cD}}

\newif\ifdtup

\def\de{{\nabla}}                                         

\newcommand{\bsubeq}{\begin{subequations}}
\newcommand{\esubeq}{\end{subequations}}

\newcommand{\mub}{{{\bar{\mu}}}}

\numberwithin{equation}{section}

\newcommand\frontmatter{\pagenumbering{roman}}
\newcommand\mainmatter{\cleardoublepage\pagenumbering{arabic}}
\begin{document}

	\begin{titlepage}
		\centering
		
		\includegraphics[width=.50\linewidth, right]{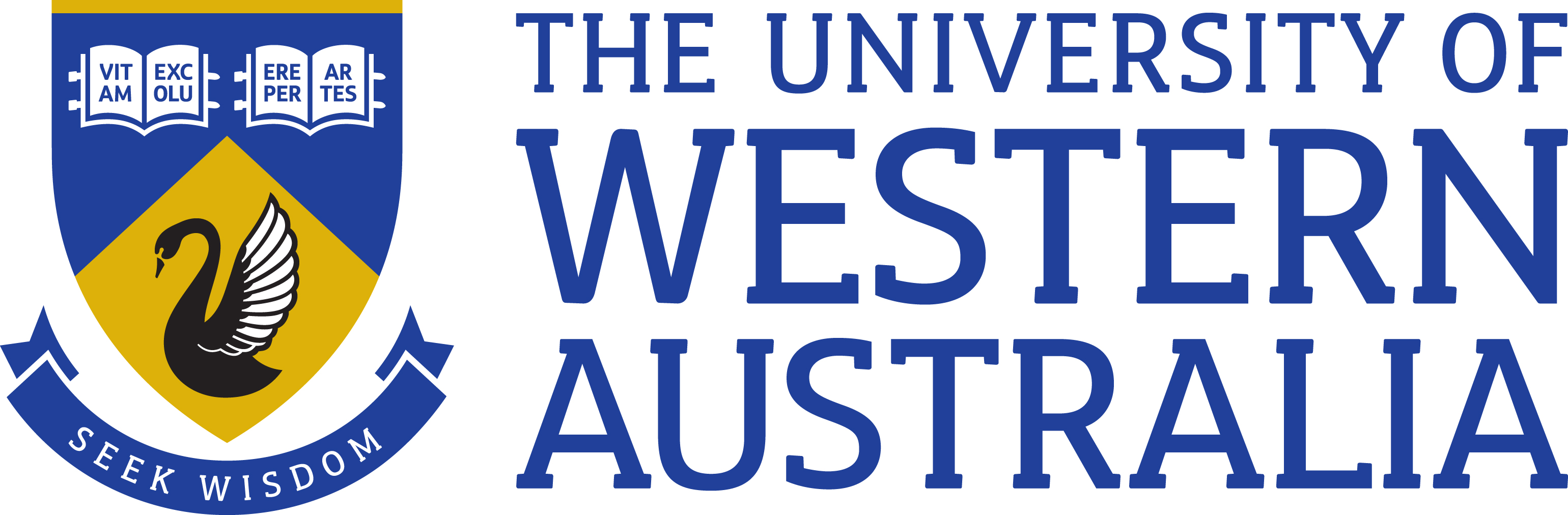}
		\vskip 1.5cm
		{\bfseries\LARGE
			Off-shell higher-spin gauge supermultiplets \\and conserved supercurrents\\} 
		
		\vskip 1.5cm
		
		{\Large Jessica Hutomo}
		
		\vskip 2cm
		\begin{center}
		{\large
		This thesis is presented for the degree of Doctor of Philosophy \\of The University of Western Australia\\
		Department of Physics\\
		March 2020}
		\end{center}
		\vskip 1.5cm
		
\begin{flushleft}
{\large {Supervisor:} \qquad\,\,\,
 		    Prof. Sergei M. Kuzenko\\
 		{Co-supervisor(s):}\,\,
 			A/Prof. Evgeny I. Buchbinder\\
 			\qquad \qquad \qquad \quad \,\,\,\,\,\,\,\,Prof. Ian N. McArthur\\ 
 		\vskip 1.1cm
 		{Examiners:} \\
 		Prof. Evgeny A. Ivanov \hfill{ (Joint Institute for Nuclear Research, Russia)}\\
Prof. Ulf Lindstr\"{o}m \hfill{(Uppsala University, Sweden)}\\
Prof. Rikard von Unge \hfill{(Masaryk University, Czech Republic)} \\	}
 \end{flushleft}

	\end{titlepage}

  \frontmatter

\begin{abstract}
\thispagestyle{plain}
\setcounter{page}{1}
This thesis presents the general structure of non-conformal higher-spin supercurrent multiplets in three and four spacetime dimensions. Such supercurrents are in one-to-one correspondence with off-shell massless higher-spin gauge supermultiplets, some of which are constructed in this thesis for the first time. Explicit realisations of these conserved current multiplets in various supersymmetric theories are worked out in detail. 

In the first part of the thesis, we begin by reviewing the key properties of known massless higher-spin ${\cN}=1$ supermultiplets in four-dimensional (4D) Minkowski and anti-de Sitter (AdS) backgrounds. We then propose a new off-shell gauge formulation for the massless integer superspin multiplet. Its novel feature is that the gauge-invariant action involves an unconstrained complex superconformal prepotential, in conjunction with two types of compensators. Its dual version is obtained by applying a superfield Legendre transformation.
Next, we deduce the structure of consistent non-conformal higher-spin ${\cN}=1$ supercurrents associated with these massless supersymmetric gauge theories. Explicit closed-form expressions for such supercurrents are derived for various supersymmetric theories in 4D ${\cN}=1$ Minkowski and AdS superspaces. These include a model of $N$ massive chiral superfields with an arbitrary mass matrix, along with free theories of tensor and complex linear multiplets. 

The second part of the thesis is devoted to a detailed study of ${\cN}=1$ and ${\cN}=2$ supersymmetric higher-spin theories in three-dimensional AdS space.
By analogy with our 4D ${\cN}=1$ constructions, we derive two dually equivalent off-shell Lagrangian formulations for the massless multiplets of \textit{arbitrary} superspin in (1,1) AdS superspace. These formulations allow us to determine the most general higher-spin supercurrent multiplets and provide their examples for models of chiral superfields. 
With regards to (2,0) AdS supersymmetry, 
our approach is to first identify a multiplet of conserved higher-spin currents in simple models for a chiral superfield. This is then used to construct two series of a massless \textit{half-integer} superspin multiplet in (2,0) AdS superspace.
Finally, our (2,0) AdS higher-spin supermultiplets are reduced to (1,0) AdS superspace, which yield four series of ${\cN}=1$ supersymmetric massless higher-spin models. 
We illustrate the duality transformations relating some of these dynamical systems. We also perform the component reduction of two new ${\cN}=1$ higher-spin actions in flat superspace. Further applications of these off-shell ${\cN}=1$ models are discussed, one of which is related to the construction of two new off-shell formulations for the massive ${\cN}=1$ gravitino supermultiplet in AdS. 
\end{abstract}
\cleardoublepage

\allowdisplaybreaks

\chapter*{Acknowledgements}
\thispagestyle{plain}
\setcounter{page}{2}
First and foremost, I would like to sincerely thank my supervisor, Prof.~Sergei Kuzenko, with whom I have been working for the past 3.5 years. His expertise and passion in theoretical physics are truly inspiring, and I feel very fortunate to be one of his PhD students.  
The work presented in this thesis would not have been possible without his guidance and constant support at every stage of my research.

I would also like to thank my co-supervisors, A/Prof.~Evgeny Buchbinder and Prof. Ian McArthur, for their helpful suggestions. I am really grateful to Evgeny for his collaborations on our papers throughout my PhD candidature. 
His guidance and availability for discussions have made it a wonderful learning experience. 

I wish to thank my thesis examiners: Prof.~Evgeny Ivanov, Prof.~Ulf Lindstr\"{o}m and Prof.~Rikard von Unge, for their valuable comments and suggestions. 

I am grateful to Dr.~Gabriele Tartaglino-Mazzucchelli, Daniel Ogburn and Daniel Hutchings for their collaborations on our joint papers.

To my fellow PhD colleagues in Field Theory and Quantum Gravity group: Michael, Daniel H, Emmanouil, James, Ben. Thank you for a wonderful time and many interesting discussions. In particular, I would like to thank Daniel, Emmanouil and Michael for their comments on some parts of this thesis. 

Last, but certainly not least, my gratitude goes out to my family, especially my parents. Thank you very much for the endless support, love and prayers throughout my studies. 

This research was supported by an Australian Government Research Training Program (RTP) Scholarship, and a University Postgraduate Award for International Students. 

\newpage

\chapter*{Authorship declaration}
\thispagestyle{plain}
This thesis is based on six published papers \cite{HK1, HK2, BHK, HKO, HK18, HK19}. The details are as follows:
\vskip 0.5cm
\begin{enumerate}
\setlength{\itemsep}{0.5cm}
\item
  J.~Hutomo and S.~M.~Kuzenko,
  ``Non-conformal higher spin supercurrents,''
  Phys.\ Lett.\ B {\bf 778}, 242 (2018)
   [arXiv:1710.10837 [hep-th]].\\
{\bf Location in thesis:} Chapter \ref{ch3}
%

\item
  J.~Hutomo and S.~M.~Kuzenko,
  ``The massless integer superspin multiplets revisited,''
  JHEP {\bf 1802}, 137 (2018)
  [arXiv:1711.11364 [hep-th]]. \\
{\bf Location in thesis:} Chapter \ref{ch3}
%

\item E.~I.~Buchbinder, J.~Hutomo and S.~M.~Kuzenko,
``Higher spin supercurrents in anti-de Sitter space,''
JHEP {\bf 1809}, 027 (2018)
[arXiv:1805.08055 [hep-th]].\\
{\bf Location in thesis:} Chapter \ref{ch4}, appendix \ref{AppendixC}

%
%
\item J.~Hutomo, S.~M.~Kuzenko and D.~Ogburn,
  ``${\cal N}=2$ supersymmetric higher spin gauge theories and current multiplets in three dimensions,''
  Phys.\ Rev.\ D {\bf 98}, no. 12, 125004 (2018)  [arXiv:1807.09098 [hep-th]].\\
{\bf Location in thesis:} Chapter \ref{ch5}
%
%
%
%
%
\item J.~Hutomo and S.~M.~Kuzenko,
  ``Higher spin supermultiplets in three dimensions: (2,0) AdS supersymmetry,''
  Phys.\ Lett.\ B {\bf 787}, 175 (2018)
  [arXiv:1809.00802 [hep-th]].\\
{\bf Location in thesis:} Chapter \ref{ch5}
%
%
%
\item J.~Hutomo and S.~M.~Kuzenko,
  ``Field theories with (2,0) AdS supersymmetry in ${\cal N}=1$ AdS superspace,''
  Phys.\ Rev.\ D {\bf 100}, no. 4, 045010 (2019)
  [arXiv:1905.05050 [hep-th]].\\
{\bf Location in thesis:} Chapter \ref{ch6}, appendices \ref{AppA2} and \ref{AppendixBB}
%
\end{enumerate}

The following published papers are not related to the subject of the thesis, but were produced during the candidature of my PhD.
\begin{enumerate}
\item E.~I.~Buchbinder, J.~Hutomo, S.~M.~Kuzenko and G.~Tartaglino-Mazzucchelli,
  ``Two-form supergravity, superstring couplings, and Goldstino superfields in three dimensions,''
  Phys.\ Rev.\ D {\bf 96}, no. 12, 126015 (2017)
  [arXiv:1710.00554 [hep-th]].
  
\item E.~I.~Buchbinder, D.~Hutchings, J.~Hutomo and S.~M.~Kuzenko,
  ``Linearised actions for $ \mathcal{N} $-extended (higher-spin) superconformal gravity,''
  JHEP {\bf 1908}, 077 (2019)
  [arXiv:1905.12476 [hep-th]].
\end{enumerate}
\vskip 2cm

Permission has been granted to use this work. 
\vskip 1cm
Jessica Hutomo
\vskip 0.5cm
Sergei Kuzenko
\vskip 0.5cm
Evgeny Buchbinder
\vskip 0.5cm
Daniel Ogburn


    
	\setcounter{tocdepth}{2}
	\tableofcontents
	\mainmatter


\chapter{Introduction}
Relativistic quantum field theory gives a powerful framework to describe all known elementary particles to a good accuracy (see \cite{Wein1, Wein2} for reviews).
At present, our theoretical understanding of the fundamental interactions of Nature is based on the Standard Model of particle physics and Einstein's theory of gravity.
The Standard Model accounts for three of the four fundamental interactions, \textit{i.e. }the electromagnetic, weak and strong forces. It is a renormalisable quantum field theory, and is described in terms of the Yang-Mills gauge fields coupled to some matter sector, with the gauge group $\rm SU(3) \times SU(2) \times U(1)$. All the fundamental particles encoded in such a formulation have spin $s \leq 1$. Gravity is described by general relativity at the classical level, which is a non-Abelian gauge theory of a spin-2 field possessing a diffeomorphism invariance. While the gravitational interaction is irrelevant at the energy scale of the Standard Model, quantum gravitational effects are not negligible at the Planck scale, which is of the order of $10^{19}$ GeV. 
Since general relativity is non-renormalisable, one of the main challenges of modern theoretical physics is to reconcile quantum field theory and general relativity in order to arrive at a consistent quantum theory of gravity.
Given that the regions associated to the Planck energy are not directly accessible using existing experiments, the development of this subject has been mainly driven by symmetry principles and other theoretical ideas. 

Supersymmetry in four dimensions was discovered by Golfand and Likhtman \cite{GL}, Volkov and Akulov \cite{VA}, and Wess and Zumino \cite{WZ1}. It is a symmetry relating two types of particles in Nature, bosons and fermions, which were previously unrelated in field theories.
The generators of supersymmetry transformations are fermionic, thus they obey anticommutation relations. 
The supersymmetric extension of the Poincar\'e group
is known as ${\cN}$-extended super-Poincar\'e group. 
This involves the addition of ${\cN}$ Majorana spinor generators to the generators of the Poincar\'e algebra, where $\cN$ is a positive integer. 

Despite the lack of experimental evidence, supersymmetry remains attractive for its phenomenological and theoretical applications.
It offers some solutions to long standing problems, such as the hierarchy problem, gauge unification, etc, see e.g. \cite{Dine} for a review. On the theoretical side, it is interesting to study a supersymmetric extension of general relativity
\footnote{As discovered by Volkov and Soroka \cite{VSor}, gauging the ${\cN}=1$ super-Poincar\'e group leads to the supergravity action with nonlinearly realised local supersymmetry.}, which is supergravity \cite{FvNF, DZ-sugra}. The latter is the gauge theory of supersymmetry. It arises if one makes supersymmetry transformations local, with the gravitino (spin 3/2) being the corresponding gauge field. Its superpartner, the graviton (spin 2), is associated to the diffeomorphism invariance. 
Supergravity multiplet consists of the graviton and ${\cN}$ gravitino fields (for $\cN >1$, the multiplet also contains some vector and scalar fields).

Supergravity has some remarkable properties, see \cite{Wein3,FSrev} for reviews. Most notably, it is a low-energy limit of superstring theory (see \cite{GSWbook, Pol} for reviews), which is currently the leading candidate to give a unified description of the four fundamental forces. 
Superstring theory includes in its spectrum an infinite tower of massive higher-spin excitations, which lead to improved ultraviolet behaviour of the theory.
In recent years, there has also been an incredibly productive area of research to understand the relationship between supergravity and the dynamics of such higher-spin states. 
In the past, most of the studies were restricted to $\cN \leq 8$ case. Theories with ${\cN} > 8$ would include massless fields with higher-spin (helicity) $s >2 $, and at that time it was not known how to couple such fields consistently to gravity \cite{Vas04}. 

The study of higher-spin fields has been carried out quite independently of string theory, initiated in the works of Dirac \cite{Dirac36}, Fierz and Pauli \cite{FPauli}, Rarita and Schwinger \cite{RS41} and many others. 
As follows from Wigner's classification \cite{Wigner}, free elementary particles are associated with the unitary irreducible representations of the Poincar\'e group, the latter are classified by mass and spin. In any physical system, the spin can take arbitrary integer or half-integer values. The term ``higher-spin" refers to fields with spin $s > 2$, which are higher-rank tensor representations of the Poincar\'e group. Of particular interest are massless higher-spin fields and their gauge symmetries underlying their dynamics, which pose challenges in the construction of interaction vertices. We now provide a brief historical overview of higher-spin theory, see e.g. \cite{Vas04, Sorokin, FTrev, BBS, BCIV} for complete reviews. 

Despite the pioneering works of Dirac \cite{Dirac36}, Fierz and Pauli \cite{FPauli} who formulated relativistic wave equations for free massive fields of arbitrary spin,
it took over thirty years until the corresponding Lagrangian description
were constructed by Singh and Hagen \cite{SH1, SH2}.
In the massless limit, consistent Lagrangians were constructed in 4D flat and (anti)-de Sitter ((A)dS) spaces. 
They were proposed by Fronsdal \cite{Fronsdal, FronsdalAdS} in the bosonic case, and by Fang and Fronsdal for fermionic fields \cite{FF, FFAdS}. 
Section 6.9 of \cite{Ideas} contains a pedagogical review of the (Fang-)Fronsdal actions in Minkowski space \cite{Fronsdal, FF} in the two-component spinor formalism.\footnote{Such a formalism will be used in this thesis. Not only is this formalism useful for higher-spin calculations, but it is also well adopted to the framework of supersymmetry.}

Constructing consistent deformations of Fronsdal's Lagrangian, which should lead to fully interacting higher-spin gauge theories, proved to be very challenging.
This was due to a large number of highly restrictive no-go theorems (e.g. \cite{CM, WT1, WW}) which, under certain assumptions, rule out any gravitational interactions of massless higher-spin fields in flat space. 
In the 1980s, several cubic vertices for higher-spin fields interacting with each other were constructed in flat space \cite{BBB83,BBvD2, Metsaev93}. It was also observed that any consistent interacting theory must involve fields of all spins and higher derivative terms.

The first successful result on higher-spin gravitational interactions was achieved by Fradkin and Vasiliev in 1987 \cite{FV87}. They constructed a Lagrangian describing consistent cubic vertices in the presence of a cosmological term, \textit{i.e.} in an (A)dS background where the no-go results can be evaded. 
This line of research culminated in Vasiliev's unfolded formulation \cite{Vas90}, in which
the fully interacting theories were described in terms of consistent nonlinear equations of motion for an infinite spectrum of higher-spin fields. This was later extended to arbitrary spacetime dimensions \cite{Vas03}.
However, one of the issues is that a \textit{Lagrangian} formulation for these nonlinear theories is still unknown, thus constraining our understanding of their quantum properties.

Since the early 1990s, higher-spin gauge theory has been an intense field of research in modern theoretical and mathematical physics.
That AdS space is the natural setup for interacting higher-spin fields has motivated further studies in the context of the AdS/CFT correspondence, leading to some conjectures relating higher-spin theories to weakly coupled conformal field theories \cite{GG, Klebanov}.
There are many other important research directions in higher-spin theory, such as the unfolded formulation, (topologically) massive higher-spin, BRST approach, etc (see e.g. \cite{Vas04, Sorokin, FTrev, BBPT, Zinoviev, KO, KT, KP1, Buch2} and the references therein). 
As mentioned previously, a major motivation for current research stems from its close connection with (super)string theory. It is conjectured that the latter is a spontaneously broken phase of a massless higher-spin theory. 
In this respect, it is of interest to study supersymmetric extensions of the massless higher-spin fields of \cite{FronsdalAdS,FFAdS}. 

A powerful approach to construct supersymmetric field theories makes use of the concept of superspace. Volkov and Akulov \cite{VAsuperspace} introduced an ${\cN}$-extended superspace in the framework of nonlinear realisations. Salam and Strathdee \cite{Salam} proposed to use superspace and superfields, as tools to construct and study supersymmetric theories. Superspace is an extension of spacetime by anticommuting (Grassmann) coordinates, while superfields are functions of the superspace coordinates. 
One has to impose certain constraints on superfields to describe irreducible representations of the ${\cN}$-extended super-Poincar\'e algebra, known as supermultiplets. A series expansion of a superfield in the Grassmann coordinates will be finite due to their anticommuting nature. The coefficients in such an expansion correspond to component fields of the superfield, which are ordinary bosonic and fermionic functions of spacetime coordinates. A supersymmetric theory is written compactly in terms of superfields, thus supersymmetry is kept manifest.
Throughout this thesis, we will employ the superspace approach. For a thorough introduction to this formalism, the reader is referred to \cite{Ideas, WB, GGRS}.

In general, the superspace approach is not widely used by higher-spin practitioners. This thesis is primarily devoted to higher-spin multiplets of conserved currents in supersymmetric field theories.
As first shown by Ferrara and Zumino \cite{FZ}, the conserved energy-momentum tensor and spin-vector supersymmetry current(s) associated with the supertranslations are embedded in a supermultiplet, called the supercurrent. It should be pointed out that this multiplet of currents is always \textit{off-shell}\footnote{A supermultiplet is called off-shell if the algebra of supersymmetry transformations closes off the mass shell, {\it i.e.}
without imposing the equations of motion. Otherwise, the supermultiplet under consideration is called on-shell.} by construction \cite{BdeRdeW}. 
Given an on-shell superfield, one may construct another multiplet by taking bilinear combinations of the component fields of the superfield. One then considers their variations under supersymmetry transformations to find the other components of the composite multiplet. The on-shell condition is not preserved when taking such products. The resulting multiplet of bilinears thus forms an off-shell multiplet. An example for this is the supercurrent, which is a composite of the underlying on-shell matter superfield. 
Off-shell supermultiplets require auxiliary fields in their description. 
In order to efficiently formulate off-shell supersymmetric theories, superspace techniques are absolutely essential, since all the appropriate auxiliary fields are included automatically.\\

\vspace{0.1cm}

\noindent \textbf{Off-shell massless higher-spin supermultiplets}\\
In 4D $\cN=1$  supersymmetric field theory,
a massless superspin-$\hat s$ ($\hat{s}=\hf,1,\dots$) multiplet describes two ordinary 
massless fields of spin $\hat s$ and $\hat s + \hf$. Such a supermultiplet is often denoted by $(\hat s, \hat s +\hf)$. The three lowest superspin values, 
$\hat s=\hf, 1$ and $ \frac{3}{2}$, 
correspond to the vector, gravitino  and supergravity multiplets, respectively. 

It follows from first principles that the sum of two actions 
for free massless spin-$\hat s$ and spin-$(\hat s+\hf)$ fields 
should possess an on-shell supersymmetry. Thus, there is no problem of constructing on-shell massless higher-spin supermultiplets ($\hat s > \frac{3}{2}$), for one only needs to work out the supersymmetry transformations leaving invariant the pair of (Fang-)Fronsdal actions \cite{Fronsdal,FF}. In four dimensions,
this task was completed first by Curtright  \cite{ Curtright} who made use of the (Fang-)Fronsdal actions, and soon after by Vasiliev \cite{Vas80} with his frame-like reformulation of the (Fang-)Fronsdal models. The nontrivial problem, however, is to construct off-shell massless higher-spin supermultiplets. Early attempts to construct 
such off-shell realisations \cite{BO, BO2} were unsuccessful, as explained in 
detail in \cite{KS94}.

The problem of constructing gauge \textit{off-shell} superfield realisations for free massless higher-spin supermultiplets
was solved in the early 1990s by Kuzenko, Postnikov and Sibiryakov
\cite{KPS,KS}.
For each superspin value $\hat s > \frac{3}{2} $, 
half-integer \cite{KPS} and integer \cite{KS},
these publications provided two dually equivalent off-shell
actions formulated in 4D ${\cal N }= 1$ Minkowski superspace.
At the component level, each of the
two superspin-$\hat s$ actions \cite{KPS,KS} reduces, upon imposing a Wess-Zumino-type
gauge and eliminating the auxiliary fields,
to a sum of the spin-$\hat s$ and  spin-$(\hat s+\hf)$ actions \cite{Fronsdal,FF}.
The models \cite{KPS, KS} thus provided the first manifestly supersymmetric extensions of the (Fang-)Fronsdal actions for massless higher-spin fields. A pedagogical review of the results obtained
in \cite{KPS,KS} can be found in section 6.9 of \cite{Ideas}. 
In \cite{KS94}, the massless higher-superspin theories of \cite{KPS,KS} were generalised 
to 4D ${\cN}=1$ AdS superspace, ${\rm AdS}^{4|4}$, and their quantisation was carried out in \cite{BKS}. Building on the ${\cN}=1$ analysis, off-shell massless ${\cN} = 2$ supermultiplets were proposed in \cite{GKS1}. 
Models describing off-shell ${\cN}=1$ superconformal higher-spin multiplets were constructed in \cite{KMT}, which made use of the gauge prepotentials introduced in \cite{HST}. 

The structure of the (Fang-)Fronsdal gauge-invariant actions and their ${\cN}=1$ supersymmetric counterparts of half-integer superspin share one common feature. For each of them, the action is written in terms of two multiplets: a (super)conformal gauge (super)field coupled to certain compensators. 
The (Fang-)Fronsdal actions \cite{Fronsdal, FF} can be interpreted as gauge-invariant models described by the Fradkin-Tseytlin conformal gauge fields \cite{FT, FL} and an appropriate set of compensators.\footnote{See the beginning of chapter \ref{ch3} for the details.} The massless half-integer superspin actions of \cite{KPS} involve not only the {\it real} superconformal gauge prepotential \cite{KMT}, but also some complex compensating superfields. Such a description was previously unknown for the massless multiplet of \textit{integer} superspin \cite{KS}. 
This was the motivation for the recent work \cite{HK2}, where we proposed a new off-shell formulation for the massless multiplet of \textit{integer} superspin. Its properties are: (i) the gauge freedom matches that of the {\it complex} superconformal integer superspin multiplet introduced in \cite{KMT}; and (ii) the
action involves two compensating multiplets, in addition to the superconformal integer superspin multiplet. Upon imposing a partial gauge fixing, this action
reduces to the so-called longitudinal formulation for the integer superspin \cite{KS}. This construction was later lifted to ${\cN}=1$ AdS supersymmetry \cite{BHK}.\\

\noindent \textbf{(Higher-spin) supercurrents}\\
As previously pointed out, the concept of supercurrent was introduced by Ferrara and Zumino \cite{FZ} in the context of ${\cN}=1$ supersymmetry. This was later extended to 4D ${\cN}=2$ supersymmetry by Sohnius \cite{Sohnius}. 

The multiplet of currents in superconformal field theories has a simpler structure. The ${\cN}=1$ conformal supercurrent multiplet contains the symmetric traceless energy-momentum tensor $T_{ab}$,
the spin-vector supersymmetry current $S_a$ and the $R$-symmetry ({\it i.e.} ${\rm U(1)}_R$) current $j_a$.

In the non-superconformal case (e.g. ${\cN}=1$ Poincar\'e supersymmetric theories), the supercurrent multiplet also includes the trace multiplet containing the trace of the energy-momentum tensor and the $\gamma$-trace of the supersymmetry current. In some cases, the trace multiplet also contains the divergence of the  ${\rm U(1)}_R$ current, $\pa_a j^a$.
Different supersymmetric theories may possess different trace multiplets. This means that the problem of classifying inequivalent non-conformal supercurrent multiplets needs to be addressed. Ten years ago, there appeared numerous papers devoted to studying consistent ${\cN}=1$ supercurrents in four dimensions \cite{MSW, KS-Fayet, SMK-Fayet, KS2, K-var,K-Noet, DS}. 

Supercurrent can be viewed as the source of supergravity \cite{OS1, FZ2}, in complete analogy with the energy-momentum tensor as the source of gravity. This idea proves to be powerful in deriving various consistent supercurrents. Given a linearised off-shell supergravity action, the supercurrent conservation equation is obtained by coupling the supergravity prepotentials to external sources and demanding invariance of the resulting action under the linearised supergravity gauge transformations. Using this procedure, the general structure of consistent supercurrents are presented in \cite{K-var, K-Noet, BK_supercurrent}
for $\cN=1$ and $\cN=2$ super-Poincar\'e cases in four 
dimensions,  
and in \cite{KT-M11} for $\cN=2$ supersymmetric theories in three dimensions. 

On the other hand, supercurrent can be used to deduce the off-shell structure of a massless supermultiplet which is associated to it. 
The procedure to follow is concisely described
by Bergshoeff et {\it al.} \cite{BdeRdeW}:
``One first assigns a field to each component of the current multiplet,
and forms a generalised inner product of field and current components.''
Indeed, this approach has been used in the past to construct off-shell supergravity multiplets in diverse dimensions \cite{BdeRdeW,Sohnius, SW,SohniusW2,SohniusW3,HL,Howe5Dsugra}.
The point is that the currents always form an off-shell multiplet, thus the fields to which they couple must also be off-shell. 
For example, in \cite{Sohnius} the ${\cN}=2$ supercurrent for the massive hypermultiplet model was used to derive the minimal multiplet of ${\cN}=2$ Poincar\'e supergravity.

All off-shell formulations for ${\cN}=1$ Poincar\'e supergravity are described in terms of the real gravitational superfield $H^{\a \ad}$ \cite{OS1} and a compensator. The gravitational superfield couples to the supercurrent $J_{\a \ad}$, while the source associated with the compensator is the trace multiplet. Thus, different choices of compensator lead to variant non-conformal supercurrents. 
Since the linearised off-shell $\cN=1$ supergravity actions have been classified \cite{GKP},
all minimal consistent supercurrents are readily derivable \cite{K-var}.
Reducible supercurrents, such as the $S$-multiplet introduced by Komargodski and Seiberg \cite{KS2}, 
can be obtained by combining some of the minimal ones.
  
Various aspects of field theories with ${\cN}=1$ AdS supersymmetry have been studied 
in detail over the last forty years, see e.g. 
\cite{FS,Keck,Zumino77,IS1, IS2,BKsigma,AJKL,Burges:1985qq,Aharony:2015hix} and references therein. The works of Ivanov and Sorin \cite{IS1,IS2} are fundamental in the development of superfield techniques. They classified off-shell superfield representations of the $\rm OSp(1|4)$ group (\textit{i.e.} the isometry group of ${\rm AdS}^{4|4}$). Furthermore, they constructed $\rm OSp(1|4)$-invariant actions generalising the Wess-Zumino model and ${\cN}=1$ super Yang-Mills theory.

The structure of consistent supercurrent multiplets in ${\rm AdS}^{4|4}$ \cite{BK11, BK12} considerably differs from that in the $\cN = 1$ super-Poincar\'e case \cite{K-var, MSW}. There exist three minimal supercurrents with $(12 + 12)$ degrees of freedom in Minkowski superspace \cite{K-var}. As discussed in \cite{BK12}, there are only two
irreducible AdS supercurrents: \textit{minimal} $(12+12)$ and \textit{non-minimal} $(20+20)$, which are related via
a well-defined improvement transformation \cite{BK12}.
The minimal supercurrent is the AdS extension of the Ferrara-Zumino supercurrent \cite{FZ}.

We remark that these consistent 
AdS supercurrents are closely related to two  classes of 
supersymmetric gauge theories: (i) the known off-shell formulations for pure ${\cN}=1$ AdS supergravity, minimal (see e.g. \cite{GGRS,Ideas} for reviews) and non-minimal \cite{BK12}; and (ii)
the two
dually equivalent series of massless higher-spin supermultiplets 
in AdS \cite{KS94}. More precisely,
the minimal supercurrent is associated with the longitudinal action
$S^{||}_{(3/2)}$ for a massless superspin-3/2 multiplet in AdS. The non-minimal supercurrent is associated
with the dual formulation $S^{\perp}_{(3/2)}$.
The functional $S^{||}_{(3/2)}$ proves to be the linearised action for
minimal $\cN=1$ AdS supergravity. The dual action  $S^{\perp}_{(3/2)}$
results from the linearisation around the AdS background of non-minimal 
$\cN=1$ AdS supergravity.
Both actions
represent the lowest superspin limits of the off-shell massless supermultiplets of half-integer superspin in AdS \cite{KS94}.


Higher-spin supercurrent multiplet is a higher-spin extension of the ordinary supercurrent. In its component expansion, it contains conserved bosonic and fermionic currents. Conserved higher-spin currents 
for scalar and spinor fields in 4D Minkowski space
have been studied in numerous publications. 
To the best of our knowledge, the first construction of currents for both scalar and spinor fields was given by Kibble \cite{Kibble}. Migdal \cite{Migdal}
and Makeenko \cite{Makeenko} later also described the spinor case. Further examples of conserved higher-spin currents were given in
\cite{Makeenko,CDT,BBvD,Anselmi,Anselmi2,KVZ}. 
Higher-spin extension (in the half-integer superspin case) of the conformal supercurrent \cite{FZ} was proposed more than thirty years ago by Howe, Stelle and Townsend \cite{HST}. 
Recently in Ref.\cite{KMT}, the structure of such a supercurrent was described in more detail, and examples for supercurrents in some superconformal models were also given. In 3D Minkowski space, explicit constructions of conserved higher-spin supercurrents in free superconformal theories were obtained in \cite{NSU}. 

As regards non-conformal higher-spin supercurrents, their properties had not been analysed.
The primary goal of this thesis is to study the general structure of non-conformal higher-spin supercurrent multiplets in three and four dimensions from the viewpoint of off-shell higher-spin gauge supermultiplets. As demonstrated in \cite{K-var, BK12, KT-M11}, the general structure of the supercurrents in AdS differ significantly from their counterparts in Minkowski space. This motivated us to look for realisations of higher-spin supercurrents in field theories with Poincar\'e and AdS supersymmetry. 

For this purpose, we developed a higher-spin extension of the general superfield approach advocated in \cite{K-var, KT-M11}. Here the task was simpler in four dimensions since such off-shell actions already existed \cite{KPS,KS,KS94}. 
In three dimensions, off-shell massless ${\cN}=1$ \cite{KT} and ${\cN}=2$ \cite{KO} supermultiplets were constructed in Minkowski space, but only for the half-integer superspin case. In \cite{KP1}, there appeared two off-shell actions corresponding to half-integer and integer ${\cN}=1$ supermultiplets in AdS. More general off-shell massless higher-spin supermultiplets in ${\cN}=1$ and ${\cN}=2$ AdS superspaces were presented in \cite{HKO, HK18, HK19}.

In four dimensions, we only concentrated on off-shell massless higher-spin multiplets and their associated conserved currents with ${\cN}=1$ Poincar\'e and AdS supersymmetry \cite{HK1, HK2, BHK}. In three dimensions, however, we considered both ${\cN}=1$ and ${\cN}=2$ AdS cases \cite{HKO, HK18, HK19}. In contrast to the four-dimensional case where pure ${\cN}=1$ AdS supergravity \cite{Townsend} is unique on-shell,
the specific feature of three dimensions is 
the existence of two distinct $\cN=2$ AdS supergravity theories \cite{AT, BILS}. They are known as the (1,1) and (2,0) AdS supergravity theories, originally constructed as Chern-Simons theories \cite{AT}. In Ref.\cite{KT-M11}, various aspects of (1,1) and (2,0) AdS supergravity theories (including the general structure of supercurrents) were studied in detail, using the superspace formalism developed by Kuzenko, Lindstr\"{o}m and Tartaglino-Mazzucchelli \cite{KLT-M11}.  
Thus, it is natural to extend the analysis of \cite{KT-M11} to the higher-spin case \cite{HKO, HK18}. 

Let us discuss some applications of our results presented in this thesis. In accordance with the standard Noether method
(see e.g. \cite{VanNieuwenhuizen:1981ae} for a review), construction of conserved higher-spin supercurrents 
for various supersymmetric theories is equivalent to generating consistent 
cubic vertices of the type $\int H  J$. Here $H$ denotes
some off-shell higher-spin gauge multiplet, 
and $J=\cD^p \F \cD^q\J$ is the higher-spin conserved current multiplet, constructed in terms of some matter multiplets $\F$ and $\J$, and superspace covariant derivatives $\cD$. 
In 4D Minkowski superspace, several cubic vertices involving the off-shell higher-spin multiplets of \cite{KPS,KS}
were constructed recently \cite{BGK1,KKvU,BGK-sigma,BGK2, BGK3,GK1} using the superfield Noether procedure \cite{MSW}. For instance, conserved supercurrents and cubic interactions between massless higher-spin supermultiplets and a single chiral superfield were constructed by Buchbinder, Gates and Koutrolikos \cite{BGK1}.
This analysis was soon extended by Koutrolikos, Ko\v{c}i and von Unge to study cubic vertices in the case of a free complex linear superfield \cite{KKvU}. The corresponding component higher-spin currents were also computed \cite{KKvU}.


Making use of the gauge off-shell formulations for massless higher-spin 
supermultiplets of \cite{KS94}, higher-spin extensions of the AdS supercurrents \cite{BK_supercurrent} were formulated in 4D ${\cN}=1$ AdS superspace \cite{BHK} for the first time. Their realisations for various supersymmetric theories in AdS were also presented, including a model of $N$ massive chiral scalar superfields with an arbitrary mass matrix. Such a program was a natural extension of the earlier flat-space results \cite{HK1, HK2}, in which we built on the structure of 
higher-spin supercurrent multiplets in models for  superconformal chiral superfields \cite{KMT}.

In the non-supersymmetric case, conserved higher-spin currents for scalar fields
in AdS were studied, e.g. in \cite{MR,MM,FIPT,FTsulaia,BekaertM}. 
The nonvanishing curvature of AdS space makes explicit calculations of conserved higher-spin currents much harder than in Minkowski space. 
Refs. \cite{MR,MM} studied only the conformal scalar, and 
only the first order correction to the flat-space expression was given explicitly. 
The construction presented in \cite{BekaertM} is more complete since all conserved higher-spin currents were computed exactly for a free massive scalar field using the so-called ambient space formulation.
All these works dealt with bosonic currents. 
The important feature of supersymmetric theories is that they also possess
fermionic currents. The conserved higher-spin supercurrents computed in \cite{BHK} can readily be reduced to components. This leads to closed-form expressions for conserved higher-spin bosonic and fermionic currents in models
with massive scalar and spinor fields.

\vspace{0.5cm}
\noindent\textbf{Thesis outline}\\
The purpose of chapter \ref{ch2} is to introduce various technical aspects and essential background materials. First, a brief account of the ${\cN}=1 $ superspace formalism of \cite{Ideas} is given. We then recall the structure of the non-conformal supercurrent multiplets in 4D ${\cal N}=1$ Minkowski and AdS superspaces following \cite{K-var,BK_supercurrent}. 
Finally, we briefly review the two dually equivalent off-shell Lagrangian formulations for massless multiplets of arbitrary superspin in 4D ${\cN}=1$ Minkowski superspace \cite{KPS, KS}. 

Chapter \ref{ch3}
presents a new off-shell formulation for the massless superspin-$s$ multiplet in 4D ${\cN}=1$ Minkowski superspace, where $s = 2, 3, \dots$ and for the massless gravitino multiplet ($s=1$). 
The non-conformal higher-spin supercurrent multiplets associated with the massless (half-)integer superspin gauge theories are derived. In addition, we compute higher-spin supercurrents that originate in the models for a single massless and massive chiral superfield, as well as the massive ${\cN}=2$ hypermultiplet.
This chapter is based on the original works \cite{HK1, HK2}.

Chapter \ref{ch4} is concerned with the extension of the flat-space results in chapter \ref{ch3} to the case of 4D ${\cN}=1$ AdS supersymmetry. 
The dual formulations for massless (half-)integer superspin multiplets in AdS \cite{KS94} are reviewed and a novel formulation for the massless integer superspin is proposed. Making use of these gauge off-shell models, higher-spin supercurrent multiplets are formulated.  
Their explicit constructions are presented for various supersymmetric in AdS, including the case of $N$ chiral scalar superfields with an arbitrary mass matrix $M$. 
We further elaborate on several nontrivial applications of the construction of higher-spin supercurrents. 
This chapter is based on the original work \cite{BHK}.

In chapter \ref{ch5}, we turn our attention to ${\cN}=2$ supersymmetric higher-spin theories in 3D AdS space. First, some of the important facts concerning (1,1) and (2,0) AdS superspaces, including superfield representations of the corresponding isometry groups, are reviewed. By analogy with our 4D ${\cN}=1$ analysis, we construct two dually equivalent off-shell Lagrangian formulations for every massless higher-spin supermultiplet in (1,1) AdS superspace, and subsequently generate consistent higher-spin supercurrents. 
In the context of (2,0) AdS supersymmetry, we begin with some simple supersymmetric models in (2,0) AdS superspace to deduce a multiplet of conserved higher-spin currents, from which the corresponding supermultiplet of higher-spin fields can be determined. This results in two off-shell gauge formulations for a massless multiplet of half-integer superspin $(s+ \hf)$, for arbitrary integer $s > 0$. This chapter is based on the original works \cite{HKO, HK18}.

In chapter \ref{ch6}, a manifestly supersymmetric setting to reduce every field theory in (2,0) AdS superspace to ${\cN}=1$ AdS superspace is developed. 
As nontrivial examples, we consider supersymmetric nonlinear sigma models described in terms of ${\cN}=2$ chiral and linear supermultiplets.  
This (2,0) $\to$ (1,0) AdS reduction technique is then applied to our off-shell massless higher-spin supermultiplets described in chapter \ref{ch5}. This results in four series of ${\cN}=1$ supersymmetric higher-spin models in AdS, two of which are new gauge theories.
This chapter is based on the original work \cite{HK19}.

Finally, in chapter \ref{ch7} we conclude this thesis by summarising its key outcomes.

There are three appendices. Our notation and conventions are summarised in appendix \ref{AppA}. Appendix \ref{AppendixC} reviews the conserved higher-spin currents 
for free $N$ scalars and Majorana spinors with arbitrary mass matrices. In appendix \ref{AppendixBB}, we analyse the component structure of the two new ${\cN}=1$ supersymmetric higher-spin models constructed in chapter \ref{ch6}.



\chapter{Field theories in ${\cN}=1$ Minkowski and AdS superspaces} \label{ch2}
\addtocontents{toc}{\protect\setcounter{tocdepth}{3}}
In this chapter we collect some technical background materials required to understand subsequent chapters. We begin with a brief introduction to some aspects of field theories in 4D ${\cN}=1$ Minkowski superspace following\cite{Ideas}.\footnote{Although we only give a review of 4D ${\cN}=1$ supersymmetry, most of its structure can be readily generalised to 3D ${\cN}=2$ super-Poincar\'e case.} The next two sections are intended to illustrate the differences between supercurrent multiplets with ${\cN}=1$ Poincar\'e and AdS supersymmetry \cite{K-var,BK12, BK11}. Finally, we review the off-shell formulations for massless higher-spin ${\cN}=1$ supermultiplets in Minkowski space, which were developed in \cite{KPS,KS}. 

\section{Field theories in ${\cN}=1$ Minkowski superspace} 
A more detailed and pedagogical introduction to various aspects covered in this section can be found in \cite{Ideas, WB, GGRS}. Our 4D notation and conventions are essentially those of \cite{Ideas} and are summarised in appendix \ref{AppA}. 

\subsection{The Poincar\'e superalgebra}

The simplest supersymmetric extension of the Poincar\'e group in four dimensions is the ${\cN}=1$ super-Poincar\'e group. Associated to this super-Lie group is the ${\cN}=1$ Poincar\'e superalgebra \cite{GL} with the following (anti-)commutation relations 
\bea \label{N1susyalg}
&&[P_a, P_b]=0~, \qquad [M_{ab}, P_c] = \ri (\eta_{ac} P_b - \eta_{bc}P_a)~, \non\\
&&[M_{ab}, M_{cd}] = \ri(\eta_{ac} M_{bd}- \eta_{ad} M_{bc} + \eta_{bd}M_{ac} - \eta_{bc} M_{ad})~, \non\\
&&[M_{ab}, Q_{\a}]= \ri (\sigma_{ab})_{\a}{}^{\b}Q_{\b}~, \qquad [P_a, Q_\a] = 0~, \non\\
&&[M_{ab}, \bar Q_{\ad}]= \ri (\tilde\sigma_{ab})_{\ad}{}^{\bd} \bar Q_{\bd}~, \qquad [P_a, \bar Q_\ad] = 0~, \\
&& \{Q_\a, Q_\b\} = 0~, \qquad \{\bar Q_\ad, \bar Q_\bd \} = 0~, \non\\
&&\{Q_{\a}, \bar Q_{\ad} \} = 2 (\sigma^a)_{\a \ad} P_{a}~. \non
\eea
Here $P_a$ and $M_{ab}$ denote the generators of the translation and Lorentz group, respectively. 
The supersymmetry generators $Q_{\a}, \bar Q_{\ad} \,(\a, \ad = 1,2)$ are Weyl spinors which transform respectively as $(1/2,0)$ and $(0,1/2)$ of the Lorentz group. 
The automorphism group ($R$-symmetry group) of the ${\cN}=1$ Poincar\'e superalgebra \eqref{N1susyalg} is ${\rm U(1)}$, which act on the supercharges in the following way
\bea
Q_{\a}^{'} = \re^{\ri\tau} Q_{\a}~, \qquad  \bar Q_{\ad}^{'} = \re^{-\ri\tau} \bar Q_{\ad}~, \qquad \tau \in \mathbb{R}~.
\eea
In the extended (${\cN} >1$) supersymmetry case, the $R$-symmetry group is ${\rm U(\cN)}$. 


\subsection{${\cN}=1$ Minkowski superspace}
We denote the 4D ${\cN}=1$ Minkowski superspace \cite{VAsuperspace, Salam} by $\mathbb{M}^{4|4}$. It can be identified with the coset space
\bea \label{MinkowskiSS}
\mathbb{M}^{4|4} = \cS\Pi / {\rm SL}(2, \mathbb{C})~.
\eea
Here $\cS\Pi$ is the ${\cN}=1$ super-Poincar\'e group, and ${\rm SL}(2, \mathbb{C})$ is the double cover of the restricted Lorentz subgroup ${\rm SO}_{0} (3,1)$.
Any element of the supergroup $\cS\Pi$ can be represented in an exponential form
\bea
{\rm exp}\big(-\ri x^a P_a + \ri (\theta^{\a} Q_{\a}+ \bar \theta_{\ad} \bar Q^{\ad})\big) \,{\rm exp} \bigg( \frac{\ri}{2} \o^{ab} M_{ab}\bigg)~.
\eea
The points of the coset space $\mathbb{M}^{4|4}$ are
\bea
{\rm exp}\big(-\ri x^a P_a + \ri (\theta^{\a} Q_{\a}+ \bar \theta_{\ad} \bar Q^{\ad})\big)~.
\eea
Thus, $\mathbb{M}^{4|4}$ can be parametrised by the local coordinates $z^A = (x^a, \theta^{\a}, \bar \theta_{\ad})$, where $x^a$ are real commuting numbers and $(\theta^{\a})^* = \bar \theta^{\ad}$ are complex anticommuting numbers. 

The action of supersymmetry transformations on the superspace coordinates can be determined using the algebra \eqref{N1susyalg}. It is given by
\bea
&&{\rm exp}\big(\ri (\e^{\a} Q_{\a}+ \bar \e_{\ad} \bar Q^{\ad})\big) \,{\rm exp}\big(-\ri x^a P_a + \ri (\theta^{\a} Q_{\a}+ \bar \theta_{\ad} \bar Q^{\ad})\big)\non\\
&=& {\rm exp} \bigg(-\ri (x^a + \ri \theta \sigma^a \bar \e -  \ri \e \sigma^a \bar \theta) P_a + \ri \big[( \theta^{\a}+ \e^{\a}) Q_{\a} +  (\bar \theta_{\ad} + \bar \e_{\ad}) \bar Q^{\ad}\big] \bigg)~,
\eea
from which we can read off
\bea 
x'^{a} = x^a + \ri (\theta \sigma^a \bar \e - \e \sigma^a \bar \theta)~, \qquad
\theta'^{\a} = \theta^\a + \e^{\a}~.
\eea
One may also compute the action of translations, ${\rm exp}\big(-\ri b^a P_a )$, and Lorentz transformations, $\L = {\rm exp} \bigg( \frac{\ri}{2} \o^{ab} M_{ab}\bigg)$, in a similar manner. The result is
\bea
x'^{a} = (\L(N))^a{}_{c}x^c + b^a ~, \qquad
\theta'^{\a} &=& \theta^{\b} (N^{-1})_{\b}{}^{\a}~,
\eea
where $\L : {\rm SL} (2, \mathbb{C}) \rightarrow {\rm SO}_0 (3,1)$ is the well-known homomorphism given by
\bea
(\L(N))^a{}_{c} = -\hf {\rm tr}(\tilde{\sigma}^a N \sigma_c N^{\dagger})~, \qquad N \in {\rm SL} (2, \mathbb{C})~.
\eea



\subsection{Superfields}
Supersymmetric field theories on superspace are naturally formulated in terms of tensor superfields. 
A tensor superfield $V$ of Lorentz type $(\frac{n}{2}, \frac{m}{2})$ is a superfield carrying $n$ undotted and $m$ dotted spinor indices, which are separately symmetrised. 
%
Furthermore, it transforms in the following way under the action of an
infinitesimal ${\cN}=1$ super-Poincar\'e group (note that we have suppressed its tensor indices):
\bea
\d V = \ri \Big(-b^a P_a + \hf \o^{ab} J_{ab} + \e^{\a} Q_{\a} + \bar \e_{\ad} \bar Q^{\ad} \Big) V~.
\eea
The generators take the form
\bsubeq \label{4dsusygen}
\bea
{P}_{a} &=& -\ri \pa_{a}~, \\
{J}_{ab} &=& \ri (x_b \pa_a - x_a \pa_b + (\sigma_{ab})^{\a \b} \q_{\a} \pa_{\b} -  (\tilde{\sigma}_{ab})^{\ad \bd} \bar \q_{\ad} \bar \pa_{\bd} - M_{ab})~, \\
{Q}_{\a} &=& \ri \pa_{\a} + \bar \q^{\ad} (\sigma^a)_{\a \ad} \pa_{a} =\ri \pa_{\a} + \bar \q^{\ad} \pa_{\a \ad} ~, \\
\bar {Q}_{\ad} &=& -\ri \bar \pa_{\ad} - \q^{\a} \pa_{\a \ad}~. 
\eea
\esubeq
We have also introduced the following notation:
\bea
\pa_{\a \ad} := (\sigma^a)_{\a\ad} \pa_a~,\qquad 
\pa_{\a} := \frac{\pa}{\pa \q^\a}~, \qquad \bar \pa_{\ad} := \frac{\pa}{\pa \bar \q^\ad}~.
\eea

We denote the set of covariant derivatives of ${\cN}=1$ Minkowski superspace by \\
$D_{A} = (\pa_a, D_\a, \bar D^{\ad})$, which have the form
\bea
D_{\a} = \pa_{\a} + \ri \bar \q^{\ad} \pa_{\a \ad}~, \qquad
\bar D^{\ad} = \bar \pa^{\ad} + \ri \q_{\a} \pa^{\a \ad}~, \qquad \bar \pa^{\ad} = -\ve^{\ad\bd} \bar \pa_{\bd}.
\eea
They obey the (anti)commutation relations\footnote{See appendix \ref{AppA} for some important properties of the covariant derivatives.}
\bea
&\{D_{\a}, D_{\b}\} = \{\bar D^{\ad}, \bar D^{\bd}\} = [D_{\a}, \pa_a] = [\bar D^{\ad}, \pa_a] = 0~, \non\\
&\{D_{\a}, \bar D_{\ad} \} = -2 \ri \pa_{\a \ad}~.
\eea
The latter indicates that flat superspace has non-vanishing torsion. 


Expanding a tensor superfield $V(x, \q, \bar \q)$ with respect to its fermionic coordinates $(\q, \bar \q)$, one obtains its corresponding component fields as the coefficients of the series. Due to the property $\q_{\a} \q_{\b} \q_{\g} = \bar \q_{\ad} \bar \q_{\bd} \bar \q_{\gd} =0 $, such a series will be finite. 
As an example, consider a Taylor expansion of a real scalar superfield, $\bar V(z) = V(z)$, but otherwise unconstrained:
\bea
V(x, \q, \bar \q) &=& A(x) + \q^{\a} \psi_{\a}(x) + \bar \q_{\ad}\bar \psi^{\ad}(x) + \q^2 F(x) + \bar \q^2 \bar F(x)  \non\\
&&+  \q^{\a} \bar \q^{\ad}C_{\a \ad}(x) +\bar \q^2 \q^{\a}\l_{\a}(x) + \q^2 \bar \q_{\ad}\bar \l^{\ad}(x) + \q^2 \bar \q^2 D(x)~.
\eea

To get insight into the physical content of the superfield $V(z)$, a more systematic and convenient way is to use space projection (also often called bar-projection in some literature) and covariant differentiation. By space projection we mean the zeroth order term in the power series expansion in $\theta$ and $\bar \theta$:
\bea
V| := V( x, \q=0, \bar \q= 0)~.
\eea
In the case of a real scalar superfield above, we may define the components using the bar-projection:
\bea
&A(x) = V(z)|~, \,\,\, \psi_{\a}(x) = D_{\a} V(z)|~, \,\,\, \bar \psi_{\ad} (x) = \bar D_{\ad} V(z) |~, \non\\
& F(x) = -\frac{1}{4}D^2 V(z)|~, \,\,\, \bar F(x) = -\frac{1}{4}\bar D^2 V(z)|~, \,\,\, C_{\a \ad}(x) = \hf [D_{\a}, \bar D_{\ad}]V(z)|~, \non\\
&\l_{\a}(x) = -\frac{1}{4}D_{\a}\bar D^2 V(z)|~, \,\,\, \bar \l_{\ad}(x) = -\frac{1}{4}\bar D_{\ad} D^2 V(z)|~, \non\\
&H(x) = \frac{1}{32} \{D^2, \bar D^2 \} V(z)|~.
\eea
In addition, one may work out how the component fields transform under the  infinitesimal supersymmetry transformations
\bea \label{inf-susy}
\d V(z) = \ri (\e^{\a} { Q}_{\a} + \bar \e_{\ad} \bar {Q}^{\ad}) V(z)~, 
\eea
by taking various numbers of covariant derivatives of \eqref{inf-susy} and then bar-project them. One should also note that the supersymmetry generator anticommutes with the spinor covariant derivatives, which implies $[D_{\a}, \e Q + \bar \e \bar Q]= [\bar D_{\ad}, \e Q + \bar \e \bar Q]= 0$.

In appendix \ref{AppendixBB} we will describe in more detail the 3D ${\cN}=1$ analogue of the above component reduction. Such a procedure is useful to study the field contents of some supersymmetric higher-spin models. 

An unconstrained superfield is a reducible representation of supersymmetry. In order to obtain an irreducible representation, we need to impose certain constraints on the superfield which are also consistent with the supersymmetry transformations. This can be done with the help of the covariant derivatives $D_{\a}, \bar D_{\ad}$. The simplest example is the chirality constraint,
\bea \label{constr-ch}
\bar D_{\ad}\F(x, \q, \bar \q) &=& 0~, 
\eea
with $\F(x, \q, \bar \q)$ being a complex scalar superfield. 
It can be shown that the above constraint implies that 
\bea \label{2122}
\bar \pa_{\ad} \big(\re^{-\ri \q \sigma^a \bar \q \pa_a} \F(x, \q, \bar \q) \big) = 0~.
\eea
In order to see this, one may act on both sides of \eqref{constr-ch} with the operator $\re^{-\ri \q \sigma^a \bar \q \pa_a}$, and make use of the identity 
\bea
\re^{-\ri \q \sigma^a \bar \q \pa_a}\bar D_{\ad} \re^{\ri \q \sigma^a \bar \q \pa_a} = -\bar \pa_{\ad}~.
\eea
It follows from \eqref{2122} that $\re^{-\ri \q \sigma^a \bar \q \pa_a} \F(x, \q, \bar \q)$ is independent of $\bar \q$, so it can be written as
\bea
\re^{-\ri \q \sigma^a \bar \q \pa_a} \F(x, \q, \bar \q) = \F(x, \q)~,
\eea
for an arbitrary superfield $\F(x, \q)$. Thus, the solution to the chirality constraint \eqref{constr-ch} is given by
\bea \label{sol-ch}
\F(x, \q, \bar \q) = \re^{\ri \q \sigma^a \bar \q \pa_a}\F(x, \q) = \F(x^a + \ri \q \sigma^a \bar \q, \q)~.
\eea
A superfield of the form \eqref{sol-ch}, which depends only on $(x,\q)$, is called chiral superfield. 

Analogously, one can also impose the anti-chirality constraint
\bea
D_{\a} \bar \F(x, \q, \bar \q) &=& 0~,
\eea
which is solved by
\bea
\bar \F(x, \q, \bar \q) = \re^{-\ri \q \sigma^a \bar \q \pa_a} \bar \F(x, \bar \q) = \bar \F(x^a - \ri \q \sigma^a \bar \q, \bar \q)~.
\eea
In contrast to the chiral superfield \eqref{sol-ch}, we see that the anti-chiral superfield essentially depends on $(x, \bar\q) $ only. Any function of a chiral superfield only is also chiral, that is
\bea
\bar D_{\ad} F(\F) = F'(\F) \bar D_{\ad} \F = 0~.
\eea
The same goes with anti-chiral superfield. 

The component structure of a chiral superfield can be studied by first decomposing $\F(x, \q)$ in terms of $\q$:
\bea
\F(x, \q) = A(x) + \q^\a \psi_\a (x) + \q^2 F(x)~,
\eea 
or, equivalently 
\bea \label{2127}
A(x) = \F|~, \qquad \psi_{\a}(x) = D_{\a} \F|~, \qquad F(x) = -\frac{1}{4} D^2 \F|~. 
\eea
As a result, we have that
\bea
\F(x, \q, \bar \q) &=&\re^{\ri \q \sigma^a \bar \q \pa_a}\F(x, \q) \non\\
&=& A(x) + \q^\a \psi_{\a} + \q^2 F(x) + \ri \q \sigma^a \bar \pa_a A(x)\non\\
&+& \frac{\ri}{2} \q^2 \bar \q \bar \sigma^a \pa_a \psi(x) + \frac{1}{4} \q^2 \bar \q^2 \Box A(x)~. \label{comp-ch}
\eea
The component fields of an anti-chiral superfield can be worked out from \eqref{comp-ch} by conjugation. 

There are other types of constrained superfields, such as complex linear and real linear superfields. They will be described in section \ref{SCflat}.

\subsection{Supersymmetric action principle}\label{rev213}
In superspace formalism, any supersymmetric field theory is described by a set of superfields, with the corresponding action functional written as an integral over the superspace of a Lagrangian superfield $\cL$. The (classical) superfield equations of motion can be obtained using a supersymmetric action principle. 

Let us first understand some basics of integration over the anticommuting coordinates ($\q, \bar \q$), which was first given by Berezin \cite{Ber}. The Berezin integral is equivalent to differentiation. We further note some properties:
\bea
\int \rd\q_{\a}\q^\b = \pa_\a \q^\b = \d_{\a}{}^{\b}~, \non
\eea
\bea
\rd^2\q = \frac{1}{4}\ve^{\a \b}\rd\q_{\a} \rd\q_{\b} \quad \Longrightarrow \quad \int \rd^2 \q = \frac{1}{4}\pa^{\a}\pa_{\a}~,\,\,\, \int \rd^2\q \,\q^2 = 1~. 
\eea
Similarly,
\bea
\int \rd \bar \q^{\ad} \bar \q_{\bd} = \bar \pa^{\ad} \bar \q_{\bd} = \d^{\ad}{}_{\bd}~,
\non
\eea
\bea
\rd^2 \bar \q = \frac{1}{4}\ve_{\ad \bd}\rd\bar \q^{\ad} \rd \bar \q^{\bd} \quad \Longrightarrow \quad \int \rd^2 \bar \q = \frac{1}{4}\bar \pa_{\ad}\bar \pa^{\ad}~,\,\,\, \int \rd^2 \bar \q \, \bar \q^2 = 1~.
\eea
The measure of full ${\cN}=1$ Minkowski superspace is denoted $\rd^8z = \rd^4x \rd^2 \q \rd^2 \bar \q$, while the measures on chiral and antichiral subspaces are given by $\rd^6z = \rd^4x \rd^2 \q$  and $ \rd^6 \bar z = \rd^4x \rd^2 \bar \q$ respectively.

There are many useful properties of integration in full superspace or chiral subspace. First, for an arbitrary superfield $V(z)$, we have that
\bea
\int \rd^8z \,\, D_{A}(V(z)) = 0~.
\eea
Integration in an (anti-)chiral subspace can be reduced to Minkowski space,
\bsubeq
\bea
\int \rd^6z \,\, V(z) &=& -\frac{1}{4} \int \rd^4x \,\,  D^2 V(z) \Big|~, \\
\int \rd^6 \bar z \,\, V(z) &=& -\frac{1}{4} \int \rd^4x \,\, \bar D^2 V(z) \Big|~.
\eea
\esubeq
Given an integration in full superspace, it can be written either in (anti-)chiral subspace or Minkowski space:
\bea
\int \rd^8z V(z) &=& -\frac{1}{4} \int \rd^6z \,\, \bar D^2 V(z)= \frac{1}{16} \int \rd^4x \,\, D^2 \bar D^2 V(z) \Big|~,\\
&=& -\frac{1}{4} \int \rd^6 \bar z \,\,  D^2 V(z) = \frac{1}{16} \int \rd^4x \,\, \bar D^2 D^2 V(z) \Big| = \frac{1}{16} \int \rd^4x \,\, D^2 \bar D^2 V(z) \Big|~.\non
\eea 

The most general supersymmetric action functional takes the form
\bea \label{susyaction}
S = \int \rd^4x \rd^2 \q \rd^2 \bar \q \cL + \int \rd^4x \rd^2 \q \cL_c + \int \rd^4x  \rd^2 \bar \q \bar \cL_c~.
\eea
Here $\cL$ is a real scalar superfield, while $\cL_c$ and $\bar \cL_c$ are chiral and anti-chiral scalar superfields, respectively.  
Performing integration over all the Grassmann variables in \eqref{susyaction} results in component form of the action, which is expressed as an integral over the 4D Minkowski space. This procedure yields
\bea
S = \int \rd^4x \Big( \frac{1}{16} D^2 \bar D^2 \cL - \frac{1}{4} D^2 \cL_c - \frac{1}{4} \bar D^2 \bar \cL_c \Big) \Big|~.
\eea

For completeness, let us prove the invariance of \eqref{susyaction} under the ${\cN}=1$ super-Poincare transformations. We will explicitly show this for the first term. The invariance of the (anti-)chiral action can also be proved in a similar way. For this we vary the Lagrangian $\cL$, which is a scalar superfield according to the rule \eqref{inf-susy}:
\bea
\d_{\e}S &=& \int \rd^4x \rd^2\q \rd^2 \bar \q \,\ri (\e^\a Q_{\a} + \bar \e_{\ad} \bar Q^{\ad}) \cL = \frac{1}{16} \int \rd^4x D^2 \bar D^2 \,\ri (\e^\a Q_{\a} + \bar \e_{\ad} \bar Q^{\ad}) \cL \Big| \non\\
&=&  \frac{1}{16} \int \rd^4x \, \ri (\e^\a Q_{\a} + \bar \e_{\ad} \bar Q^{\ad})D^2 \bar D^2 \cL \Big|  = -\frac{1}{16} \int \rd^4x \,  (\e^\a D_{\a} + \bar \e_{\ad} \bar D^{\ad})D^2 \bar D^2 \cL \Big| \non\\
&=& -\frac{1}{16} \int \rd^4x  \, \bar \e_{\ad} [\bar D^{\ad}, D^2 ]\bar D^2 \cL \Big|  = \frac{\ri}{4} \int \rd^4x \,\pa_{\a \ad} (\bar \e^{\ad} D^\a \bar D^2 \cL) \Big|~,
\eea
{\it i.e.} the Lagrangian changes by a total spacetime derivative. 

Consider a simple superfield model for a free massless chiral scalar superfield $\F$. Its action is given by
\bea
S = \int \rd^4x \rd^2 \q \rd^2 \bar \q\, \bar \F \F~, \qquad \bar D_{\ad} \F = 0~.
\eea
The superfield equations of motion can be derived by varying the action with respect to $\F$, which is defined as an integral over the chiral subspace. This leads to $\bar D^2 \bar \F = 0$. Similarly, one gets $D^2 \F = 0$ by varying the anti-chiral superfield $\bar \F$.
Recalling the definition of the components of $\F$ given in \eqref{2127}, the corresponding component action is easily found to be
\bea
S = \int \rd^4x \bigg( -\pa^a \bar A \pa_a A-\frac{\ri}{2} \psi^{\a} \pa_{\a \ad} \bar \psi^{\ad} + \bar F F \bigg)~.
\eea
The component fields $F$ and $\bar F$ are auxiliary fields. They enter the action without derivatives (or kinetic terms), thus they have no non-trivial dynamics. 
One further finds that the component fields $A, \psi, F$ (and their conjugates) transform \textit{linearly} under the infinitesimal supersymmetry transformations:
\bea
\d A &=& -\e^{\a} \psi_{\a}~,\non\\
\d \psi_{\a} &=& -2 \e_{\a} F-2 \ri \bar \e^{\ad} \pa_{\a \ad} A~, \\
\d F &=& -\ri \bar \e^{\ad} \pa_{\a \ad} \psi^{\a}~. \non
\eea
Suppose the auxiliary fields $F, \bar F$ are eliminated through their equations of motion (in this case it is $F = \bar F = 0$). Computing the commutators of two infinitesimal supersymmetry transformations, one finds that the supersymmetry algebra is broken when the auxiliary fields are eliminated. More precisely, the result is of the form
\bea
[\d_{\e_1}, \d_{\e_2}]A = c^m\pa_m A~, \qquad 
[\d_{\e_1}, \d_{\e_2}] \psi_{\a}=
c^m \pa_m \psi_{\a} 
+ \ri c_{\a \ad} \frac{\d S}{\d \bar \psi_{\ad}}~,
\eea
with $c^m = 2 \ri (\e^{\a}_1 \sigma^m_{\a \ad} \bar \e^{\ad}_2 - \e^{\a}_2 \sigma^m_{\a \ad} \bar \e^{\ad}_1)$. 
This algebra is closed only on the equations of motion for spinor fields ${\d S}/{\d \bar \psi_{\ad}} = 0$. Therefore, the
supersymmetry algebra in the theory without auxiliary fields is closed only on-shell (on
equations of motion). This explains the role of auxiliary fields, which are to ensure  off-shell closure of the supersymmetry algebra on the component fields. 

\section{Linearised ${\cN}=1$ Poincar\'e supergravity and variant supercurrents} \label{SCflat}
Given an ${\cN}=1$ supersymmetric theory, one can derive the conserved spin-vector current associated to rigid supersymmetry. By computing Noether currents in the massive Wess-Zumino model, Ferrara and Zumino demonstrated that the conserved energy-momentum tensor $T_{ab}$ and the spin-vector current $S_a$ belong to a supermultiplet, called the supercurrent \cite{FZ}. 
Additionally, the supercurrent contains the axial ${\rm U(1)}_R$  current, $j_{a}$, which is only conserved for a theory with ${\rm U(1)}_R$ symmetry. The  trace of the energy-momentum tensor, the gamma-trace of the spinor current $\gamma^a S_a$, and the axial current divergence $\pa^{a}j_{a}$ form a smaller supermultiplet, \textit{i.e.} the trace supermultiplet (also called supertrace). 

For 4D ${\cN}=1$ supersymmetric theories in Minkowski space, the most general supercurrent multiplet is subject to the following conservation law \cite{K-var, K-Noet}
\bea \label{PoinSC}
&{\bar D}^{\ad}{J} _{\a \ad} = { \c}_\a  +{\rm i}\,\eta_\a +D_\a T~, 
\\
&{\bar D}_\ad {\c}_\a  = {\bar D}_\ad \eta_\a= {\bar D}_\ad {T}=0~, 
\quad D^\a {\c}_\a - {\bar D}_\ad {\bar {\c}}^\ad
=D^\a {\eta}_\a - {\bar D}_\ad {\bar {\eta}}^\ad = 0~.
\non
\eea
Here the real vector superfield $J_{\a\ad} = \bar J_{\a \ad}$ is the supercurrent. The chiral superfields $T$, $\c_\a$ and $\eta_{\a}$ are the trace supermultiplets.

Depending on supersymmetric theories, some of the trace supermultiplets might vanish. In the case of superconformal theories, we can set all of them to zero. 
The three terms on the right-hand side of \eqref{PoinSC} correspond to the fact that there exist exactly three linearised off-shell formulations for \textit{minimal} (12+12) $\cN=1$ Poincar\'e supergravity, which have been studied in \cite{GKP}.  These off-shell models are related by duality transformations, \textit{i.e.} they are equivalent on-shell. 
More precisely, the authors of \cite{GKP} classified the following off-shell ${\cN}=1$ superfield models for linearised supergravity: (i) three minimal formulations with (12+12) off-shell degrees of freedom; (ii) three reducible realisations with (16+16) components; and (iii) one non-minimal formulation with (20+20) components. 
Each formulation corresponds to a different way of gauge-fixing ${\cN}=1$ conformal supergravity to describe Poincar\'e supergravity (see \cite{Ideas, GGRS} for reviews). These seven supergravity models give rise to variant supercurrent multiplets. 

We recall that ${\cN}=1$ Poincar\'e supergravity \cite{FvNF, DZ-sugra} describes interacting spin-2 $h_{mn}$ (the graviton) and spin-3/2 fields $\psi_{m}^{\a}, \bar \psi_{m}^{\ad}$ (the gravitino), with local translational and supersymmetry invariance. 
All off-shell formulations for linearised ${\cN}=1$ Poincar\'e supergravity are described by two types of dynamical superfields: the real gravitational gauge superfield $H_{\a \ad} = \bar H_{\a \ad}$ \cite{OS1} and a compensating superfield. 
Each off-shell description contains the graviton and gravitino as dynamical fields, but differs in the set of auxiliary fields. 
The graviton and gravitino fields can be identified with the components of $H_{\a \ad}$. Switching to the two-component spinor notation (see appendix \ref{AppA}), they read
\bea
h_{\a\ad \b \bd} \sim [D_{(\b}, \bar D_{( \bd}] H_{\a) \ad)} \big|~, \qquad \psi_{\b \a \ad} \sim \bar D^2 D_{(\b}H_{\a)\ad}\big|~.
\eea

The real gravitational superfield $H_{\a \ad}$  has the following linearised gauge transformation
\bea \label{confsugra}
\d H_{\a \ad} = \bar D_{\ad}L_{\a} - D_{\a}\bar L_{\ad}~, 
\eea
where $L_{\a}$ is an unconstrained spinor superfield. Upon imposing the Wess-Zumino gauge, the remaining transformations correspond to superconformal transformations \cite{Kaku-sugra}, see also \cite{Ideas,GGRS} for reviews.
A compensating superfield is required in order to remove this extra symmetry, and thus describing Poincar\'e supergravity. The difference between the off-shell models is thus encoded in the choice of the compensators.

Let us now focus on the structure of irreducible\footnote{A supercurrent multiplet is called irreducible 
if it is associated with an off-shell formulation for pure supergravity.} supercurrents and the linearised off-shell minimal supergravity formulations that they correspond to. There are three irreducible supercurrent multiplets with (12+12) off-shell degrees of freedom.
\begin{itemize}
\item Setting $\chi_{\a}= \eta_{\a}=0$ leads to the well-known Ferrara-Zumino multiplet \cite{FZ}, which corresponds to the old minimal formulation for ${\cN}=1$ supergravity \cite{WZ78, old1, old2}, with the following supergravity gauge transformation law:
\bea \label{om}
\d H_{\a \ad} &= & \bar D_{\ad}L_{\a} - D_{\a}\bar L_{\ad}~, \non\\
\d \sigma &=& -\frac{1}{12} \bar D^2 D^{\a}L_{\a}~.
\eea
Here $\sigma$ is the chiral compensator, $\bar D_{\ad} \sigma = 0$. 
\item The case $T=\eta_{\a}=0$ is known as the $R$-multiplet, which exists if the model has an $R$-symmetry. This multiplet corresponds to the new minimal supergravity \cite{SW}. New minimal supergravity uses a real linear superfield $G$ as a compensator, $\bar G - G = \bar D^2 G = 0$. The constrained superfield $G$ is the gauge-invariant field strength of a chiral spinor potential $\J_{\a}$
\bea
G = D^{\a}\J_{\a} + \bar D_{\ad} \bar \J^{\ad}~, \qquad \bar D_{\ad}\J_{\b} = 0~,
\eea
which is defined modulo gauge freedom
\bea
\d \J_{\a} = \ri \bar D^2 D_{\a}K~, \qquad K = \bar K~.
\eea
The gauge transformation law is
\bea
\d H_{\a \ad} &= & \bar D_{\ad}L_{\a} - D_{\a}\bar L_{\ad}~, \non\\
\d G &=& \frac{1}{4}(D^{\a} \bar D^2  L_{\a} + \bar D_{\ad} D^2 \bar L^{\ad}) \qquad \Longrightarrow \qquad \d \J_{\a} = \frac{1}{4} \bar D^2 L_{\a}~.
\eea

\item The third option with $T = \chi_{\a} =0$, corresponds to the Virial multiplet, which was studied quite recently \cite{Nakayama}. It corresponds to another minimal supergravity theory introduced in \cite{BGLP}. It might have fewer applications because it is known only at linearised level, unlike the old and new minimal theories. This theory also makes use of a real linear compensator superfield $F$, which is the gauge-invariant field strength of a chiral spinor potential $\rho_{\a}$
\bea
F = D^{\a}\rho_{\a} + \bar D_{\ad} \bar \rho^{\ad}~, \qquad \bar D_{\ad}\rho_{\b} = 0~.
\eea  
The supergravity transformation is given by
\bea
\d H_{\a \ad} &= & \bar D_{\ad}L_{\a} - D_{\a}\bar L_{\ad}~, \non\\
\d F &=& \frac{\ri}{12}(D^{\a} \bar D^2  L_{\a} - \bar D_{\ad} D^2 \bar L^{\ad})\qquad \Longrightarrow \qquad \d \rho_{\a} = \frac{\ri}{12} \bar D^2 L_{\a}~.
\eea
\end{itemize}

If only one of the trace multiplets is zero, the supercurrent multiplet contains bigger (16+16) components and is said to be reducible. The most famous one is the so-called $S$-multiplet, introduced by Komargodski and Seiberg \cite{KS}. The $S$-multiplet is subject to the conservation equation
\bea
&\bar D^{\ad}J_{\a \ad} = D_{\a}T + \chi_{\a}~, \non\\
&\bar D_{\ad} T = \bar D_{\ad} \chi_{\a} = 0~, \quad D^{\a}\chi_{\a} -\bar D_{\ad} \bar \chi^{\ad} = 0~.
\eea
It has been shown in \cite{GKP} that such models with 16+16 off-shell degrees of freedom can be written as a sum of two of the three minimal models discussed above. 

Let us show how to derive a supercurrent multiplet and its conservation equation, starting from a linearised off-shell formulation for ${\cN}=1$ supergravity, for instance the old minimal supergravity. This approach is based on \cite{K-var}, in which variant ${\cN}=1$ supercurrent multiplets are derived. The analysis for the other formulations should  be analogous.  First, the following linearised action \cite{GGRS}
\bea
S^{(I)}[H, \sigma] &=& \int \rd^4x \rd^2\q \rd^2 \bar \q \bigg\{ -\frac{1}{16} H^{\a \ad} D^{\b} \bar D^2 D_{\b} H_{\a \ad} - \frac{1}{4} (\pa_{\a \ad} H^{\a\ad})^2 + \frac{1}{48} ([D_{\a}, \bar D_{\ad}] H^{\a\ad})^2
\non \\
&&- \ri (\sigma - \bar \sigma)\pa^{\a \ad} H_{\a \ad}- 3 \bar{\sigma} \sigma \bigg\} 
\eea
is invariant under the linearised gauge transformations \eqref{om} of the supergravity prepotentials. Next, we consider the following coupling of the dynamical variables to external sources
\bea
S^{(I)} \rightarrow S^{(I)}[H, \sigma] - \hf \int \rd^4x \rd^2\q \rd^2 \bar \q \,\, H^{\a \ad} J_{\a \ad} - \frac{3}{2} \bigg\{  \int \rd^4x \rd^2 \q \,\, \sigma T + \rm{c.c.} \bigg\}~.
\eea
Demanding invariance of the above action under \eqref{om}, it is straightforward to show that the sources $J_{\a \ad}$ and $T$ must satisfy the conservation equation \eqref{PoinSC}, that is
\bea 
\bar D^{\ad} J_{\a \ad} = D_{\a} T~, \qquad \bar D_{\ad} T = 0~.
\eea

In the case of conformal supergravity, the coupling becomes very simple
\bea
S_{\rm source} = \int \rd^4x \rd^2\q \rd^2 \bar \q \,\, H^{\a \ad} J_{\a \ad}~,
\eea
which leads to the conservation condition
\bea \label{confcurrent-eq}
\bar D^{\ad}J_{\a \ad} = 0 \quad \Longleftrightarrow \quad D^{\a} J_{\a \ad} = 0~,
\eea
as a consequence of imposing invariance under \eqref{confsugra}. 
The independent components of the conformal supercurrent $J_{\a \ad}$ are 
\bea
j_{\a\ad}:= J_{\a\ad}|~, \quad S_{\a \b \ad} := D_\b J_{\a\ad}| =S_{(\a\b)\ad}~, 
\quad T_{\a\b \ad \bd } := [D_{(\b}, \bar D_{( \bd}] J_{\a) \ad)} |~.
\non
\eea
Here $j_{\a\ad}$ is the $R$-symmetry current, which is not always conserved. The supersymmetry current
$S_{\a \b \ad}, \bar S_{\a \ad \bd}$   and the energy-momentum tensor $T_{\a\b \ad \bd }$ are conserved,
\bea
\pa^{\a \ad}S_{\a \b \ad} = 0~, \qquad \pa^{\a \ad}T_{\a \b \ad \bd} = 0~.
\eea

Let us look at some simple supersymmetric theories in which the above (non-)conformal supercurrents are realised. We first consider a superconformal model for a massless chiral scalar $\F, \,\bar D_{\ad} \F= 0$, with action
\bea
S = \int \rd^4x \rd^2 \q  \rd^2 \bar \q \, \bar \F \F
\eea
This theory is characterised by the conformal supercurrent \cite{FZ}
\bea
J_{\a \ad} = D_{\a}\F \bar D_{\ad} \bar \F + 2 \ri (\F \pa_{\a\ad} \bar \F - \bar \F \pa_{\a\ad}\F )~,
\label{mc}
\eea
which obeys the conservation equation 
\eqref{confcurrent-eq},
provided the matter superfield is put on-shell: $D^2 \F =0 ~, \bar D^2 \bar \F =0$. 

A single massive chiral superfield can be coupled to the old minimal supergravity, which is reflected in the existence of the Ferrara-Zumino supercurrent \cite{FZ}. A massive chiral superfield is described by the action
\bea
S = \int \rd^4x \rd^2 \q  \rd^2 \bar \q \, \bar \F \F
+\Big\{ \frac{m}{2} \int \rd^4x \rd^2 \q\, \F^2 +{\rm c.c.} \Big\}~.
\eea
For this model, $J_{\a \ad}$ can be chosen to have the same functional form as in the massless case, eq.~\eqref{mc}\footnote{This follows from the fact that the gravitational superfield does not couple to the superpotential \cite{Siegel}.}. The trace multiplet is given by
\bea
T = m \F^2~. \label{1.6-1}
\eea
It may be shown that the conservation equation
\bea
\bar D^{\ad}J_{\a \ad} = D_{\a}T~,\quad  \bar D_{\ad}T =0
\eea
holds on the use of the massive equations of motion 
\bea
-\frac{1}{4}\bar D^2 \bar \F + m \F  =0~, \quad -\frac{1}{4}D^2 \F + m \bar \F  =0~.
\eea

As explained in \cite{DS}, the Ferrara-Zumino supercurrent multiplet is not well defined in some supersymmetric theories. On the other hand, the $S$-multiplet always exists in all known rigid supersymmetric theories in Minkowski space. 

Given real scalar superfields $U$ and $V$,  the non-conformal supercurrent multiplets \eqref{PoinSC} can be transformed by the rule \cite{K-Noet}
\begin{subequations}
\bea
J_{\a \ad} &\rightarrow & J_{\a \ad} + [D_{\a}, \bar D_{\ad}]V - 2 \pa_{\a\ad}U~, \\
T &\rightarrow & T + \hf \bar D^2 (V-\ri U)~, \\
\chi_{\a} & \rightarrow & \chi_{\a} + \frac{3}{2}\bar D^2 D_{\a}V~,\\
\eta_{\a} & \rightarrow & \eta_{\a} + \hf \bar D^2 D_{\a}U~,
\eea
\end{subequations}
while keeping the conservation equation \eqref{PoinSC} unchanged. Such a transformation is called an improvement.

\section{Field theories in ${\cN}=1$ AdS superspace} \label{sumAdS}
In four dimensions, ${\cN}=1$ supersymmetry in anti-de Sitter space was first studied by Keck \cite{Keck}, and analysis of its nonlinear realisations was given by Zumino \cite{Zumino77}. Ivanov and Sorin \cite{IS1} extensively developed the concept of 4D ${\cN}=1$ AdS superspace and superfield techniques.
The $\cN=1$ AdS superspace, ${\rm AdS}^{4|4}$, is the simplest member of the family of ${\cN}$-extended AdS superspaces
\bea
{\rm AdS}^{4|4{\cN}} =\frac{{\rm OSp} ({\cN}|4)}{{\rm SO}(3,1) \times {\rm SO}({\cN})}~.
\eea

In the following we give a summary of the results which are 
absolutely essential for constructing $\cN=1$ supersymmetric field theory in 
AdS in a manifestly  ${\rm OSp(1|4)}$-invariant way.
We mostly follow the presentation in \cite{KS94}. 
Our notation and two-component spinor conventions agree with \cite{Ideas}, 
except for the superspace integration measures.

Let $z^M =(x^m , \q^\mu, \bar \q_{\dot \mu} ) $ be local coordinates
for ${\rm AdS}^{4|4}$. 
The geometry  of ${\rm AdS}^{4|4}$
may be  described in terms of covariant derivatives
of the form
\bea
\cD_A = (\cD_a , \cD_\a ,\bar \cD^\ad ) = E_A + \O_A~, \qquad
E_A = E_A{}^M \partial_M  ~,
\eea
where $E_A{}^M $ is the inverse superspace vielbein, and 
\bea
\O_A = \frac{1}{2}\,\O_A{}^{bc} M_{bc}
= \O_A{}^{\b \g} M_{\b \g}
+\bar \O_A{}^{\bd \gd} \bar M_{\bd \gd} 
\eea	
is the Lorentz connection. 
The Lorentz generators $M_{bc} \Leftrightarrow (M_{\b\g},{\bar M}_{\bd\gd})$ act on two-component spinors, see appendix \ref{AppA}. In particular, they act on the spinor covariant derivatives by the rule
\bea
[M_{\a \b}, \cD_\g]= \ve_{\g(\a}\cD_{\b)}~, \quad [\bar M_{\ad \bd}, \bar \cD_{\gd}]= \ve_{\gd(\ad}\bar \cD_{\bd)}~,
\eea
while $[M_{\a \b}, \bar \cD_{\gd}]= [\bar M_{\ad \bd}, \cD_{\g}]  = 0 $. 
The covariant derivatives of ${\rm AdS}^{4|4}$ satisfy the following algebra
\begin{subequations}  \label{1.2}
\bea
&& \qquad \{ \cD_\a , \bar \cD_\ad \} = -2\rm i \cD_{\a \ad} ~, \\
&& \qquad \{\cD_\a, \cD_\b \} = -4\bar \m\, M_{\a \b}~, \qquad
\{ {\bar \cD}_\ad, {\bar \cD}_\bd \} = 4\m\,\bar M_{\ad \bd}~, \\
&& \qquad [ \cD_\a , \cD_{ \b \bd }] 
=\rm i \bar \m\,\ve_{\a \b} \bar \cD_\bd~,  \qquad
\,\,[{ \bar \cD}_{\ad} , \cD_{ \b \bd }] 
=-\rm i \m\,\ve_{\ad \bd} \cD_\b~,    \\
&&\quad \,\,\,\,[ \cD_{\a \ad} , \cD_{ \b \bd } ] = -2 \bar \m \m \Big({\ve}_{\a \b} \bar M_{\ad \bd }+ \ve_{\ad \bd} M_{\a \b}\Big)~,  
\eea
\end{subequations} 
with $\m\neq 0$ being a  complex parameter, which is related to the scalar curvature 
$\cR$ of  AdS space by the rule $\cR = -12 |\m|^2$.

The isometry group of $\cN=1$ AdS superspace is ${\rm OSp(1|4)}$.
The  isometries transformations of AdS$^{4|4}$ are generated by the Killing vector fields
$\L^A E_A$ which are defined to solve the Killing equation
\bea
\big[\L+\hf \o^{bc}M_{bc},\cD_{A} \big]=0~,\qquad
\L:=\l^B \cD_B =
\l^{b} \cD_{b}+\l^\b \cD_\b+{\bar \l}_\bd {\bar \cD}^\bd~, 
\label{N=1-killings-0}
\eea
for some Lorentz superfield parameter   $\o^{bc}= -\o^{cb}$. 
As shown in \cite{Ideas}, 
the equations in (\ref{N=1-killings-0}) are equivalent to 
\begin{subequations}
\bea
\cD_{(\a}\l_{\b)\bd}&=&0~, \qquad  {\bar \cD}^\bd\l_{\a\bd}   + 8\ri\l_\a=0~,\\
\cD_\a\l^\a&=&0~,
\qquad
{\bar \cD}_\ad\l_\a  + \frac{\ri}{ 2}{\mu}\l_{\a\ad}  =0~, \\
\o_{\a\b}&=&\cD_\a\l_\b~.
\eea
\end{subequations}
The solution to these equations is given in \cite{Ideas}.
If $T$ is a tensor superfield (with  suppressed indices), 
its infinitesimal ${\rm OSp(1|4)}$ transformation is 
\bea
\d T = \big( \L+\hf \o^{bc}M_{bc} \big) T ~.
\eea

In Minkowski space, we have seen that there are two ways to generate supersymmetric invariants, 
one of which corresponds to the 
integration over the full superspace and the other over its chiral subspace. 
In AdS superspace, every chiral integral can be always recast as 
a full superspace integral.
Associated with a scalar superfield $\cL$ is the following ${\rm OSp(1|4)}$ invariant
\bea
\int \rd^4x \rd^2 \q  \rd^2 \bar \q
\,E\,{\cal L} &=& 
-\frac{1}{4} \int
\rd^4x \rd^2 \q  
\, \cE\, {({\bar \cD}^2 - 4 \m)} {\cal L} ~, \qquad
E^{-1}= {\rm Ber}\, (E_{\rm A}{}^M)~,
\eea
where 
$\cE$ denotes the chiral integration measure.\footnote{In the chiral 
representation \cite{Ideas,GGRS}, the chiral measure
is $\cE= \vf^3$, where $\vf$ is the chiral compensator of old minimal 
supergravity \cite{Siegel78}.} 
Let $\cL_{\rm c}$ be a chiral scalar, $\bar \cD_\ad \cL_{\rm c} =0$. 
It generates the supersymmetric invariant 
$
\int \rd^4x \rd^2 \q  \, \cE \,{\cal L}_{\rm c}. 
$
The specific feature
of AdS superspace is that the chiral action can equivalently
be written as an integral over the full superspace \cite{Siegel78}
\bea
\int \rd^4x \rd^2 \q  \, \cE \,{\cal L}_{\rm c} 
= \frac{1}{\m} \int \rd^4x \rd^2 \q  \rd^2 \bar \q
\,{E}\, {\cal L}_{\rm c} ~.
\eea
Unlike the flat superspace case, the integral on the right does not vanish in AdS.


\section{Variant supercurrents in AdS space} \label{VarAdS}
We now turn to describing the structure of ${\cN}=1$ supercurrent multiplets in AdS.
In contrast to the variant supercurrents in Minkowski space, there exist only two irreducible AdS supercurrents, 
with $(12+12)$ and $(20+20)$ degrees of 
freedom \cite{BK11}.
The former is associated with the old minimal AdS supergravity (see e.g. \cite{Ideas,GGRS} for reviews). This supercurrent is the AdS extension of the Ferrara-Zumino multiplet satisfying the conservation equation
\bea
\bar \cD^\ad J_{\a\ad} = \cD_\a T~, \qquad \bar \cD_\ad T =0~.
\label{FZsupercurrent}
\eea
The latter corresponds to non-minimal AdS supergravity \cite{BK12}, with the following conservation law
\bea
\label{nonminCurrent}
\bar \cD^\ad {\mathbb J}_{\alpha \ad} = -\frac{1}{4} \bar \cD^2 \zeta_\alpha~,\qquad
\cD_{(\beta} \zeta_{\alpha)} = 0~.
\eea
The vector superfields $J_a$ and ${\mathbb J}_a$ are real.

The non-minimal supercurrent \eqref{nonminCurrent} is equivalent to the Ferrara-Zumino multiplet \eqref{FZsupercurrent}, since there exists 
a well-defined improvement 
transformation that turns \eqref{nonminCurrent} into \eqref{FZsupercurrent}, 
as demonstrated in \cite{BK12}. In AdS superspace, the constraint on the 
trace multiplet $\z_\a$, $\cD_{(\b} \z_{\a)}$, can always be solved as 
\bea
\zeta_\alpha 
	= \cD_\alpha (V + \ri \,U)~,
\eea
for well-defined real operators $V$ and $U$.\footnote{This follows from the properties of linear superfields which will be discussed further in section \ref{s41}.}
If we now introduce 
\bea
J_{\alpha \ad} := {\mathbb J}_{\alpha \ad} 
+ \frac{1}{6} [\cD_\alpha, \bar \cD_\ad] V
	- \cD_{\alpha \ad} U~,
	\qquad
T := \frac{1}{12} (\bar \cD^2 - 4 \mu) (V - 3 \ri U)~,
\eea
then the operators $J_{\a\ad} $ and $T$ prove to satisfy the conservation equation 
 \eqref{FZsupercurrent}.
 
 For the Ferrara-Zumino supercurrent  \eqref{FZsupercurrent}, 
 there exists an improvement transformation that is 
 generated by a chiral scalar operator $\O$.
 Specifically, using the operator $\O$ allows one to  
 introduce new supercurrent $\widetilde J_{\a\ad}$ 
 and chiral trace multiplet $\widetilde T$ defined by 
 \begin{subequations}\label{A.55}
 \bea
 \widetilde J_{\a\ad} &=& J_{\a\ad} +\ri \cD_{\a\ad} \big( \O -\bar \O \big)~, 
 \qquad \bar \cD_\ad \O=0~,
 \\
 \widetilde T &=& T +2\m \O +\frac 14 (\bar \cD^2 -4\m) \bar \O~.
 \eea
 \end{subequations}
The operators  $\widetilde J_{\a\ad}$ and $\widetilde T$ obey the conservation equation  \eqref{FZsupercurrent}.

\section{Off-shell higher-spin multiplets: a brief review} \label{s22}
In later chapters, we are going to construct higher-spin extensions of supercurrents in three and four dimensions. For this we require off-shell formulations for massless higher-spin supermultiplets. In the framework of 4D ${\cN}=1$ Poincar\'e and AdS supersymmetry, such gauge theories have already been developed in a series of papers \cite{KPS, KS, KS94}. Here we briefly review the constructions in Minkowski space \cite{KPS, KS}. Their AdS counterparts \cite{KS94} will be reviewed in the beginning of chapter \ref{ch4}.
\subsection{Massless half-integer superspin multiplets}
There exist two dually equivalent off-shell formulations for a free massless superspin-$(s+\hf)$ multiplet, with $s=1, 2, \ldots$. They are referred to as transverse and longitudinal formulations \cite{KPS}. 
In these two off-shell gauge models, the main feature is the use of the so-called transverse and longitudinal linear superfields as one of the dynamical variables. 
Both are complex superfields and subject to different constraints. More generally, a complex tensor superfield\footnote{All Lorentz tensor (super)fields considered in this thesis are completely symmetric in their undotted spinor indices and separately in their dotted ones. We use the shorthand notation
$V_{\a(m)\ad(n)} := V_{\a_1 \cdots \a_m \ad_1 \cdots \ad_n} (z)
= V_{(\a_1 \cdots \a_m)(\ad_1 \cdots \ad_n)}(z) $
and
$V^{\a(m) \ad(n)} U_{\a(m)\ad(n)} := V^{\a_1 \cdots \a_m \ad_1 \cdots \ad_n} 
U_{\a_1 \cdots \a_m \ad_1 \cdots \ad_n}$.
Parentheses denote symmetrisation of indices; 
the undotted and dotted spinor indices are symmetrised independently. 
Indices sandwiched between vertical  bars 
(for instance,  $|\g|$) are not subject to symmetrisation. } 
$\G_{\a(m) \ad(n)}$ 
is called \textit{transverse linear}, if it obeys the constraint 
\bea
\bar D^\bd \G_{ \a(m) \bd \ad(n - 1) } = 0~, \qquad n>0~.
\label{transverse1}
\eea
A \textit{longitudinal linear} superfield $G_{\a(m) \ad(n)}$ is defined to satisfy the constraint
\bea
\bar D_{(\ad_1} G_{\a(m)\ad_2 \dots \ad_{n+1})} = 0~.
\label{longitudinal1}
\eea
The above constraints imply that $\G_{\a(m) \ad(n)}$ and $G_{\a(m) \ad(n)}$ are linear superfields,
\bea
\bar D^2 \G_{\a(m) \ad(n)} = \bar D^2 G_{\a(m) \ad(n)} = 0~.
\eea
In the case $n=0$, the constraint \eqref{transverse1} has to be replaced 
with the standard linear 
constraint  $\bar D^2 \G_{ \a(m) } = 0 $. The constraint \eqref{longitudinal1} for $n=0$ is the chirality condition
$\bar D_\bd G_{ \a(m) } = 0$. 

In the case of 4D $\cN=1$ AdS supersymmetry, 
longitudinal linear and transverse linear superfields 
were first described in \cite{IS1} to realise the irreducible representations of the AdS isometry group
${\rm OSp}(1|4)$ (see \cite{West} for a nice review of the results of \cite{IS1}). In the framework of 4D $\cN=1$ conformal supergravity,
primary longitudinal linear and transverse linear supermultiplets were
introduced for the first time by Kugo and Uehara \cite{KU}.
Such superfields were used in \cite{KS94,KPS,KS,KO,BHK, HKO}
for the description of off-shell massless gauge theories in three and four dimensions.

\subsubsection{Transverse formulation}
The transverse formulation is realised in terms of the following dynamical variables:
\bea
\cV^\bot_{s+1/2}& = &\Big\{H_{\a(s)\ad(s)}~, ~
\G_{\a(s-1) \ad(s-1)}~,
~ \bar{\G}_{\a(s-1) \ad(s-1)} \Big\} ~.
\label{4dtr}
\eea
Here $H_{\a(s) \ad (s)}$ is a real unconstrained superfield. The complex superfield $\G_{\a(s-1) \ad(s-1)}$ is transverse linear,
\bea
\bar D^\bd \G_{ \a(s-1) \bd \ad(s - 2) } = 0 \quad
&\Longrightarrow & \quad \bar D^2 \G_{\a(s-1) \ad(s-1)}=0~.
\label{tr}
\eea
The constraint \eqref{tr} can be solved in terms of an \textit{unconstrained} prepotential:
\bea
 \G_{\a(s-1) \ad(s-1)}&=& \bar D^\bd 
{ \Phi}_{\a(s-1)\,(\bd \ad_1 \cdots \ad_{s-1}) } ~.
\eea
The prepotential is defined modulo gauge transformation of the form
\bea
\d_\x \Phi_{\a(s-1)\, \ad (s)} 
&=&  \bar D^\bd 
{ \x}_{\a(s-1)\, (\bd \ad_1 \cdots \ad_{s}) } ~,
\label{tr-prep-gauge1}
\eea
with the gauge parameters $ {\x}_{\a(s-1)\,  \ad (s+1) } $
being unconstrained.

It was postulated in \cite{KPS} that the linearised gauge transformations for the superfields $H_{\a(s) \ad (s)}$ and $\G_{\a(s-1) \ad(s-1)}$ are
\begin{subequations} 
\bea \d_\L H_{\a_1 \dots \a_s \ad_1  \dots \ad_s} 
&= &\bar D_{(\ad_1} \L_{\a_1\dots  \a_s \ad_2 \dots \ad_s )} 
- D_{(\a_1} \bar{\L}_{\a_2 \dots \a_s)\ad_1  \dots \ad_s} \ , \label{trgaugeH} \\
\d_\L \G_{\a_1 \dots \a_{s-1}\ad_1 \dots \ad_{s-1}} &= &
-\frac{1}{4} \bar D^\bd D^2 \bar{\L}_{\a_1 \dots \a_{s-1}\bd \ad_1 
\dots \ad_{s-1}} \ , \label{trgaugeGG}
\eea
\end{subequations}
where the complex gauge parameter $\L_{\a_1 \dots \a_s \ad_1 \dots \ad_{s-1}}
=\L_{(\a_1 \dots \a_s )(\ad_1 \dots \ad_{s-1})}$ 
is unconstrained. 
It follows from \eqref{trgaugeGG} that the transformation law of the prepotential $\F_{\a(s-1) \ad(s)}$ is
\bea
\d_\L \F_{\a_1 \dots \a_{s-1}\ad_1 \dots \ad_s} &= &
-\frac{1}{4} D^2 \bar{\L}_{\a_1 \dots \a_{s-1} \ad_1 
\dots \ad_s} ~.
\label{trgaugePhi}
\eea
The action invariant 
under the gauge transformations \eqref{trgaugeH} and
\eqref{trgaugeGG} is 
\bea
S^{\bot}_{(s+\hf)}[H, \G ,\bar \G]&=&
\Big( - \frac{1}{2}\Big)^s  \int \rd^4x \rd^2 \q  \rd^2 \bar \q \,
\Big\{ \frac{1}{8} H^{ \a(s) \ad(s) }  D^\b {\bar D}^2 D_\b 
H_{\a(s) \ad(s) }  \non \\
&+& H^{ \a(s) \ad(s) }
\left( D_{\a_s}  {\bar D}_{\ad_s} \G_{\a(s-1) \ad(s-1) }
- {\bar D}_{\ad_s}  D_{\a_s} 
{\bar \G}_{\a (s-1) \ad (s-1) } \right) \non \\
&+&\Big( {\bar \G}^{\a(s-1) \ad(s-1)} \G_{\a(s-1) \ad(s-1)} \non\\
&+& \frac{s+1} {s} \, \G^{\a(s-1) \ad(s-1)} \G_{\a(s-1) \ad(s-1)} + {\rm c.c.} \Big)
\Big\} ~.
\label{hi-t}
\eea

We now briefly comment on the limiting $s=1$ case which should correspond to linearised supergravity. The transverse linear constraint \eqref{tr} cannot be used
for $s=1$, however its corollary $\bar D^2 \G_{\a(s-1) \ad(s-1)}=0$ can be used,
\bea
\bar D^2 \G =0~.
\eea 
This constraint defines a complex linear superfield. In accordance with \eqref{trgaugeGG}, 
the gauge transformation of $\G$ is 
\bea
\d_\L \G = \frac{1}{4} \bar D_\bd D^2 \bar{\L}^\bd ~.
\eea
The action \eqref{hi-t} for $s=1$ coincides with the linearised action 
for the $n=-1$ non-minimal supergravity, see \cite{GKP,Ideas} for reviews.

\subsubsection{Longitudinal formulation} \label{2212}
The longitudinal formulation is described in terms of the following dynamical variables:
\bea
\cV^{\|}_{s+1/2}& = &\Big\{H_{\a(s)\ad(s)}~, ~
G_{\a(s-1) \ad(s-1)}~,
~ \bar{G}_{\a(s-1) \ad(s-1)} \Big\} ~,
\label{4dlon}
\eea
where the real superfield $H_{\a(s) \ad (s)}$ is unconstrained, and the compensating superfield $G_{\a(s-1) \ad(s-1)}$ is longitudinal linear,
\bea
\bar D_{(\ad_1}G_{ \a(s-1) \ad_2 \cdots \ad_s) } = 0 \quad
&\Longrightarrow & \quad \bar D^2 G_{\a(s-1) \ad(s-1)}=0~.
\label{lon}
\eea
The constraint \eqref{lon} can be solved in terms of an \textit{unconstrained} prepotential
\bea
 G_{\a(s-1) \ad(s-1)} &=& {\bar D}_{( \ad_1 }
 \Psi_{ \a(s-1) \, \ad_2 \cdots \ad_{s-1}) } ~.
\eea
The prepotential is defined modulo gauge transformations of the form
\bea
\d_\z  \Psi_{ \a(s-1) \, \ad {(s-2}) } &=&  {\bar D}_{( \ad_1 }
 \z_{ \a(s-1) \, \ad_2 \cdots \ad_{s-2}) } ~,
\label{lon-prep-gauge1}
\eea
with the gauge parameter $ \z_{ \a(s-1) \, \ad (s-3)}$ 
being unconstrained.\footnote{For $s=2$
the gauge transformation law \eqref{lon-prep-gauge1} has to be replaced with 
 $\d \J_{\a} = \z_{\a}$, with the gauge parameter $\z_{\a}$ being chiral, $\bar D_\bd \z_{\a}=0$.}

The gauge transformations for the dynamical superfields are given by
\begin{subequations} 
\bea 
\d_\L H_{\a_1 \dots \a_s \ad_1  \dots \ad_s} 
&= &\bar D_{(\ad_1} \L_{\a_1\dots  \a_s \ad_2 \dots \ad_s )} 
- D_{(\a_1} \bar{\L}_{\a_2 \dots \a_s)\ad_1  \dots \ad_s}~,  \label{longaugeH} \\ 
\d_\L G_{\a_1 \dots \a_{s-1}\ad_1 \dots \ad_{s-1}} &= & - \hf \bar D_{(\ad_1} 
\bar D^{|\bd|} D^\b \L_{\b\a_1 \dots \a_{s-1} \ad_2 \dots \ad_{s-1}) \bd} \non\\
&&+ \ri (s-1) \bar D_{(\ad_1} \pa^{\b |\bd|} 
\L_{\b \a_1 \dots \a_{s-1} \ad_2 \dots \ad_{s-1} ) \bd} \ . \label{longaugeG}
\eea
\end{subequations}
The symmetrisation in \eqref{longaugeG} is extended only to the indices 
$\ad_1, \ad_2, \dots,  \ad_{s-1}$. 
It follows from \eqref{longaugeG} that the transformation law of the prepotential $\J_{\a(s-1) \ad(s-2)}$ is
\bea
\d_\L \J_{\a_1 \dots \a_{s-1}\ad_1 \dots \ad_{s-2}} &= & - \hf 
\Big( \bar D^{\bd} D^\b -2\ri (s-1) \pa^{\b \bd} \Big)
\L_{\b\a_1 \dots \a_{s-1} \bd \ad_1 \dots \ad_{s-2}} ~.
\label{longaugeJ}
\eea
The action invariant 
under the gauge transformations \eqref{longaugeH} and \eqref{longaugeG} 
is
\bea
S^{\|}_{(s+\hf)}[H,G,\bar G]&=&
\Big(-\hf \Big)^s  \int \rd^4x \rd^2 \q  \rd^2 \bar \q \, \Big\{
\frac 18 
H^{\a (s) \ad (s) }   D^\b \bar D^2
 D_\b H_{\a (s) \ad (s) }  \non \\
&-& \frac{1}{8} \, \frac{s}{2s+1} \, \Big( \,
\big[ D_{\g}, \bar D_{\gd}\big] H^{\g \a (s-1)\gd  \ad (s-1)}
\, \Big)  \,
\big[ D^\b, \bar D^{\bd}\big]
H_{\b\a(s-1)\bd\ad(s-1)} \,  \non \\
&+& \frac{s}{2}\, \Big( \partial_{\g \gd} 
H^{\g \a (s-1) \gd \ad (s-1)}  \Big) \,
\partial^{\b\bd}
H_{\b\a(s-1)\bd\ad(s-1)} 
\non \\
&+& 2{\rm i} \, \frac{ s}{2s+1}  \,  \pa_{\g \gd } 
H^{\g \a (s-1) \gd \ad (s-1)}
\Big( G_{\a(s-1) \ad(s-1)} - \bar G_{\a(s-1) \ad(s-1)} \Big)  \non \\
&+& \frac{1}{2s+1} \Big( \bar G^{\a(s-1) \ad(s-1)} G_{\a(s-1) \ad(s-1)} \non\\
&-& \frac{s+1}s G^{\a(s-1) \ad(s-1)} G_{\a(s-1) \ad(s-1)}
+ {\rm c.c.}\Big)\Big\}   ~.
\label{hi-l}
\eea
The models (\ref{hi-t}) and (\ref{hi-l}) are dually equivalent in a sense that they are related by a superfield Legendre transformation described in \cite{KPS}.
In general, the procedure works as follows. 
Let us start with the transverse theory \eqref{hi-t} 
and associate with it the following first-order action
\bea
S[H, V, \bar V, G, \bar G] &=& S^{\perp}_{(s+\hf)}[H, V, \bar V] \non\\
&&+ \int  \rd^4x \rd^2 \q  \rd^2 \bar \q  \,
 \Big(V^{\a(s-1) \ad(s-1)} G_{\a(s-1) \ad(s-1)} 
  +{\rm c.c.} \Big) ~,
\label{2220}
\eea
with $V_{\a(s-1) \ad(s-1)}$ being \textit{unconstrained} complex. Here $S^{\perp}_{(s+\hf)}[H, V, \bar V]$ is obtained from $S^{\perp}_{(s+\hf)}[H, \G, \bar \G]$ by the replacement $\G_{\a(s-1) \ad(s-1)} \to V_{\a(s-1) \ad(s-1)}$. The Lagrange multiplier $G_{\a(s-1)\ad(s-1)}$ is longitudinal linear.
Varying the first-order action $S[H, V, \bar V, G, \bar G]$ with respect to the Lagrange multiplier $G_{\a(s-1) \ad(s-1)}$ gives $V_{\a(s-1) \ad(s-1)} = \G_{\a(s-1) \ad(s-1)}$, thus $S[H, V, \bar V, G, \bar G]$ reduces to the original action $S^{\perp}_{(s+\hf)}[H, \G, \bar \G]$.
On the other hand, one can integrate out $V_{\a(s-1) \ad(s-1)}$ using its equation of motion
\bea
\frac{\d}{\d V^{\a(s-1) \ad(s-1)}} S^{\perp}_{(s+\hf)} [H, V, \bar V] + G_{\a(s-1) \ad(s-1)} =0~. 
\label{d-eq}
\eea
We assume that \eqref{d-eq} can be solved to express $V_{\a(s-1) \ad(s-1)}$ in terms of $G_{\a(s-1) \ad(s-1)}$ and its conjugate. Plugging this solution back into \eqref{2220} leads to the dual action $S_{D} [H, G, \bar G]$, given by the expression \eqref{hi-l}.

The constraint  
\eqref{lon} 
is the chirality condition for $s=1$, $\bar D_\ad G=0$~.
The gauge transformation law \eqref{longaugeG} cannot directly be used for $s=1$. 
Nevertheless,  it can be rewritten in the form 
\bea
\d_\L G_{\a_1 \dots \a_{s-1}\ad_1 \dots \ad_{s-1}} &= &
 - \frac 14  \bar D^2  
 D^\b \L_{\b\a_1 \dots \a_{s-1} \ad_1 \dots \ad_{s-1}} \non\\
&&+ \ri (s-1)  \pa^{\b \bd} \bar D_{(\ad_1}
\L_{\b \a_1 \dots \a_{s-1} \ad_2 \dots \ad_{s-1} ) \bd} ~, 
\eea
which is well defined for  $s=1$:
\bea
\d_\L G &= & - \frac 14  \bar D^2   D^\b \L_\b~.
\eea
The action \eqref{hi-l} for $s=1$ coincides with the linearised action 
for the old minimal supergravity, see \cite{GKP,Ideas} for reviews.

\subsection{Massless integer superspin multiplets} \label{222}
We now recall the two off-shell gauge models for a massless multiplet of integer superspin $s \geq 2$, which were originally constructed in \cite{KS}. In each of the formulations, the dynamical variables consist of a real unconstrained prepotential $H_{\a(s-1) \ad(s-1)}$ in conjunction with some compensating supermultiplets. 

\subsubsection{Longitudinal formulation}
The longitudinal theory is described by the following set of superfields
\bea
\cV^{\|}_{s} &=& 
\Big\{H_{\a(s-1)\ad(s-1)}~, ~
G_{\a(s) \ad(s)}~,
~ \bar{G}_{\a(s) \ad(s)} \Big\}~.
\eea
The superfield $H_{\a(s-1)  \ad(s-1)}$ is unconstrained real, while the compensator $G_{\a(s)\ad(s)}$ is longitudinal linear. The latter is a field strength associated 
with a complex unconstrained 
prepotential $\J_{\a(s) \ad(s-1)}$,
\bea
G_{\a_1 \dots \a_s \ad_1 \dots \ad_s} := 
\bar D_{(\ad_1} \J_{\a_1 \dots \a_s \ad_2 \dots \ad_s)}
\quad \Longrightarrow \quad
\bar D_{(\ad_1} G_{\a_1 \dots \a_s \ad_2 \dots \ad_{s+1})}=0~.
\eea

The dynamical superfields are defined modulo gauge transformations of the form
\bsubeq \label{2226}
\bea
\d_L H_{\a(s-1) \ad(s-1)} 
&= &D^{\b} L_{\b \a(s-1)\ad(s-1) )} 
- \bar D^{\bd} \bar L_{\a(s-1)\bd \ad(s-1) )}~, \\
\d_{L, \z} \J_{\a(s) \ad(s-1)} &=& \hf D_{(\a_1} D^{|\g|} L_{\a_2 \dots \a_s) \g \ad(s-1)}+ \bar D_{(\ad_1}\z_{\a_1 \dots \a_s \ad_2 \dots \ad_{s-1})}~, \label{lon-gauge-Jint}
\eea
\esubeq
with the gauge parameter $L_{\a(s) \ad(s-1)}$ being complex unconstrained.
The action functional which is quadratic in the superfields $H, G, \bar G$ and invariant under the gauge transformations \eqref{2226} is given by 
\bea
S^{\|}_{(s)} &=&
\Big( - \frac{1}{2}\Big)^s  \int 
 \rd^4x \rd^2 \q  \rd^2 \bar \q
\,
\left\{ \frac{1}{8} H^{ \a (s-1) \ad (s-1) }  D^\b {\bar D}^2 D_\b 
H_{\a (s-1) \ad (s-1)} \right. \non \\
&+& \frac{s}{s+1}H^{ \a(s-1) \ad(s-1) }
\Big( D^{\b}  {\bar D}^{\bd} G_{\b\a(s-1) \bd\ad(s-1) }
- {\bar D}^{\bd}  D^{\b} 
{\bar G}_{\b \a (s-1) \bd \ad (s-1) } \Big) \non \\
&+& 2 \bar G^{ \a (s) \ad (s) } G_{ \a (s) \ad (s) } 
+ \frac{s}{s+1}\Big( G^{ \a (s) \ad (s) } G_{ \a (s) \ad (s) } 
+ \bar G^{ \a (s) \ad (s) }  \bar G_{ \a (s) \ad (s) } \Big) \bigg\}~.~~~~~
\label{i-l}
\eea

\subsubsection{Transverse formulation}
The transverse formulation is realised by the following set of superfields
\bea
\cV^\bot_{s}& = &\Big\{H_{\a(s-1)\ad(s-1)}~, ~
\G_{\a(s)\, \ad (s)} ~,
~ \bar \G_{\a(s)\, \ad (s)}  \Big\} ~.
\eea
Here the compensating multiplet is described by a transverse linear superfield $\G_{\a(s) \ad(s)}$ (and its conjugate $\bar \G_{\a(s) \ad(s)}$) constrained by 
\bea
\bar D^{\bd} \G_{\a(s) \bd \ad(s-1)} = 0 \qquad \Longrightarrow \qquad \bar D^2 \G_{\a(s) \ad(s)} = 0~.
\eea
The dynamical superfields are defined modulo gauge transformations of the form
\begin{subequations} \label{trgaugeint}
\bea 
\d_L H_{\a(s-1) \ad(s-1)} 
&= &D^{\b} L_{\b \a(s-1)\ad(s-1) )} 
- \bar D^{\bd} \bar L_{\a(s-1)\bd \ad(s-1) )}~,\\
\d_L \G_{\a(s) \ad(s)} &=& \frac{s+1}{2(s+2)} \bar D^\bd \Big\{ 
\bar D_{(\bd} D_{(\a_1} 
+ 2\ri (s+2) \pa_{(\a_1(\bd}\Big\}
 \bar{L}_{\a_2 \dots \a_{s})\ad_1 \dots \ad_{s})}~.~~~~~
\eea
\esubeq
The gauge-invariant action is given by
\bea
S^{\perp}_{(s+ \hf)} &=& -\Big( - \frac{1}{2}\Big)^s  \int 
 \rd^4x \rd^2 \q  \rd^2 \bar \q
\, \Big\{ -\frac{1}{8}H^{\a(s-1)\ad(s-1)} D^{\b} \bar D^2 D_{\b} H_{\a(s-1) \ad(s-1)} \non\\
&&+ \frac{1}{8} \frac{s^2}{(s+1)(2s+1)} [D^{\b}, \bar D^{\bd}]H^{\a_1 \dots \a_{s-1} \ad_1 \dots \ad_{s-1}}  [D_{(\b}, \bar D_{(\bd}]H_{\a_1 \dots \a_{s-1}) \ad_1 \dots \ad_{s-1})} \non \\
&&+ \hf \frac{s^2}{s+1} \pa^{\b\bd} H^{\a_1 \dots \a_{s-1} \ad_1 \dots \ad_{s-1}} \pa_{(\b ( \bd} H_{\a_1 \dots \a_{s-1}) \ad_1 \dots \ad_{s-1})} \non\\
&&+ \frac{2\ri s}{2s+1}H^{\a(s-1) \ad(s-1)} \pa^{\b \bd} \big(\G_{\b\a(s-1) \bd \ad(s-1)} - \bar \G_{\b \a(s-1) \ad(s-1)}\big)\non\\
&&+ \frac{1}{2s+1} \big( \bar \G^{\a(s) \ad(s) } \G_{\a(s) \ad(s)} - \frac{s}{s+1} \G^{\a(s) \ad(s)} \G_{\a(s) \ad(s)} + \rm c.c. \big) \Big\}~. 
\label{i-t}
\eea

As demonstrated in \cite{KS}, the two actions \eqref{i-l} and \eqref{i-t} are classically equivalent, for they are related by a superfield Legendre transformation.


\chapter{Non-conformal higher-spin supercurrents in Minkowski space} \label{ch3}
In the previous chapter, the off-shell structure of the massless superspin-$\hat{s}$ ${\cN}=1$ multiplets was described. In the case of half-integer superspin $\hat{s}= s+\hf$, with $s=2,3,\dots$ 
both models involve one and the same gauge superfield $H_{\a(s) \ad(s)}$, but differ in the compensating multiplets used, {\it i.e.} transverse or longitudinal superfield. The real unconstrained prepotential $H_{\a(s) \ad(s)}$ is the higher-spin superconformal gauge multiplet introduced in \cite{HST}. In the $s=1$ case, the gauge transformation \eqref{trgaugeH} corresponds to linearised conformal supergravity \cite{FZ2}.
It is important to note that the non-supersymmetric higher-spin theories proposed by (Fang-)Fronsdal \cite{Fronsdal, FF} and their supersymmetric counterparts of half-integer superspin share one common feature. Specifically, for each of them, the gauge-invariant action is formulated in terms of a (super)conformal gauge (super)field coupled to certain compensators. Such a description was not known for the massless multiplet of {\it integer} superspin until recently \cite{HK2}, in which a reformulation of the integer superspin action \eqref{i-l} was given.
We now make these points more precise. 

\section{Conformal gauge (super)fields and compensators}

Given an integer $s\geq 2$, the conformal spin-$s$ field \cite{FT,FL}
is described by a real potential
$h_{\a_1 \dots \a_{s} \ad_1 \dots \ad_{s} } 
=h_{(\a_1 \dots \a_{s}) (\ad_1 \dots \ad_{s} )} 
\equiv h_{\a(s) \ad(s)}$
with the gauge freedom
\begin{subequations}
\bea
 \d h_{\a_1 \dots \a_{s} \ad_1 \dots \ad_{s} } 
&=& \pa_{(\a_1 (\ad_1} \l_{\a_2\dots \a_{s}) \ad_2 \dots \ad_{s})}~,
\eea
for an arbitrary real gauge parameter 
$\l_{\a_1 \dots \a_{s-1}  \ad_1 \dots \ad_{s-1}}
=\l_{(\a_1 \dots \a_{s-1} )( \ad_1 \dots \ad_{s-1})} \equiv \l_{\a(s-1) \ad(s-1)}$. 
In addition to the gauge field $h_{\a(s) \ad(s)}$,
the massless spin-$s$ action \cite{Fronsdal} also involves  a real compensator
$h_{\a(s-2) \ad(s-2)}$ with the gauge transformation
\bea
 \d h_{\a_1 \dots \a_{s-2} \ad_1 \dots \ad_{s-2} } 
&=& \pa^{\b \bd} \l_{\b \a_1\dots \a_{s-2} \bd \ad_1 \dots \ad_{s-2}}~.
\eea
\end{subequations}
In the fermionic case, the conformal spin-$(s+\hf)$ field \cite{FT,FL}
is described by a potential
$\j_{\a(s+1) \ad(s)}$ and its conjugate $\bar \j_{\a(s) \ad(s+1)}$
with the gauge freedom
\begin{subequations}
\bea
 \d \j_{\a_1 \dots \a_{s+1} \ad_1 \dots \ad_{s} } 
&=& \pa_{(\a_1 (\ad_1} \x_{\a_2\dots \a_{s+1}) \ad_2 \dots \ad_{s})}~,
\eea
for an arbitrary gauge parameter 
$\x_{\a(s) \ad(s-1)}$.
In addition to the gauge fields $ \j_{\a(s+1) \ad(s)}$ and $\bar \j_{\a(s) \ad(s+1)}$, 
the massless spin-$(s+\hf)$ action \cite{FF} also involves two 
compensators $ \j_{\a(s-1) \ad(s)}$ and $ \j_{\a(s-1) \ad(s-2)}$
and their conjugates, with the following gauge transformations
\bea
 \d \j_{\a_1 \dots \a_{s-1} \ad_1 \dots \ad_{s} } 
&=& \pa^\b{}_{ (\ad_1} \x_{\b\a_1\dots \a_{s-1} \ad_2 \dots \ad_{s})}~, \\
 \d \j_{\a_1 \dots \a_{s-1} \ad_1 \dots \ad_{s-2} } 
&=& \pa^{\b \bd} \x_{\b \a_1\dots \a_{s-1} \bd \ad_1 \dots \ad_{s-2}}~.
\eea
\end{subequations}

%
In the case of an integer superspin $\hat{s}= s$, with $s = 2,3,\dots$ the superconformal multiplet introduced in \cite{KMT} 
is described in terms of an unconstrained prepotential $\J_{\a(s)\ad(s-1)} $ and its complex conjugate with the gauge freedom
\bea
 \d_{ {\frak V} ,\z} \J_{\a_1 \dots \a_s \ad_1 \dots \ad_{s-1}} 
 &=& \hf D_{(\a_1}  {\frak V}_{\a_2 \dots \a_s)\ad_1 \dots \ad_{s-1}}
+  \bar D_{(\ad_1} \z_{\a_1 \dots \a_s \ad_2 \dots \ad_{s-1} )}  ~.~~~~ \label{conf-int}
\eea
Here the gauge parameters ${\frak V}_{\a(s-1) \ad(s-1)}$ and $\z_{\a(s) \ad(s-2)}$ are both \textit{unconstrained}.
As shown in subsection \ref{222} (see also \cite{KS}), the prepotential  $\J_{\a(s)\ad(s-1)} $ 
naturally appears as one of the dynamical variables in the \textit{longitudinal} formulation for the massless 
superspin-$s$ multiplet, in addition to the real unconstrained prepotential $H_{\a(s-1)\ad(s-1)}$. However, the gauge transformation of
$\J_{\a(s)\ad(s-1)} $ given in eq. \eqref{lon-gauge-Jint} differs from eq. \eqref{conf-int}. 
The difference is that the parameter ${\frak V}_{\a(s-1) \ad(s-1)}$ in \cite{KS} is not unconstrained, but instead takes the form ${\frak V}_{\a(s-1) \ad(s-1)} = D^{\b} L_{(\b \a_1 \cdots \a_{s-1}) \ad(s-1)}$. 
Furthermore, the prepotential $\J_{\a(s)\ad(s-1)}$ enters the action functional \eqref{i-l} only via the constrained field strength $G_{\a(s) \ad(s)} := \bar D_{(\ad_1} \J_{\a(s) \ad_2 \dots \ad_s)} $, which is longitudinal linear. 
It is then natural to look for a new formulation by properly generalising the off-shell supersymmetric action \eqref{i-l}. 

In this chapter we propose a new off-shell realisation for the massless superspin-$s$ multiplet with the following properties: (i) the gauge freedom of $\J_{\a(s)\ad(s-1)} $ is given by \eqref{conf-int}; and (ii) the original longitudinal formulation \eqref{i-l} emerges upon imposing a gauge condition. 
The new model is shown to possess a dual formulation obtained by applying a superfield Legendre transformation. 
We then introduce non-conformal higher-spin  supercurrents associated to the off-shell actions for the massless ${\cN}=1$ supermultiplets in 4D Minkowski space.
Explicit realisations for these conserved higher-spin supercurrents are given for models for a single massless and massive chiral superfield, as well as the massive ${\cN}=2$ hypermultiplet. 


\section{The massless integer superspin multiplets revisited} \label{s31}
Here we present a new off-shell gauge formulation for the massless superspin-$s$ multiplet, as well as for the massless gravitino multiplet ($s=1$) which requires special consideration. 
\subsection{Reformulation of the longitudinal theory}
Given a positive integer $s \geq 2$, we propose to describe the massless 
superspin-$s$ multiplet in terms of the following 
 superfield variables: 
(i) an unconstrained prepotential
 $\J_{\a(s)\ad(s-1)}  $ 
and its complex conjugate $\bar \J_{\a(s-1)\ad(s)}$; 
(ii) a real  superfield 
 $H_{\a(s-1)\ad(s-1)}  =\bar H_{\a(s-1)\ad(s-1)}  $; and (iii)
a complex superfield $\S_{\a(s-1) \ad (s-2) }$ 
and its conjugate $\bar \S_{\a (s-2) \ad(s-1)}$, 
where $\S_{\a(s-1) \ad (s-2) }$ is constrained 
to be transverse linear,
\bea
 \bar D^\bd \S_{\a(s-1) \bd \ad(s-3)} =0~.
 \label{a1}
 \eea
 In the $s=2$ case, 
 for which \eqref{a1} is not defined, 
 $\S_{\a} $ is constrained to be complex linear, 
 \bea
 \bar D^2 \S_{\a } =0~.
 \label{a2}
 \eea
 The constraint \eqref{a1}, or its counterpart \eqref{a2} for $s=2$, 
  can be solved in terms of a complex unconstrained
 prepotential $Z_{\a(s-1) \ad (s-1)}$ by the rule
 \bea
 \S_{\a(s-1) \ad (s-2)} = \bar D^\bd Z_{\a(s-1) (\bd \ad_1 \dots \ad_{s-2} )} ~.
 \label{a3}
 \eea
 This prepotential is defined modulo 
 gauge transformations 
 \bea
 \d_\x Z_{\a(s-1) \ad (s-1)}=  \bar D^\bd \x_{\a(s-1) (\bd \ad_1 \dots \ad_{s-1} )} ~,
 \label{a4}
 \eea
 with the gauge parameter  $\x_{\a(s-1) \ad (s) }$ being  unconstrained complex.

The gauge freedom of $\J_{\a_1 \dots \a_s \ad_1 \dots \ad_{s-1}} $ is
chosen to coincide with that of the superconformal superspin-$s$ multiplet 
\cite{KMT}, which is
\begin{subequations} \label{a5}
\bea
 \d_{ {\frak V} ,\z} \J_{\a_1 \dots \a_s \ad_1 \dots \ad_{s-1}} 
 &=& \hf D_{(\a_1}  {\frak V}_{\a_2 \dots \a_s)\ad_1 \dots \ad_{s-1}}
+  \bar D_{(\ad_1} \z_{\a_1 \dots \a_s \ad_2 \dots \ad_{s-1} )}  ~ , \label{a5a}
\eea
with \textit{unconstrained} complex gauge parameters ${\frak V}_{\a(s-1) \ad(s-1)}$ 
and $\z_{\a(s) \ad(s-2)}$. 
The $\frak V$-transformation is defined to act on the superfields $H_{\a(s-1) \ad(s-1)}$
and $\S_{\a(s-1) \ad(s-2) }$ as follows
\bea 
\d_{\frak V} H_{\a(s-1) \ad(s-1)}&=& {\frak V}_{\a(s-1) \ad(s-1)} +\bar {\frak V}_{\a(s-1) \ad(s-1)}
~, \label{a5b}\\
\d_{\frak V} \S_{\a(s-1) \ad(s-2) }&=&  \bar D^\bd \bar {\frak V}_{\a(s-1) \bd \ad(s-2)}
\,\, \Longrightarrow \,\, \d_{\frak V} Z_{\a(s-1) \ad (s-1)}
=\bar  {\frak V}_{\a(s-1) \ad (s-1)}~.~~~
\label{a5c}
\eea
\end{subequations}
The longitudinal linear superfield 
\bea
G_{\a_1 \dots \a_s \ad_1 \dots \ad_s} := 
\bar D_{(\ad_1} \J_{\a_1 \dots \a_s \ad_2 \dots \ad_s)}~, 
\qquad \bar D_{(\ad_1} G_{\a_1 \dots \a_s \ad_2 \dots \ad_{s+1})}=0
\label{a6}
\eea
is invariant under the $\z$-transformation \eqref{a5a} 
and  varies under the $\frak V$-transformation as 
\bea
 \d_{ {\frak V} } G_{\a_1 \dots \a_s \ad_1 \dots \ad_{s}} 
 &=& \hf \bar D_{(\ad_1} D_{(\a_1}  {\frak V}_{\a_2 \dots \a_s)\ad_2 \dots \ad_{s})}~.
 \eea

It may be checked that the following action
\bea
S^{\|}_{(s)} &=&
\Big( - \frac{1}{2}\Big)^s  \int 
 \rd^4x \rd^2 \q  \rd^2 \bar \q
\,
\left\{ \frac{1}{8} H^{ \a (s-1) \ad (s-1) }  D^\b {\bar D}^2 D_\b 
H_{\a (s-1) \ad (s-1)} \right. \non \\
&&+ \frac{s}{s+1}H^{ \a(s-1) \ad(s-1) }
\Big( D^{\b}  {\bar D}^{\bd} G_{\b\a(s-1) \bd\ad(s-1) }
- {\bar D}^{\bd}  D^{\b} 
{\bar G}_{\b \a (s-1) \bd \ad (s-1) } \Big) \non \\
&&+ 2 \bar G^{ \a (s) \ad (s) } G_{ \a (s) \ad (s) } 
+ \frac{s}{s+1}\Big( G^{ \a (s) \ad (s) } G_{ \a (s) \ad (s) } 
+ \bar G^{ \a (s) \ad (s) }  \bar G_{ \a (s) \ad (s) } 
 \Big) \non \\
 &&+ \frac{s-1}{4s}H^{ \a(s-1) \ad(s-1) }
\Big( D_{\a_1} \bar D^2 \bar \S_{\a_2 \dots \a_{s-1}\ad(s-1)}
 - {\bar D}_{\ad_1}  D^2 \S_{\a(s-1) \ad_2 \dots \ad_{s-1} } \Big)  \non \\
&&+\frac{1}{s} \J^{\a(s) \ad(s-1)} \Big( 
D_{\a_1} \bar D_{\ad_1} -2\ri (s-1) \pa_{\a_1 \ad_1} \Big)
\S_{\a_2 \dots \a_s \ad_2 \dots \ad_{s-1} }\non  \\
&&+\frac{1}{s} \bar \J^{\a(s-1) \ad(s)} \Big( 
 \bar D_{\ad_1} D_{\a_1}-2\ri (s-1) \pa_{\a_1 \ad_1} \Big)
\bar \S_{\a_2 \dots \a_{s-1} \ad_2 \dots \ad_{s} }\non \\
&&+ \frac{s-1}{8s} \Big( \S^{\a(s-1) \ad(s-2) } D^2 \S_{\a(s-1) \ad(s-2)} 
- \bar \S^{\a(s-2) \ad(s-1) }\bar D^2 \bar \S_{\a(s-2) \ad(s-1)} \Big)
\non \\
&& \left.
- \frac{1}{s^2}\bar \S^{\a(s-2) \ad(s-2)\bd } \Big( \hf (s^2 +1) D^\b \bar D_\bd 
+\ri  {(s-1)^2} \pa^\b{}_\bd \Big) \S_{\b \a(s-2) \ad(s-2)} \right\}
\label{action}
\eea
is invariant under the gauge transformations \eqref{a5}. By construction, the action is also invariant under \eqref{a4}. It should be pointed out that in constructing the new action \eqref{action}, one has to check explicitly its invariance under the $\z$-gauge transformation \eqref{a5a}. In contrast, the original action \eqref{i-l} is formulated in terms of $H_{\a(s-1) \ad(s-1)}$ and the field strength $G_{\a(s) \ad(s)}$. The latter is manifestly invariant under the $\z$-transformation.

It is important to keep in mind the following identity, which is used quite often in proving the gauge-invariance (and also for other higher-spin calculations in this thesis)
\bea
D_{\b} U_{\a_1 \dots \a_m \ad_1 \dots \ad_n} &=& D_{(\b} U_{\a_1 \dots \a_m)\ad_1 \dots \ad_n} + \frac{1}{m+1} \sum_{k=1}^{m} \ve_{\b \a_k} D^{\g} U_{|\g| \a_1 \dots \hat{\a_k} \dots \a_{m} \ad_1 \dots \ad_n} \non\\
&=& D_{(\b} U_{\a_1 \dots \a_m)\ad_1 \dots \ad_n} + \frac{m}{m+1} \ve_{\b \a_1} D^{\g} U_{|\g| \a_2 \dots \a_{m} \ad_1 \dots \ad_n}
\eea
The reader is referred to appendix \ref{AppA} for the symmetrisation convention used in this work.

The $\frak V$-gauge freedom \eqref{a5} may be used to impose the condition 
\bea
 \S_{\a(s-1) \ad(s-2) }=0~.
\label{a9}
\eea
In this gauge, the action \eqref{action} reduces to 
that describing the {\it longitudinal} formulation for the massless superspin-$s$ multiplet  \eqref{i-l}. The gauge condition \eqref{a9} does not 
fix completely the $\frak V$-gauge freedom. There remains a residual gauge transformation generated by 
\bea
{\frak V}_{\a(s-1) \ad(s-1)} = D^\b L_{(\b \a_1 \dots \a_{s-1}) \ad(s-1)}~,
\label{a99}
\eea
with the parameter $L_{\a(s) \ad(s-1)}$ being an unconstrained superfield. 
With this expression for ${\frak V}_{\a(s-1) \ad(s-1)}$, the gauge transformations \eqref{a5a}  and \eqref{a5b} coincide with those given in \eqref{2226}.
Our consideration implies that the action \eqref{action} indeed provides an off-shell formulation for the massless superspin-$s$ multiplet.

Instead of choosing the condition \eqref{a99},
one can impose an alternative gauge fixing 
\bea
H_{\a(s-1) \ad(s-1)} =0~.
\label{a10}
\eea
In accordance with \eqref{a5b}, in this gauge
the residual gauge freedom is described by 
\bea
{\frak V}_{\a(s-1) \ad(s-1)} = \ri {\frak R}_{\a(s-1) \ad(s-1)}~, \qquad 
\bar{\frak R}_{\a(s-1) \ad(s-1)}={\frak R}_{\a(s-1) \ad(s-1)}~.
\eea

The action \eqref{action} includes a single term which involves the `naked' 
gauge field $\J_{\a(s)\ad(s-1)} $ and not the field strength $G_{\a(s)\ad(s)} $.
This is actually a BF term, for it can be written in two different forms
\bea
\frac{1}{s}  \int  \rd^4x \rd^2 \q  \rd^2 \bar \q
 \,
 \J^{\a(s) \ad(s-1)} \Big( 
D_{\a_1} \bar D_{\ad_1} &-&2\ri (s-1) \pa_{\a_1 \ad_1} \Big)
\S_{\a_2 \dots \a_s \ad_2 \dots \ad_{s-1} } \non \\
=- \frac{1}{s+1}  \int 
 \rd^4x \rd^2 \q  \rd^2 \bar \q
\,
 G^{\a(s) \ad(s)} \Big( \bar D_{\ad_1}D_{\a_1}  
&+&2\ri (s+1) \pa_{\a_1 \ad_1} \Big)
Z_{\a_2 \dots \a_s \ad_2 \dots \ad_{s} }~.
\label{a12}
\eea
The former makes the $\xi$-gauge symmetry \eqref{a4} manifestly realised, 
while the latter
turns the $\z$-transformation \eqref{a5a} into a manifest symmetry.
Making use of \eqref{a12} allows us to write the action \eqref{action} in the following form:
\bea
S^{\|}_{(s)} &=&
\Big( - \frac{1}{2}\Big)^s  \int 
 \rd^4x \rd^2 \q  \rd^2 \bar \q
\,
\left\{ \frac{1}{8} H^{ \a (s-1) \ad (s-1) }  D^\b {\bar D}^2 D_\b 
H_{\a (s-1) \ad (s-1)} \right. \non \\
&&+ \frac{s}{s+1}H^{ \a(s-1) \ad(s-1) }
\Big( D^{\b}  {\bar D}^{\bd} G_{\b\a(s-1) \bd\ad(s-1) }
- {\bar D}^{\bd}  D^{\b} 
{\bar G}_{\b \a (s-1) \bd \ad (s-1) } \Big) \non \\
&&+ 2 \bar G^{ \a (s) \ad (s) } G_{ \a (s) \ad (s) } 
+ \frac{s}{s+1}\Big( G^{ \a (s) \ad (s) } G_{ \a (s) \ad (s) } 
+ \bar G^{ \a (s) \ad (s) }  \bar G_{ \a (s) \ad (s) } 
 \Big) \non \\
 &&+ \frac{s-1}{4s}H^{ \a(s-1) \ad(s-1) }
\Big( D_{\a_1} \bar D^2 \bar \S_{\a_2 \dots \a_{s-1}\ad(s-1)}
 - {\bar D}_{\ad_1}  D^2 \S_{\a(s-1) \ad_2 \dots \ad_{s-1} } \Big)  \non \\
&&
- \frac{1}{s+1}  G^{\a(s) \ad(s)} \Big( \bar D_{\ad_1}D_{\a_1}  
+2\ri (s+1) \pa_{\a_1 \ad_1} \Big)
Z_{\a_2 \dots \a_s \ad_2 \dots \ad_{s} }
\non  \\
&&
+ \frac{1}{s+1}  \bar G^{\a(s) \ad(s)} \Big(  D_{\a_1}\bar D_{\ad_1}  
+2\ri (s+1) \pa_{\a_1 \ad_1} \Big)
\bar Z_{\a_2 \dots \a_s \ad_2 \dots \ad_{s} }
\non \\
&&+ \frac{s-1}{8s} \Big( \S^{\a(s-1) \ad(s-2) } D^2 \S_{\a(s-1) \ad(s-2)} 
- \bar \S^{\a(s-2) \ad(s-1) }\bar D^2 \bar \S_{\a(s-2) \ad(s-1)} \Big)
\non \\
&& \left.
- \frac{1}{s^2}\bar \S^{\a(s-2) \ad(s-2)\bd } \Big( \hf (s^2+1)D^\b \bar D_\bd 
+\ri  {(s-1)^2} \pa^\b{}_\bd \Big) \S_{\b \a(s-2) \ad(s-2)} \right\}~.~~~
\label{action2}
\eea

\subsection{Dual formulation}
The theory with action \eqref{action2} possesses a dual formulation 
that can be obtained by applying the duality transformation described in subsection \ref{2212}. 
We now associate with our theory \eqref{action2} 
the following first-order action
\bea
S_{\text{first-order} }&=&S^{\|}_{(s)}[U, \bar U, H , Z, \bar Z]  \non \\
&&+ \Big( \frac{-1}{2} \Big)^s\int  \rd^4x \rd^2 \q  \rd^2 \bar \q\,
 \Big(\frac{2}{s+1} \G^{\a(s) \ad(s)} U_{\a(s) \ad(s)} 
  +{\rm c.c.} \Big)~,~~
\label{action3}
\eea
where $S^{\|}_{(s)}[U, \bar U, H , Z, \bar Z] $ is obtained from the action
\eqref{action2} by replacing $G_{\a(s) \ad(s)} $ with an \textit{unconstrained} 
complex superfield $U_{\a(s) \ad(s)} $. The Lagrange multiplier $\G_{\a(s) \ad(s)} $
is a transverse linear superfield, 
\bea
\bar D^\bd \G_{\a(s) \bd \ad_1 \dots \ad_{s-1}} =0 ~.
\eea
The specific normalisation of the Lagrange multiplier in \eqref{action3} is chosen to match that of \cite{KS94,KS}.

The first-order model introduced 
is equivalent to the original theory 
 \eqref{action2}, which can be seen by varying $S_{\text{first-order}}$ with respect to the Lagrange multiplier. The action  \eqref{action3} is invariant under the gauge $\x$-transformation 
 \eqref{a4} which acts on $U_{\a (s) \ad (s)}$ and
 $\G_{\a(s) \ad(s)}$ by the rule
\begin{subequations}\label{b7}
 \bea
 \d_\x U_{\a (s) \ad (s)} &=&0~,\\
 \d_\x \G_{\a(s) \ad(s)} &=& \bar D^\bd \Big\{  \frac{s+1}{2(s+2)}
\bar D_{(\bd} D_{(\a_1} \x_{\a_2 \dots \a_s) \ad_1 \dots \ad_s) } 
+\ri (s+1)\pa_{(\a_1 (\bd } \x_{\a_2 \dots \a_s) \ad_1 \dots \ad_s) } \Big\}~.~~~~~~~~~~~
\eea
\end{subequations}
The first-order action  \eqref{action3} is also invariant under the gauge $\frak V$-transformation \eqref{a5b} and \eqref{a5c}, which acts on $U_{\a (s) \ad (s)}$ and
$\G_{\a(s) \ad(s)} $ as
\begin{subequations}
\bea
 \d_{ {\frak V} } U_{\a (s) \ad (s)} 
 &=& \hf \bar D_{(\ad_1} D_{(\a_1}  {\frak V}_{\a_2 \dots \a_s)\ad_2 \dots \ad_{s})}~, \\
\d_{\frak V} \G_{\a(s) \ad(s)} &=&0~.
\eea 
\end{subequations}

On the other hand, eliminating the auxiliary superfields  $U_{\a(s) \ad(s)} $ and  $\bar U_{\a(s) \ad(s)} $ 
from \eqref{action3} using their equations of motion leads to 
\bea
S^{\perp}_{(s)} &=& - \Big( - \hf \Big)^s 
  \int  \rd^4x \rd^2 \q  \rd^2 \bar \q\,
\Bigg\{ - \frac{1}{8} H^{\a(s-1) \ad(s-1)} D^\b \bar D^2  D_\b H_{\a(s-1)\ad(s-1)} \non\\
&&+ \frac{1}{8} \frac{s^2}{(s+1)(2s+1)} [D^\b, \bar D^\bd] H^{\a(s-1)\ad(s-1)} 
[D_{(\b}, \bar D_{(\bd}] H_{\a(s-1))\ad(s-1))} \non\\
&&+\hf \frac{s^2}{s+1} \pa^{\b\bd} H^{\a(s-1)\ad(s-1)} \pa_{(\b(\bd} H_{\a(s-1))\ad(s-1))}
 \non\\
&&+ \frac{2 \ri s}{2s+1} H^{\a(s-1)\ad(s-1)} \pa^{\b\bd} 
\Big({\bm \G}_{\b\a(s-1)\bd\ad(s-1)} 
- \bar{\bm \G}_{\b\a(s-1)\bd\ad(s-1)}\Big) \non\\
&& 
+ \frac{2}{2s+1} \bar{\bm \G}^{ \a (s) \ad (s) } {\bm \G}_{ \a (s) \ad (s) } 
- \frac{s}{(s+1)(2s+1)} \Big({\bm \G}^{ \a (s) \ad (s) }  {\bm \G}_{ \a (s) \ad (s) } 
+ \bar{\bm \G}^{ \a (s) \ad (s) } \bar{\bm \G}_{ \a (s) \ad (s) }\Big) \non\\
 &&- \frac{s-1}{2(2s+1)} H^{ \a(s-1) \ad(s-1) }
\Big( D_{\a_1} \bar D^2 \bar \S_{\a_2 \dots \a_{s-1}\ad(s-1)}
 - {\bar D}_{\ad_1}  D^2 \S_{\a(s-1) \ad_2 \dots \ad_{s-1} } \Big)  \non \\
&&+ \frac{1}{2(2s+1)} H^{ \a(s-1) \ad(s-1) }
\Big(  D^2 {\bar D}_{\ad_1} \S_{\a(s-1) \ad_2 \dots \ad_{s-1} }
 - \bar D^2 D_{\a_1} \bar \S_{\a_2 \dots \a_{s-1}\ad(s-1)} \Big)  \non \\
&&- \ri \frac{(s-1)^2}{s(2s+1)} H^{ \a(s-1) \ad(s-1) }
 \pa_{\a_1 \ad_1} \Big( D^\b \S_{ \b \a_2 \dots \a_{s-1} \ad_2 \dots \ad_{s-1}}
+ \bar D^\bd \bar \S_{ \a_2 \dots \a_{s-1} \bd \ad_2 \dots \ad_{s-1}} \Big)  \non \\
&&- \frac{s-1}{8s} \Big( \S^{\a(s-1) \ad(s-2) } D^2 \S_{\a(s-1) \ad(s-2)} 
- \bar \S^{\a(s-2) \ad(s-1) }\bar D^2 \bar \S_{\a(s-2) \ad(s-1)} \Big)
\non \\
&&
+ \frac{1}{s^2}\bar \S^{\a(s-2) \ad(s-2)\bd } \Big( \hf (s^2 +1) D^\b \bar D_\bd 
+\ri  {(s-1)^2} \pa^\b{}_\bd \Big) \S_{\b \a(s-2) \ad(s-2)}
\Bigg\} ~,
\label{action4}
\eea
 where we have defined
\bea
{\bm \G}_{ \a (s) \ad (s) } = \G_{ \a (s) \ad (s) }
-\hf \bar D_{(\ad_1} D_{(\a_1} Z_{\a_2 \dots \a_s) \ad_2 \dots \ad_s) } 
-\ri (s+1)\pa_{(\a_1 (\ad_1 } Z_{\a_2 \dots \a_s) \ad_2 \dots \ad_s) } ~.~~~
\label{shifted}
\eea
 
In accordance with \eqref{a5c}, the $\frak V$-gauge freedom may be used to
impose the condition 
\bea
Z_{\a(s-1) \ad(s-1)} =0~.
\label{b11}
\eea
In this gauge the action \eqref{action4} reduces to the one defining 
the {\it transverse} formulation for the massless superspin-$s$ multiplet, eq.~\eqref{i-t}.
The gauge condition \eqref{b11} is preserved by residual local 
$\frak V$- and $\x$-transformations of the form 
\bea
  \bar D^\bd \x_{\a(s-1) \bd \ad (s-1 )}  +
  \bar {\frak V}_{\a(s-1)\ad (s-1 )} =0~.
\eea
 Making use of the parametrisation \eqref{a99}, the residual gauge freedom is
\begin{subequations}
\bea
\d H_{\a(s-1)\ad(s-1)} &=& D^\b L_{\b \a(s-1) \ad(s-1)} - \bar D^\bd \bar{L}_{\a(s-1)\bd\ad(s-1)} \ ,\\
\d \G_{\a(s) \ad(s)} &=& \frac{s+1}{2(s+2)} \bar D^\bd \Big\{ 
\bar D_{(\bd} D_{(\a_1} 
+ 2\ri (s+2) \pa_{(\a_1(\bd}\Big\}
 \bar{L}_{\a_2 \dots \a_{s})\ad_1 \dots \ad_{s})} ~,~~~~
\eea
\end{subequations}
which is exactly the gauge symmetry of the transverse formulation for the massless superspin-$s$ multiplet, eq.~\eqref{trgaugeint}.

We note that the action \eqref{action} involves the transverse linear compensator 
$\S_{\a(s-1) \ad (s-2) }$ and its conjugate $\bar \S_{\a (s-2) \ad(s-1)}$. 
These superfields cannot be dualised into a longitudinal linear supermultiplet
without destroying the locality of the theory, for the action  \eqref{action} 
contains terms with derivatives of 
$\S_{\a(s-1) \ad (s-2) }$ and $\bar \S_{\a (s-2) \ad(s-1)}$. 

\subsection{Massless gravitino multiplet}
The original longitudinal and transverse actions for the 
massless superspin-$s$ multiplet, given by \eqref{i-l} and \eqref{i-t} respectively, are well defined 
for $s=1$, in which case they describe two off-shell formulations for the 
massless gravitino multiplet. However, the new action functional \eqref{action} is not defined in the $s=1$ case. The point is that 
the gauge transformation law \eqref{a5a} is not defined for $s=1$.
Instead, one should replace the gauge transformation \eqref{a5a} with 
\begin{subequations}
\bea
\d \J_\a = \hf D_\a {\frak V} + \z_\a ~, \qquad \bar D_\bd \z_\a =0~,
\label{3124a}
\eea
in accordance with the superconformal gravitino model \cite{KMT}.
This transformation law of $\J_\a$ coincides with the one occurring in the 
off-shell model for the massless gravitino multiplet proposed in \cite{GS80}.
In addition to the gauge superfield $\J_\a$, 
this model also involves two compensators: a real scalar $H$ and a chiral scalar $\F$, 
$\bar D_\ad \F=0$, with the gauge transformations
\bea 
\d H &=& {\frak V} + \bar {\frak V} ~, \label{3124b}\\
\d \F&= & -\hf \bar D^2 \bar {\frak V} ~.
\eea
\end{subequations}
The gauge-invariant action of \cite{GS80} takes the form (see also \cite{Ideas} for a review):
\bea
S^{\rm (I)}_{\rm GM}= 
S^{\|}_{(1, \frac 32 )} [\J, \bar \J, H] - \hf \int  \rd^4x \rd^2 \q  \rd^2 \bar \q\,
\Big( \bar \F \F +\F D^\a \J_\a +\bar \F \bar D_\ad \bar \J^\ad \Big)~,
\label{6.4}
\eea
where $S^{\|}_{(1, \frac 32 )} [\J, \bar \J, H] $ denotes the longitudinal action for the gravitino multiplet, 
which is obtained from \eqref{action} by choosing the gauge \eqref{a9}
and setting $s=1$. At the component level, this model corresponds to the Fradkin-Vasiliev-de Wit-van Holten formulation for the gravitino multiplet \cite{FV79,deWvanH}.

There exists a dual formulation for \eqref{6.4}. This is  obtained by performing a superfield Legendre transformation \cite{LR2}, which gives the dual action \cite{LR2} 
\bea
S^{\rm (II)}_{\rm GM}= 
S^{\|}_{(1, \frac 32 )} [\J, \bar \J, H] +\frac 14 \int  \rd^4x \rd^2 \q  \rd^2 \bar \q\,
\Big( G + D^\a \J_\a + \bar D_\ad \bar \J^\ad \Big)^2~,
\eea
where $G=\bar G$ is a real linear superfield, $\bar D^2 G = D^2G=0$.
The gauge freedom in this theory is given by eqs. \eqref{3124a}, \eqref{3124b} and 
\bea
\d G = -D^{\a}\z_{\a} - \bar D_{\ad} \bar \z^{\ad}~.
\eea 
It may be used 
to impose two conditions $H =0$ and $G=0$. We then end up 
with 
the Ogievetsky-Sokatchev formulation for the gravitino multiplet 
\cite{OS} (see section 6.9.5 of the book \cite{Ideas} for the technical details).

There exists one more dual formulation  for \eqref{6.4} that is  obtained
by performing the complex linear-chiral duality transformation. 
 It leads to 
\bea
S^{\rm (III)}_{\rm GM}= 
S^{\|}_{(1, \frac 32 )} [\J, \bar \J, H] +\frac 12 \int  \rd^4x \rd^2 \q  \rd^2 \bar \q\,
( \S + D^\a \J_\a)(\bar \S  + \bar D_\ad \bar \J^\ad )~,
\label{3128}
\eea
where $\S$ is a complex linear superfield constrained by $\bar D^2 \S=0$. The gauge freedom in this theory is given by eqs. \eqref{3124a}, \eqref{3124b} and
\bea
\d \S = - D^{\a} \z_{\a}~.
\eea
This gauge freedom does not allow one to gauge away $\S$ off the mass shell. To the best of our knowledge, the supersymmetric gauge theory  \eqref{3128} is a new off-shell realisation for the massless gravitino multiplet. 

Let us also remark that all the constructions considered in this section can naturally be lifted to the case of anti-de Sitter supersymmetry to extend the results of \cite{KS94}. This will be studied in chapter \ref{ch4}.

\section{Higher-spin multiplets of conserved currents}
This section is devoted to the study of non-conformal higher-spin supercurrent multiplets in Minkowski space, as an extension of the superconformal case which was first described in \cite{HST} and further elaborated in \cite{KMT}. Our approach will be a higher-spin extension of that used in subsection \ref{SCflat} to derive consistent supercurrents associated with a linearised off-shell formulation for ${\cN}=1$ Poincar\'e supergravity. Here we will demonstrate that the off-shell actions for the massless half-integer superspin multiplet described in section \ref{s22}, along with the new integer superspin action \eqref{action2}, allow us to formulate ${\cN}=1$ non-conformal higher-spin supercurrents in 4D Minkowski space \cite{HK1, HK2}. 
\subsection{Non-conformal supercurrents: Half-integer superspin} \label{ss321}
Let us proceed with the massless half-integer superspin case and derive the current multiplet corresponding to the longitudinal formulation \eqref{hi-l}. The first step is to add a source (or coupling) term to the action $S^{\|}_{(s+\hf)}[H,G,\bar G]$, eq.~\eqref{hi-l}
\bea
S^{\|}_{(s+\hf)}[H,G,\bar G]&-&\int \rd^4x \rd^2 \q  \rd^2 \bar \q \, \Big\{ 
H^{ \a (s) \ad (s) } J_{ \a (s) \ad (s) } \non\\
&&+ \Big(\J^{ \a (s-1) \ad (s-2) } T_{ \a (s-1) \ad (s-2) } + {\rm c.c.} \Big) 
 \Big\}~.~~~~~~~~
\label{hic1}
\eea
Next, requiring the above to be invariant under the $\z$-transformation \eqref{lon-prep-gauge1} 
\bea
\d_\z  \Psi_{ \a(s-1) \, \ad {(s-2}) } &=&  {\bar D}_{( \ad_1 }
 \z_{ \a(s-1) \, \ad_2 \cdots \ad_{s-2}) } ~,
\non
\eea
implies that  $T_{ \a (s-1) \ad (s-2) } $ is a transverse linear superfield,
\bea
\bar D^\bd T_{\a(s-1) \bd \ad_1 \dots \ad_{s-3}} =0~.
\label{hic2}
\eea
The action \eqref{hic1} should also respect the $\L$-gauge freedom given in
(\ref{longaugeH}) and (\ref{longaugeJ}): 
\bea 
\d_\L H_{\a_1 \dots \a_s \ad_1  \dots \ad_s} 
&= &\bar D_{(\ad_1} \L_{\a_1\dots  \a_s \ad_2 \dots \ad_s )} 
- D_{(\a_1} \bar{\L}_{\a_2 \dots \a_s)\ad_1  \dots \ad_s}~,\non\\
\d_\L \J_{\a_1 \dots \a_{s-1}\ad_1 \dots \ad_{s-2}} &= & - \hf 
\Big( \bar D^{\bd} D^\b -2\ri (s-1) \pa^{\b \bd} \Big)
\L_{\b\a_1 \dots \a_{s-1} \bd \ad_1 \dots \ad_{s-2}} ~.
\non
\eea
This demands the sources to satisfy the following conservation equation
\bsubeq
\bea
\bar D^\bd J_{\a_1 \dots \a_s \bd \ad_1 \dots \ad_{s-1}} 
+\hf \Big( D_{(\a_1} \bar D_{(\ad_1}
-2\ri (s-1) \pa_{ (\a_1 (\ad_1 } \Big)  T_{\a_2\dots \a_s) \ad_2 \dots \ad_{s-1})} =0~.
\label{hic3a}
\eea
For completeness, we also give the conjugate equation
\bea
D^\b J_{\b \a_1 \dots \a_{s-1}  \ad_1 \dots \ad_{s}} 
-\hf \Big( \bar D_{(\ad_1}  D_{(\a_1}
-2\ri (s-1) \pa_{ (\a_1 (\ad_1 } \Big)  \bar T_{\a_2\dots \a_{s-1}) \ad_2 \dots \ad_{s})} 
=0~. \label{hic3c}
\eea
\end{subequations}

Associated with the transverse model \eqref{hi-t} is the following non-conformal supercurrent multiplet
\begin{subequations}
\bea
\bar D^\bd J_{\a_1 \dots \a_s \bd \ad_1 \dots \ad_{s-1}} 
-\frac{1}{4} \bar D^2 F_{\a_1 \dots \a_s \ad_1 \dots \a_{s-1}} &=&0~,\\
D_{(\a_1 } F_{\a_2 \dots \a_{s+1} )\ad_1 \dots \a_{s-1}}&=&0  ~.
\eea
\end{subequations}
Thus  the trace multiplet $\bar F_{\a(s-1) \ad(s)}$ is longitudinal linear. 


When working with higher-spin supercurrents, it proves to be convenient to make use of a condensed notation. Let us introduce 
auxiliary commuting complex variables $\z^\a \in {\mathbb C}^2$ and their conjugates 
$\bar \z^\ad$. Given a tensor superfield $U_{\a(p) \ad(q)}$, we associate with it 
the following index-free field on ${\mathbb C}^2$ 
\bea
U_{(p,q)} (\z, \bar \z):= \z^{\a_1} \dots \z^{\a_p} \bar \z^{\ad_1} \dots \bar \z^{\ad_q}
U_{\a_1 \dots \a_p \ad_1 \dots \ad_q}~,
\eea
which is a homogeneous polynomial of degree $(p,q)$ in $\z^\a$ and $\bar \z^\ad$.
Furthermore, we make use of the bosonic variables $(\z^\a, \bar \z^\ad)$ and their corresponding partial derivatives $({\pa}/{\pa \z^\a}, \pa /{ \pa \bar \z^\ad})$ to convert the spinor and vector covariant derivatives into index-free operators. 
We introduce operators that  increase the degree 
of homogeneity in $\z^\a$ and $\bar \z^\ad$: 
\bea
{D}_{(1,0)} := \z^\a D_\a~, \qquad 
{\bar D}_{(0,1)} := \bar \z^\ad \bar D_\ad~, 
\qquad 
{\pa}_{(1,1)} := 2\ri \z^\a \bar \z^\ad \pa_{\a\ad}~, \label{326S}
\eea
and their descendants 
\bea
A_{(1,1)} := -D_{(1,0)} \bar D_{(0,1)} +(s-1) \pa_{(1,1)} ~, \quad 
\bar A_{(1,1)} := \bar D_{(0,1)}  D_{(1,0)} -(s-1) \pa_{(1,1)} ~. 
\eea
The fermionic operators ${D}_{(1,0)} $ and 
${\bar D}_{(0,1)} $ are \textit{nilpotent}, ${D}_{(1,0)}^2=0 $ and 
${\bar D}_{(0,1)}^2=0 $~.
Additionally, we also have the following {\it nilpotent} operators which decrease the degree 
of homogeneity in $\z^\a$ and $\bar \z^\ad$:
\begin{subequations}
\bea
D_{(-1,0)} &:=& D^\a \frac{\pa}{\pa \z^\a}~, \qquad D_{(-1,0)}^2 =0~,\\
\bar D_{(0,-1)}& :=& \bar D^\ad \frac{\pa}{\pa \bar \z^\ad}~ \qquad 
\bar D_{(0,-1)}^2 =0 
~.
\eea
\end{subequations}

Making use of the above notation, 
the transverse linearity condition \eqref{hic2} and its conjugate become
\begin{subequations} \label{hicc1}
\bea
\bar D_{(0,-1)} T_{(s-1,s-2)} &=&0~,  \label{hic8a}\\
D_{(-1,0)} \bar T_{(s-2,s-1)} &=&0~.
\eea
\end{subequations}
The conservation equations \eqref{hic3a} and \eqref{hic3c} turn into 
\begin{subequations} \label{hicc2}
\bea
\bar D_{(0,-1)} J_{(s,s)} -\hf A_{(1,1)} T_{(s-1, s-2)}&=&0~, \label{hic9a}\\
D_{(-1,0)} J_{(s,s)} -\hf \bar A_{(1,1)} \bar T_{(s-2, s-1)}&=&0~.
\eea
\end{subequations}
Since the operator $\bar D_{(0,-1)} J_{(s,s)} $ is nilpotent, 
the conservation equation \eqref{hic9a} is consistent provided
\bea
\bar D_{(0,-1)}  A_{(1,1)} T_{(s-1, s-2)}=0~.
\eea
This is indeed true, as a consequence of
\eqref{hic8a}. 

\subsubsection{Examples of higher-spin supercurrents} \label{ss3211}
Consider a free massless chiral scalar superfield $\F$ with the action
\bea \label{flatchiral}
S = \int \rd^4x \rd^2 \q  \rd^2 \bar \q \, \bar \F \F ~,\quad \bar D_\ad \F =0~.
\eea
The conserved higher-spin supercurrent multiplet associated to the model \eqref{flatchiral} was first constructed in \cite{KMT}. It is 
\bea
J_{\a(s) \ad (s)} &=&   (2\ri)^{s-1} \sum_{k=0}^{s} (-1)^k \binom{s}{k}\non \\
&&\times  \left\{ \binom{s}{k+1} 
 \pa_{(\a_1 ( \ad_1} \dots \pa_{\a_k \ad_k} D_{\a_{k+1}} \F \,
\bar D_{\ad_{k+1} } \pa_{\a_{k+2} \ad_{k+2} }\dots \pa_{ \a_s) \ad_s  ) } \bar \F 
\right.\non \\
&& \qquad \left.+ 2\ri \binom{s}{k} 
\pa_{(\a_1  (\ad_1} \dots \pa_{\a_k \ad_k}  \F \,
 \pa_{\a_{k+1} \ad_{k+1} }\dots \pa_{ \a_s) \ad_s  ) } \bar \F \right\}~.
\eea
Using our notation, it reads
\bea
J_{(s,s)} &=& \sum_{k=0}^s (-1)^k
\binom{s}{k}
\left\{ \binom{s}{k+1} 
{\pa}^k_{(1,1)}
 D_{(1,0)} \F \,
{\pa}^{s-k-1}_{(1,1)}
\bar D_{(0,1)} 
\bar \F  
\right. \non \\ 
&& \left.
 \qquad \qquad
+ \binom{s}{k} 
{\pa}^k_{(1,1)}
  \F \,
{\pa}^{s-k}_{(1,1)}
\bar \F \right\}~,	
\label{hic12}
\eea
which obeys the conservation equations on-shell
\bea \label{conf-CE}
D_{(-1,0)} J_{(s,s)} = 0 \quad \Longleftrightarrow \quad 
\bar D_{(0,-1)} J_{(s,s)} = 0~. 
\eea
It is useful to understand the construction of the conformal higher-spin supercurrent \eqref{hic12}. For this we need to discuss a few important notions of ${\cN}=1$ superconformal multiplet following the presentation of \cite{KMT}. 

A tensor superfield $T$ (with suppressed indices) is called superconformal primary of weight $(p,q)$ if it transforms as
\bea
\d_{\x} T = \Big( \x + \hf \o^{bc}[\x] M_{bc}\Big)T + \big( p \sigma[\x] + q \bar \sigma[\x] \big)T~,
\eea
for some parameters $p$ and $q$. Here $\x = \x^A D_A$ is the ${\cN}=1$ conformal Killing real supervector field generating superconformal transformations in Minkowski space. The superfields $\o^{bc}[\x]$ and $\sigma[\x]$ denote some local Lorentz and super-Weyl parameters, respectively. The dimension of $T$ is $(p+q)$ and its $R$-symmetry charge is proportional to $(p-q)$. If $T$ is chiral, $q=0$, and we say that $T$ is superconformal primary of dimension $p$.

For example, by requiring that both $H_{\a(s) \ad(s)}$ and the gauge parameter $\L_{\a(s) \ad(s-1)}$ in \eqref{longaugeH} to be superconformal primary, the superconformal transformation law for $H_{\a(s) \ad(s)}$ can be derived \cite{KMT}. It is
\bea \label{sctlawH}
\d_{\x} H_{\a(s) \ad(s) } =  \Big( \x + \hf \o^{bc}[\x] M_{bc}\Big) H_{\a(s) \ad(s)}- \frac{s}{2} \big(  \sigma[\x] +  \bar \sigma[\x] \big)H_{\a(s) \ad(s)}~.
\eea

Given a real scalar $\cL$, the action functional over the full superspace,
\bea
S = \int \rd^4x \rd^2\q \rd^2 \bar\q \,\cL
\eea
is invariant under the superconformal transformations if $\cL$ is superconformal primary of weight $(1,1)$. On the other hand, the chiral action 
\bea
S = \int \rd^4x \rd^2\q  \,\cL_c~,\qquad \bar D_{\ad} \cL_c = 0
\eea
is superconformally invariant provided $\cL_c$ is superconformal primary of dimension +3. For instance, the massless model \eqref{flatchiral} is superconformal provided the chiral scalar superfield is superconformal primary of dimension +1.

Now, in order to describe the structure of $J_{\a(s) \ad(s)}$, the authors of \cite{KMT} first consider coupling of the form
\bea
S^{(s+\hf)}_{{\rm source}} = \int \rd^4x \rd^2\q \rd^2\bar\q H^{\a(s) \ad(s)}J_{\a(s) \ad(s)}
\eea
and require invariance under the superconformal transformations \eqref{sctlawH}. From \eqref{sctlawH}, one sees that $H_{\a(s) \ad(s)}$ is superconformal primary of weight $(-\frac{s}{2}, -\frac{s}{2})$, thus the real superfield $J_{\a(s) \ad(s)}$ must be of weight $(1+\frac{s}{2},\,1+\frac{s}{2} )$.
Next, the requirement of gauge-invariance under \eqref{longaugeH} leads to the conservation equations \eqref{conf-CE}. Since $\F$ is superconformal primary of dimension +1, the following ansatz for $J_{\a(s) \ad(s)}$ as composites of $\F$ and $\bar \F$ was considered \cite{KMT}:
\bea
J_{(s,s)} &=& \sum_{k=0}^s\bigg\{ a_k\,
{\pa}^k_{(1,1)}
 D_{(1,0)} \F \,
{\pa}^{s-k-1}_{(1,1)}
\bar D_{(0,1)} 
\bar \F  
+ b_k\,
{\pa}^k_{(1,1)}
  \F \,
{\pa}^{s-k}_{(1,1)}
\bar \F \bigg\}~.
\eea
The coefficients $a_k$ and $b_k$ can be fixed uniquely by imposing two conditions: (i) $J_{\a(s) \ad(s)}$ must be real; and (ii) it must obey the conservation equation \eqref{conf-CE}. 
Indeed, setting $s=1$ leads to the Ferrara-Zumino supercurrent \cite{FZ} which we reviewed in subsection \ref{SCflat}: 
\bea
J_{\a \ad} = D_{\a}\F \bar D_{\ad} \bar \F + 2 \ri (\F \pa_{\a \ad} \bar \F - \bar \F \pa_{\a \ad} \F)~.
\eea

Our aim is to construct non-conformal higher-spin supercurrent
arising in the model for a massive chiral superfield
\bea
S = \int \rd^4x \rd^2 \q  \rd^2 \bar \q \, \bar \F \F
+\Big\{ \frac{m}{2} \int \rd^4x \rd^2 \q\, \F^2 +{\rm c.c.} \Big\}~.
\label{massive-flat}
\eea
As will be demonstrated below, it is the longitudinal 
higher-spin supercurrent multiplet described by \eqref{hicc1} and \eqref{hicc2}, which naturally arises in \eqref{massive-flat}. Guided by the structure of  
the Ferrara-Zumino supercurrent for the model \eqref{massive-flat},
we assume that $J_{(s,s)} $ has the same functional form as in the massless
case, eq. \eqref{hic12}. 
We first compute the left-hand side of \eqref{hic9a} and use the massive equation of motion, $-\frac{1}{4} \bar D^2 \bar \F +m \F =0~$.
This gives
\begin{subequations}
\bea
\bar D_{(0,-1)} J_{(s,s)} &=& F_{(s,s-1)}~, \label{hic12a}
\eea
where we have denoted 
\bea
F_{(s,s-1)} &=& 2m(s+1) \sum_{k=0}^s (-1)^{s-1+k} \binom{s}{k} \binom{s}{k+1}
\non \\ 
&& 
\times \left\{1+(-1)^s \frac{k+1}{s-k+1}\right\} 
 {\pa}^k_{(1,1)} \F \,{\pa}^{s-k-1}_{(1,1)}
 D_{(1,0)} \F ~.
\eea
\end{subequations}

Let us now determine the trace multiplet $T_{(s-1,s-2)}$. For this we consider a general ansatz in the form
\bea
T_{(s-1, s-2)} = (-1)^s m \sum_{k=0}^{s-2} c_k 
{\pa}^k_{(1,1)} \F\,
{\pa}^{s-k-2}_{(1,1)}
 D_{(1,0)} \F ~.
 \label{T3.15}
\eea
This ansatz is chosen based on the following requirements: (i) $T_{(s-1,s-2)}$ must be transverse linear \eqref{hic8a}; and 
(ii) it solves the equation \eqref{hic9a},
\bea
F_{(s,s-1)} = \hf A_{(1,1)} T_{(s-1, s-2)}~. 
\eea
For $k = 1,2,...s-2$, the first condition implies that
the coefficients $c_k$ must satisfy
\begin{subequations}\label{hic16}
\begin{align}
kc_k = (s-k-1) c_{s-k-1}~.\label{hic16a}
\end{align}
Imposing condition (ii) leads to
\begin{align}
c_{s-k-1} + s c_k + (s-1) c_{k-1} &= -4(s+1)(-1)^k \binom{s}{k} \binom{s}{k+1} 
\non \\
& \qquad \qquad \times \left\{ 1+ (-1)^s \frac{k+1}{s-k+1} \right\} ~.\label{hic16b}
\end{align}
In addition, it also follows from (ii) that
\begin{align}
(s-1) c_{s-2} +c_0 &= 4(-1)^s s(s+1)\left\{1+(-1)^s \frac{s}{2}\right\}~, \label{hic16c}\\
c_0 &= -4(s+1+(-1)^s)~. \label{hic16d}
\end{align}
\end{subequations}
We find that the set of equations \eqref{hic16} 
leads to a unique expression for $c_k$,
\bea\label{hic17}
c_k &=& -\frac{4(s+1)(s-k-1)}{s-1}
\sum_{l=0}^k \frac{(-1)^k}{s-l} \binom{s}{l} \binom {s}{l+1} \left\{ 1+(-1)^s \frac{l+1}{s-l+1} \right\}  ~,~~~~  \\
&& \qquad \qquad \qquad  k=0,1,\dots s-2~. \non 
\eea
If the parameter $s$ is odd, $s=2n+1$, with  $n=1,2,\dots$, 
one can check that the equations \eqref{hic16a}--\eqref{hic16c} are identically 
satisfied. 
However, if the parameter $s$ is even, $s=2n$, with $n=1,2,\dots$, 
there appears an inconsistency: 
 the right-hand side of \eqref{hic16c} is positive, while the left-hand side 
is negative, $(s-1) c_{s-2} + c_0 < 0$. As a result, our solution \eqref{hic17} is only consistent for $s=2n+1, n=1,2,\dots$.

Relations \eqref{hic12}, \eqref{T3.15} and \eqref{hic17} determine the non-conformal higher-spin supercurrent 
in the massive chiral model \eqref{massive-flat}, with the trace multiplet
$T_{(s-1, s-2)}$
being the higher-spin extension of \eqref{1.6-1}.
Unlike the conformal higher-spin supercurrent \eqref{hic12},
the non-conformal one exists only for the odd values of $s$,
$s=2n+1$, with  $n=1,2,\dots$. The same conclusion was also reached by the authors of \cite{BGK1} who employed the superfield Noether procedure.

\subsection{Non-conformal supercurrents: Integer superspin} \label{ss322}
Having derived a new off-shell gauge formulation for the massless superspin-$s$ multiplet, we turn to describing the structure of the non-conformal higher-spin supercurrents associated to the model \eqref{action2}.

As in the half-integer superspin case, let us 
couple the prepotentials 
$H_{ \a (s-1) \ad (s-1) } $, $Z_{ \a (s-1) \ad (s-1) }$ and $\Psi_{ \a (s) \ad (s-1) } $ to external sources
\bea
S^{(s)}_{\rm source} &=& \int \rd^4x \rd^2 \q  \rd^2 \bar \q \, \Big\{ 
\Psi^{ \a (s) \ad (s-1) } J_{ \a (s) \ad (s-1) }
-\bar \Psi^{ \a (s-1) \ad (s) } \bar J_{ \a (s-1) \ad (s) }
\non \\
&&+H^{ \a (s-1) \ad (s-1) } S_{ \a (s-1) \ad (s-1) } \non \\
&&+ Z^{ \a (s-1) \ad (s-1) } T_{ \a (s-1) \ad (s-1) } 
+ \bar Z^{ \a (s-1) \ad (s-1) } \bar T_{ \a (s-1) \ad (s-1) }
 \Big\}~.
\label{ic-s}
\eea
In order for the source term $S^{(s)}_{\rm source}$ to be invariant under the $\z$-transformation 
in \eqref{a5a}, the source  $J_{ \a (s) \ad (s-1) }$ must obey
\bea
\bar D^\bd J_{\a(s) \bd \ad(s-2)} =0 \quad \Longleftrightarrow \quad
D^\b \bar J_{\b \a(s-2)  \ad(s)} =0 ~.
\label{ic2a}
\eea
Next, in order for $S^{(s)}_{\rm source}$ to be invariant under the transformation \eqref{a4}, we require 
the superfield
$T_{ \a (s-1) \ad (s-1) }$ to satisfy
\bea
\bar D_{(\ad_1} T_{\a(s-1) \ad_2 \dots \ad_{s})} =0
 \quad \Longleftrightarrow \quad
 D_{(\a_1} \bar T_{\a_2 \dots \a_{s})  \ad (s-1)} =0~.
\label{ic2b}
\eea
We see that  the superfields $J_{ \a (s) \ad (s-1) }$ and $T_{ \a (s-1) \ad (s-1) } $ are transverse linear and longitudinal linear, respectively.
Finally, requiring $S^{(s)}_{\rm source}$ to be invariant under the 
$\frak V$-transformation 
\eqref{a5} gives the following conservation equation
\begin{subequations} \label{ic4}
\bea
-\hf D^\b J_{\b \a(s-1) \ad(s-1)} 
+S_{\a(s-1) \ad(s-1)} + \bar T_{\a(s-1) \ad(s-1)} =0~. 
\label{ic2c}
\eea
and its conjugate
\bea
\hf \bar D^\bd \bar J_{ \a(s-1) \bd \ad(s-1)} 
+S_{\a(s-1) \ad(s-1)} + T_{\a(s-1) \ad(s-1)} =0~. 
\label{ic4b}
\eea
\end{subequations}

As a consequence of  \eqref{ic2b}, from \eqref{ic2c} we deduce
\bea
\frac{1}{4} D^2 J_{ \a(s) \ad(s-1)} + D_{(\a_1} S_{\a_2 \dots \a_s) \ad(s-1) } =0~.
\label{ic5}
\eea
The equations \eqref{ic2a} and \eqref{ic5} describe the conserved 
current supermultiplet which corresponds to our theory in the gauge \eqref{a9}.

Taking the sum of \eqref{ic2c} and \eqref{ic4b}
leads to
\bea
\hf D^\b J_{\b \a(s-1) \ad(s-1)} 
+\hf \bar D^\bd \bar J_{\a(s-1) \bd \ad(s-1)}
+ T_{\a(s-1) \ad(s-1)}-\bar T_{\a(s-1) \ad(s-1)} =0~. 
\label{ic3}
\eea
The equations \eqref{ic2a}, \eqref{ic2b} and \eqref{ic3} describe the conserved 
current supermultiplet which corresponds to our theory in the gauge \eqref{a10}.
As a consequence of \eqref{ic2b}, the conservation equation \eqref{ic3} 
implies
\bea
\hf D_{(\a_1} \left\{D^{|\b|} J_{\a_2 \dots \a_s ) \b\ad(s-1)} 
+ \bar D^\bd \bar J_{\a_2 \dots \a_s ) \bd \ad(s-1)}\right\}
+D_{(\a_1} T_{\a_2 \dots \a_s ) \ad(s-1)} =0~. 
\label{ic44}
\eea

Using our condensed notation,
the transverse linear condition \eqref{ic2a}  turns into 
\bea
\bar D_{(0,-1)} J_{(s,s-1)} &=& 0~,  \label{ic5a}
\eea
while the longitudinal linear condition \eqref{ic2b} takes the form
\bea
\bar D_{(0,1)} T_{(s-1,s-1)} &=& 0~. \label{ic5b}
\eea
The conservation equation \eqref{ic2c} becomes
\bea
-\frac{1}{2s} D_{(-1,0)} J_{(s,s-1)} + S_{(s-1,s-1)} + \bar T_{(s-1,s-1)} = 0~.
\label{ic6}
\eea
and \eqref{ic44} takes the form
\bea
\frac{1}{2s} D_{(1,0)} \left\{D_{(-1,0)} J_{(s,s-1)} + \bar D_{(0,-1)} \bar J_{(s-1,s)}\right\}
+D_{(1,0)} T_{(s-1,s-1)} =0~. 
\label{ic7}
\eea
\subsubsection{Examples of higher-spin supercurrents} \label{ss3221}
Let us consider the Fayet-Sohnius model \cite{Fayet,Soh-central}
for a free massive hypermultiplet
\bea
S = \int \rd^4x \rd^2 \q  \rd^2 \bar \q \, \Big( \bar \F_+ \F_+
+\bar \F_- \F_-\Big)
+\Big\{ {m} \int \rd^4x \rd^2 \q\, \F_+ \F_- +{\rm c.c.} \Big\}~,
\label{icc1}
\eea
where the superfields $\F_\pm$ are chiral, $\bar D_\ad \F_\pm =0$,
and  the mass parameter $m$ is chosen to be positive.

In the massless case, $m=0$, 
the fermionic higher-spin supercurrent $J_{\a(s) \ad(s-1)}$ 
was first constructed in 
\cite{KMT}. In our notation it reads
\bea
J_{(s,s-1)} &=& \sum_{k=0}^{s-1} (-1)^k
\binom{s-1}{k}
\left\{ \binom{s}{k+1} 
{\pa}^k_{(1,1)}
 D_{(1,0)} \F_{+} \,\,
{\pa}^{s-k-1}_{(1,1)}
 \F_{-}  
\right. \non \\ 
&& \left.
 \qquad \qquad
- \binom{s}{k} 
{\pa}^k_{(1,1)}
  \F_{+} \,\,
{\pa}^{s-k-1}_{(1,1)}
D_{(1,0)} \F_{-} \right\}~.
\label{ic8}
\eea
One may check that $J_{(s,s-1)}$ obeys, 
for $s > 1$,  the conservation equations
\bea
D_{(-1,0)} J_{(s,s-1)} = 0, \qquad
\bar D_{(0,-1)} J_{(s,s-1)} = 0 ~,~
\label{ic9}
\eea
are satisfied as a consequence of the massless equations of motion,  $D^2 \F_\pm = 0$. 

Let us construct conserved fermionic supercurrent
corresponding to  the massive model \eqref{icc1}.
Assuming that $J_{(s,s-1)}$ has the same functional form as in the massless case, 
eq. \eqref{ic8}, and making use of the equations of motion 
\bea
-\frac{1}{4} D^2 \F_+ +m \bar \F_{-} =0, \qquad
-\frac{1}{4} D^2 \F_- +m \bar \F_{+} =0,
\eea
we obtain 
\bea
D_{(-1,0)} J_{(s,s-1)} &=& 2m (s+1) \sum_{k=0}^{s-1} (-1)^{k+1} \binom{s-1}{k} \binom{s}{k} \non \\
&&\qquad \times \left\{ -\frac{s-k}{k+1} {\pa}^k_{(1,1)} \bar \F_- \,{\pa}^{s-k-1}_{(1,1)} \F_-
+ {\pa}^k_{(1,1)} \F_+ \,{\pa}^{s-k-1}_{(1,1)} \bar \F_+ \right\}\non \\
&&+ 2m(s+1) \sum_{k=1}^{s-1} (-1)^{k+1} \binom{s-1}{k} \binom{s}{k} \frac{k}{k+1} 
\non \\
&& \qquad \times {\pa}^{k-1}_{(1,1)} \bar D_{(0,1)} \bar \F_- \,{\pa}^{s-k-1}_{(1,1)} D_{(1,0)} \F_- 
\non \\
&&+ 2m(s+1)\sum_{k=0}^{s-2} (-1)^{k+1} \binom{s-1}{k} \binom{s}{k} \frac{s-1-k}{k+1} 
\non \\
&& \qquad \times {\pa}^{k}_{(1,1)} D_{(1,0)} \F_+ \,{\pa}^{s-k-2}_{(1,1)} \bar D_{(0,1)} \bar \F_+ ~.~ \label{ic10}
\eea
It can be shown that the massive supercurrent $J_{(s,s-1)}$ also obeys \eqref{ic5a}. 

As the next step, we need to construct a superfield $T_{(s-1,s-1)}$, which has the following properties: (i) it is longitudinal linear \eqref{ic5b}; and (ii) it satisfies \eqref{ic7}, which is a consequence of the conservation equation \eqref{ic6}. 
Within these conditions, our ansatz takes the form
\bea
T_{(s-1, s-1)} &=& 
\sum_{k=0}^{s-1} c_k {\pa}^k_{(1,1)} \F_-\, {\pa}^{s-k-1}_{(1,1)} \bar \F_-  \non \\
&&+ \sum_{k=0}^{s-1} d_k {\pa}^k_{(1,1)} \F_+ \,
{\pa}^{s-k-1}_{(1,1)} \bar \F_+  \non \\
&&+ \sum_{k=1}^{s-1} f_k {\pa}^{k-1}_{(1,1)} D_{(1,0)} \F_-\, {\pa}^{s-k-1}_{(1,1)} \bar D_{(0,1)} \bar \F_-  \non \\
&&+ \sum_{k=1}^{s-1} g_k {\pa}^{k-1}_{(1,1)}  D_{(1,0)} \F_+\,\, {\pa}^{s-k-1}_{(1,1)} \bar D_{(0,1)} \bar \F_+ ~.
\eea
Imposing the first condition, we find that the coefficients must be related by
\begin{subequations} 
\bea
c_0 = d_0 = 0~, \qquad f_k = c_k~, \qquad g_k = d_k~. 
\eea
On the other hand, for $k=1,2, \dots s-2$, condition  (ii) yields the following recurrence relations:
\bea 
c_k + c_{k+1} &=& 
\frac{m(s+1)}{s} (-1)^{s+k} \binom{s-1}{k} \binom{s}{k} 
\non \\
&& \times \frac{1}{(k+2)(k+1)} \Big\{(2k+2-s)(s+1)-k-2\Big\}~, \\
d_k + d_{k+1} &=& \frac{m(s+1)}{s} (-1)^{k} \binom{s-1}{k} \binom{s}{k} 
\non \\
&& \times \frac{1}{(k+2)(k+1)} \Big\{(2k+2-s)(s+1)-k-2\Big\}~.
\eea
Condition (ii) also implies that
\bea
c_1 = -(-1)^s \frac{m(s^2-1)}{2}~, \qquad c_{s-1} &=&- \frac{m(s^2-1)}{s}~;\\
d_1 =- \frac{m(s^2-1)}{2}~, \qquad d_{s-1} &=& -(-1)^s \frac{m(s^2-1)}{s}~.
\eea
\end{subequations}
The above relations lead to simple expressions for $c_k$ and $d_k$:
\begin{subequations} \label{qqq21}
\bea
d_k &=& \frac{m(s+1)}{s} \frac{k}{k+1} (-1)^{k} \binom{s-1}{k} \binom{s}{k}~,\\
c_k &=& -(-1)^s d_k ~,
\eea
\end{subequations}
where $ k=1,2,\dots s-1$.
Now that we have already derived an expression for the trace multiplet $T_{(s-1,s-1)}$, the superfield $S_{(s-1,s-1)}$ can be computed using the conservation equation \eqref{ic6}. This leads to
\bea
S_{(s-1,s-1)} &=& -m (s+1)\sum_{k=0}^{s-1} (-1)^{k+1}
\binom{s-1}{k}\binom{s}{k} \frac{1}{k+1}
 \non \\
 && \qquad \qquad
\times \left\{ 
{\pa}^k_{(1,1)}
\bar \F_{-} \,
{\pa}^{s-k-1}_{(1,1)}
 \F_{-}  
+ (-1)^s  
{\pa}^k_{(1,1)}
 \bar \F_{+} \,
{\pa}^{s-k-1}_{(1,1)}
 \F_{+} \right\}~.
\eea
One may verify that $S_{(s-1,s-1)}$ is a real superfield. 

\section{Discussion}
A novel off-shell formulation for the massless superspin-$s$ multiplet has been proposed in this chapter. In addition, we derived consistent higher-spin supercurrents associated with the off-shell gauge theories of massless $\cN = 1$ supermultiplets in Minkowski space. Several supercurrents were constructed explicitly, paying particular attention to models of free chiral scalar superfields.

Actually, the theory of a free massive chiral superfield \eqref{massive-flat} proves to possess conserved fermionic higher-spin supercurrents only for \textit{even} integer superspin $s=2,4, \dots$.
Indeed, one can extract from eq.~\eqref{ic8} (by setting $\F_{+} = \F_{-} = \F$) the following supercurrent $J_{(s,s-1)}$, which is a complex fermionic superfield:
\bea
J_{(s,s-1)} &=& \sum_{k=0}^{s-1} (-1)^k
\binom{s-1}{k}
\left\{ \binom{s}{k+1} 
{\pa}^k_{(1,1)}
 D_{(1,0)} \F \,\,
{\pa}^{s-k-1}_{(1,1)}
\F  
\right. \non \\ 
&& \left.
 \qquad \qquad
- \binom{s}{k} 
{\pa}^k_{(1,1)}
  \F \,\,
{\pa}^{s-k-1}_{(1,1)} D_{(1,0)}
\F \right\}~.
\label{3246}
\eea
The above expression can be further simplified by changing the index of summation of the second term (\textit{i.e.} let $k'=s-k-1$). We obtain
\bea
J_{(s,s-1)} &=& \sum_{k=0}^{s-1} (-1)^k
\binom{s-1}{k}
\binom{s}{k+1}  \big\{ 1+ (-1)^s \big\}
{\pa}^k_{(1,1)}
 D_{(1,0)} \F \,\,
{\pa}^{s-k-1}_{(1,1)}
\F  ~.
\eea
This implies that $J_{(s,s-1)} = 0$ if $s$ is odd. Thus, for even values of $s$, we have
\bea \label{JJint}
J_{(s,s-1)} &=& 2 \sum_{k=0}^{s-1} (-1)^k
\binom{s-1}{k}
\binom{s}{k+1}  {\pa}^k_{(1,1)}
 D_{(1,0)} \F \,\,
{\pa}^{s-k-1}_{(1,1)}
\F  ~.
\eea
The corresponding 
trace multiplet $T_{(s-1,s-1)}$ is given by
\bea
T_{(s-1,s-1)} &=& \sum_{k=0}^{s-1} c_k {\pa}^k_{(1,1)} \F \,\, {\pa}^{s-k-1}_{(1,1)} \bar \F \non \\
&&  + \sum_{k=1}^{s-1} d_k {\pa}^{k-1}_{(1,1)} {D}_{(1,0)} \F \,\, {\pa}^{s-k-1}_{(1,1)}  \bar {D}_{(0,1)} \bar \F ~,
\eea
with the coefficients $c_k$ and $d_k$ given by \eqref{qqq21}. 
It may be checked that the conservation equations
\bea
&&\bar D_{(0,-1)} J_{(s,s-1)} =0~, \qquad 
\bar D_{(1,0)} T_{(s-1,s-1)} =0~,\non \\
&&\frac{1}{2s}D_{(1,0)} \Big( D_{(-1,0)} J_{(s,s-1)} + \bar D_{(0,-1)} \bar J_{(s-1,s)} \Big) = -D_{(1,0)}T_{(s-1,s-1)}~,
\eea
are satisfied for the \textit{even} values of $s$, $s=2n$, with $n=1,2,\dots$. 
An alternative approach based on the superfield Noether procedure \cite{MSW} was recently developed in \cite{BGK1,KKvU,BGK-sigma,BGK2, BGK3,GK1} to study supercurrents and cubic vertices between various matter and massless higher-spin multiplets in 4D Minkowski superspace. 

An interesting open question is to classify all non-conformal deformations 
of the higher-spin supercurrents \eqref{ic9}, along the lines of the recent 
analysis of non-conformal $\cN=(1,0)$ supercurrents in six dimensions
\cite{KNT}.
Our results provide the setup required for developing a program to derive higher-spin supersymmetric models from quantum correlation functions,  as an extension of the non-supersymmetric approaches pursued, e.g., in \cite{Bonora1,Bonora2,Bonora3}.
Another interesting project would be to study ${\cN}=2$ supercurrents corresponding to the off-shell massless higher-spin ${\cN} = 2$ supermultiplets in 4D Minkowski space constructed in \cite{GKS1}.



\chapter{Higher-spin supercurrents in AdS space} \label{ch4}

An interesting feature of our results in the previous chapter is the existence of a selection rule for higher-spin supercurrents in ${\cN}=1$ supersymmetric theories. We recall that in the case of a massless half-integer superspin multiplet, the bosonic supercurrent $J_{\a(s) \ad(s)}$ for a massless chiral superfield is defined for \textit{all} values of $s$, while those corresponding to the massive chiral model exists only for {\it{odd}} $s$. The situation turns out to be different for the integer superspin case. For a single (massless or massive) chiral superfield, the fermionic supercurrent $J_{\a(s) \ad(s-1)}$ exists only for \textit{even} values of $s$, yet it is defined for {\it arbitrary} $s$ in the massive hypermultiplet model. It is thus natural to look for a generalisation of these flat space results to various supersymmetric theories in 4D ${\cN}=1$ AdS superspace ${\rm AdS}^{4|4}$, for instance a model of $N$ massive chiral scalar superfields with an arbitrary mass matrix. A large part of this chapter will be devoted to this analysis.

This chapter is organised as follows. In section \ref{s41} we review the general properties of transverse and longitudinal linear superfields. Novel off-shell gauge formulations for the massless integer superspin multiplet in AdS are presented in section \ref{s42}. They are shown to reduce to those proposed in \cite{KS94} upon partially fixing the gauge freedom. We also describe off-shell formulations (including a novel one) for the massless gravitino multiplet in AdS. In section \ref{s43}
we introduce higher-spin supercurrent multiplets in AdS and describe improvement 
transformations for them. Sections \ref{s44} and \ref{s45} are devoted to the explicit constructions
of higher-spin supercurrents for $N$ chiral superfields. 
Several nontrivial applications of the results obtained are given in section \ref{s46}.
 
\section{Linear superfields} \label{s41}
Before we describe superfield formulations for off-shell massless higher-spin gauge multiplets in ${\rm AdS}^{4|4}$ \cite{KS94}, it is important to first recall the notion of transverse and longitudinal superfields \cite{IS1}.
Complex tensor superfields 
$\G_{\a(m) \ad(n)} :=\G_{\a_1 \dots \a_m \ad_1 \dots \ad_n}
=\G_{(\a_1 \dots \a_m)( \ad_1 \dots \ad_n)}$ and 
$G_{\a(m) \ad(n)}$ 
are called transverse linear and longitudinal linear respectively, if the constraints\footnote{Our 4D AdS notation and two-component spinor conventions correspond to \cite{Ideas}. For concise results concerning field theories in ${\rm AdS}^{4|4}$, see subsection \ref{sumAdS}.}
\begin{subequations} \label{1.6}
\bea
&& \bar \cD^\bd \G_{ \a(m) \bd \ad(n - 1) } = 0 ~,  \qquad n \neq 0~,   \label{1.6a}
\\
&& \bar \cD_{(\ad_1} G_{\a(m)\ad_2 \dots \ad_{n+1} )} = 0  \label{1.6b}
\eea
\end{subequations}
are satisfied. 
For $n=0$ the latter constraint 
coincides with  the condition of covariant chirality, 
 $\bar \cD_\bd G_{ \a(m) } = 0$. 
 The relations \eqref{1.6} lead to the linearity conditions
\begin{subequations}
\bea
(\bar \cD^2-2(n+2)\m)\,\G_{\a(m) \ad(n)} &=& 0~, \label{1.7a}\\
(\bar \cD^2+2n\m)\,G_{\a(m) \ad(n)} &=& 0~. \label{1.7b}
\eea
\end{subequations}
The transverse condition \eqref{1.6a} is not defined for $n=0$. 
However, its corollary \eqref{1.7a} remains consistent for the choice $n=0$
and corresponds to complex linear superfields $\G_{\a(m)}$
 constrained by 
\bea
(\bar \cD^2-4\m)\,\G_{\a(m) } = 0~.
\eea
In the family of constrained superfields $\G_{\a(m)}$ introduced, 
the scalar multiplet, $m=0$, is used most often in applications. 
One can define projectors $P^{\perp}_{n}$ and $P^{||}_{n}$
on the spaces of  transverse linear and longitudinal linear superfields
respectively: 
\begin{subequations}
\bea
P^{\perp}_{n}&=& \frac{1}{4 (n+1)\m} (\bar \cD^2+2n\m) ~,\\
P^{||}_{n}&=&- \frac{1}{4 (n+1)\m} (\bar \cD^2-2(n+2)\m ) ~,
\eea
\end{subequations} 
with the properties 
\bea
\big(P^{\perp}_{n}\big)^2 =P^{\perp}_{n} ~, \quad 
\big(P^{||}_{n}\big)^2=P^{||}_{n}~,
\quad P^{\perp}_{n} P^{||}_{n}=P^{||}_{n}P^{\perp}_{n}=0~,
\quad P^{\perp}_{n} +P^{||}_{n} ={\mathbbm 1}~.
\eea

Given a complex tensor superfield $V_{\a(m)  \ad(n)} $ with $n \neq 0$, 
it can be represented
as a sum of transverse linear and longitudinal linear multiplets, 
\bea
V_{\a(m) \ad(n)} = &-& 
\frac{1}{2 \mu (n+2)} \cDB^\gd \cDB_{(\gd} V_{\a(m) \ad_1 \dots  \ad_n)} 
- \frac{1}{2 \mu (n+1)} \cDB_{(\ad_1} \cDB^{|\gd|} V_{\a(m) \ad_2 \dots \ad_{n} ) \gd} 
~ . ~~~
\eea
Choosing $V_{\a(m) \ad(n)} $ to be  transverse linear ($\G_{\a(m) \ad(n)} $)
or longitudinal linear ($G_{\a(m) \ad(n)} $), the above relation
gives
\begin{subequations}
\bea
 \G_{\a(m) \ad(n)}&=& \bar \cD^\bd 
{ \Phi}_{\a(m)\,(\bd \ad_1 \cdots \ad_{n}) } ~,
 \\
 G_{\a(m) \ad(n)} &=& {\bar \cD}_{( \ad_1 }
 \Psi_{ \a(m) \, \ad_2 \cdots \ad_{n}) } ~,
\eea
\end{subequations}
for some 
prepotentials $ \F_{\a(m) \ad(n+1)}$ and $  \J_{\a(m) \ad(n-1)}$.
The constraints \eqref{1.6} hold for unconstrained $ \F_{\a(m) \ad(n+1)}$ and $  \J_{\a(m) \ad(n-1)}$.
These prepotentials  are defined modulo gauge transformations of the form:
\begin{subequations}
\bea
\d_\x \Phi_{\a(m)\, \ad (n+1)} 
&=&  \bar \cD^\bd 
{ \x}_{\a(m)\, (\bd \ad_1 \cdots \ad_{n+1}) } ~,
\\
\d_\z  \Psi_{ \a(m) \, \ad {(n-1}) } &=&  {\bar \cD}_{( \ad_1 }
 \z_{ \a(m) \, \ad_2 \cdots \ad_{n-1}) } ~,
\eea
\end{subequations}
with the gauge parameters $ { \x}_{\a(m)\,  \ad (n+2) } $
and $ \z_{ \a(m) \, \ad (n-2)}$ being unconstrained.

\section{Massless integer superspin multiplets} \label{s42}

Let $s$ be a positive integer. The longitudinal formulation for the massless
superspin-$s$ multiplet in AdS was realised in \cite{KS94} in terms
of the following dynamical variables 
\bea
\cV^{||}_{(s)} &=& \left\{ H_{\a(s-1) \ad(s-1)}, G_{\a(s)\ad(s)}, \bar{G}_{\a(s)\ad(s)} \right\} \ . 
\eea
Here $H_{\a(s-1)  \ad(s-1)}$ is an unconstrained real superfield,  
while $G_{\a(s)\ad(s)}$ is a longitudinal linear superfield.
The latter is the field strength associated with a complex unconstrained 
prepotential $\J_{\a(s) \ad(s-1)}$,
\bea
G_{\a_1 \dots \a_s \ad_1 \dots \ad_s} := 
\bar \cD_{(\ad_1} \J_{\a_1 \dots \a_s \ad_2 \dots \ad_s)}
\quad \Longrightarrow \quad
\bar \cD_{(\ad_1} G_{\a_1 \dots \a_s \ad_2 \dots \ad_{s+1})}=0~.
\label{g2.5S}
\eea
The gauge freedom postulated in \cite{KS94} is given by 
\begin{subequations} \label{oldgaugefreedomS}
\bea
\d H_{\a(s-1)\ad(s-1)} &=& \cD^\b L_{\b \a(s-1) \ad(s-1)} - \cDB^\bd \bar{L}_{\a(s-1)\bd\ad(s-1)} \ ,
\\
\d G_{\b \a(s-1) \bd \ad(s-1)} &=& \hf \cDB_{(\bd} \cD_{(\b} \cD^{|\g|} L_{\a(s-1))\g\ad(s-1))} \ , 
\eea
\end{subequations}
where the gauge parameter is $L_{\a(s)\ad(s-1)}$ is unconstrained.

The goal of this section is to reformulate the longitudinal theory by enlarging the gauge freedom \eqref{oldgaugefreedomS}
at the cost of 
introducing a new compensating superfield, in addition to
$H_{\a(s-1)  \ad(s-1)}$, $\J_{\a(s)  \ad(s-1)}$ and $\bar\J_{\a(s-1)  \ad(s)}$.
In such a setting, the  gauge freedom of  $\J_{\a(s)  \ad(s-1)}$ 
coincides with that of  a superconformal multiplet of superspin-$s$ \cite{KMT}.
This new formulation will be an extension of the one given in \cite{HK2} (and described in section \ref{s31})
in flat superspace case.

\subsection{New formulation} \label{ss421}

We fix an integer $s \geq 2$. Our task is to derive an AdS extension of the gauge-invariant action \eqref{action} in Minkowski superspace. The geometry of ${\rm AdS}^{4|4}$ is completely determined by the covariant derivatives algebra \eqref{1.2}. To start with, we consider the following action functional, which is a minimal lift of \eqref{action} to ${\rm AdS}^{4|4}$
\bea
S^{\|}_{(s)} &=&
\Big( - \frac{1}{2}\Big)^s  \int 
 \rd^4x \rd^2 \q  \rd^2 \bar \q
\, E
\left\{ \frac{1}{8} H^{ \a (s-1) \ad (s-1) }  \cD^\b {\bar \cD}^2 \cD_\b 
H_{\a (s-1) \ad (s-1)} \right. \non \\
&&+ \frac{s}{s+1}H^{ \a(s-1) \ad(s-1) }
\Big( \cD^{\b}  {\bar \cD}^{\bd} G_{\b\a(s-1) \bd\ad(s-1) }
- {\bar \cD}^{\bd}  \cD^{\b} 
{\bar G}_{\b \a (s-1) \bd \ad (s-1) } \Big) \non \\
&&+ 2 \bar G^{ \a (s) \ad (s) } G_{ \a (s) \ad (s) } 
+ \frac{s}{s+1}\Big( G^{ \a (s) \ad (s) } G_{ \a (s) \ad (s) } 
+ \bar G^{ \a (s) \ad (s) }  \bar G_{ \a (s) \ad (s) } 
 \Big) \non \\
 &&+ \frac{s-1}{4s}H^{ \a(s-1) \ad(s-1) }
\Big( \cD_{\a_1} \bar \cD^2 \bar \S_{\a_2 \dots \a_{s-1}\ad(s-1)}
 - {\bar \cD}_{\ad_1}  \cD^2 \S_{\a(s-1) \ad_2 \dots \ad_{s-1} } \Big)  \non \\
&&+\frac{1}{s} \J^{\a(s) \ad(s-1)} \Big( 
\cD_{\a_1} \bar \cD_{\ad_1} -2\ri (s-1) \cD_{\a_1 \ad_1} \Big)
\S_{\a_2 \dots \a_s \ad_2 \dots \ad_{s-1} }\non  \\
&&+\frac{1}{s} \bar \J^{\a(s-1) \ad(s)} \Big( 
 \bar \cD_{\ad_1} \cD_{\a_1}-2\ri (s-1) \cD_{\a_1 \ad_1} \Big)
\bar \S_{\a_2 \dots \a_{s-1} \ad_2 \dots \ad_{s} }\non \\
&&+ \frac{s-1}{8s} \Big( \S^{\a(s-1) \ad(s-2) } \cD^2 \S_{\a(s-1) \ad(s-2)} 
- \bar \S^{\a(s-2) \ad(s-1) }\bar \cD^2 \bar \S_{\a(s-2) \ad(s-1)} \Big) \non \\
&&- \frac{1}{s^2}\bar \S^{\a(s-2) \ad(s-2)\bd } \Big( \hf (s^2 +1) \cD^\b \bar \cD_\bd 
+\ri  {(s-1)^2} \cD^\b{}_\bd \Big) \S_{\b \a(s-2) \ad(s-2)}\Big\}
+\dots~~~~~
\label{2.7}
\eea
In accordance with section \ref{s31}, our dynamical superfields consist of a complex unconstrained prepotential $\J_{\a(s)\ad(s-1)}$, a real superfield $H_{\a(s-1)\ad(s-1)}$ and a complex superfield $\S_{\a(s-1) \ad (s-2) }$ constrained to be transverse linear,  
\bea
\bar \cD^\bd \S_{\a(s-1) \bd \ad(s-3)} =0~.
\label{2.1}
\eea
In the $s=2$ case, 
 for which \eqref{2.1} is not defined, 
 $\S_{\a} $ is instead constrained by
 \bea
( \bar \cD^2 - 4 \m) \S_{\a } =0~.
 \label{21S}
 \eea
The constraint \eqref{2.1}, or its counterpart \eqref{21S} for $s=2$, can be solved in terms of a complex unconstrained prepotential $Z_{\a(s-1) \ad (s-1)}$,
\bea
\S_{\a(s-1) \ad (s-2)} = \bar \cD^\bd Z_{\a(s-1) (\bd \ad_1 \dots \ad_{s-2} )} ~,
\label{2.2}
\eea
which is defined modulo gauge shifts 
\bea
\d_\x Z_{\a(s-1) \ad (s-1)}=  \bar \cD^\bd \x_{\a(s-1) (\bd \ad_1 \dots \ad_{s-1} )} ~.
\label{2.3}
\eea
Here the gauge parameter  $\x_{\a(s-1) \ad (s) }$ is unconstrained. 

The gauge-invariant action in AdS is expected to differ from \eqref{2.7} by some $\m$-dependent terms. These are required to ensure invariance under the linearised gauge transformations which we postulate to be of the form
\begin{subequations} \label{2.4}
\bea
 \d_{ {\mathfrak V} ,\z} \J_{\a_1 \dots \a_s \ad_1 \dots \ad_{s-1}} 
 &=& \hf \cD_{(\a_1}  {\mathfrak V}_{\a_2 \dots \a_s)\ad_1 \dots \ad_{s-1}}
+  \bar \cD_{(\ad_1} \z_{\a_1 \dots \a_s \ad_2 \dots \ad_{s-1} )}  ~ , \label{2.4a}
\\
\d_{\mathfrak V} H_{\a(s-1) \ad(s-1)}&=& {\mathfrak V}_{\a(s-1) \ad(s-1)} +\bar {\mathfrak V}_{\a(s-1) \ad(s-1)}
~, \label{2.4b}\\
\d_{\mathfrak V} \S_{\a(s-1) \ad(s-2) }&=&  \bar \cD^\bd \bar {\mathfrak V}_{\a(s-1) \bd \ad(s-2)}
\quad \Longrightarrow \quad \d_{\mathfrak V} Z_{\a(s-1) \ad (s-1)}
=\bar  {\mathfrak V}_{\a(s-1) \ad (s-1)}~,~~~~~~~
\label{2.4c}
\eea
\end{subequations}
with unconstrained gauge parameters ${\mathfrak V}_{\a(s-1) \ad(s-1)}$ 
and $\z_{\a(s) \ad(s-2)}$. We note that the gauge freedom of $\J_{\a_1 \dots \a_s \ad_1 \dots \ad_{s-1}} $ is
chosen to coincide with that of the superconformal superspin-$s$ multiplet 
\cite{KMT}.
The longitudinal linear superfield $G_{\a(s) \ad(s)}$ defined by \eqref{g2.5S}
is invariant under the $\z$-transformation \eqref{2.4a} 
and  varies under the $\mathfrak V$-transformation as 
\bea
 \d_{ {\mathfrak V} } G_{\a_1 \dots \a_s \ad_1 \dots \ad_{s}} 
 &=& \hf \bar \cD_{(\ad_1} \cD_{(\a_1}  {\mathfrak V}_{\a_2 \dots \a_s)\ad_2 \dots \ad_{s})}~.
\eea

Let us compute the variation of \eqref{2.7} under \eqref{2.4} and iteratively add certain $\m$-dependent terms to achieve a gauge-invariant action. The following identities are derived from the covariant derivatives algebra \eqref{1.2} and prove to be useful in carrying out such calculations:
\begin{subequations} 
\label{1.4}
\bea 
\cD_\a\cD_\b
\!&=&\!\frac{1}{2}\ve_{\a\b}\cD^2-2{\bar \m}\,M_{\a\b}~,
\quad\qquad \,\,\,
{\bar \cD}_\ad{\bar \cD}_\bd
=-\frac{1}{2}\ve_{\ad\bd}{\bar \cD}^2+2\m\,{\bar M}_{\ad\bd}~,  \label{1.4a}\\
\cD_\a\cD^2
\!&=&\!4 \bar \m \,\cD^\b M_{\a\b} + 4{\bar \m}\,\cD_\a~,
\quad\qquad
\cD^2\cD_\a
=-4\bar \m \,\cD^\b M_{\a\b} - 2\bar \m \, \cD_\a~, \label{1.4b} \\
{\bar \cD}_\ad{\bar \cD}^2
\!&=&\!4 \m \,{\bar \cD}^\bd {\bar M}_{\ad\bd}+ 4\m\, \bar \cD_\ad~,
\quad\qquad
{\bar \cD}^2{\bar \cD}_\ad
=-4 \m \,{\bar \cD}^\bd {\bar M}_{\ad\bd}-2\m\, \bar \cD_\ad~,  \label{1.4c}\\
\left[\bar \cD^2, \cD_\a \right]
\!&=&\!4\rm i \cD_{\a\bd} \bar \cD^\bd +4 \m\,\cD_\a = 
4\rm i \bar \cD^\bd \cD_{\a\bd} -4 \m\,\cD_\a~,
 \label{1.4d} \\
\left[\cD^2,{\bar \cD}_\ad\right]
\!&=&\!-4\rm i \cD_{\b\ad}\cD^\b +4\bar \m\,{\bar \cD}_\ad = 
-4\rm i \cD^\b \cD_{\b\ad} -4 \bar \m\,{\bar \cD}_\ad~,
 \label{1.4e}
\eea
\end{subequations} 
where $\cD^2=\cD^\a\cD_\a$ and ${\bar \cD}^2={\bar \cD}_\ad{\bar \cD}^\ad$. 

This procedure leads to the following action in AdS, which is invariant under  \eqref{2.4} and, by construction, \eqref{2.3}:
\bea
S^{\|}_{(s)} &=&
\Big( - \frac{1}{2}\Big)^s  \int 
 \rd^4x \rd^2 \q  \rd^2 \bar \q
\, E
\left\{ \frac{1}{8} H^{ \a (s-1) \ad (s-1) }  \cD^\b ({\bar \cD}^2- 4\mu) \cD_\b 
H_{\a (s-1) \ad (s-1)} \right. \non \\
&&+ \frac{s}{s+1}H^{ \a(s-1) \ad(s-1) }
\Big( \cD^{\b}  {\bar \cD}^{\bd} G_{\b\a(s-1) \bd\ad(s-1) }
- {\bar \cD}^{\bd}  \cD^{\b} 
{\bar G}_{\b \a (s-1) \bd \ad (s-1) } \Big) \non \\
&&+ \frac{(s+1)^2}{2} \bar \mu \mu 
H^{\a (s-1) \ad (s-1)} H_{\a (s-1) \ad (s-1)} \non \\
&&+ 2 \bar G^{ \a (s) \ad (s) } G_{ \a (s) \ad (s) } 
+ \frac{s}{s+1}\Big( G^{ \a (s) \ad (s) } G_{ \a (s) \ad (s) } 
+ \bar G^{ \a (s) \ad (s) }  \bar G_{ \a (s) \ad (s) } 
 \Big) \non \\
 &&+ \frac{s-1}{4s}H^{ \a(s-1) \ad(s-1) }
\Big( \cD_{\a_1} \bar \cD^2 \bar \S_{\a_2 \dots \a_{s-1}\ad(s-1)}
 - {\bar \cD}_{\ad_1}  \cD^2 \S_{\a(s-1) \ad_2 \dots \ad_{s-1} } \Big)  \non \\
&&+\frac{1}{s} \J^{\a(s) \ad(s-1)} \Big( 
\cD_{\a_1} \bar \cD_{\ad_1} -2\ri (s-1) \cD_{\a_1 \ad_1} \Big)
\S_{\a_2 \dots \a_s \ad_2 \dots \ad_{s-1} }\non  \\
&&+\frac{1}{s} \bar \J^{\a(s-1) \ad(s)} \Big( 
 \bar \cD_{\ad_1} \cD_{\a_1}-2\ri (s-1) \cD_{\a_1 \ad_1} \Big)
\bar \S_{\a_2 \dots \a_{s-1} \ad_2 \dots \ad_{s} }\non \\
&&- \mu \frac{s^2+4s-1}{2s} H^{\a (s-1) \ad (s-1)} \cD_{\a_1} 
\bar \S_{\a_2 \dots \a_{s-1} \ad(s-1)} \non \\
&&+ \bar \mu \frac{s^2+4s-1}{2s} H^{\a (s-1) \ad (s-1)} \bar \cD_{\ad_1} 
 \S_{ \a(s-1) \ad_2 \dots \ad_{s-1}} \non \\
&&+ \frac{s-1}{8s} \Big( \S^{\a(s-1) \ad(s-2) } \cD^2 \S_{\a(s-1) \ad(s-2)} 
- \bar \S^{\a(s-2) \ad(s-1) }\bar \cD^2 \bar \S_{\a(s-2) \ad(s-1)} \Big) \non \\
&&- \frac{1}{s^2}\bar \S^{\a(s-2) \ad(s-2)\bd } \Big( \hf (s^2 +1) \cD^\b \bar \cD_\bd 
+\ri  {(s-1)^2} \cD^\b{}_\bd \Big) \S_{\b \a(s-2) \ad(s-2)} \non\\
&&+ \mu \frac{s^2+4s-1}{4s} \bar \S^{\a(s-2) \ad(s-1)} \bar \S_{\a(s-2) \ad(s-1)} \non \\
&&+ \bar \mu \frac{s^2+4s-1}{4s} \S_{\a(s-1) \ad(s-2)} \S^{\a(s-1) \ad(s-2)} \Big\}
~.~~~
\label{actionS}
\eea
The above action is real due to the identity \bea
\cD^\a (\cDB^2- 4 \mu) \cD_\a = \cDB_\ad (\cD^2 - 4 \mub) \cDB^{\ad}~.
\eea
In the limit of vanishing curvature of the AdS superspace ($\m\rightarrow 0$), we see that \eqref{actionS} reduces to \eqref{action}.

The $\mathfrak V$-gauge freedom \eqref{2.4} allows us to gauge away $\S_{\a(s-1) \ad(s-2) }$,
\bea
 \S_{\a(s-1) \ad(s-2) }=0~.
\label{2.10}
\eea
In this gauge, the action \eqref{actionS} reduces to 
that describing the longitudinal formulation for the massless superspin-$s$ multiplet  \cite{KS94}. The gauge condition \eqref{2.10} does not 
fix completely the $\mathfrak V$-gauge freedom. The residual gauge transformations  
are generated by 
\bea
{\mathfrak V}_{\a(s-1) \ad(s-1)} = \cD^\b L_{(\b \a_1 \dots \a_{s-1}) \ad(s-1)}~,
\label{2.11}
\eea
with $L_{\a(s) \ad(s-1)}$ being an unconstrained superfield. 
With this expression for ${\mathfrak V}_{\a(s-1) \ad(s-1)}$, the gauge transformations \eqref{2.4a}  and \eqref{2.4b} coincide with 
 \eqref{oldgaugefreedomS}.
Thus, the action \eqref{actionS} indeed provides an off-shell formulation for the massless superspin-$s$ multiplet in ${\rm AdS}^{4|4}$. 

Alternatively, one can impose a gauge fixing 
\bea
H_{\a(s-1) \ad(s-1)} =0~.
\label{2.12ads}
\eea
In accordance with \eqref{2.4b}, in this gauge
the residual gauge freedom is 
\bea
{\mathfrak V}_{\a(s-1) \ad(s-1)} = \ri {\mathfrak R}_{\a(s-1) \ad(s-1)}~, \qquad 
\bar{\mathfrak R}_{\a(s-1) \ad(s-1)}={\mathfrak R}_{\a(s-1) \ad(s-1)}~.
\eea

The gauge-invariant action \eqref{actionS} includes a single term which involves the prepotential $\J_{\a(s)\ad(s-1)} $ and not the field strength $G_{\a(s)\ad(s)} $, 
the latter being 
defined by \eqref{g2.5S} and invariant under the $\z$-transformation \eqref{2.4a}.
This is actually a BF term, for it can be written in two different forms
\bea
\frac{1}{s}  \int  \rd^4x \rd^2 \q  \rd^2 \bar \q \, E
 \,
 \J^{\a(s) \ad(s-1)} \Big( 
\cD_{\a_1} \bar \cD_{\ad_1} &-&2\ri (s-1) \cD_{\a_1 \ad_1} \Big)
\S_{\a_2 \dots \a_s \ad_2 \dots \ad_{s-1} } \non \\
=- \frac{1}{s+1}  \int 
 \rd^4x \rd^2 \q  \rd^2 \bar \q \, E
\,
 G^{\a(s) \ad(s)} \Big( \bar \cD_{\ad_1} \cD_{\a_1}  
&+&2\ri (s+1) \cD_{\a_1 \ad_1} \Big)
Z_{\a_2 \dots \a_s \ad_2 \dots \ad_{s} }~.~~~
\label{2.14}
\eea
The former makes the gauge symmetry \eqref{2.3} manifestly realised, 
while the latter
turns the $\z$-transformation \eqref{2.4a} into a manifest symmetry.
Making use of \eqref{2.14} leads to a different representation 
for the action \eqref{actionS}. It is 
\bea
S^{\|}_{(s)} &=&
\Big( - \frac{1}{2}\Big)^s  \int 
 \rd^4x \rd^2 \q  \rd^2 \bar \q
\, E
\left\{ \frac{1}{8} H^{ \a (s-1) \ad (s-1) }  \cD^\b ({\bar \cD}^2- 4\mu) \cD_\b 
H_{\a (s-1) \ad (s-1)} \right. \non \\
&&+ \frac{s}{s+1}H^{ \a(s-1) \ad(s-1) }
\Big( \cD^{\b}  {\bar \cD}^{\bd} G_{\b\a(s-1) \bd\ad(s-1) }
- {\bar \cD}^{\bd}  \cD^{\b} 
{\bar G}_{\b \a (s-1) \bd \ad (s-1) } \Big) \non \\
&&+ \frac{(s+1)^2}{2} \bar \mu \mu 
H^{\a (s-1) \ad (s-1)} H_{\a (s-1) \ad (s-1)} \non \\
&&+ 2 \bar G^{ \a (s) \ad (s) } G_{ \a (s) \ad (s) } 
+ \frac{s}{s+1}\Big( G^{ \a (s) \ad (s) } G_{ \a (s) \ad (s) } 
+ \bar G^{ \a (s) \ad (s) }  \bar G_{ \a (s) \ad (s) } 
 \Big) \non \\
 &&+ \frac{s-1}{4s}H^{ \a(s-1) \ad(s-1) }
\Big( \cD_{\a_1} \bar \cD^2 \bar \S_{\a_2 \dots \a_{s-1}\ad(s-1)}
 - {\bar \cD}_{\ad_1}  \cD^2 \S_{\a(s-1) \ad_2 \dots \ad_{s-1} } \Big)  \non \\
&&-\frac{1}{s+1} G^{\a(s) \ad(s-1)} \Big( 
\bar \cD_{\ad_1} \cD_{\a_1} +2\ri (s+1) \cD_{\a_1 \ad_1} \Big)
Z_{\a_2 \dots \a_s \ad_2 \dots \ad_s }\non  \\
&&+ \frac{1}{s+1} G^{\a(s) \ad(s-1)} \Big( 
\cD_{\a_1} \bar \cD_{\ad_1} +2\ri (s+1) \cD_{\a_1 \ad_1} \Big)
\bar Z_{\a_2 \dots \a_s \ad_2 \dots \ad_s }\non \\
&&- \mu \frac{s^2+4s-1}{2s} H^{\a (s-1) \ad (s-1)} \cD_{\a_1} 
\bar \S_{\a_2 \dots \a_{s-1} \ad(s-1)} \non \\
&&+ \bar \mu \frac{s^2+4s-1}{2s} H^{\a (s-1) \ad (s-1)} \bar \cD_{\ad_1} 
 \S_{ \a(s-1) \ad_2 \dots \ad_{s-1}} \non \\
&&+ \frac{s-1}{8s} \Big( \S^{\a(s-1) \ad(s-2) } \cD^2 \S_{\a(s-1) \ad(s-2)} 
- \bar \S^{\a(s-2) \ad(s-1) }\bar \cD^2 \bar \S_{\a(s-2) \ad(s-1)} \Big) \non \\
&&- \frac{1}{s^2}\bar \S^{\a(s-2) \ad(s-2)\bd } \Big( \hf (s^2 +1) \cD^\b \bar \cD_\bd 
+\ri  {(s-1)^2} \cD^\b{}_\bd \Big) \S_{\b \a(s-2) \ad(s-2)} \non\\
&&+ \mu \frac{s^2+4s-1}{4s} \bar \S^{\a(s-2) \ad(s-1)} \bar \S_{\a(s-2) \ad(s-1)} \non \\
&&+ \bar \mu \frac{s^2+4s-1}{4s} \S_{\a(s-1) \ad(s-2)} \S^{\a(s-1) \ad(s-2)} \Big\}
~.~~~
\label{action2S}
\eea


\subsection{Dual formulation}

By analogy with the flat superspace case, the action \eqref{action2S} can be reformulated in terms of a transverse linear superfield by applying the duality transformation \cite{KS94}.
Let us associate with our theory \eqref{action2S} 
the following first-order action
\bea
S_{\text{first-order} }&=&S^{\|}_{(s)}[U, \bar U, H , Z, \bar Z]  \non \\
&&+ \Big( \frac{-1}{2} \Big)^s\int  \rd^4x \rd^2 \q  \rd^2 \bar \q \,E \,
 \Big(\frac{2}{s+1} \G^{\a(s) \ad(s)} U_{\a(s) \ad(s)} 
  +{\rm c.c.} \Big)~,~~
\label{action3S}
\eea
where $S^{\|}_{(s)}[U, \bar U, H , Z, \bar Z] $ is obtained from the action
\eqref{action2S} by replacing $G_{\a(s) \ad(s)} $ with an {\it unconstrained} 
complex superfield $U_{\a(s) \ad(s)} $. The Lagrange multiplier $\G_{\a(s) \ad(s)} $
is transverse linear, 
\bea \label{2.17}
\bar \cD^\bd \G_{\a(s) \bd \ad_1 \dots \ad_{s-1}} =0 ~.
\eea
We note that the specific normalisation of the Lagrange multiplier in \eqref{action3S} is chosen to match that of \cite{KS94}.
Varying \eqref{action3S} with respect to the Lagrange multiplier and taking into account the constraint \eqref{2.17} yields $U_{\a(s) \ad(s)} = G_{\a(s) \ad(s)} $. As a result, $S_{\text{first-order} }$ turns into the original action \eqref{action2S}. On the other hand, we can eliminate the auxiliary superfields  $U_{\a(s) \ad(s)}$ and  $\bar U_{\a(s) \ad(s)}$ from \eqref{action3S} using their equations of motion. This leads to the dual action
\bea
S^{\perp}_{(s)} &=& - \Big( - \hf \Big)^s 
  \int  \rd^4x \rd^2 \q  \rd^2 \bar \q \, E \,
\Bigg\{ - \frac{1}{8} H^{\a(s-1) \ad(s-1)} \cD^\b ({\bar \cD}^2- 4\mu) \cD_\b 
H_{\a (s-1) \ad (s-1)} \non\\
&&+ \frac{1}{8} \frac{s^2}{(s+1)(2s+1)} [\cD^\b, \bar \cD^\bd] H^{\a(s-1)\ad(s-1)} 
[\cD_{(\b}, \bar \cD_{(\bd}] H_{\a(s-1))\ad(s-1))} \non\\
&&+\hf \frac{s^2}{s+1} \cD^{\b\bd} H^{\a(s-1)\ad(s-1)} \cD_{(\b(\bd} H_{\a(s-1))\ad(s-1))} \non \\
&&- \frac{(s+1)^2}{2} \bar \mu \mu 
H^{\a (s-1) \ad (s-1)} H_{\a (s-1) \ad (s-1)} \non \\
&&+ \frac{2 \ri s}{2s+1} H^{\a(s-1)\ad(s-1)} \cD^{\b\bd} 
\Big({\bm \G}_{\b\a(s-1)\bd\ad(s-1)} 
- \bar{\bm \G}_{\b\a(s-1)\bd\ad(s-1)}\Big) \non\\
&& 
+ \frac{2}{2s+1} \bar{\bm \G}^{ \a (s) \ad (s) } {\bm \G}_{ \a (s) \ad (s) } 
- \frac{s}{(s+1)(2s+1)} \Big({\bm \G}^{ \a (s) \ad (s) }  {\bm \G}_{ \a (s) \ad (s) } 
+ \bar{\bm \G}^{ \a (s) \ad (s) } \bar{\bm \G}_{ \a (s) \ad (s) }\Big) \non\\
 &&- \frac{s-1}{2(2s+1)} H^{ \a(s-1) \ad(s-1) }
\Big( \cD_{\a_1} \bar \cD^2 \bar \S_{\a_2 \dots \a_{s-1}\ad(s-1)}
 - {\bar \cD}_{\ad_1}  \cD^2 \S_{\a(s-1) \ad_2 \dots \ad_{s-1} } \Big)  \non \\
&&+ \frac{1}{2(2s+1)} H^{ \a(s-1) \ad(s-1) }
\Big( \cD^2 {\bar \cD}_{\ad_1} \S_{\a(s-1) \ad_2 \dots \ad_{s-1} }
 - \bar \cD^2 \cD_{\a_1} \bar \S_{\a_2 \dots \a_{s-1}\ad(s-1)} \Big)  \non \\
&&- \ri \frac{(s-1)^2}{s(2s+1)} H^{ \a(s-1) \ad(s-1) }
 \cD_{\a_1 \ad_1} \Big( \cD^\b \S_{ \b \a_2 \dots \a_{s-1} \ad_2 \dots \ad_{s-1}}
+ \bar \cD^\bd \bar \S_{ \a_2 \dots \a_{s-1} \bd \ad_2 \dots \ad_{s-1}} \Big)  \non \\
&&+  \mu \frac{(s+2)(s+1)}{2s+1} H^{\a (s-1) \ad (s-1)}  \cD_{\a_1} 
\bar \S_{\a_2 \dots \a_{s-1} \ad(s-1)} \non \\
&&- \bar \mu \frac{(s+2)(s+1)}{2s+1} H^{\a (s-1) \ad (s-1)} \bar \cD_{\ad_1} 
 \S_{ \a(s-1) \ad_2 \dots \ad_{s-1}} \non \\
&&- \frac{s-1}{8s} \Big( \S^{\a(s-1) \ad(s-2) } \cD^2 \S_{\a(s-1) \ad(s-2)} 
- \bar \S^{\a(s-2) \ad(s-1) }\bar \cD^2 \bar \S_{\a(s-2) \ad(s-1)} \Big) \non \\
&&+ \frac{1}{s^2}\bar \S^{\a(s-2) \ad(s-2)\bd } \Big( \hf (s^2 +1) \cD^\b \bar \cD_\bd 
+\ri  {(s-1)^2} \cD^\b{}_\bd \Big) \S_{\b \a(s-2) \ad(s-2)} \non\\
&&- \mu \frac{s^2+4s-1}{4s} \bar \S^{\a(s-2) \ad(s-1)} \bar \S_{\a(s-2) \ad(s-1)} \non \\
&&- \bar \mu \frac{s^2+4s-1}{4s} \S_{\a(s-1) \ad(s-2)} \S^{\a(s-1) \ad(s-2)}
\Bigg\} ~,
\label{action4S}
\eea
 where we have defined
\bea
{\bm \G}_{ \a (s) \ad (s) } = \G_{ \a (s) \ad (s) }
-\hf \bar \cD_{(\ad_1} \cD_{(\a_1} Z_{\a_2 \dots \a_s) \ad_2 \dots \ad_s) } 
-\ri (s+1)\cD_{(\a_1 (\ad_1 } Z_{\a_2 \dots \a_s) \ad_2 \dots \ad_s) } ~.~~~
\label{shiftedS}
\eea

The first-order model introduced is equivalent to the original theory \eqref{action2S}. The action \eqref{action3S} is invariant under the gauge $\x$-transformation 
 \eqref{2.3} which acts on $U_{\a (s) \ad (s)}$ and
 $\G_{\a(s) \ad(s)}$ by the rule
\begin{subequations}
 \bea
 \d_\x U_{\a (s) \ad (s)} &=&0~,\\
 \d_\x \G_{\a(s) \ad(s)} &=& \bar \cD^\bd \Big\{  \frac{s+1}{2(s+2)}
\bar \cD_{(\bd} \cD_{(\a_1} \x_{\a_2 \dots \a_s) \ad_1 \dots \ad_s) } 
+\ri (s+1)\cD_{(\a_1 (\bd } \x_{\a_2 \dots \a_s) \ad_1 \dots \ad_s) } \Big\}~.~~~~~~~~~~~ \label{2.20}
\eea
\end{subequations}
Here we point out that ${\bm \G}_{ \a (s) \ad (s) }$ is invariant under the gauge transformations \eqref{2.3} and \eqref{2.20}.
The first-order action  \eqref{action3S} is also invariant under the gauge $\mathfrak V$-transformation \eqref{2.4b} and \eqref{2.4c}, which acts on $U_{\a (s) \ad (s)}$ and
$\G_{\a(s) \ad(s)} $ as
\begin{subequations}
\bea
 \d_{ {\mathfrak V} } U_{\a (s) \ad (s)} 
 &=& \hf \bar \cD_{(\ad_1} \cD_{(\a_1}  {\mathfrak V}_{\a_2 \dots \a_s)\ad_2 \dots \ad_{s})}~, \\
\d_{\mathfrak V} \G_{\a(s) \ad(s)} &=&0~.
\eea 
\end{subequations}
 
The $\mathfrak V$-gauge freedom in \eqref{2.4c} may be used to impose the condition 
\bea
Z_{\a(s-1) \ad(s-1)} =0~.
\label{3.11}
\eea
As a result, the action \eqref{action4S} reduces to that describing 
the transverse formulation for the massless superspin-$s$ multiplet \cite{KS94}.
The gauge condition \eqref{3.11} is preserved by residual local 
$\mathfrak V$- and $\x$-transformations of the form 
\bea
  \bar \cD^\bd \x_{\a(s-1) \bd \ad (s-1 )}  +
  \bar {\mathfrak V}_{\a(s-1)\ad (s-1 )} =0~.
\eea
Making use of the parametrisation \eqref{2.11}, the residual gauge freedom is
\begin{subequations}
\bea
\d H_{\a(s-1)\ad(s-1)} &=& \cD^\b L_{\b \a(s-1) \ad(s-1)} - \bar \cD^\bd \bar{L}_{\a(s-1)\bd\ad(s-1)} \ ,\\
\d \G_{\a(s) \ad(s)} &=& \frac{s+1}{2(s+2)} \bar \cD^\bd \Big\{ 
\bar \cD_{(\bd} \cD_{(\a_1} 
+ 2\ri (s+2) \cD_{(\a_1(\bd}\Big\}
 \bar{L}_{\a_2 \dots \a_{s})\ad_1 \dots \ad_{s})}~.~~~~~
\eea
\end{subequations}
This is exactly the gauge symmetry of the transverse formulation for the massless superspin-$s$ multiplet \cite{KS94}.

\subsection{Models for the massless gravitino multiplet in AdS} \label{ss423}

The massless gravitino multiplet ({\it i.e.} the massless superspin-1 multiplet)
was excluded from the above consideration. Here we will fill the gap. 
 
The (generalised) longitudinal formulation for the gravitino multiplet is described by the action
\begin{subequations}\label{3.26}
\bea
S^{\|}_{\rm GM} &=&
 - \int  \rd^4x \rd^2 \q  \rd^2 \bar \q\, E
\left\{ \frac{1}{16} H  \cD^\a ({\bar \cD}^2- 4\mu) \cD_\a H 
+ \frac{1}{4} H
\big( \cD^{\a}  {\bar \cD}^{\ad} G_{\a \ad}
- {\bar \cD}^{\ad}  \cD^{\a} {\bar G}_{\a \ad} \big) 
\right. 
\non \\
&& \qquad +  \bar G^{ \a  \ad  } G_{ \a  \ad } 
+ \frac 14 \big( G^{ \a  \ad } G_{ \a  \ad } 
+ \bar G^{ \a  \ad  }  \bar G_{ \a  \ad  } 
\big)   \non\\
&& 
\qquad
 \left. 
+|\m|^2 \Big( H - \frac{\F}{\m} -\frac{\bar \F}{\bar \m} \Big)^2
+\Big( \frac{\F}{\m} +\frac{\bar \F}{\bar \m}\Big)
\Big( \m \cD^\a \J_\a + \bar \m \bar \cD_\ad \bar \J^\ad \Big) 
\right\}~,
\eea
where $\F$ is a chiral scalar superfield, $\bar \cD_\ad \F=0$, and
\bea
G_{\a\ad} = \bar \cD_\ad \J_\a ~, \qquad \bar G_{\a\ad} = - \cD_\a \bar \J_\ad~.
\eea
\end{subequations}
This action is invariant under gauge transformations of the form 
\begin{subequations}
\bea
\d H&=& {\mathfrak V} +\bar {\mathfrak V}~,  \label{3.28a} \\
\d \J_\a &=& = \hf \cD_\a  {\mathfrak V}+ \eta_\a~, \qquad \bar \cD_\ad \eta_\a =0~,
\label{3.28b} \\
\d \F &=& -\frac 14 (\bar \cD^2 -4\m) \bar  {\mathfrak V}~.
\label{3.28c} 
\eea
\end{subequations}
This is one of the two models for the massless gravitino multiplet in AdS introduced in 
\cite{BK11}.  In a flat superspace limit, the action reduces to that given in \cite{GS80}.
Imposing the gauge condition $\F=0$ reduces the action \eqref{3.26}
to the original longitudinal formulation for the massless gravitino multiplet in AdS \cite{KS94}.

The action \eqref{3.26} involves the chiral scalar $\F$ and its conjugate only in the combination $(\vf + \bar \vf)$, where $\vf = \F /\m$. This means that the model \eqref{3.26} 
possesses a dual formulation realised in terms of a real linear superfield $L$, 
\bea
\big( \bar \cD^2 -4\m\big) L =0~, \qquad \bar L =L~.
\eea
The dual model is described by the action \cite{BK11}
\bea
S_{\rm GM} &=&
 - \int  \rd^4x \rd^2 \q  \rd^2 \bar \q\, E
\left\{ \frac{1}{16} H  \cD^\a ({\bar \cD}^2- 4\mu) \cD_\a H 
+ \frac{1}{4} H
\big( \cD^{\a}  {\bar \cD}^{\ad} G_{\a \ad}
- {\bar \cD}^{\ad}  \cD^{\a} {\bar G}_{\a \ad} \big) 
\right. 
\non \\
&& \qquad +  \bar G^{ \a  \ad  } G_{ \a  \ad } 
+ \frac 14 \big( G^{ \a  \ad } G_{ \a  \ad } 
+ \bar G^{ \a  \ad  }  \bar G_{ \a  \ad  } 
\big)  + |\m|^2H^2 \non\\
&& 
\qquad
 \left. 
-\frac 14  \Big( 2 |\m|H + L- \frac{\m}{|\m|} \cD^\a \J_\a - \frac{\bar \m}{|\m|} 
\bar \cD_\ad \bar \J^\ad \Big)^2
\right\}~.
\label{3.299}
\eea
This action is invariant under the gauge transformations \eqref{3.28a}, \eqref{3.28b}
and
\bea
\d L =\frac{1}{|\m|} \big( {\m} \cD^\a \eta_\a +{\bar \m} \bar \cD_\ad \bar \eta^\ad \big)~.
\eea
In a flat superspace limit, the action \eqref{3.299} reduces to that given in \cite{LR2}.

In Minkowski superspace, there exists one more dual realisation for the massless gravitino 
multiplet model \cite{HK2} which is obtained by performing a Legendre transformation 
converting $\F$ into a complex linear superfield. This formulation cannot be lifted to 
the AdS case, the reason being the fact that 
the action \eqref{3.26} involves the chiral scalar $\F$ and its conjugate only in the combination $(\vf + \bar \vf)$, where $\vf = \F /\m$. 

The dependence on $\J_\a $ and $\bar \J_\ad$
in  the last term of \eqref{3.26} 
can be expressed in terms of $G_{\a\ad}$ and $\bar G_{\a\ad}$
if we introduce a complex unconstrained prepotential $U$
for $\F$ in the standard way
\bea
\F = -\frac 14 (\bar \cD^2 -4\m) U~.
\eea
Then making use of \eqref{1.4d} gives
\bea
 \int  \rd^4x \rd^2 \q  \rd^2 \bar \q\, E \, \F \cD^\a \J_\a 
 = - \int  \rd^4x \rd^2 \q  \rd^2 \bar \q\, E\,G^{\a\ad}
 \Big( \frac 14 \bar \cD_\ad \cD_\a  +\ri \cD_{\a\ad}  \Big)U~.
 \eea
 Since the resulting action depends on $G_{\a\ad}$ and $\bar G_{\a\ad}$, 
we can introduce a dual formulation for the theory that is obtained
turning $G_{\a\ad}$ and $\bar G_{\a\ad}$ into a transverse linear superfield
\bea
\G_{\a\ad} = \bar \cD^\bd 
{ \Phi}_{\a\,\ad \bd } ~, \qquad { \Phi}_{\a\,\bd \ad } ={ \Phi}_{\a\,\ad \bd } 
\eea
and its conjugate using the scheme described in \cite{KS94}.
The resulting action is 
\bea
S^{\perp}_{\rm GM} &=&
\int  \rd^4x \rd^2 \q  \rd^2 \bar \q\, E
\left\{ -\frac{1}{16} H  \cD^\a ({\bar \cD}^2- 4\mu) \cD_\a H \right.
\non \\
&&+ \frac{1}{96} [\cD^\a, \bar \cD^\ad] H \, [\cD_\a, \bar \cD_\ad] H + \frac{1}{8} {\cD}^{\a \ad} H\, {\cD}_{\a \ad}H 
\non\\
&&+ \frac{1}{3} \bar {\bm \G}^{\a \ad} {\bm \G}_{\a \ad} - \frac{1}{12} \Big( {\bm \G}^{\a \ad} {\bm \G}_{\a \ad} + \bar {\bm \G}^{\a \ad} \bar {\bm \G}_{\a \ad} \Big)
+ \frac{\ri}{3}   \Big(\bar {\bm \G}^{\a \ad} - {\bm \G}^{\a \ad} \Big)
{\cD}_{\a \ad} H
\non\\
&&- \frac{1}{6} \F {\cD}^2 H- \frac{1}{6} {\bar \F} {\bar \cD}^2 H -|\m|^2 \Big( H - \frac{\F}{\m} - \left. \frac{\bar \F}{\bar \m} \Big)^2 \right\}~,
\label{Tspin1}
\eea
where we have defined
\bea
{\bm \G}_{\a \ad} := \G_{\a \ad} -\hf {\bar \cD}_\ad {\cD}_\a U -2\ri\, {\cD}_{\a \ad} U ~.
\label{shifted.s1}
\eea
The action \eqref{Tspin1} is invariant under the following gauge transformations 
\begin{subequations}
\bea
\d_\x U &=& {\bar \cD}_\ad \bar \x^\ad ~,\\
\d_\x \G_{\a \ad} &=& -\frac{1}{3} \bar \cD^\bd \Big\{\bar \cD_{(\bd} \cD_{\a} \bar \x_{\ad) } 
+6 \ri \, \cD_{\a (\bd } \bar \x_{\ad) } \Big\}~.~~~ \label{xi-spin1}
\eea
\end{subequations}
Both $\F$ and ${\bm \G}_{\a \ad}$ are invariant under $\x$-gauge transformations.
The action \eqref{Tspin1} is also invariant under the gauge transformations \eqref{3.28a}, \eqref{3.28c} and
\begin{subequations}
\bea
\d_{\mathfrak V} U &=& \bar {\mathfrak V}~, \\
\d_{\mathfrak V} \G_{\a \ad} &=& 0~.
\eea 
\end{subequations}
Imposing the gauge condition $U=0$ reduces the action \eqref{Tspin1}
to the original  transverse formulation for the massless gravitino multiplet in AdS \cite{KS94}.


\section{Higher-spin supercurrents} \label{s43}

In this section we introduce higher-spin supercurrent multiplets in AdS. 
First, we recall the structure of the gauge superfields in terms of which
the massless superspin-$(s+1/2)$ multiplet ($s=2, 3, \ldots$) are described \cite{KS94}.

\subsection{Massless half-integer superspin multiplets} \label{ss431}

For a  massless superspin-$(s+1/2)$ multiplet in AdS, there exist two dually equivalent off-shell formulations ({\it i.e.}\,transverse and longitudinal), which were first constructed in \cite{KS94}.
The corresponding dynamical variables are \cite{KS94}
\begin{subequations}
\bea
\cV^\bot_{s+1/2}& = &\Big\{H_{\a(s)\ad(s)}~, ~
\G_{\a(s-1) \ad(s-1)}~,
~ \bar{\G}_{\a(s-1) \ad(s-1)} \Big\} ~,  \label{10}  \\
\cV^{\|}_{s+1/2} &=& 
\Big\{H_{\a(s)\ad(s)}~, ~
G_{\a(s-1) \ad(s-1)}~,
~ \bar{G}_{\a(s-1) \ad(s-1)} \Big\} \label{10-lon}
~.
\eea
\end{subequations}
Here $H_{\a(s) \ad (s)}$ is a real unconstrained superfield.
The complex superfields 
$\G_{\a (s-1) \ad (s-1)} $ and 
$G_{\a (s-1) \ad (s-1)}$ are  transverse linear and 
longitudinal linear, respectively,
\begin{subequations}
\bea
{\bar \cD}^\bd \,\G_{\a(s-1) \bd \ad(s-2)} &=&  0 ~,
\label{transverse}
\\
{\bar \cD}_{ (\ad_1} \,G_{\a(s-1) \ad_2 \dots \ad_{s})}&=&0 ~.
\label{longitudinal}
\eea
\end{subequations}
These constraints are solved in terms of unconstrained prepotentials
as follows:
\begin{subequations}
\bea
 \G_{\a(s-1) \ad(s-1)}&=& \bar \cD^\bd 
{ \Phi}_{\a(s-1)\,(\bd \ad_1 \cdots \ad_{s-1}) } ~,
\label{6.3a}
 \\
 G_{\a(s-1) \ad(s-1)} &=& {\bar \cD}_{( \ad_1 }
 \Psi_{ \a(s-1) \, \ad_2 \cdots \ad_{s-1}) } ~.
\label{6.3b}
\eea
\end{subequations}
The prepotentials  are defined modulo gauge transformations of the form:
\begin{subequations}
\bea
\d_\x \Phi_{\a(s-1)\, \ad (s)} 
&=&  \bar \cD^\bd 
{ \x}_{\a(s-1)\, (\bd \ad_1 \cdots \ad_{s}) } ~,
\label{tr-prep-gauge}
\\
\d_\z  \Psi_{ \a(s-1) \, \ad {(s-2}) } &=&  {\bar \cD}_{( \ad_1 }
 \z_{ \a(s-1) \, \ad_2 \cdots \ad_{s-2}) } ~,
\label{lon-prep-gauge}
\eea
\end{subequations}
with the gauge parameters $ { \x}_{\a(s-1)\,  \ad (s+1) } $
and $ \z_{ \a(s-1) \, \ad (s-3)}$ being unconstrained.

The gauge transformations of the superfields $H$, $\G$ and $G$ are 
\begin{subequations} \label{6.5}
\bea 
\d_\L H_{\a_1 \dots \a_s \ad_1  \dots \ad_s} 
&= &\bar \cD_{(\ad_1} \L_{\a_1\dots  \a_s \ad_2 \dots \ad_s )} 
- \cD_{(\a_1} \bar{\L}_{\a_2 \dots \a_s)\ad_1  \dots \ad_s} \ , \label{6.5a} \\
\d_\L \G_{\a_1 \dots \a_{s-1} \ad_1  \dots \ad_{s-1}} 
&= & -\frac{s}{2(s+1)} {\bar \cD}^\bd {\cD}^\b {\cD}_{(\b} {\bar \L}_{\a(s-1)) \bd \ad(s-1)} \non\\
&=& -\frac{1}{4} \bar \cD^\bd \cD^2 \bar{\L}_{\a_1 \dots \a_{s-1}\bd \ad_1 
\dots \ad_{s-1}} \non\\
&&-\hf {\bar \m} (s-1) {\bar \cD}^\bd \bar{\L}_{\a_1 \dots \a_{s-1}\bd \ad_1 
\dots \ad_{s-1}} \ ,\label{6.5b} \\
\d_\L G_{\a_1 \dots \a_{s-1}\ad_1 \dots \ad_{s-1}} &= & - \hf \bar \cD_{(\ad_1} 
\bar \cD^{|\bd|} \cD^\b \L_{\b\a_1 \dots \a_{s-1} \ad_2 \dots \ad_{s-1}) \bd} \non\\
&&+ \ri (s-1) \bar \cD_{(\ad_1} \cD^{\b |\bd|} 
\L_{\b \a_1 \dots \a_{s-1} \ad_2 \dots \ad_{s-1} ) \bd} \ . \label{6.5c}
\eea
\end{subequations}
Here the gauge parameter $\L_{\a_1 \dots \a_s \ad_1 \dots \ad_{s-1}}
=\L_{(\a_1 \dots \a_s )(\ad_1 \dots \ad_{s-1})}$ 
is unconstrained. 
The symmetrisation in \eqref{6.5c} is extended only to the indices 
$\ad_1, \ad_2, \dots,  \ad_{s-1}$. 
It follows from \eqref{6.5b} and \eqref{6.5c} that the transformation laws of the prepotentials $\Phi_{\a(s-1) \ad(s)}$ and $\J_{\a(s-1) \ad(s-2)}$ are
\begin{subequations} \label{6.6}
\bea
\d_\L \Phi_{\a_1 \dots \a_{s-1}\ad_1 \dots \ad_{s}} &= &  -\frac{1}{4} \cD^2 \bar{\L}_{\a_1 \dots \a_{s-1} \ad_1 
\dots \ad_{s}} 
-\hf {\bar \m} (s-1) \bar{\L}_{\a_1 \dots \a_{s-1} \ad_1 
\dots \ad_{s}}\ , \label{6.6a}\\
\d_\L \J_{\a_1 \dots \a_{s-1}\ad_1 \dots \ad_{s-2}} &= & - \hf 
\Big( \bar \cD^{\bd} \cD^\b -2\ri (s-1) \cD^{\b \bd} \Big)
\L_{\b\a_1 \dots \a_{s-1} \bd \ad_1 \dots \ad_{s-2}} \ . \label{6.6b}
\eea
\end{subequations}


\subsection{Non-conformal supercurrents: Half-integer superspin} \label{ss432}

In the framework of the longitudinal formulation \eqref{10-lon}, 
let us couple the prepotentials 
$H_{ \a (s) \ad (s) } $, $\J_{ \a (s-1) \ad (s-2) }$ and $\bar \J_{ \a (s-2) \ad (s-1) }  $,
to external sources
\bea
S^{(s+\hf)}_{\rm source}&=&\int \rd^4x \rd^2 \q  \rd^2 \bar \q \, E\, \Big\{ 
H^{ \a (s) \ad (s) } J_{ \a (s) \ad (s) }
+ \J^{ \a (s-1) \ad (s-2) } T_{ \a (s-1) \ad (s-2) } \non \\
&&+ \bar \J_{ \a (s-2) \ad (s-1) } \bar T^{ \a (s-2) \ad (s-1) } \Big\}~.
\label{7.1}
\eea
Requiring $S^{(s+\hf)}_{\rm source}$ to be invariant under 
\eqref{lon-prep-gauge} gives
\begin{subequations} \label{7.3}
\bea
\bar \cD^\bd T_{\a(s-1) \bd \ad_1 \dots \ad_{s-3}} =0~,
\label{7.2}
\eea
and therefore $T_{ \a (s-1) \ad (s-2) } $ is a transverse linear superfield. 
Requiring $S^{(s+\hf)}_{\rm source}$ to be invariant under the gauge transformations
(\ref{6.5a}) and (\ref{6.6b}) gives the following conservation equation:
\bea
\bar \cD^\bd J_{\a_1 \dots \a_s \bd \ad_1 \dots \ad_{s-1}} 
+\hf \Big( \cD_{(\a_1} \bar \cD_{(\ad_1}
-2\ri (s-1) \cD_{ (\a_1 (\ad_1 } \Big)  T_{\a_2\dots \a_s) \ad_2 \dots \ad_{s-1})} =0~.
\label{7.3a}
\eea
For completeness, we also give the conjugate equation
\bea
\cD^\b J_{\b \a_1 \dots \a_{s-1}  \ad_1 \dots \ad_{s}} 
-\hf \Big( \bar \cD_{(\ad_1}  \cD_{(\a_1}
-2\ri (s-1) \cD_{ (\a_1 (\ad_1 } \Big)  \bar T_{\a_2\dots \a_{s-1}) \ad_2 \dots \ad_{s})} 
=0~. \label{7.3c}
\eea
\end{subequations}

Similar considerations for the transverse formulation \eqref{10} lead to the following non-conformal supercurrent multiplet
\begin{subequations}\label{TSupercurrent}
\bea
\bar \cD^\bd {\mathbb J}_{\a_1 \dots \a_s \bd \ad_1 \dots \ad_{s-1}} 
-\frac{1}{4} (\bar \cD^2 + 2 \m (s-1)) {\mathbb F}_{\a_1 \dots \a_s \ad_1 \dots \ad_{s-1}} &=&0~,\\
\cD_{(\a_1 } {\mathbb F}_{\a_2 \dots \a_{s+1} )\ad_1 \dots \ad_{s-1}}&=&0  ~. \label{TSupercurrentAdS}
\eea
\end{subequations}
It follows from \eqref{TSupercurrentAdS} that the trace multiplet $\bar {\mathbb F}_{\a(s-1) \ad(s)}$ is longitudinal linear.
In the flat-superspace limit, the higher-spin supercurrent multiplets  
\eqref{7.3} and \eqref{TSupercurrent}
reduce to those described in subsection \ref{ss321}.

Let us recall our condensed notation used in subsection \ref{ss321}. Associated with any tensor superfield $U_{\a(m) \ad(n)}$ is the following index-free field on ${\mathbb C}^2$ 
\bea
U_{(m,n)} (\z, \bar \z):= \z^{\a_1} \dots \z^{\a_m} \bar \z^{\ad_1} \dots \bar \z^{\ad_n}
U_{\a_1 \dots \a_m \ad_1 \dots \ad_n}~,
\label{4.100}
\eea
We also introduce the AdS analogues of the operators \eqref{326S}:
\begin{subequations}
\bea
{\cD}_{(1,0)} &:=& \z^\a \cD_\a~,\\
{\bar \cD}_{(0,1)} &:=& \bar \z^\ad \bar \cD_\ad~, \\
{\cD}_{(1,1)} &:=& 2\ri \z^\a \bar \z^\ad \cD_{\a\ad}
= -\big\{ {\cD}_{(1,0)} , \bar {\cD}_{(0,1)} \big\}
~.
\eea
\end{subequations}
The following operators decrease the degree 
of homogeneity in the variables $\z^\a$ and $\bar \z^\ad$, specifically
\begin{subequations}
\bea
\cD_{(-1,0)} &:=& \cD^\a \frac{\pa}{\pa \z^\a}~,\\
\bar \cD_{(0,-1)}& :=& \bar \cD^\ad \frac{\pa}{\pa \bar \z^\ad}~ 
~.
\eea
\end{subequations}

Making use of the above notation, the transverse linear condition \eqref{7.2} and its conjugate become
\begin{subequations}
\bea
\bar \cD_{(0,-1)} T_{(s-1,s-2)} &=&0~,  \label{7.8a}\\
\cD_{(-1,0)} \bar T_{(s-2,s-1)} &=&0~.  \label{7.8b}
\eea
\end{subequations}
The conservation equations \eqref{7.3a} and \eqref{7.3c} turn into 
\begin{subequations}
\bea
\frac{1}{s}\bar \cD_{(0,-1)} J_{(s,s)} -\hf A_{(1,1)} T_{(s-1, s-2)}&=&0~, \label{7.9a}\\
\frac{1}{s}\cD_{(-1,0)} J_{(s,s)} -\hf \bar A_{(1,1)} \bar T_{(s-2, s-1)}&=&0~. \label{7.9b}
\eea
\end{subequations}
where 
\bea
A_{(1,1)} := -\cD_{(1,0)} \bar \cD_{(0,1)} +(s-1) \cD_{(1,1)} ~, \quad
\bar A_{(1,1)} := \bar \cD_{(0,1)}  \cD_{(1,0)} -(s-1) \cD_{(1,1)} ~. ~~~
\eea
Since 
$\bar \cD_{(0,-1)} ^2 J_{(s,s)} =0$,
the conservation equation \eqref{7.9a} is consistent provided
\bea
\bar \cD_{(0,-1)}  A_{(1,1)} T_{(s-1, s-2)}=0~.
\eea
This is indeed true, as a consequence of the transverse linear condition
\eqref{7.8a}. 


\subsection{Improvement transformations}

The conservation equations \eqref{7.3} and \eqref{TSupercurrent}
define two consistent higher-spin supercurrents in AdS. 
Similar to the two 
irreducible AdS supercurrents \cite{BK12}, with $(12+12)$ and $(20+20)$ degrees of 
freedom (see also the review in section \ref{VarAdS}), the  higher-spin supercurrents \eqref{7.3} and \eqref{TSupercurrent}
are equivalent in the sense that there always exists a well-defined improvement 
transformation that converts \eqref{7.3} into \eqref{TSupercurrent}.
Such an improvement transformation is constructed below.

Since the trace multiplet $T_{\a(s-1) \ad (s-2)}$ is transverse, eq. \eqref{7.2}, 
there exists a well-defined complex tensor operator  $X_{\a(s-1) \ad (s-1)}$ such that 
\bea
T_{\a(s-1) \ad (s-2)}
= \bar \cD^\bd X_{\a(s-1) (\bd \ad_1 \dots \ad_{s-2})} ~.
\eea
Let us introduce the real $U_{\a(s-1) \ad (s-1)}$ and imaginary $V_{\a(s-1) \ad (s-1)}$
parts of $X_{\a(s-1) \ad (s-1)}$, 
\bea
X_{\a(s-1) \ad (s-1)} = U_{\a(s-1) \ad (s-1)} + \ri V_{\a(s-1) \ad (s-1)}~.
\eea
Then it may be checked that the operators
\begin{subequations}
\bea
{\mathbb J}_{\a(s) \ad (s)} &:=& J_{\a(s) \ad (s)}
+\frac{s}{2} \big[ \cD_{(\a_1}, \bar \cD_{(\ad_1} \big]
U_{\a_2 \dots \a_s) \ad_2 \dots \ad_s ) }
+ s \cD_{(\a_1 (\ad_1 } V_{\a_2 \dots \a_s) \ad_2 \dots \ad_s)}~, ~~~~~~\\
{\mathbb F}_{\a(s) \ad(s-1)} &:=& 
\cD_{(\a_1} \Big\{ (2s+1) U_{\a_2 \dots \a_s) \ad(s-1)}
- \ri V_{\a_2 \dots \a_s) \ad(s-1)}\Big\} \eea
\end{subequations}
enjoy the conservation equation \eqref{TSupercurrent} and the constraint \eqref{TSupercurrentAdS}. It is also not difficult to construct an inverse improvement transformation 
converting the higher-spin supercurrent \eqref{TSupercurrent} to \eqref{7.3}. 

In accordance with the result obtained, for all applications it suffices to work with the 
longitudinal supercurrent \eqref{7.3}. This is why in the integer superspin case, 
which will be studied in the next subsection, we will introduce only 
a higher-spin supercurrent corresponding to the new gauge formulation \eqref{action2S}.

There exists an improvement transformation\footnote{One may compare this to \eqref{A.55} in the lower-spin case.} for the supercurrent multiplet \eqref{7.3}
Given a chiral scalar superfield $\O$, we introduce
\begin{subequations} \label{improvement4.20}
\bea
\widetilde{J}_{(s, s)} &:=& J_{(s,s)} 
+ \cD^s_{(1,1)} \big( \O +(-1)^s \bar \O\big)~, \qquad \bar \cD_\ad \O=0~,\\
\widetilde{T}_{(s-1,s-2)} &:=& T_{(s-1,s-2)} 
+ \frac{2  (-1)^s }{s(s-1)} \bar \cD_{(0,-1)} \cD^{s-1}_{(1,1)} \bar \O \non \\
&& \phantom{T_{\a(s-1) \ad (s-2)}}
+ \frac{4(s+1)}{s} \m \cD^{s-2}_{(1,1)} \cD_{(1,0)} \O~. 
\eea
\end{subequations}
The operators $\widetilde{J}_{(s,s)} $ and 
$\widetilde{T}_{(s-1,s-2)} $ prove to obey the conservation 
equation \eqref{7.3}. 


\subsection{Non-conformal supercurrents: Integer superspin} \label{subsection4.4}

We now make use of the new gauge formulation \eqref{actionS}, 
or  equivalently \eqref{action2S}, for the integer superspin-$s$ multiplet to derive the AdS analogue of the non-conformal higher-spin supercurrents in subsection \ref{ss322}.

Let us couple the prepotentials 
$H_{ \a (s-1) \ad (s-1) } $, $Z_{ \a (s-1) \ad (s-1) }$ and $\Psi_{ \a (s) \ad (s-1) } $ to external sources
\bea
S^{(s)}_{\rm source} &=& \int \rd^4x \rd^2 \q  \rd^2 \bar \q \, E\, \Big\{ 
\Psi^{ \a (s) \ad (s-1) } J_{ \a (s) \ad (s-1) }
-\bar \Psi^{ \a (s-1) \ad (s) } \bar J_{ \a (s-1) \ad (s) }
\non \\
&&+H^{ \a (s-1) \ad (s-1) } S_{ \a (s-1) \ad (s-1) } \non \\
&&+ Z^{ \a (s-1) \ad (s-1) } T_{ \a (s-1) \ad (s-1) } 
+ \bar Z^{ \a (s-1) \ad (s-1) } \bar T_{ \a (s-1) \ad (s-1) }
 \Big\}~.
\label{4.1}
\eea
In order for $S^{(s)}_{\rm source}$ to be invariant under the $\z$-transformation 
in \eqref{2.4a}, the source  $J_{ \a (s) \ad (s-1) }$ must satisfy
\bea
\bar \cD^\bd J_{\a(s) \bd \ad(s-2)} =0 \quad \Longleftrightarrow \quad
\cD^\b \bar J_{\b \a(s-2)  \ad(s)} =0 ~.
\label{4.2a}
\eea
Next, requiring $S^{(s)}_{\rm source}$ to be invariant under the transformation \eqref{2.3} leads to
\bea
\bar \cD_{(\ad_1} T_{\a(s-1) \ad_2 \dots \ad_{s})} =0
 \quad \Longleftrightarrow \quad
 \cD_{(\a_1} \bar T_{\a_2 \dots \a_{s})  \ad (s-1)} =0~.
\label{4.2b}
\eea
We see that  the superfields $J_{ \a (s) \ad (s-1) }$ and $T_{ \a (s-1) \ad (s-1) } $ are transverse linear and longitudinal linear, respectively.
Finally, requiring $S^{(s)}_{\rm source}$ to be invariant under the 
$\mathfrak V$-transformation 
\eqref{2.4} gives the following conservation equation
\begin{subequations} \label{3.4}
\bea
-\hf \cD^\b J_{\b \a(s-1) \ad(s-1)} 
+S_{\a(s-1) \ad(s-1)} + \bar T_{\a(s-1) \ad(s-1)} =0
\label{4.2c}
\eea
as well as its conjugate
\bea
\hf \bar \cD^\bd \bar J_{ \a(s-1) \bd \ad(s-1)} 
+S_{\a(s-1) \ad(s-1)} + T_{\a(s-1) \ad(s-1)} =0~. 
\label{3.4b}
\eea
\end{subequations}

As a consequence of  \eqref{4.2b}, from \eqref{4.2c} we deduce
\bea
\frac{1}{4} \cD^2 J_{ \a(s) \ad(s-1)} -\hf \bar \m(s+2)J_{\a(s) \ad(s-1)} + \cD_{(\a_1} S_{\a_2 \dots \a_s) \ad(s-1) } =0~.
\label{3.5}
\eea
The equations \eqref{4.2a} and \eqref{3.5} describe the conserved 
current supermultiplet which corresponds to our theory in the gauge \eqref{2.12ads}.

Taking the sum of \eqref{4.2c} and \eqref{3.4b}
leads to
\bea
\hf \cD^\b J_{\b \a(s-1) \ad(s-1)} 
+\hf \bar \cD^\bd \bar J_{\a(s-1) \bd \ad(s-1)}
+ T_{\a(s-1) \ad(s-1)}-\bar T_{\a(s-1) \ad(s-1)} =0~. 
\label{4.3}
\eea
The equations \eqref{4.2a}, \eqref{4.2b} and \eqref{4.3} describe the conserved 
current supermultiplet which corresponds to our theory in the gauge \eqref{2.10}.
As a consequence of \eqref{4.2b}, the conservation equation \eqref{4.3} 
implies
\bea
\hf \cD_{(\a_1} \left\{\cD^{|\b|} J_{\a_2 \dots \a_s ) \b\ad(s-1)} 
+ \bar \cD^\bd \bar J_{\a_2 \dots \a_s ) \bd \ad(s-1)}\right\}
+\cD_{(\a_1} T_{\a_2 \dots \a_s ) \ad(s-1)} =0~. 
\label{4.4}
\eea

Written in the condensed notation, the transverse linear condition \eqref{4.2a} turns into 
\bea
\bar \cD_{(0,-1)} J_{(s,s-1)} &=& 0~,  \label{4.5a}
\eea
while the longitudinal linear condition \eqref{4.2b} takes the form
\bea
\bar \cD_{(0,1)} T_{(s-1,s-1)} &=& 0~. \label{4.5b}
\eea
The conservation equation \eqref{4.2c} becomes
\bea
-\frac{1}{2s} \cD_{(-1,0)} J_{(s,s-1)} + S_{(s-1,s-1)} + \bar T_{(s-1,s-1)} = 0
\label{4.6}
\eea
and \eqref{4.4} takes the form
\bea
\frac{1}{2s} \cD_{(1,0)} \left\{\cD_{(-1,0)} J_{(s,s-1)} + \bar \cD_{(0,-1)} \bar J_{(s-1,s)}\right\}
+\cD_{(1,0)} T_{(s-1,s-1)} =0~. 
\label{4.7}
\eea


\subsection{Improvement transformation}

There exist an improvement transformation for the supercurrent multiplet \eqref{3.4}. 
Given a chiral scalar superfield $\O$, we introduce
\begin{subequations}
\bea
\widetilde{J}_{(s,s-1)} &:=& J_{(s,s-1)} 
+ \cD^{s-1}_{(1,1)} \cD_{(1,0)} \O ~, \qquad \bar \cD_\ad \O=0~,\\
\widetilde{\bar T}_{(s-1, s-1)} &:=& \bar T_{(s-1, s-1)} 
+ \frac{s-1}{4s} \cD^{s-1}_{(1,1)} (\cD^2 -4 \bar \m) \O \non \\
&&+ (-1)^s (s-1) \Big(\bar \m + \frac{\m}{s}\Big) \cD^{s-1}_{(1,1)} \bar \O~,\\
\widetilde{S}_{(s-1,s-1)} &:=& S_{(s-1,s-1)} 
+ \m (s-1) \cD^{s-1}_{(1,1)} \O + (-1)^{s-1} \bar \m (s-1)\cD^{s-1}_{(1,1)} \bar \O \non \\
&&+ \bar \m \frac{s-1}{s} \cD^{s-1}_{(1,1)} \O + (-1)^{s-1} \m \frac{s-1}{s} \cD^{s-1}_{(1,1)} \bar \O~.
\eea
\end{subequations}
It may be checked that the operators $\widetilde{J}_{(s,s-1)} $, $\widetilde{\bar T}_{(s-1,s-1)} $ and $\widetilde{S}_{(s-1,s-1)}$ obey the conservation 
equation \eqref{4.6}, as well as \eqref{4.2b} and \eqref{4.5a}.


\section{Higher-spin supercurrents for chiral superfields: Half-integer superspin} \label{s44}

In the remainder of this chapter we will study explicit realisations
of the higher spin supercurrents introduced above in various supersymmetric 
field theories in AdS. 

\subsection{Superconformal model for a chiral superfield} \label{1chiral}


Let us consider the superconformal theory of a single chiral scalar superfield 
\bea
S = \int \rd^4x \rd^2 \q  \rd^2 \bar \q \,E\, \bar \F \F ~,
\label{chiral}
\eea
where  $\F$ is covariantly chiral, $\bar \cD_\ad \F =0$.
We can define the conformal supercurrent $J_{(s, s)}$ 
in direct analogy with the flat superspace case~\cite{KMT,  HK1}
\bea
J_{(s,s)} &=& \sum_{k=0}^s (-1)^k
\binom{s}{k}
\left\{ \binom{s}{k+1} 
{\cD}^k_{(1,1)}
 \cD_{(1,0)} \F \,\,
{\cD}^{s-k-1}_{(1,1)}
\bar \cD_{(0,1)} 
\bar \F  
\right. \non \\ 
&& \left.
 \qquad \qquad
+ \binom{s}{k} 
{\cD}^k_{(1,1)}
  \F \,\,
{\cD}^{s-k}_{(1,1)}
\bar \F \right\}~.~
\label{7.15}
\eea
Making use of
the massless equations of motion,   $(\cD^2-4\bar \m)\, \F = 0$, 
one may check that $J_{(s,s)}$ satisfies the conservation equation
\bea
\cD_{(-1,0)} J_{(s,s)} = 0 \quad \Longleftrightarrow \quad 
\bar \cD_{(0,-1)} J_{(s,s)} = 0 ~.~
\label{7.9}
\eea
The calculation of \eqref{7.9} in AdS is much more complicated than in flat superspace due to the fact that the algebra of covariant derivatives \eqref{1.2}
is nontrivial. 
Let us sketch the main steps in evaluating the left-hand side of eq.~\eqref{7.9}
with $J_{(s,s)} $ given by \eqref{7.15}.
We start with the obvious relations
\begin{subequations}
\bea
\frac{\pa}{\pa \z^\a} {\cD}_{(1,1)} &=&2\ri {\bar \z}^\ad {\cD}_{\a \ad}~, \\
\frac{\pa}{\pa \z^\a} {\cD}^k_{(1,1)} &=& 
\sum_{n=1}^k\,{\cD}^{n-1}_{(1,1)} \,\,  2\ri \, {\bar \z}^\ad {\cD}_{\a \ad}\,\, {\cD}^{k-n}_{(1,1)} ~, \qquad k>1
~.\label{eq1}
\eea
\end{subequations}
To simplify eq.~\eqref{eq1}, we may push ${\bar \z}^\ad{\cD}_{\a \ad}$, say,  to the left 
provided that we take into account its commutator with ${\cD}_{(1,1)}$:
\bea
[{\bar \z}^\ad {\cD}_{\a \ad}\,, {\cD}_{(1,1)}] = -4\ri \,\bar \m \m \,\z_\a  {\bar \z}^\ad {\bar \z}^\bd {\bar M}_{\ad \bd}~.
\label{555}
\eea
Associated with the Lorentz generators are the operators
\begin{subequations}
\bea
{\bar M}_{(0,2)} &:=& {\bar \z}^\ad {\bar \z}^\bd {\bar M}_{\ad \bd}~,\\
{M}_{(2,0)} &:=& {\z}^\a {\z}^\b {M}_{\a \b}~,
\eea
where ${\bar M}_{(0,2)}$ appears in the right-hand side of \eqref{555}.
These operators annihilate every superfield $U_{(m,n)}(\z, \bar \z) $ of the form 
\eqref{4.100},\footnote{These properties are analogous to those 
that play a fundamental role for the consistent definition of covariant projective supermultiplets
 in 5D $\cN=1$ \cite{KT-M08} and 4D $\cN=2$ \cite{KLRT-M} supergravity theories.}
\bea
{\bar M}_{(0,2)} U_{(m,n)}=0~, \qquad  M_{(2,0)} U_{(m,n)} =0~.
\eea
\end{subequations}
From the above consideration, it follows that
\begin{subequations}
\bea
[{\bar \z}^\ad {\cD}_{\a \ad}\,, {\cD}^k_{(1,1)}]\, U_{(m,n)} &=& 0 ~, \\
\Big(\frac{\pa}{\pa \z^\a} {\cD}^k_{(1,1)}\Big)U_{(m,n)} &=& 2\ri k\, {\bar \z}^\ad {\cD}_{\a \ad}\, {\cD}^{k-1}_{(1,1)}U_{(m,n)}~.
\eea
\end{subequations}
We also state some other properties which we often use throughout our calculations
\begin{subequations}
\bea
{\cD}^2_{(0,1)} &=& -2\bar \m M_{(2,0)} ~,\\
\big[ {\cD}_{(1,0)}\,, {\cD}_{(1,1)} \big] 
&=& 
\big[ \bar \cD_{(0,1)}\,, \cD_{(1,1)} \big] = 0~,\\
\big[ \cD^\a, \cD_{(1,1)} \big] &=& -2 \bar \m \, \z^\a \bar {\cD}_{(0,1)} ~,\\
\big[\cD^\a, \cD^k_{(1,1)}\big] &=& -2 \bar \m \,k \,\z^\a \cD^{k-1}_{(1,1)} \bar {\cD}_{(0,1)}~,\\
\big[\cD^\a, \bar \z^\bd \cD_{\b \bd}\big] &=& \ri \bar \m \,\d^\a_\b \, \bar \cD_{(0,1)}~.
\eea
\end{subequations}
The above identities suffice to prove that the supercurrent  \eqref{7.15}
does obey the conservation equation \eqref{7.9}.


\subsection{Non-superconformal model for a chiral superfield} \label{subsection5.2}

Let us now add the mass term to~\eqref{chiral} and consider the following action
\bea
S = \int \rd^4x \rd^2 \q  \rd^2 \bar \q \,E\, \bar \F \F
+\Big\{ \frac{m}{2} \int \rd^4x \rd^2 \q \, \cE \, \F^2 +{\rm c.c.} \Big\}~,
\label{chiral-massive}
\eea
with $m$ a complex mass parameter. 
The real supercurrent $J_{(s,s)}$ takes the same form as in the massless case, \eqref{7.15}. However, 
in the massive case $J_{(s,s)}$ satisfies a more general conservation equation~\eqref{7.9a}
for some superfield $T_{(s-1, s-2)}$, which we need to determine.
Indeed, making use of the equations of motion
\bea
-\frac{1}{4} (\cD^2-4\bar\m) \F  +\bar m \bar \F =0~, \qquad
-\frac{1}{4} (\bar \cD^2-4\m) \bar \F +m \F =0~,
\eea
we obtain 
\begin{subequations}
\bea
\bar \cD_{(0,-1)} J_{(s,s)} &=& F_{(s,s-1)}~, \label{7.12a}
\eea
where we have denoted 
\bea
F_{(s,s-1)} &=& 2m(s+1) \sum_{k=0}^s (-1)^{s-1+k} \binom{s}{k} \binom{s}{k+1}
\non \\ 
&& 
\times \left\{1+(-1)^s \frac{k+1}{s-k+1}\right\} 
 {\cD}^k_{(1,1)} \F \,\,{\cD}^{s-k-1}_{(1,1)}
 \cD_{(1,0)} \F ~.
\eea
\end{subequations}

We now  look for a superfield $T_{(s-1, s-2)}$
such that (i) it obeys the transverse linear constraint \eqref{7.8a}; and 
(ii) it satisfies the equation
\bea
F_{(s,s-1)} = \frac{s}{2} A_{(1,1)} T_{(s-1, s-2)}~. 
\eea
Our analysis will be similar to the one performed in~\ref{ss3211} in flat superspace. 
We consider a general ansatz
\bea
T_{(s-1, s-2)} = (-1)^s m \sum_{k=0}^{s-2} c_k 
{\cD}^k_{(1,1)} \F\,
{\cD}^{s-k-2}_{(1,1)}
 \cD_{(1,0)} \F 
 \label{T7.15}
\eea
with some coefficients $c_k$ which have to be determined. 
For $k = 1,2,...s-2$, condition (i) implies that 
the coefficients $c_k$ must satisfy
\begin{subequations}\label{7.16}
\begin{align}
kc_k = (s-k-1) c_{s-k-1}~,\label{7.16a}
\end{align}
while (ii) gives the following equation
\begin{align}
c_{s-k-1} + s c_k + (s-1) c_{k-1} &= -4(-1)^k \frac{s+1}{s} \binom{s}{k} \binom{s}{k+1} 
\non \\
& \qquad \qquad \times \left\{ 1+ (-1)^s \frac{k+1}{s-k+1} \right\} ~.\label{7.16b}
\end{align}
Condition (ii) also implies that 
\begin{align}
(s-1) c_{s-2} +c_0 &= 4(-1)^s (s+1)\left\{1+(-1)^s \frac{s}{2}\right\}~, \label{7.16c}\\
c_0 &= -\frac{4}{s}(s+1+(-1)^s)~. \label{7.16d}
\end{align}
\end{subequations}
It turns out that the equations \eqref{7.16} 
lead to a unique expression for $c_k$ given by 
\bea\label{7.17}
c_k &=& -\frac{4(s+1)(s-k-1)}{s(s-1)}
\sum_{l=0}^k \frac{(-1)^k}{s-l} \binom{s}{l} \binom {s}{l+1} \left\{ 1+(-1)^s \frac{l+1}{s-l+1} \right\}  ~,~~~~  \\
&& \qquad \qquad \qquad  k=0,1,\dots s-2~. \non 
\eea

If the parameter $s$ is odd, $s=2n+1$, with  $n=1,2,\dots$, 
one can check that the equations \eqref{7.16a}--\eqref{7.16c} are identically 
satisfied. 
However, if the parameter $s$ is even, $s=2n$, with $n=1,2,\dots$, 
there appears an inconsistency: 
 the right-hand side of \eqref{7.16c} is positive, while the left-hand side 
is negative, $(s-1) c_{s-2} + c_0 < 0$. Therefore, our solution \eqref{7.17} is only consistent for $s=2n+1, n=1,2,\dots$. 

Relations \eqref{7.15}, \eqref{T7.15} and \eqref{7.17} determine the non-conformal higher-spin supercurrents 
in the massive chiral model \eqref{chiral-massive}.
Unlike the conformal higher-spin supercurrents \eqref{7.15},
the non-conformal ones exist only for the odd values of $s$,
$s=2n+1$, with  $n=1,2,\dots$.
In the flat superspace limit, the above results reduce to those derived in \ref{ss3211}
and in Ref. \cite{BGK1}.


\subsection{Superconformal model with $N$ chiral superfields}

We now  generalise
the  superconformal model \eqref{chiral}
to the case of $N$ covariantly 
chiral scalar superfields $\F^i$, 
$i=1,\dots N$,
\bea
S = \int \rd^4x \rd^2 \q  \rd^2 \bar \q\,E \,{\bar \F}^i \F^i ~,
\qquad {\bar \cD}_\ad \F^i = 0 ~.~
\label{Nchiral-4D}
\eea
The novel feature of the $N>1$ case is that there exist two different types of
conformal supercurrents, which are:
\bea
J^+_{(s,s)} &=& S^{ij}\sum_{k=0}^s (-1)^k
\binom{s}{k}
\left\{ \binom{s}{k+1} 
{\cD}^k_{(1,1)}
 \cD_{(1,0)} \F^i \,\,
{\cD}^{s-k-1}_{(1,1)}
\bar \cD_{(0,1)} 
\bar \F^j  
\right. \non \\ 
&& \left.
 \qquad \qquad
+ \binom{s}{k} 
{\cD}^k_{(1,1)}
  \F^i \,\,
{\cD}^{s-k}_{(1,1)}
\bar \F^j \right\}~, \qquad S^{ij}= S^{ji} 
\label{A.1}
\eea
and
\bea
J^-_{(s,s)} &=& \ri \, A^{ij}\sum_{k=0}^s (-1)^k
\binom{s}{k}
\left\{ \binom{s}{k+1} 
{\cD}^k_{(1,1)}
 \cD_{(1,0)} \F^i \,\,
{\cD}^{s-k-1}_{(1,1)}
\bar \cD_{(0,1)} 
\bar \F^j  
\right. \non \\ 
&& \left.
 \qquad \qquad
+ \binom{s}{k} 
{\cD}^k_{(1,1)}
  \F^i \,\,
{\cD}^{s-k}_{(1,1)}
\bar \F^j \right\}~, \qquad A^{ij}= -A^{ji} ~.~
\label{A.2}
\eea
Here $S$ and $A$ are arbitrary real symmetric and antisymmetric 
constant matrices, respectively. 
We have put an overall factor  $\sqrt{-1} $ in eq.~\eqref{A.2} in order 
to make $J^-_{(s,s)}$  real. 
One can show that the currents \eqref{A.1} and  \eqref{A.2} are conserved on-shell:
\bea
\cD_{(-1,0)} J^\pm_{(s,s)} = 0 \quad \Longleftrightarrow \quad 
\bar \cD_{(0,-1)} J^\pm_{(s,s)} = 0 ~.~
\label{A.3}
\eea

The above results can be recast in terms of the matrix conformal supercurrent
$J_{(s,s)} =\big(J^{ij}_{(s,s)} \big)$ with components
\bea
J^{ij}_{(s,s)} &:=& \sum_{k=0}^s (-1)^k
\binom{s}{k}
\left\{ \binom{s}{k+1} 
{\cD}^k_{(1,1)}
 \cD_{(1,0)} \F^i \,\,
{\cD}^{s-k-1}_{(1,1)}
\bar \cD_{(0,1)} 
\bar \F^j  
\right. \non \\ 
&& \left.
 \qquad \qquad
+ \binom{s}{k} 
{\cD}^k_{(1,1)}
  \F^i \,\,
{\cD}^{s-k}_{(1,1)}
\bar \F^j \right\}
\label{520-4D}~,
\eea
which is  Hermitian, $J_{(s,s)}{}^\dagger = J_{(s,s)}$. 
The chiral action  \eqref{Nchiral-4D}
possesses rigid ${\rm U}(N)$ symmetry acting on the chiral column-vector $\F = (\F^i$) 
by $\F \to g \F$, with $g \in {\rm U}(N)$, which implies that 
the supercurrent \eqref{520-4D} transforms
as $J_{(s,s)} \to gJ_{(s,s)} g^{-1}$.

\subsection{Massive model with $N$ chiral superfields} \label{ss444}
Consider a  theory of $N$ massive chiral 
multiplets with action
\bea
S = \int \rd^4x \rd^2 \q  \rd^2 \bar \q \,E\, \bar \F^i \F^i
+\Big\{\hf \int \rd^4x \rd^2 \q  \, \cE \, M^{ij} \F^i \F^j +{\rm c.c.} \Big\}~,
\label{Nchiralm}
\eea
where $M^{ij}$ is a constant symmetric $N\times N$ mass matrix.
The corresponding equations of motion are
\bea
-\frac{1}{4} (\cD^2-4\bar\m) \F^i  + {\bar M}^{ij} \bar \F^j =0~, \qquad
-\frac{1}{4} ({\bar \cD}^2-4\m) {\bar \F }^i + M^{ij} \F^j =0~.
\eea

First we will consider the case where $S$ is a real and symmetric matrix. 
Making use of the equations of motion,
we obtain 
\bea
\cD_{(-1,0)} J_{(s,s)} &=& 2(s+1) (S\bar M)^{ji} \sum_{k=0}^s (-1)^{k} \binom{s}{k} \binom{s}{k} 
\non \\ 
&& 
\times  \frac{k}{k+1}
 {\cD}^{k-1}_{(1,1)} \,{\bar \cD}_{(0,1)} {\bar \F}^i \,\,{\cD}^{s-k}_{(1,1)}
 {\bar \F}^j 
 \non \\
 &&
 +2(s+1) (S\bar M)^{ji} \sum_{k=0}^{s-1} (-1)^{k} \binom{s}{k} \binom{s}{k+1} 
\non \\ 
&& 
\times
 {\cD}^{k}_{(1,1)} {\bar \F}^i \,\,{\cD}^{s-k-1}_{(1,1)} {\bar \cD}_{(0,1)} {\bar \F}^j ~.
\label{A.4}
\eea
Now, suppose the product ${S \bar M}$ is symmetric, which implies $[S,\bar M]=0$. Then, \eqref{A.4} becomes
\bea
\cD_{(-1,0)} J_{(s,s)} &=& 2(s+1) (S\bar M)^{ij} \sum_{k=0}^{s-1} (-1)^{k} \binom{s}{k} \binom{s}{k+1} 
\non \\ 
&& 
\times \left\{1+(-1)^s \frac{k+1}{s-k+1}\right\} 
 {\cD}^{k}_{(1,1)} {\bar \F}^i \,\,{\cD}^{s-k-1}_{(1,1)} {\bar \cD}_{(0,1)} {\bar \F}^j ~.
\label{A.4a}
\eea
We now look for a superfield $\bar T_{(s-2, s-1)}$
such that (i) it obeys the transverse antilinear constraint \eqref{7.8b}; and 
(ii) it satisfies the conservation equation \eqref{7.9b}:
\bea
\cD_{(-1,0)} J_{(s,s)} = \frac{s}{2} \bar A_{(1,1)} \bar T_{(s-2, s-1)}~. 
\eea
As in the single field case we consider a general ansatz
\bea
\bar T_{(s-2, s-1)} = (S\bar M)^{ij}\sum_{k=0}^{s-2} c_k 
{\cD}^k_{(1,1)} \bar \F^i\,
{\cD}^{s-k-2}_{(1,1)}
 \bar \cD_{(0,1)} \bar \F^j ~.
\label{A.5}
\eea
Then for $k = 1,2,...s-2$, condition (i) implies that 
the coefficients $c_k$ must satisfy
\begin{subequations}\label{A.6}
\begin{align}
kc_k = (s-k-1) c_{s-k-1}~,\label{A.6a}
\end{align}
while (ii) gives the following equation
\begin{align}
c_{s-k-1} + s c_k + (s-1) c_{k-1} &= -4(-1)^k \frac{s+1}{s} \binom{s}{k} \binom{s}{k+1} 
\non \\
& \qquad \qquad \times \left\{ 1+ (-1)^s \frac{k+1}{s-k+1} \right\} ~.\label{A.6b}
\end{align}
Condition (ii) also implies that 
\begin{align}
(s-1) c_{s-2} +c_0 &= 4(-1)^s (s+1)\left\{1+(-1)^s \frac{s}{2}\right\}~, \label{A.6c}\\
c_0 &= -\frac{4}{s}(s+1+(-1)^s)~. \label{A.6d}
\end{align}
\end{subequations}
The above conditions coincide with eqs.\eqref{7.16a}--\eqref{7.16d} in the case of a single, massive chiral superfield, which are satisfied only for $s= 2n+1, n=1,2,\dots$.
Hence, the solution for the coefficients $c_k$ is given by~\eqref{7.17} for odd values of $s$ and there is no solution for even $s$.

On the other hand, if $S\bar M$ is antisymmetric (which is equivalent to $\{S,\bar M\}=0$), eq.~\eqref{A.4a} is slightly modified
\bea
\cD_{(-1,0)} J_{(s,s)} &=& 2(s+1) (S\bar M)^{ij} \sum_{k=0}^{s-1} (-1)^{k} \binom{s}{k} \binom{s}{k+1} 
\non \\ 
&& 
\times \left\{-1+(-1)^s \frac{k+1}{s-k+1}\right\} 
 {\cD}^{k}_{(1,1)} {\bar \F}^i \,\,{\cD}^{s-k-1}_{(1,1)} {\bar \cD}_{(0,1)} {\bar \F}^j ~.
\eea
Starting with a  general ansatz 
\bea
\bar T_{(s-2, s-1)} = (S\bar M)^{ij}\sum_{k=0}^{s-2} d_k 
{\cD}^k_{(1,1)} \bar \F^i\,
{\cD}^{s-k-2}_{(1,1)}
 \bar \cD_{(0,1)} \bar \F^j 
 \label{TA.7}
\eea
and imposing conditions (i) and (ii) yield the following equations for the coefficients $d_k$
\begin{subequations}\label{A.7}
\begin{align}
kd_k = -(s-k-1) d_{s-k-1}~.\label{A.7a}
\end{align}
\begin{align}
-d_{s-k-1} + s d_k + (s-1) d_{k-1} &= -4(-1)^k \frac{s+1}{s} \binom{s}{k} \binom{s}{k+1} 
\non \\
& \qquad \qquad \times \left\{ -1+ (-1)^s \frac{k+1}{s-k+1} \right\} ~.\label{A.7b}
\end{align}
\begin{align}
(s-1) d_{s-2} -d_0 &= 4(-1)^s (s+1)\left\{-1+(-1)^s \frac{s}{2}\right\}~. \label{A.7c}\\
d_0 &= \frac{4}{s}(s+1+(-1)^{s-1})~. \label{A.7d}
\end{align}
\end{subequations}
The equations \eqref{A.7} 
lead to a unique expression for $d_k$ given by 
\bea\label{A.8}
d_k &=& -\frac{4(s+1)(s-k-1)}{s(s-1)}
\sum_{l=0}^k \frac{(-1)^k}{s-l} \binom{s}{l} \binom {s}{l+1} \left\{ -1+(-1)^s \frac{l+1}{s-l+1} \right\}  ~,~~~~  \\
&& \qquad \qquad \qquad  k=0,1, \dots s-2~. \non 
\eea
If the parameter $s$ is even, $s=2n$, with  $n=1,2,\dots$, 
one can check that the equations \eqref{A.7a}--\eqref{A.7d} are identically 
satisfied. 
However, if the parameter $s$ is odd, $s=2n+1$, with $n=1,2,\dots$, 
there appears an inconsistency: 
 the right-hand side of \eqref{A.7c} is positive, while the left-hand side 
is negative, $(s-1) d_{s-2} - d_0 < 0$. Therefore, our solution \eqref{A.8} is only consistent for $s=2n, n=1,2,\dots$. 

Finally, we consider $A^{ij}=-A^{ji}$ with the corresponding $J_{(s,s)}$ given by~\eqref{A.2}. 
The analysis in this case is similar to the one presented above and we will simply state the results. 
If $s$ is odd the non-conformal higher-spin supercurrents exist if  $\{A,\bar M\}=0$. The trace supercurrent $\bar T_{(s-2, s-1)}$ is given by~\eqref{A.5}
with the coefficients $c_k$ given by 
\bea\label{eee1S}
c_k &=& \ri \frac{4(s+1)(s-k-1)}{s(s-1)}
\sum_{l=0}^k \frac{(-1)^k}{s-l} \binom{s}{l} \binom {s}{l+1} \left\{ 1+(-1)^s \frac{l+1}{s-l+1} \right\}  ~,~~~~  \\
&& \qquad \qquad \qquad  k=0,1,\dots s-2~. \non 
\eea
If $s$ is even the non-conformal higher-spin supercurrents exist if  $[A,\bar M]=0$. The trace supercurrent $\bar T_{(s-2, s-1)}$ is given by~\eqref{TA.7}
with the coefficients $d_k$ given by
\bea\label{eee2}
d_k &=& \ri \frac{4(s+1)(s-k-1)}{s(s-1)}
\sum_{l=0}^k \frac{(-1)^k}{s-l} \binom{s}{l} \binom {s}{l+1} \left\{ -1+(-1)^s \frac{l+1}{s-l+1} \right\}  ~,~~~~  \\
&& \qquad \qquad \qquad  k=0,1, \dots s-2~. \non 
\eea

Note that the coefficients $c_k$ in~\eqref{eee1} differ from similar coefficients in~\eqref{7.17} by a factor of $- \ri$. This means that for odd $s$ we can define a more general supercurrent
\bea
J_{(s,s)} &=& H^{ij}\sum_{k=0}^s (-1)^k
\binom{s}{k}
\left\{ \binom{s}{k+1} 
{\cD}^k_{(1,1)}
 \cD_{(1,0)} \F^i \,\,
{\cD}^{s-k-1}_{(1,1)}
\bar \cD_{(0,1)} 
\bar \F^j  
\right. \non \\ 
&& \left.
 \qquad \qquad
+ \binom{s}{k} 
{\cD}^k_{(1,1)}
  \F^i \,\,
{\cD}^{s-k}_{(1,1)}
\bar \F^j \right\} ~,
\label{eee3}
\eea
where $H^{ij}$ is a generic matrix which can be split into the symmetric and antisymmetric parts $H^{ij} = S^{ij} + \ri A^{ij}$. Here both $S$ and $A$ are real 
and we put an $\ri$ in front of $A$ because $J_{(s,s)} $ must be real. From the above consideration it then follows that the corresponding more general solution for $\bar T_{(s-2, s-1)}$ reads
\bea
\bar T_{(s-2, s-1)} = (\bar H\bar M)^{ij}\sum_{k=0}^{s-2} c_k 
{\cD}^k_{(1,1)} \bar \F^i\,
{\cD}^{s-k-2}_{(1,1)}
 \bar \cD_{(0,1)} \bar \F^j ~,
\label{A.5a}
\eea
where $[S, \bar M]=0$, $\{A, \bar M\}=0$ and $c_k$ are, as before, given by eq.~\eqref{7.17}.
Similarly, the coefficients $d_k$ in~\eqref{eee2} differ from similar coefficients in~\eqref{A.8} by a factor of $- \ri$. This means that for even $s$ we can define a more general 
supercurrent~\eqref{eee3}, where $H^{ij}$ is a generic matrix which we can split as before into the symmetric and antisymmetric parts,  $H^{ij} = S^{ij} + \ri A^{ij}$. 
From the above consideration it then follows that the corresponding more general solution for $\bar T_{(s-2, s-1)}$ reads
\bea
\bar T_{(s-2, s-1)} = (\bar H\bar M)^{ij}\sum_{k=0}^{s-2} d_k 
{\cD}^k_{(1,1)} \bar \F^i\,
{\cD}^{s-k-2}_{(1,1)}
 \bar \cD_{(0,1)} \bar \F^j ~,
\label{A.5b}
\eea
where $\{S, \bar M\}=0$, $[A, \bar M]=0$ and $d_k$ are given by eq.~\eqref{A.8}.

\section{Higher-spin supercurrents for chiral superfields: Integer superspin} \label{s45}

In this section we provide explicit realisations for the fermionic higher-spin supercurrents 
(integer superspin) in models described by chiral scalar superfields. 


\subsection{Massive hypermultiplet model}

Consider a free massive hypermultiplet in AdS
\bea
S = \int \rd^4x \rd^2 \q  \rd^2 \bar \q \,E\, \Big( \bar \J_+ \J_+
+\bar \J_- \J_-\Big)
+\Big\{ {m} \int \rd^4x \rd^2 \q  \,\cE\, \J_+ \J_- +{\rm c.c.} \Big\}~,
\label{hyper1}
\eea
where the superfields $\J_\pm$ are covariantly chiral, $\bar \cD_\ad \J_\pm =0$ and $m$ is a complex mass parameter.\footnote{This model possesses
off-shell $\cN=2$ AdS supersymmetry \cite{BKsigma,KT-M-ads}.}
By a change of variables it is possible to make $m$ real. 
Let us introduce another set of fields $\F_\pm$, $\bar \cD_\ad \F_\pm =0$,  related to $\J_\pm$ by the following transformations
\bea \label{phase}
\F_\pm = e^{\ri \a/2} \J_\pm~, \qquad
m = M e^{\ri \a}~.
\eea
Under the transformations \eqref{phase}, the action \eqref{hyper1} turns into
\bea
S = \int \rd^4x \rd^2 \q  \rd^2 \bar \q \,E \, \Big( \bar \F_+ \F_+
+\bar \F_- \F_-\Big)
+\Big\{ {M} \int \rd^4x \rd^2 \q \,\cE \, \F_+ \F_- +{\rm c.c.} \Big\}~,
\label{hyper21}
\eea
where the mass parameter $M$ is now real.
In the massless case, $M=0$, 
the conserved fermionic supercurrent $J_{\a(s) \ad(s-1)}$ was constructed in~\cite{KMT} and 
is given by
\bea
J_{(s,s-1)} &=& \sum_{k=0}^{s-1} (-1)^k
\binom{s-1}{k}
\left\{ \binom{s}{k+1} 
{\cD}^k_{(1,1)}
 \cD_{(1,0)} \F_{+} \,\,
{\cD}^{s-k-1}_{(1,1)}
 \F_{-}  
\right. \non \\ 
&& \left.
 \qquad \qquad
- \binom{s}{k} 
{\cD}^k_{(1,1)}
  \F_{+} \,\,
{\cD}^{s-k-1}_{(1,1)}
\cD_{(1,0)} \F_{-} \right\}~.
\label{4.8}
\eea
Making use of
the massless equations of motion,  $-\frac{1}{4}(\cD^2-4\bar \m)\, \F_\pm = 0$, 
one may check that $J_{(s,s-1)}$ obeys, 
for $s > 1$,  the conservation equations
\bea
\cD_{(-1,0)} J_{(s,s-1)} = 0, \qquad
\bar \cD_{(0,-1)} J_{(s,s-1)} = 0 ~.~
\label{4.9}
\eea

We will construct fermionic higher-spin supercurrents 
corresponding to the massive model \eqref{hyper21}.
Making use of the massive equations of motion 
\bea
-\frac{1}{4} (\cD^2-4\bar\m) \F_+ +M \bar \F_{-} =0, \qquad
-\frac{1}{4} (\cD^2-4\bar\m)\F_- +M \bar \F_{+} =0,
\eea
we obtain 
\bea
\cD_{(-1,0)} J_{(s,s-1)} &=& 2M (s+1) \sum_{k=0}^{s-1} (-1)^{k+1} \binom{s-1}{k} \binom{s}{k} \non \\
&&\qquad \times \left\{ -\frac{s-k}{k+1} {\cD}^k_{(1,1)} \bar \F_- \,{\cD}^{s-k-1}_{(1,1)} \F_-
+ {\cD}^k_{(1,1)} \F_+ \,{\cD}^{s-k-1}_{(1,1)} \bar \F_+ \right\}\non \\
&&+ 2M(s+1) \sum_{k=1}^{s-1} (-1)^{k+1} \binom{s-1}{k} \binom{s}{k} \frac{k}{k+1} 
\non \\
&& \qquad \times {\cD}^{k-1}_{(1,1)} \bar \cD_{(0,1)} \bar \F_- \,\,{\cD}^{s-k-1}_{(1,1)} \cD_{(1,0)} \F_- 
\non \\
&&+ 2M(s+1)\sum_{k=0}^{s-2} (-1)^{k+1} \binom{s-1}{k} \binom{s}{k} \frac{s-1-k}{k+1} 
\non \\
&& \qquad \times {\cD}^{k}_{(1,1)} \cD_{(1,0)} \F_+ \,\,{\cD}^{s-k-2}_{(1,1)} \bar \cD_{(0,1)} \bar \F_+ ~.~ \label{4.10}
\eea
It can be shown that the massive supercurrent $J_{(s,s-1)}$ also obeys \eqref{4.5a}. 

We now look for a superfield $T_{(s-1,s-1)}$ such that (i) it obeys the longitudinal linear constraint \eqref{4.5b}; and 
(ii) it satisfies \eqref{4.7}, which is a consequence of the conservation equation \eqref{4.6}. 
For this we consider a general ansatz 
\bea
T_{(s-1, s-1)} &=& 
\sum_{k=0}^{s-1} c_k \,{\cD}^k_{(1,1)} \F_-\,\, {\cD}^{s-k-1}_{(1,1)} \bar \F_-  \non \\
&&+ \sum_{k=0}^{s-1} d_k \,{\cD}^k_{(1,1)} \F_+ \,\,
{\cD}^{s-k-1}_{(1,1)} \bar \F_+  \non \\
&&+ \sum_{k=1}^{s-1} f_k \,{\cD}^{k-1}_{(1,1)} \cD_{(1,0)} \F_-\,\, {\cD}^{s-k-1}_{(1,1)} \bar \cD_{(0,1)} \bar \F_-  \non \\
&&+ \sum_{k=1}^{s-1} g_k \,{\cD}^{k-1}_{(1,1)}  \cD_{(1,0)} \F_+\,\, {\cD}^{s-k-1}_{(1,1)} \bar \cD_{(0,1)} \bar \F_+ ~.
\label{T4.11}
\eea
Condition (i) implies that the coefficients must be related by
\begin{subequations} 
\bea
c_0 = d_0 = 0~, \qquad f_k = c_k~, \qquad g_k = d_k~, 
\label{qqq1}
\eea
while for $k=1,2, \dots s-2$, condition  (ii) gives the following recurrence relations:
\bea 
c_k + c_{k+1} &=& 
\frac{M(s+1)}{s} (-1)^{s+k} \binom{s-1}{k} \binom{s}{k} 
\non \\
&& \times \frac{1}{(k+2)(k+1)} \Big\{(2k+2-s)(s+1)-k-2\Big\}~, \\
d_k + d_{k+1} &=& \frac{M(s+1)}{s} (-1)^{k} \binom{s-1}{k} \binom{s}{k} 
\non \\
&& \times \frac{1}{(k+2)(k+1)} \Big\{(2k+2-s)(s+1)-k-2\Big\}~.
\eea
Condition (ii) also implies that
\bea
c_1 = -(-1)^s \frac{M(s^2-1)}{2}~, \qquad c_{s-1} &=&- \frac{M(s^2-1)}{s}~;\\
d_1 =- \frac{M(s^2-1)}{2}~, \qquad d_{s-1} &=& -(-1)^s \frac{M(s^2-1)}{s}~.
\eea
\end{subequations}
The above conditions lead to simple expressions for $c_k$ and $d_k$:
\begin{subequations}
\bea
d_k &=& \frac{M(s+1)}{s} \frac{k}{k+1} (-1)^{k} \binom{s-1}{k} \binom{s}{k}~,\\
c_k &=& (-1)^s d_k ~,
\eea
\label{qqq2}
\end{subequations}
where $ k=1,2,\dots s-1$.


\subsection{Superconformal model with $N$ chiral superfields}


In this subsection we will generalise the above results for $N$ chiral superfields $\Phi^i$, $i=1, \dots N$. We first consider the superconformal model \eqref{Nchiral-4D}.
Let us construct the following fermionic supercurrent
\bea
J_{(s,s-1)} &=& C^{ij}\sum_{k=0}^{s-1} (-1)^k
\binom{s-1}{k}
\left\{ \binom{s}{k+1} 
{\cD}^k_{(1,1)}
 \cD_{(1,0)} \F^i \,\,
{\cD}^{s-k-1}_{(1,1)} \F^j  
\right. \non \\ 
&& \left.
 \qquad \qquad
- \binom{s}{k} 
{\cD}^k_{(1,1)}
  \F^i \,\,
{\cD}^{s-k-1}_{(1,1)} \cD_{(1,0)} \F^j \right\}~, 
\label{B.1}
\eea
where $C^{ij}$ is a constant complex matrix. By changing the summation index it is not hard to show that $J_{(s,s-1)} =0$ if  (i) $s$ is odd and 
$C^{ij}$ is symmetric; and 
 (ii) 
 $s$ is even and $C^{ij}$ is antisymmetric, that is
 \begin{subequations}
 \bea
 C^{ij} =C^{ji} ~, \quad s=1,3,\dots  \quad & \Longrightarrow \quad J_{(s,s-1)}=0~; \\
 C^{ij} =-C^{ji} ~, \quad s=2,4,\dots  \quad & \Longrightarrow \quad J_{(s,s-1)}=0~.
 \eea
 \end{subequations}
This means that we have to consider the two separate cases: the case of even $s$ with symmetric $C$, and the case of odd $s$ with antisymmetric $C$. 
Using the massless equation of motion,  $-\frac{1}{4}(\cD^2-4\bar \m)\, \F^i = 0$, one may check that $J_{(s,s-1)}$ satisfies the conservation equations \eqref{4.9}
\bea
\cD_{(-1,0)} J_{(s,s-1)} = 0~, \qquad
\bar \cD_{(0,-1)} J_{(s,s-1)} = 0~.
\label{B.2}
\eea

In the case of a single chiral superfield, the supercurrent \eqref{B.1} exists 
for even $s$, 
\bea
J_{(s,s-1)} &=& 2 \sum_{k=0}^{s-1} (-1)^k
\binom{s-1}{k} \binom{s}{k+1} 
{\cD}^k_{(1,1)}
 \cD_{(1,0)} \F \,
{\cD}^{s-k-1}_{(1,1)} \F
\label{6.14}
\eea
The flat-superspace version of \eqref{6.14} is given by eq.~\eqref{JJint} and Ref. \cite{BGK3}.
 

\subsection{Massive model with $N$ chiral superfields}

Let us turn to the massive model \eqref{Nchiralm}. As was discussed in previous subsection, to construct the conserved currents we first have to 
calculate ${\cD}_{(-1,0)} J_{(s,s-1)} $ using the equations of motion in the massive theory. The calculation depends on whether $C^{ij}$ is symmetric or antisymmetric. 

\subsubsection{Symmetric $C$}

If $C^{ij}$ is a symmetric matrix, using the massive equation of motion, we obtain
\bea
{\cD}_{(-1,0)} J_{(s,s-1)} &=& -2(s+1)(C\bar M)^{ji}\sum_{k=0}^{s-1} (-1)^{k+1} \binom{s-1}{k} \binom{s}{k} \frac{s-k}{k+1} \non \\
&&\qquad \times {\cD}^k_{(1,1)} \bar \F^i \,{\cD}^{s-k-1}_{(1,1)} \F^j \non\\
&&+ 2(s+1)(C\bar M)^{ij}\sum_{k=0}^{s-1} (-1)^{k+1} \binom{s-1}{k} \binom{s}{k} \non \\
&& \qquad \times {\cD}^k_{(1,1)} \F^i \,{\cD}^{s-k-1}_{(1,1)} \bar \F^j
\non \\
&&+ 2(s+1) (C\bar M)^{ji} \sum_{k=1}^{s-1} (-1)^{k+1} \binom{s-1}{k} \binom{s}{k} \frac{k}{k+1}
\non \\
&& \qquad \times {\cD}^{k-1}_{(1,1)} \bar \cD_{(0,1)} \bar \F^i \,\,{\cD}^{s-k-1}_{(1,1)} \cD_{(1,0)} \F^j
\non \\
&&+ 2(s+1) (C\bar M)^{ij}\sum_{k=0}^{s-2} (-1)^{k+1} \binom{s-1}{k} \binom{s}{k} \frac{s-1-k}{k+1} 
\non \\
&& \qquad \times {\cD}^{k}_{(1,1)} \cD_{(1,0)} \F^i \,\,{\cD}^{s-k-2}_{(1,1)} \bar \cD_{(0,1)} \bar \F^j  ~.~ 
\label{B.3}
\eea

Here we have two cases to consider:
\begin{enumerate}
\item $C \bar M$ is symmetric $\Longleftrightarrow [C,\bar M]=0, \,\, s$ even.
\item $C \bar M$ is antisymmetric $\Longleftrightarrow \{C,\bar M\}=0, \,\,s$ even.
\end{enumerate}
\textbf{Case 1:} Eq. \eqref{B.3} can be simplified to yield
\bea
{\cD}_{(-1,0)} J_{(s,s-1)} &=& 4(s+1)(C\bar M)^{ij}\sum_{k=0}^{s-1} (-1)^{k+1} \binom{s-1}{k} \binom{s}{k} \non \\
&&\qquad \times {\cD}^k_{(1,1)} \F^i \,{\cD}^{s-k-1}_{(1,1)} \bar \F^j \non\\
&&+ 4(s+1) (C\bar M)^{ij} \sum_{k=1}^{s-1} (-1)^{k+1} \binom{s-1}{k} \binom{s}{k} \frac{k}{k+1}
\non \\
&& \qquad \times {\cD}^{k-1}_{(1,1)} \bar \cD_{(0,1)}\bar \F^i \,\,{\cD}^{s-k-1}_{(1,1)} \cD_{(1,0)} \F^j ~.~ 
\label{B.4}
\eea
We now look for a superfield $T_{(s-1,s-1)}$ such that (i) it obeys the longitudinal linear constraint \eqref{4.5b}; and (ii) it satisfies \eqref{4.7}, 
which is a consequence of the conservation equation \eqref{4.6}.
The precise form of eq.~\eqref{4.7} in the present case is
\bea
&&\frac{1}{2s} \cD_{(1,0)} \left\{\cD_{(-1,0)} J_{(s,s-1)} + \bar \cD_{(0,-1)} \bar J_{(s-1,s)}\right\}\non\\
&&= \frac{2}{s+1} {\cD}_{(1,0)} \sum_{k=0}^{s-1}(-1)^{k+1} \binom{s-1}{k} \binom{s}{k} \non \\
&&\qquad \times \left\{\frac{s}{k+1} (C \bar M)^{ij}-\frac{(s+1)(s-k)}{(k+1)(k+2)} (\bar C M)^{ij} \right\} \non\\
&& \qquad \times {\cD}^k_{(1,1)} \F^i \,{\cD}^{s-k-1}_{(1,1)} \bar \F^j \non \\
&& =-\cD_{(1,0)} T_{(s-1,s-1)} ~. \label{B.4a}
\eea
To find $T_{(s-1,s-1)}$ we  consider a general ansatz 
\bea
T_{(s-1, s-1)} &=& 
\sum_{k=0}^{s-1} (c_k)^{ij} \,{\cD}^k_{(1,1)} \F^i\,\, {\cD}^{s-k-1}_{(1,1)} \bar \F^j  \non \\
&&+ \sum_{k=1}^{s-1} (d_k)^{ij} \,{\cD}^{k-1}_{(1,1)} \cD_{(1,0)} \F^j\,\, {\cD}^{s-k-1}_{(1,1)} \bar \cD_{(0,1)} \bar \F^j ~.
\label{B.5}
\eea
It is possible to show that no solution for $T_{(s-1, s-1)} $ can be found unless we impose\footnote{Since $C$ and ${\bar M}$ commute 
we can take them both to be diagonal, $C= {\rm diag} (c_1, \dots, c_N)$, $M= {\rm diag} (m_1, \dots, m_N)$. 
Then the condition~\eqref{e1} means that ${\rm arg} (c_i) - {\rm arg} (m_i) =n_i \pi$ for some integers $n_i$.} 
\be 
C \bar M = \bar C M~.
\label{e1}
\ee
Furthermore,  condition (i) implies that the coefficients must be related by
\begin{subequations} 
\bea
(c_0)^{ij} = 0~, \qquad (c_k)^{ij} = (d_k)^{ij}~,
\eea
while for $k=1,2, \dots s-2$, while condition (ii) and eq.~\eqref{e1} gives the following recurrence relations 
\bea 
(d_k)^{ij} + (d_{k+1})^{ij} &=& -2\frac{(s+1)}{s} (C\bar M)^{ij} (-1)^{k+1} \binom{s-1}{k} \binom{s}{k} 
\non \\
&& \times \frac{1}{k+1} \Big\{s-\frac{(s+1)(s-k)}{k+2}\Big\}~.
\eea
Condition (ii) also implies that
\bea
(d_1)^{ij} = (1-s^2) (C \bar M)^{ij}~, \qquad (d_{s-1})^{ij} &=& \frac{2(1-s^2)}{s} (C \bar M)^{ij}~.
\eea
\label{ee1}
\end{subequations}
The above conditions lead to simple expressions for $d_k$:
\bea
(d_k)^{ij} &=& \frac{2(s+1)}{s} (C \bar M)^{ij} \frac{k}{k+1} (-1)^{k} \binom{s-1}{k} \binom{s}{k}~,
\eea
where $ k=1,2,\dots s-1$ and $s$ is even.\\
\textbf{Case 2:}
If we take $C \bar M$ to be antisymmetric, a similar analysis shows that no solution for $T_{(s-1, s-1)}$ exists for even values of $s$. 

\subsubsection{Antisymmetric $C$}
If $C^{ij}$ is antisymmetric we get: 
\bea 
{\cD}_{(-1,0)} J_{(s,s-1)} &=& 2(s+1)(C\bar M)^{ji}\sum_{k=0}^{s-1} (-1)^{k+1} \binom{s-1}{k} \binom{s}{k} \frac{s-k}{k+1} \non \\
&&\qquad \times {\cD}^k_{(1,1)} \bar \F^i \,{\cD}^{s-k-1}_{(1,1)} \F^j \non\\
&&+ 2(s+1)(C\bar M)^{ij}\sum_{k=0}^{s-1} (-1)^{k+1} \binom{s-1}{k} \binom{s}{k} \non \\
&& \qquad \times {\cD}^k_{(1,1)} \F^i \,{\cD}^{s-k-1}_{(1,1)} \bar \F^j
\non \\
&&- 2(s+1) (C\bar M)^{ji} \sum_{k=1}^{s-1} (-1)^{k+1} \binom{s-1}{k} \binom{s}{k} \frac{k}{k+1}
\non \\
&& \qquad \times {\cD}^{k-1}_{(1,1)} \bar \cD_{(0,1)} \bar \F^i \,\,{\cD}^{s-k-1}_{(1,1)} \cD_{(1,0)} \F^j
\non \\
&&+ 2(s+1) (C\bar M)^{ij}\sum_{k=0}^{s-2} (-1)^{k+1} \binom{s-1}{k} \binom{s}{k} \frac{s-1-k}{k+1} 
\non \\
&& \qquad \times {\cD}^{k}_{(1,1)} \cD_{(1,0)} \F^i \,\,{\cD}^{s-k-2}_{(1,1)} \bar \cD_{(0,1)} \bar \F^j  ~.~ 
\label{B.3.new}
\eea
As in the symmetric $C$ case, there are also two cases to consider:
\begin{enumerate}
\item $C \bar M$ is symmetric $\Longleftrightarrow \{C,\bar M\}=0, \,\,s$ odd.
\item $C \bar M$ is antisymmetric $\Longleftrightarrow [C,\bar M]=0, \,\,s$ odd.
\end{enumerate}
\textbf{Case 1:}
Using eq.~\eqref{B.3.new} and keeping in mind that $s$ is odd, we obtain
\bea
{\cD}_{(-1,0)} J_{(s,s-1)} &=& 4(s+1)(C\bar M)^{ij}\sum_{k=0}^{s-1} (-1)^{k+1} \binom{s-1}{k} \binom{s}{k} \non \\
&&\qquad \times {\cD}^k_{(1,1)} \F^i \,{\cD}^{s-k-1}_{(1,1)} \bar \F^j \non\\
&&- 4(s+1) (C\bar M)^{ij} \sum_{k=1}^{s-1} (-1)^{k+1} \binom{s-1}{k} \binom{s}{k} \frac{k}{k+1}
\non \\
&& \qquad \times {\cD}^{k-1}_{(1,1)} \bar \cD_{(0,1)}\bar \F^i \,\,{\cD}^{s-k-1}_{(1,1)} \cD_{(1,0)} \F^j ~.~ 
\label{B.6}
\eea
Then it follows that eq.~\eqref{4.7} becomes
\bea
&&\frac{1}{2s} \cD_{(1,0)} \left\{\cD_{(-1,0)} J_{(s,s-1)} + \bar \cD_{(0,-1)} \bar J_{(s-1,s)}\right\}\non\\
&&= \frac{2}{s+1} {\cD}_{(1,0)} \sum_{k=0}^{s-1}(-1)^{k+1} \binom{s-1}{k} \binom{s}{k} \non \\
&&\qquad \times \left\{\frac{s}{k+1} (C \bar M)^{ij}-\frac{(s+1)(s-k)}{(k+1)(k+2)} (\bar C M)^{ij} \right\} \non\\
&& \qquad \times {\cD}^k_{(1,1)} \F^i \,{\cD}^{s-k-1}_{(1,1)} \bar \F^j \non \\
&& =-\cD_{(1,0)} T_{(s-1,s-1)} ~. \label{B.6a}
\eea
Note that it is the equation same as eq.~\eqref{B.4a} which means that the solution for $T_{(s-1,s-1)}$ is the same as in \textbf{Case 1}. That is, 
the matrices $C$ and $M$ must satisfy $C \bar M = \bar C M$, $T_{(s-1,s-1)}$ is given by eq.~\eqref{B.5} and the coefficients 
$(c_k)^{ij}, (d_k)^{ij}$ are given by eqs.~\eqref{ee1}.
\textbf{Case 2:}
If we take $C \bar M$ to be antisymmetric, a similar analysis shows that no solution for $T_{(s-1, s-1)}$ exists for odd values of $s$. 

\subsubsection{Massive hypermultiplet model revisited}

As a consistency check of our general method, let us reconsider the case of a hypermultiplet studied previously. 
For this we will take $N=2$, the mass matrix in the form 
\be
M=
\begin{pmatrix}
0 & m \\
m \,& 0 
\end{pmatrix} ~,
\label{ee2}
\ee
and denote $\Phi^i = (\Phi_+, \Phi_-)$. If $s$ is even we will take $C$ in the form
\be
C=
\begin{pmatrix}
0 & c \\
c \,& 0 
\end{pmatrix} ~.
\label{ee3}
\ee
Note that $C$ commutes with $M$. The condition $C \bar M= \bar C M$ is equivalent to ${\rm arg} (c)= {\rm arg} (m) +n \pi$. For simplicity, let us choose both $c$ and $m$ 
to be real. Under these conditions eq.~\eqref{B.1} for $J_{(s, s-1)}$ becomes
\bea
J_{(s,s-1)} &=& c\sum_{k=0}^{s-1} (-1)^k
\binom{s-1}{k} \binom{s}{k+1} 
\left\{{\cD}^k_{(1,1)}
 \cD_{(1,0)} \F_+ \,\,
{\cD}^{s-k-1}_{(1,1)} \F_- \right.
\non \\
&& \left. \qquad +  {\cD}^k_{(1,1)} \cD_{(1,0)} \F_- \,\,
{\cD}^{s-k-1}_{(1,1)} \F_+ \right\} 
\non \\ 
&&+ c\sum_{k=0}^{s-1} (-1)^{k+1}
\binom{s-1}{k} \binom{s}{k} 
\left\{\cD^k_{(1,1)} \F_+ \,\,
{\cD}^{s-k-1}_{(1,1)} {\cD}_{(1,0)} \F_- \right.
\non \\
&& \left. \qquad  +\cD^k_{(1,1)} \F_- \,\,
{\cD}^{s-k-1}_{(1,1)} {\cD}_{(1,0)} \F_+ \right\}~.
\label{ee4}
\eea
Introducing a new summation variable $k'= s-1-k$ for the second and fourth terms, we obtain
\bea
J_{(s,s-1)} &=& c\sum_{k=0}^{s-1} (-1)^k
\binom{s-1}{k} \binom{s}{k+1} 
\Big[(1+(-1)^s\Big]
{\cD}^k_{(1,1)}
 \cD_{(1,0)} \F_+ \,\,
{\cD}^{s-k-1}_{(1,1)} \F_- \non\\
&&- c\sum_{k=0}^{s-1} (-1)^{k}
\binom{s-1}{k} \binom{s}{k} 
\Big[(1+(-1)^s\Big]
{\cD}^{k}_{(1,1)} \F_+ {\cD}^{s-k-1}_{(1,1)} {\cD}_{(1,0)} \F_- ~.
\label{J.even}
\eea
We see that for even $s$ it coincides with the hypermultiplet supercurrent given by~\eqref{4.8} up to an overall coefficient $2c$. 
If $s$ is odd we have to choose $C$ to be antisymmetric 
\be
C=
\begin{pmatrix}
0 & c \\
-c \,& 0 
\end{pmatrix} ~.
\label{ee5}
\ee
Note that $C$ now anticommutes with $M$. For simplicity, we again choose $c$ and $m$ to be real. Now the expression~\eqref{B.1} for $J_{(s, s-1)}$ becomes
\bea
J_{(s,s-1)} &=& c\sum_{k=0}^{s-1} (-1)^k
\binom{s-1}{k} \binom{s}{k+1} 
\Big[(1- (-1)^s\Big]
{\cD}^k_{(1,1)}
 \cD_{(1,0)} \F_+ \,\,
{\cD}^{s-k-1}_{(1,1)} \F_- \non\\
&&- c\sum_{k=0}^{s-1} (-1)^{k}
\binom{s-1}{k} \binom{s}{k} 
\Big[(1- (-1)^s\Big]
{\cD}^{k}_{(1,1)} \F_+ {\cD}^{s-k-1}_{(1,1)} {\cD}_{(1,0)} \F_- ~.
\label{J.odd}
\eea
We see that for odd $s$ it coincides with the hypermultiplet supercurrent given by~\eqref{4.8} up to an overall coefficient $2c$. 
To summarise, we reproduced the hypermultiplet supercurrent~\eqref{4.8} for both even and odd values of $s$. However, for even $s$ it came 
from a symmetric matrix~\eqref{ee3} and for odd $s$ it came from an antisymmetric matrix~\eqref{ee5}.

Let us now consider $T_{(s-1, s-1)}$. First, we will note that the product $C \bar M$ is given by 
\be
C \bar M= c m 
\begin{pmatrix}
1 & 0 \\
0\,& (-1)^s 
\end{pmatrix} ~.
\label{ee6}
\ee
This means that $T_{(s-1, s-1)}$ is given by the following expression valid for all values of  $s$
\be
T_{(s-1, s-1)} = 
\sum_{k=0}^{s-1} (d_k)^{ij} \, \Big[ {\cD}^k_{(1,1)} \F^i\,\, {\cD}^{s-k-1}_{(1,1)} \bar \F^j  
+  {\cD}^{k-1}_{(1,1)} \cD_{(1,0)} \F^j\,\, {\cD}^{s-k-1}_{(1,1)} \bar \cD_{(0,1)} \bar \F^j  \Big]~, 
\label{ee7}
\ee
where the matrix $(d_k)^{ij}$ is given by 
\be
(d_k)^{ij} = 2 c m \frac{s+1}{s} \frac{k}{k+1} (-1)^k \binom{s-1}{k} \binom{s}{k} 
\begin{pmatrix}
1 & 0 \\
0\,& (-1)^s 
\end{pmatrix} ~.
\label{ee8}
\ee
It is easy to see that this expression for $T_{(s-1, s-1)}$ coincides with the one obtained for the hypermultiplet in the previous subsections 
in eqs.~\eqref{T4.11}, \eqref{qqq1}, \eqref{qqq2} up to an overall factor $2 c$.


\section{Summary and applications} \label{s46}

In this chapter, we have described higher-spin conserved supercurrents 
for ${\cal N}=1$ supersymmetric theories in four-dimensional anti-de Sitter space. 
We have explicitly constructed such supercurrents 
in the case of $N$ chiral scalar superfields with an arbitrary mass matrix $M$.  
The structure of the supercurrents depends on whether the superspin is integer or half-integer, as well as on the value of the superspin, and the mass matrix. 
Let us summarise our results. 

In the case of half-integer superspin-$(s+ \hf)$, the supercurrent has the structure $J_{(s,s)}= H^{ij} J^{ij}_{(s, s)}$, 
where $i, j =1, \dots N$ and $H^{ij}$ is a Hermitian matrix. The precise form of  $ J^{ij}_{(s, s)}$ was discussed in section \ref{s44}. In the massless theory it is conserved for all values of $s$. 
In the massive theory, the conservation 
equation involves an additional complex multiplet $T_{(s-1, s-2)}$ whose existence depends on the value of $s$ and the mass matrix. 
For odd values of $s$, it exists provided $[S, \bar M]=0$, $\{A, \bar M\}=0$, where $S$ and $A$  are the symmetric and antisymmetric parts of $H$, respectively. When $s$ is even, it exists provided $\{S, \bar M\}=0$, $[A, \bar M]=0$.

In the case of integer superspin-$s$, the fermionic supercurrent was discussed in section \ref{s45}. It has the form $J_{(s, s-1)}= C^{ij} J^{ij}_{(s, s-1)}$. In the massless 
theory it exists for even values of $s$ if $C$ is symmetric and for odd values of $s$ if $C$ is antisymmetric. In the massive theory the conservation equation involves 
an additional complex multiplet $T_{(s-1 , s-1)}$ and a real multiplet $S_{(s-1, s-1)}$. Their existence also depends on the value of $s$. 
For $s$ even they exist provided $C \bar M = \bar C M$, $[C, \bar M]=0$ and for $s$ odd provided $C \bar M = \bar C M$, $\{ C, \bar M\}=0$.

It should be mentioned that in the non-supersymmetric case, conserved higher-spin currents for scalar and spinor fields in Minkowski space have been studied extensively in the past. Appendices \ref{AppendixC-scalar} and \ref{AppendixD-spinor} review the construction of conserved higher-spin currents 
for $N$ scalars and spinors, respectively, with arbitrary mass matrices. 
These results are scattered in the literature, including 
\cite{Migdal,Makeenko,CDT,BBvD}.  

In the rest of this section, we will discuss several applications of the results obtained.

\subsection{Higher-spin supercurrents for a 
tensor multiplet}

Let us consider a special case of the non-superconformal chiral model 
\eqref{chiral-massive} with the mass parameter $m=\m$, 
\bea
S [\F, \bar \F]= \hf \int \rd^4x \rd^2 \q  \rd^2 \bar \q \,E\, ( \F +\bar \F)^2~,
\qquad \bar \cD_\ad \F=0~.
\label{7.111}
\eea
This theory is known to be dual to a tensor multiplet model \cite{Siegel79}
\bea
S [L]=- \hf \int \rd^4x \rd^2 \q  \rd^2 \bar \q \,E\, L^2~,
\label{tensor}
\eea
which is realised in terms of a real linear superfield $L=\bar L$, 
constrained by $(\bar \cD^2 -4\m) L =0$,
which is the gauge-invariant field strength
of a chiral spinor superfield 
\bea
L= \cD^\a \eta_\a +\bar \cD_\ad \bar \eta^\ad~, \qquad \bar \cD_\bd \eta_\a =0~.
\eea
We recall that the duality between \eqref{7.111} and \eqref{tensor} follows, e.g., 
from the fact the off-shell constraint
\begin{subequations}
\bea
 (\bar \cD^2 -4\m) \cD_\a (\F +\bar \F)=0
 \eea
 and the equation of motion for $\F$ 
 \bea
  (\bar \cD^2 -4\m) (\F +\bar \F) =0
  \eea
  \end{subequations}
are equivalent to the equation of motion for $\eta_\a$ 
\begin{subequations}
\bea
 (\bar \cD^2 -4\m) \cD_\a L=0
 \eea
 and the off-shell constraint
 \bea
  (\bar \cD^2 -4\m) L =0~,
  \eea
  \end{subequations}
respectively. 

Higher-spin supercurrents for the tensor model \eqref{tensor} 
can be obtained from the results derived in subsection \ref{subsection5.2} in conjunction
with an improvement transformation of the type \eqref{improvement4.20}
with  $\O = -\hf \F^2$.
Given an odd $s=3, 5 \dots$, for the supercurrent we get
\bea
{J}_{(s,s)} &=& -L\,\, {\cD}^{s-1}_{(1,1)} \, [{\cD}_{(1,0)},{\bar \cD}_{(0,1)}]L
\non \\
&&+ \sum_{k=0}^{s-1} (-1)^k
\binom{s}{k}
\binom{s}{k+1} 
{\cD}^k_{(1,1)}
 \cD_{(1,0)} L \,\,
{\cD}^{s-k-1}_{(1,1)}
\bar \cD_{(0,1)} L  
\non \\ 
&&+ \hf \sum_{k=1}^{s-1} \left\{-1+ (-1)^k
\binom{s}{k}\right\}
\binom{s}{k} 
{\cD}^{k-1}_{(1,1)}\, [\cD_{(1,0)}, {\bar \cD}_{(0,1)}] L \,\,\,
{\cD}^{s-k}_{(1,1)} L ~.
\label{currentL}
\eea
The corresponding trace multiplet proves to be
\bea
{T}_{(s-1, s-2)} &=& -\frac{4\m}{s} L \,\, {\cD}^{s-2}_{(1,1)} {\cD}_{(1,0)}L 
+ 4\m \frac{s+1}{s} {\cD}_{(1,0)}L \,\, {\cD}^{s-3}_{(1,1)} {\bar \cD}_{(0,1)} {\cD}_{(1,0)}L \non \\
&&-\frac{2}{s} {\cD}^{s-2}_{(1,1)} \left\{{\cD}_{(1,0)} {\bar \cD}_\ad L \,\,\, {\bar \cD}^\ad L \right\} \non \\
&&+ \m \sum_{k=1}^{s-2} c_k {\cD}^{k-1}_{(1,1)} {\bar \cD}_{(0,1)} {\cD}_{(1,0)}L \,\, {\cD}^{s-k-2}_{(1,1)} {\cD}_{(1,0)}L \non \\
&&+ \frac{4\m}{s} \sum_{k=1}^{s-2} \binom{s-2}{k} {\cD}^{k-1}_{(1,1)} {\cD}_{(1,0)} {\bar \cD}_{(0,1)}L \,\, {\cD}^{s-k-2}_{(1,1)} {\cD}_{(1,0)}L \non \\
&&+ 2\m \frac{s+1}{s}\sum_{k=1}^{s-3} \binom{s-2}{k} \left\{ {\cD}^{k}_{(1,1)} {\cD}_{(1,0)}L \,\, {\cD}^{s-k-3}_{(1,1)} {\bar \cD}_{(0,1)} {\cD}_{(1,0)}L \right. \non \\
&& \left. \qquad \qquad + {\cD}^{k-1}_{(1,1)} {\bar \cD}_{(0,1)} {\cD}_{(1,0)}L \,\, {\cD}^{s-k-2}_{(1,1)} {\cD}_{(1,0)}L \right\} ~. 
\label{traceL}
\eea
The coefficient $c_k$ is given by eq.~\eqref{7.17}, $s$ is odd. 
The Ferrara-Zumino supercurrent ($s=1$) for the model \eqref{tensor}
in an arbitrary supergravity background
 was derived in section 6.3 of \cite{Ideas}. Modulo normalisation, 
 the AdS supercurrent is 
 \begin{subequations}
 \bea
 J_{\a\ad} = \bar \cD_\ad L \cD_\a L+ L \big[\cD_\a, \bar \cD_\ad \big] L~,
 \eea
 and the corresponding trace multiplet is 
 \bea
 T=\frac 14 (\bar \cD^2 -4\m) L^2~.
 \eea
 \end{subequations}
 The supercurrent obeys the conservation equation  \eqref{FZsupercurrent}.


\subsection{Higher-spin supercurrents for a complex linear multiplet}
Conserved higher-spin supercurrents for a complex linear multiplet in Minkowski superspace were first studied by Koutrolikos, Ko\v{c}i and von Unge \cite{KKvU}, as an extension of the lower-spin case \cite{KKvU-sc}.
In AdS, the superconformal non-minimal scalar multiplet is described by the action 
\bea
S [\G, \bar \G]= - \int \rd^4x \rd^2 \q  \rd^2 \bar \q \,E\, \bar \G  \G ~,
\label{7.88}
\eea
where $\G$ is a complex linear scalar, $(\bar \cD^2 -4\m) \G =0$. 
This is a dual formulation for the superconformal chiral model \eqref{chiral}.
As is well known, the duality between \eqref{chiral} and \eqref{7.88} 
follows from the fact that the off-shell constraint 
\begin{subequations}
\bea
(\cD^2 -4 \bar \m) \bar \G =0~,
\eea
and the equation of motion for $\G$ 
\bea
\bar \cD_\ad \bar \G=0
\eea
\end{subequations}
are equivalent to the equation of motion for $\bar \F$, $(\cD^2 -4 \bar \m) \F=0$, 
and the off-shell constraint $\bar \cD_\ad \F=0$, respectively.
In other words, on the mass shell we can identify $\bar \G$ with $\F$.

The higher-spin supercurrents, $J_{(s,s)}$ and $J_{(s,s-1)}$, 
 for the model \eqref{7.88} are obtained from \eqref{7.15} and \eqref{6.14}, 
 respectively, by replacing $\F$ with $\bar \G$. The fermionic supercurrent 
 $J_{(s,s-1)}$ exists for even values of $s$. 
Indeed, in Minkowski superspace, the expression for  $J_{(s,s)}$ obtained coincides with the main result of Ref.\cite{KKvU}\footnote{See also \cite{koci-thesis} for the discussion of the fermionic supercurrent 
 $J_{(s,s-1)}$.}, which applied the Noether procedure to generate cubic vertices between massless higher-spin supermultiplets and the free complex linear superfield model
\bea
S [\G, \bar \G]= - \int \rd^4x \rd^2 \q  \rd^2 \bar \q  \bar \G  \G ~,
\qquad \bar D^2 \G=0~.
\eea


\subsection{Gauge higher-spin multiplets and conserved supercurrents}

For each of the two off-shell formulations for 
the massless multiplet of half-integer superspin-$(s+\hf)$, with $s=2, 3, \ldots$, 
which we reviewed in section \ref{ss431}, it was shown
in \cite{KS94} that there exists a  gauge-invariant field strength 
$W_{\a(2s+1)} $ which is covariantly chiral, $\cDB_\bd W_{\a(2s+1)} = 0$, 
and is given by the expression
\bea
W_{\a (2s + 1)} &=& -\frac 14(\cDB^2 - 4 \mu) \cD_{(\a_1}{}^{\bd_1} \cdots \cD_{(\a_s}{}^{\bd_s} \cD_{\a_{s+1}} H_{\a_{s+2} \cdots \a_{2s+1}) \bd_1 \cdots \bd_s} ~.
 \label{7.115}
\eea
It was also shown in \cite{KS94} that 
on the mass shell it holds that  (i) $W_{\a(2s+1)} $ and its conjugate $\bar W_{\ad(2s+1)} $ 
are the only  independent gauge-invariant  field strengths; and (ii) $W_{\a(2s+1)} $
obeys the irreducibility condition 
\bea
\cD^\b W_{\b \a(2s)}=0~.
\label{7.116}
\eea
The relations \eqref{7.115} and \eqref{7.116} also hold for the cases 
$s=0$ and $s=1$, which correspond to the vector multiplet and linearised
supergravity, respectively. 
In terms of $W_{\a(2s+1)} $ and  $\bar W_{\ad(2s+1)} $, 
we can define the following 
higher-spin supercurrent 
\bea
J_{\a(2s+1) \ad(2s+1)} =  W_{\a(2s+1)} \bar W_{\ad(2s+1)} ~,\qquad 
s=0,1,\dots~, 
\eea
which obeys the conservation equation
\bea
 \bar \cD_{(0,-1)} J_{(2s+1,2s+1)} =0 \quad \Longleftrightarrow \quad
 \cD_{(-1,0)} J_{(2s+1,2s+1)} =0 ~.
\eea

In the case of the longitudinal formulation for 
the massless multiplet of integer superspin-$s$, with $s=2, 3, \ldots$, 
which we described in section \ref{s42}, it was shown
in \cite{KS94} that there exists a  gauge-invariant field strength 
$W_{\a(2s)} $ which is covariantly chiral, $\cDB_\bd W_{\a(2s)} = 0$, 
and is given by the expression\footnote{The flat-superspace version 
of \eqref{7.199} is given in section 6.9 of \cite{Ideas}.}
\bea
W_{\a (2s )} &=& -\frac 14(\cDB^2 - 4 \mu) \cD_{(\a_1}{}^{\bd_1} \cdots 
\cD_{(\a_{s-1}}{}^{\bd_{s-1}} \cD_{\a_{s}} \J_{\a_{s+1} \cdots \a_{2s}) 
\bd_1 \cdots \bd_{s-1}} ~.
\label{7.199}
\eea
As demonstrated  in \cite{KS94},
on the mass shell it holds that  (i) $W_{\a(2s)} $ and its conjugate $\bar W_{\ad(2s) }$ 
are the only  independent gauge-invariant  field strengths; and (ii) $W_{\a(2s)} $
obeys the irreducibility condition 
\bea
\cD^\b W_{\b \a(2s-1) }=0~.
\label{7.120}
\eea
The relations \eqref{7.199} and \eqref{7.120} also hold for the case $s=1$, 
which corresponds to the gravitino  multiplet.
In terms of $W_{\a(2s)} $ and  $\bar W_{\ad(2s)} $, 
we can define the higher-spin supercurrent 
\bea
J_{\a (2s) \ad(2s) } =  W_{\a(2s)} \bar W_{\ad(2s)} ~, \qquad s=1,2,\dots~,
\eea
which obeys the  conservation equation
\bea
 \bar \cD_{(0,-1)} J_{(2s,2s)} =0 \quad \Longleftrightarrow \quad
 \cD_{(-1,0)} J_{(2s,2s)} =0 ~.
\eea
The conserved supercurrents $J_{\a (n) \ad(n) } =  W_{\a(n)} \bar W_{\ad(n)}$,
with $n=1,2,\dots$, are the AdS extensions of those introduced many years
ago by Howe, Stelle and Townsend \cite{HST}.

Now, for any positive integer $n>0$,
 we can try to generalise the higher-spin supercurrent \eqref{7.15}
as follows:
\bea
{\mathfrak J}_{(s+n,s+n)} &=& \sum_{k=0}^s (-1)^{k} \frac{\binom{s}{k} \binom{s+n}{k}}{\binom{n+k}{n}}
\left\{ (-1)^n \frac{s-k}{n+k+1}
{\cD}^k_{(1,1)}
 \cD_{(1,0)} W_{(n,0)} \,\,
{\cD}^{s-k-1}_{(1,1)} \bar {\cD}_{(0,1)}
\bar W_{(0,n)} 
\right. \non \\ 
&& \left.
 \qquad 
+{\cD}^k_{(1,1)}
  W_{(n,0)} \,\,
{\cD}^{s-k}_{(1,1)}
\bar W_{(0,n)} \right\}~.~
\label{currentW}
\eea
Making use of
the on-shell condition   
\bea
\cD_{(-1,0)} W_{(n,0)} = 0 \quad \Longleftrightarrow \quad ({\cD}^2- 2(n+2) \bar \m)W_{(n,0)} = 0 ~,
\eea
one may check that 
\bea
\cD_{(-1,0)} {\mathfrak J}_{(s+n,s+n)} &=& 2n \bar \m \sum_{k=0}^{s-1} (-1)^{n+k} \frac{s-k}{n+k+1} \,\, \frac{\binom{s}{k} \binom{s+n}{k}}{\binom{n+k}{n}}
\non \\
&&\qquad \quad \times 
{\cD}^k_{(1,1)} W_{(n,0)} \,\,
{\cD}^{s-k-1}_{(1,1)}
\bar {\cD}_{(0,1)} \bar W_{(0,n)} ~.~
\label{eqW}
\eea
This demonstrates that ${\mathfrak J}_{(s+n,s+n)} $ is not conserved in AdS${}^{4|4}$.

In the flat-superspace limit, $ \m\to 0$, the right-hand side of \eqref{eqW}
vanishes and ${\mathfrak J}_{(s+n,s+n)} $ becomes conserved. 
In Minkowski superspace, the conserved supercurrent 
${\mathfrak J}_{(s+n,s+n)} $ was recently constructed in \cite{BGK3}
as an extension of the non-supersymmetric approach \cite{GSV}.

As a generalisation of the  conserved supercurrents $J_{\a (n) \ad(n) } =  W_{\a(n)} \bar W_{\ad(n)}$, one can introduce 
\bea
J_{\a (n) \ad(m) } =  W_{\a(n)} \bar W_{\ad(m)}~,
\label{7.266}
\eea
with $n\neq m$. They obey the conservation equations
\bea
 \bar \cD_{(0,-1)} J_{(n,m)} =0 ~, \qquad 
 \cD_{(-1,0)} J_{(n,m)} =0 
\eea
and can be viewed as Noether currents for the generalised superconformal 
higher-spin multiplets introduced in \cite{KMT}.
Starting from the conserved supercurrents \eqref{7.266}, one can construct a generalisation of \eqref{currentW}.  We will not elaborate on a construction 
here.



\chapter{${\cN} = 2$ supersymmetric higher-spin gauge theories and current multiplets in three dimensions} \label{ch5}

In four dimensions, there exists a correspondence between 
$\cN=1$ anti-de Sitter (AdS) supergravity \cite{Townsend} and the two dually equivalent series of massless multiplets of half-integer superspin-$(s+ \hf)$, with $s=1,2,\dots$ \cite{KS94}. Specifically, there are two off-shell formulations for pure ${\cN}=1$ AdS supergravity: minimal (see e.g. \cite{GGRS,Ideas} for reviews) and non-minimal \cite{BK12}. These theories possess a single maximally supersymmetric solution, which is the ${\cN}=1$ AdS superspace ${\rm AdS}^{4|4}$. For the lowest superspin value corresponding to $s=1$, the longitudinal series yields the linearised action for minimal  AdS supergravity, while the transverse one leads to linearised non-minimal AdS supergravity.

In three dimensions, the AdS group is a product of two simple groups,
$$\rm SO(2,2) \cong \Big( SL(2, {\mathbb R}) \times SL( 2, {\mathbb R}) \Big)/{\mathbb Z}_2~,$$ 
and so are its simplest supersymmetric extensions,  
${\rm OSp} (p|2; {\mathbb R} ) \times  {\rm OSp} (q|2; {\mathbb R} )$.
This implies that $\cN$-extended AdS supergravity exists in several incarnations  \cite{AT}. These are known as the $(p,q)$ AdS supergravity theories,
where the  non-negative integers $p \geq q \geq 0$ are such that $\cN=p+q$. Superspace approach to 3D ${\cN}$-extended conformal supergravity was developed by Kuzenko, Lindstr\"{o}m and Tartaglino-Mazzucchelli \cite{KLT-M11}, and used to construct off-shell ${\cN} \leq 4$ supergravity-matter couplings. The formalism of \cite{KLT-M11} was then applied to study the geometry of $(p,q)$ AdS superspaces \cite{KLT-M12}.
The so-called $(p,q)$ AdS superspace \cite{KLT-M12}
\bea
{\rm AdS}^{(3|p,q)} = \frac{ {\rm OSp} (p|2; {\mathbb R} ) \times  {\rm OSp} (q|2; {\mathbb R} ) } 
{ {\rm SL}( 2, {\mathbb R}) \times {\rm SO}(p) \times {\rm SO}(q)} \non
\eea
can be realised as a maximally symmetric solution of $(p,q)$ AdS supergravity (see \cite{KLT-M12} for the technical details).  

In the case of 3D $\cN=2$ supersymmetry, there exist two distinct AdS superspaces, 
 ${\rm AdS}^{(3|1,1)}$ and ${\rm AdS}^{(3|2,0)}$. The former is the 3D counterpart of the 4D $\cN=1$ AdS superspace, while 
the latter has no 4D analogue. The existence of these superspaces and their superconformal flatness were studied for the first time in \cite{BILS}. Ref. \cite{KT-M11} presented superfield formulations for 3D ${\cN}=2$ AdS supergravity theories and their corresponding supercurrent multiplets.
Two off-shell formulations for (1,1) AdS supergravity have been developed: minimal \cite{RvanN86,ZupnikPak,NG,BCSS,KLT-M11,KT-M11,KLRST-M}  and non-minimal  \cite{KT-M11,KLRST-M} theories; and one for (2,0) AdS supergravity
 \cite{HIPT,KLT-M11,KT-M11,KLRST-M}. 
${\rm AdS}^{(3|1,1)}$ is the unique maximally symmetric solution 
of the two dually equivalent  (1,1) AdS supergravity theories,
minimal and non-minimal ones.
${\rm AdS}^{(3|2,0)}$ is the unique maximally symmetric solution 
of the (2,0) AdS supergravity. This supergravity theory was originally formulated in \cite{HIPT} 
in the component setting. The early superspace descriptions of
 the minimal (1,1) supergravity were given in \cite{ZupnikPak,NG}.
 
It has recently been pointed out \cite{HKO} that the correspondence between AdS supergravity theories and massless higher-spin supermultiplets in 3D anti-de Sitter space, ${\rm AdS}_3$, might occur
in the $\cN=2$ case. Since there are three off-shell $\cN=2$ AdS supergravity theories, 
one might expect the existence of three  series of massless 
higher-spin gauge supermultiplets. Two series of massless higher-spin actions associated with the minimal and the non-minimal (1,1) AdS 
supergravity theories were presented in \cite{HKO}. These generalise similar constructions in the super-Poincar\'e case \cite{KO}. As will be explained in sections \ref{s62} and \ref{s57}, the off-shell higher-spin supermultiplets in (2,0) AdS superspace \cite{HK18} were constructed using a different approach.  

Pure $\cN=2$ supergravity 
(massless superspin-3/2 multiplet) 
and its higher-spin extensions have no propagating degrees of freedom in three dimensions.
Nevertheless, there are at least two nontrivial applications of 
the massless higher-spin gauge supermultiplets.
Firstly, one can follow the pattern of topologically massive (super)gravity
\cite{DJT1,DJT2,DK,Deser84} and construct massive higher-spin supermultiplets
by combining a massless 
action with a higher-spin 
extension of the action for linearised conformal supergravity. 
This has been achieved in \cite{KO} in the $\cN=2$ super-Poincar\'e case, 
and similar ideas have been implemented  in the frameworks 
of $\cN=1$ Poincar\'e and AdS supersymmetry  \cite{KT,KP1}. Topologically massive higher-spin supermultiplets in (1,1) and (2,0) AdS superspaces have been formulated in \cite{HKO} and \cite{HK18}, respectively.
The second application is to develop a 3D extension of the higher-spin supercurrents presented in chapters \ref{ch3} and \ref{ch4}. Specifically, making use of the off-shell formulations for massless higher-spin 
supermultiplets in ${\rm AdS}_{3} $, one can define consistent higher-spin supercurrent 
multiplets that contain ordinary bosonic and fermionic conserved currents in ${\rm AdS}_{3} $. 
One can then look for explicit realisations of such 
higher-spin supercurrents in concrete supersymmetric theories in ${\rm AdS}_{3} $ \cite{HKO}. 

This chapter can be divided into two parts: sections \ref{s51} to \ref{s55} focus on rigid supersymmetric higher-spin gauge theories in (1,1) AdS superspace which were studied in \cite{HKO}, while sections \ref{s62} and \ref{s57} are concerned with the construction of off-shell massless higher-spin gauge multiplets with (2,0) AdS supersymmetry as described in \cite{HK18}. In section \ref{s51}, we review the superspace geometry of 3D ${\cN}=2$ conformal supergravity.  We then introduce primary linear supermultiplets and conformal higher-spin gauge superfields coupled to ${\cN}=2$ conformal supergravity, the latter being one of the key ingredients in constructing massless higher-superspin actions. Section \ref{s52} reviews the two inequivalent ${\cN}=2$ AdS superspaces. Two dual off-shell Lagrangian formulations for every massless higher-spin supermultiplet in (1,1) AdS superspace will be presented in sections \ref{s53} and \ref{s54}. As in the 4D AdS constructions, the two cases of half-integer and integer superspin, as well as massless gravitino multiplet have to be treated separately. Section \ref{s55} is devoted to constructing non-conformal higher-spin supercurrent multiplets in models for chiral scalar superfields. The materials presented in sections \ref{s51}, \ref{s53} and subsections \ref{subsection5.1}$-$\ref{subsect5.2} are based on the work by Kuzenko and Ogburn \cite{HKO}. Here I only include a summary of those results which are essential for constructing a new off-shell model for the massless integer superspin, as well as describing (1,1) AdS higher-spin supercurrents.

Starting with simple models for a chiral scalar supermultiplet in (2,0) AdS superspace, in section \ref{s62} we obtain the conservation equation obeyed by the multiplet of higher-spin currents. This will allow us to determine the off-shell gauge superfields which couple to the current multiplet. Two off-shell formulations for 
a massless multiplet of half-integer superspin in (2,0) AdS superspace are developed in section \ref{s57}. Our results, their implications and possible
extensions are discussed in section \ref{s58}.

\section{Superconformal higher-spin multiplets} \label{s51}
Before presenting superconformal higher-spin multiplets, let us first give a succinct review of the formulation for $\cN=2$ conformal supergravity 
following \cite{KLT-M11}. There exists a more general formulation 
for conformal supergravity  \cite{BKNT-M1}, the so-called $\cN=2$ conformal superspace. However, for our purposes it suffices to 
use the formulation of \cite{KLT-M11}, which is obtained from the $\cN=2$ conformal 
superspace by partially fixing the gauge freedom. The reader is referred to appendix \ref{AppA2} for more details on our 3D conventions.

\subsection{Conformal supergravity} \label{ss511}

All known off-shell formulations for 3D $\cN=2$ supergravity \cite{KLT-M11,KT-M11}
can be realised in a curved superspace $\cM^{3|4}$
 with the structure group  ${\rm SL}(2,{\mathbb{R}})\times {\rm U(1)}_R$. Here ${\rm SL}(2,{\mathbb{R}})$ and $ {\rm U(1)}_R$
 stand for the spin group and the $R$-symmetry group, respectively.
We parametrise the superspace by
local bosonic ($x^m$) and fermionic ($\q^\m, \bar \q_\m$)
coordinates  $z^{{\cM}}=(x^{m},\q^{\mu},{\bar \q}_{{\mu}})$,
where $m = 0,1,2~, \mu = 1,2$. The Grassmann  variables $\q^{\mu} $ and $\bar \q_{{\mu}}$
are related to each other by complex conjugation:
$\overline{\q^{\mu}}=\bar \q^{{\mu}}$.

The superspace covariant derivatives have the form
\bea
{\bcD}_{{\cA}}=(\bcD_{{a}}, \bcD_{{\a}},\bar \bcD^\a)
=E_{{\cA}}+\O_{{\cA}}+\ri \F_{{\cA}} J~.
\label{CovDev-gen}
\eea
Here $E_{\cA}$ is the inverse supervielbein, while $\O_{\cA}$ and $\F_{\cA}$ denote the Lorentz and ${\rm U(1)}_R$ connections, respectively,
\bea
E_{\cA}=E_{\cA}{}^{\cM} \frac{\pa}{\pa z^{\cM}}~,
\qquad \O_{\cA}=\hf\O_{\cA}{}^{bc} M_{bc}= -\O_{\cA}{}^b M_b
=\hf\O_{\cA}{}^{\b\g}M_{\b\g}~.
\eea
The explicit relations between Lorentz generators with two vector indices 
($M_{ab}= -M_{ba}$), one vector index ($M_a$)
and two spinor indices ($M_{\a\b} = M_{\b\a} $) 
are defined in appendix \ref{AppA2}.
The actions of the generators ${\rm SL}(2,{\mathbb{R}})\times {\rm U(1)}_R$ on the covariant derivatives are defined as
\bea
&&{[} J,{\bcD}_{\a}{]}
={\bcD}_{\a}~,
\qquad
{[} J, \bar {\bcD}^{\a}{]}
= - \bar {\bcD}^\a~,
\qquad 
{[}J, {\bcD}_a{]}=0~, \non\\
\,\,\,{[} M_{\a \b}, {\bcD}_{\g} {]} &=& \ve_{\g (\a}{\bcD}_{\b)}~,\qquad  
{[}M_{\a \b}, \bar {\bcD}_{\g} {]} = \ve_{\g (\a}\bar {\bcD}_{\b)}~, \qquad 
{[} M_{a b}, {\bcD}_{c} {]} = 2 \eta_{c [a}{\bcD}_{b]}~.~~~~~~~
\eea

The covariant derivatives obey (anti-)commutation relations
\bea
{[} \bcD_{\cA} , \bcD_{\cB} \} = T_{\cA\cB}{}^{\cC}\bcD_{\cC} + \hf R_{\cA\cB}{}^{cd}M_{cd} + \ri R_{\cA\cB}J.
\eea
In the above, $T_{\cA \cB}{}^{\cC}$ is the torsion, while $R_{\cA \cB}{}^{cd}$ and $R_{\cA \cB}$ describe the curvature. 
In order to describe $\cN=2$ conformal supergravity, the torsion 
has to obey the covariant constraints proposed in \cite{HIPT}.
Solving the constraints gives the following algebra of covariant derivatives  \cite{KLT-M11,KT-M11}
\bsubeq \label{algebra-final}
\bea
\{ \bcD_\a , \bcD_\b \} &=& - 4 \bar \cR M_{\a\b} \ , \\
\{ \bcD_\a , {\bar \bcD}_\b \} &=&
- 2 \ri (\g^c)_{\a\b} \bcD_c 
- 2 \cC_{\a\b} J
- 4 \ri \ve_{\a\b} \cS J 
+ 4 \ri \cS M_{\a\b}
- 2 \ve_{\a\b} \cC^{\g\d} M_{\g\d} \ , \label{algebra-final-b} \\
{[}\bcD_a , \bcD_\b {]}
&=& \ri \ve_{abc} (\g^b)_\b{}^\g \cC^c \bcD_\g
+ (\g_a)_\b{}^\g \cS \bcD_\g 
- \ri (\g_a)_{\b\g} \bar \cR \bar \bcD^\g
+ \ri (\g_a)_\b{}^\g \bcD_{(\g} \cC_{\d\r)} M^{\d\r} \non\\
&&- \frac{1}{3} (2 \bcD_\b \cS + \ri \bar \bcD_\b \bar \cR) M_a
- \frac{2}{3} \ve_{abc} (\g^b)_\b{}^\a (2 \bcD_\a \cS + \ri \bar \bcD_\a \bar \cR) M^c \non\\
&&+ \frac{\ri}{2} \Big( 
(\g_a)^{\a\g} \bcD_{(\a} \cC_{\b\g)}
+ \frac{1}{3} (\g_a)_\b{}^\g (8 \ri \bcD_\g \cS - \bar \bcD_\g \bar \cR)
\Big) J \ ,
\eea
\esubeq
We thus see that the algebra is parametrised by three torsion superfields: a real scalar $\cS$, a complex scalar $\cR$ and its conjugate $\bar \cR$, and a real vector $\cC_{\a \b}:= (\g^a)_{\a \b}\cC_{a}$. The ${\rm U}(1)_{R}$ charges of the torsion superfields  $\cR$, $\bar \cR$ and $\cC_{\a\b}$
are $-2$, $+2$ and 0, respectively. 
They satisfy the Bianchi identities
\bea 
\bcD_\a \bar \cR &=& 0 \ , \quad ( \bar \bcD^2 - 4\cR) \cS=0\, \quad
\bcD^\b \cC_{\a\b} =- \frac{1}{2} (\bar \bcD_{\a} \bar \cR + 4 \ri \bcD_{\a} \cS) \ .
\label{BItypeIC}
\eea
Throughout this chapter, we define
$\bcD^2:=\bcD^\a\bcD_\a$ and $\bar \bcD^2:=\bar \bcD_\a \bar \bcD^\a$.

The algebra of covariant derivatives given by \eqref{algebra-final}
is invariant under the super-Weyl transformation
\cite{KLT-M11, KT-M11}
\bsubeq  \label{super-WeylN=2}
\bea
\bcD'_\a&=&\re^{\hf\s}\Big(\bcD_\a+\bcD^{\g}\s
M_{\g\a}-\bcD_{\a} \s J\Big)~,
\\
\bar \bcD'_{\a}&=&\re^{\hf\s}\Big(\bar \bcD_{\a}+\bar \bcD^{\g}\s {M}_{\g\a}
+ \bar \bcD_{\a}\s J\Big)~,
\\
\bcD'_{a}
&=&\re^{\s}\Big(
\bcD_{a}
-\frac{\ri}{2}(\g_a)^{\g\d}\bcD_{\g}\s \bar \bcD_{\d}
-\frac{\ri}{2}(\g_a)^{\g\d} \bar \bcD_{\g}\s\bcD_{\d} 
+\ve_{abc}\bcD^b \s M^c \non\\
&&~~~~~-\frac{\ri}{2}(\bcD^{\g}\s) \bar \bcD_{\g}\s M_{a}
- \frac{\ri}{24} (\g_a)^{\g\d} \re^{- 3 \s} [\bcD_\g , \bar \bcD_\d] \re^{3 \s}
J
\Big)~,
\eea
which induces the following transformation of the torsion tensors:
\bea 
\cS'&=&\re^{\s}\Big(
\cS
+\frac{\ri}{4}\bcD^\g \bar \bcD_{\g}\s
\Big)~,
 \label{2.11d} \\
\cC'_{a}&=&
\Big( \cC_a + \frac{1}{8} (\g_a)^{\g\d} [\bcD_\g , \bar \bcD_\d] \Big) \re^\s
~,
\\
\cR' &=&
- \frac{1}{4} \re^{2 \s} (\bar \bcD^2 - 4 \cR) \re^{- \s}
~.
\label{2.11f}
\eea
\esubeq
The parameter $\s$ is an arbitrary real scalar superfield.
The super-Weyl invariance \eqref{super-WeylN=2}
is intrinsic to conformal supergravity. 
For every supergravity-matter system, 
its action is required to be a super-Weyl invariant functional  
of the supergravity multiplet coupled to certain conformal
compensators, see \cite{KLT-M11,KT-M11} for more details.

The $\cN=2$ supersymmetric extension of the Cotton tensor 
\cite{Kuzenko12} is
\bea 
\cW_{\a\b} = - \frac{\ri}{4} [\bcD^\g , \bar\bcD_\g] \cC_{\a\b}
+ \hf [\bcD_{(\a} , \bar\bcD_{\b)}] \cS + 2 \cS \cC_{\a\b}  ~.
\label{N=2covsuperspaceCotton}
\eea
It transforms homogeneously under \eqref{super-WeylN=2},
\bea 
\cW'_{\a\b} = \re^{2 \s} \cW_{\a\b} \ ,
\label{2.9}
\eea
and obeys the Bianchi identities \cite{BKNT-M1}
\bea
 \bar \bcD^\b \cW_{\a\b} = \bcD^\b \cW_{\a\b} =0~.
\eea
The curved superspace is conformally flat if and only if $\cW_{\a\b} =0$
\cite{BKNT-M1}.


\subsection{Primary superfields} \label{ss512}

Let $T_{\a(n) } := T_{\a_1 \dots \a_n}
=T_{(\a_1 \dots \a_n)}$ be a symmetric rank-$n$ spinor superfield 
of ${\rm U(1)}_R$ charge $q$, 
\bea
J T_{\a(n) }  = q T_{\a(n) } ~.
\eea
The superfield $T_{\a(n) } $
is called super-Weyl primary of dimension $d$ if it transforms under the infinitesimal super-Weyl transformation law as
\bea
\d_\s T_{\a(n) }  = d \s T_{\a(n) } ~.
\eea
As an example, the super-Cotton tensor is super-Weyl primary of dimension $+2$.  
Let us introduce several types of primary superfields which will be important 
for our subsequent analysis. 

A symmetric rank-$n$ spinor superfield $G_{\a(n)}$ is called longitudinal linear if
it obeys the following first-order constraint
\bea
 \bar \bcD_{(\a_1} G_{\a_2 \dots \a_{n+1} )} = 0  ~, \label{2.111}
 \eea
 which implies 
 \bea
 \big(\bar \bcD^2+2n\cR \big)G_{\a(n)} &=& 0~. \label{2.12}
 \eea
 If $G_{\a(n)}$ is super-Weyl primary, the constraint \eqref{2.111} 
 is consistent provided 
 the dimension $d_{G_{(n)} }$ and ${\rm U(1)}_R$ charge $q_{G_{(n)} }$ 
 of $G_{\a(n)}$ are related as
 \bea
 d_{G_{(n)} } =  -\frac{n}{2}  -q_{G_{(n)} }~. \label{2.13}
 \eea
 In the scalar case, $n=0$, the constraint \eqref{2.111} becomes the condition 
 of covariant chirality, $\bar \bcD_\a G=0$.  
 The dimension $d_{G }$ and ${\rm U(1)}_R$ charge $q_{G}$ of 
 any primary chiral scalar superfield $G$ are related as
 $ d_{G} +q_{G }=0$, in accordance with  \cite{KLT-M11}.
 

Given a positive integer $n$, a symmetric rank-$n$ spinor superfield 
$\G_{\a(n)}$ is called transverse linear if it obeys the  first-order constraint
 \bea
 \bar \bcD^\b \G_{ \b \a_1 \dots \a_{n - 1} } = 0 ~,  \qquad n \neq 0~,   
 \label{2.144-11}
 \eea
 which implies 
 \bea
 \big(\bar \bcD^2-2(n+2)\cR\big)\G_{\a(n)} = 0~. \label{2.15}
 \eea
If $\G_{\a(n)}$ is super-Weyl primary, then the constraint \eqref{2.144-11} 
 is consistent provided 
 the dimension $d_{\G_{(n)} }$ and ${\rm U(1)}_R$ charge $q_{\G_{(n)} }$ 
 of $\G_{\a(n)}$ are related to each other as follows:
 \bea
 d_{\G_{(n)} } =  1+ \frac{n}{2}-q_{\G_{(n)} }
 ~. \label{2.16}
 \eea
In the $n=0$ case, the constraint \eqref{2.144-11} is not defined.
However, its corollary \eqref{2.15} is perfectly consistent, 
\bea
 \big(\bar \bcD^2- 4\cR\big)\G = 0~, 
\label{2.18}
 \eea
 and defines a covariantly  linear scalar superfield $\G$.
 The dimension $d_{\G}$ and ${\rm U(1)}_R$ charge $q_{\G}$ of 
 any primary linear scalar  $\G$ are related as
 $ d_{\G} +q_{\G}=1$, in accordance with  \cite{KLT-M11}.

The constraints \eqref{2.111} and \eqref{2.144-11} are solved in terms of 
prepotentials $\J_{\a(n-1)}$ and $\F_{\a(n+1)}$ as follows:
\begin{subequations} \label{2.19-11}
\bea
 G_{\a(n)}&=& \bar \bcD_{(\a_1}
{ \J}_{ \a_2 \dots \a_{n}) } ~, 
\label{2.19a}\\
 \G_{\a(n)}&=& \bar \bcD^\b
{ \Phi}_{(\b \a_1 \dots \a_n )} ~.
\label{2.19b}
\eea
\end{subequations}
Provided the constraints \eqref{2.111} and \eqref{2.144-11} are the only conditions
imposed on $ G_{\a(n)}$ and $\G_{\a(n)}$ respectively, 
the prepotentials $\J_{\a(n-1)}$ and $\F_{\a(n+1)}$ can be chosen 
to be  unconstrained complex, and  
are defined modulo
gauge transformations of the form:
\begin{subequations} 
\bea
\d_\z \J_{\a(n-1)} 
&=&  \bar \bcD_{(\a_1 }
{ \z}_{\a_2 \dots \a_{n-1})} ~, \label{2.20a}\\
\d_\x \Phi_{\a(n+1)} &=&  \bar \bcD^\g
{ \x}_{(\g \a_1 \dots \a_{n+1})} ~,
\eea
\end{subequations}
with the gauge parameters ${\z_{\a(n-2)}}$ and $\x_{\a(n+2)}$
being unconstrained. If the linear superfields $ G_{\a(n)}$
and  $\G_{\a(n)}$ are super-Weyl primary, then their prepotentials 
$\J_{\a(n-1)}$ and $\F_{\a(n+1)}$ can also be chosen to be super-Weyl 
primary. 


In the $n=0$ case, the prepotential solution \eqref{2.19b} 
is still valid. The prepotential $\F_\a$ can be chosen to be unconstrained complex  
provided the constraint \eqref{2.18} is the only condition imposed on $\G$. 
However, if we are dealing with a real linear superfield, 
\bea
 \big(\bar \bcD^2- 4\cR\big)L = 0~,  \qquad \bar L =L~,
\label{2.22}
 \eea
then the constraints are solved \cite{KLT-M12} in terms of 
an unconstrained real prepotential $V$,  
\bea
L= \ri {\bar \bcD}^\a \bcD_\a V~,  \qquad \bar V =V~,
\label{G-prep}
\eea
which is defined modulo gauge transformations of the form:
\bea 
\d V = \l + \bar \l~, \qquad  J \l =0~, \quad {\bar \bcD}_\a \l =0~.
\eea
If $L$ is super-Weyl primary, then eq. \eqref{2.16} tells us
that the dimension of $L$ is $+1$. 
In this case it is consistent to consider the gauge prepotential $V$ 
to be inert under the super-Weyl transformations \cite{KLT-M11}, 
$\d_\s V =0$.



\subsection{Conformal gauge superfields}\label{ss513}

Let $n$ be a positive integer.
A real symmetric rank-$n$ spinor superfield ${\mathfrak H}_{\a(n) } $ 
is said to be a conformal 
gauge supermultiplet
if (i)  it is super-Weyl primary of dimension $(-{n}/{2})$, 
\bea
\d_\s {\mathfrak H}_{\a(n)} = -\frac{n}{2} \s {\mathfrak H}_{\a(n)}~;
\label{2.27}
\eea
and (ii) it is defined modulo gauge transformations of the form
\bea
\d_\l {\mathfrak H}_{\a(n) } =\bar  \bcD_{(\a_1} \l_{\a_2 \dots \a_n) }
-(-1)^n\bcD_{(\a_1} \bar \l_{\a_2 \dots \a_n) }~,
\label{2.28-11}
\eea
with the  gauge parameter $\l_{\a(n-1)}$
being unconstrained complex.
The dimension of ${\mathfrak H}_{\a(n) } $   in \eqref{2.27}
is uniquely fixed by requiring the longitudinal 
linear superfield $g_{\a(n)} =\bar  \bcD_{(\a_1} \l_{\a_2 \dots \a_n) }$
 in the right-hand side of \eqref{2.28-11} to be super-Weyl primary. Indeed,  
the gauge parameter $g_{\a(n)}$ must be neutral with respect to the 
$R$-symmetry  group ${\rm U(1)}_R$ since ${\mathfrak H}_{\a(n) } $ is real. Hence, the dimension of  $g_{\a(n)}$ is equal to $(-n/2)$, in accordance with
\eqref{2.13}.

\section{Geometry of ${\cN}=2$ AdS superspaces} \label{s52}
Let us briefly discuss maximally supersymmetric backgrounds 
in the off-shell $\cN=2$ supergravity theories, since the superspaces 
 ${\rm AdS}^{(3|1,1)}$ and ${\rm AdS}^{(3|2,0)}$ are special examples 
 of such supermanifolds.
The most general maximally supersymmetric backgrounds 
are characterised by several conditions \cite{KLRST-M}
on the  torsion superfields $\cR$, $\cS$ and $\cC_a$, 
which parametrise the superspace geometry of $\cN=2$ conformal supergravity, see \ref{ss511}. 
These requirements are as follows:
\begin{subequations}
\bea
\cR \cS&=&0~, \qquad \cR \,{\cC}_a = 0 ~, \\
{\bcD}_{\cA}  {\cR}&=& 0~, \qquad
{\bcD}_{\cA}  {\cS}= 0~,
\qquad \bcD_\a \cC_b =0 \quad \Longrightarrow \quad
{\bcD}_{a} {\cC}_b=
2\ve_{abc}{\cC}^c {\cS} ~.~~~
\eea
\end{subequations}
The (1,1) AdS superspace is singled out by 
the conditions $\cS=0$ and ${\cC}_a = 0$,
 with $\cR$ and its conjugate $\bar \cR$
having non-zero constant values \cite{KT-M11}.
On the other hand, the solution with ${\cR}=0, {\cC}_a =0$ and $\cS\neq 0$ corresponds to the (2,0) AdS superspace
\cite{KT-M11}. It may be shown that the ${\rm U(1)}_R$ connection is flat if and only if 
$\cS =0$ \cite{KLT-M12}. 
The non-vanishing ${\rm U(1)}_R$ curvature 
is the main reason why 
the structure of massless higher-spin gauge supermultiplets in (2,0) AdS superspace
\cite{HK18}
considerably differs from their counterparts with (1,1) AdS supersymmetry. This will be the subject of section \ref{s57}. 

\subsection{(1,1) AdS superspace} 

In this subsection we collect salient facts about the geometry of (1,1) AdS superspace \cite{KT-M11}, ${\rm AdS}^{(3|1,1)}$,
as well as elaborate on superfield representations of the isometry group.

The geometry of ${\rm AdS}^{(3|1,1)}$ is characterised by covariant derivatives 
\bea
{\frak{D}}_{{\cA}}=(\frak D_{{a}}, \frak D_{{\a}},\bar {\frak D}^\a)
=E_{\cA}{}^{\cM}\frac{\pa}{\pa z^{\cM}}+ \hf \O_{{\cA}}{}^{cd} M_{cd}
\eea
obeying the following graded commutation relations\cite{KT-M11}:
\begin{subequations}  \label{AdS11}
\bea
&& \qquad \{ \frak{D}_{\a} , \bar {\frak D}_{\b} \} = -2\rm i \frak{D}_{\a \b} ~, \\
&& \qquad \{\frak{D}_{\a}, \frak{D}_{\b} \} = -4\bar \m\, M_{\a \b}~, \qquad
\{ \bar {\frak D}_\a, \bar {\frak{D}}_\b \} = 4\m\,M_{\a \b}~, \\
&& \qquad [ \frak{D}_{ \a \b }, \frak{D}_\g ] = -2 \rm i \bar \m\,\ve_{\g (\a} \bar {\frak{D}}_{\b)}~,  \qquad
\,\,[\frak{D}_{ \a \b },  \bar {\frak{D}}_{\g} ] = 2 \rm i \m\,\ve_{\g (\a} \frak{D}_{\b)}~,   \\
&&\quad \,\,\,\,[ \frak{D}_{\a \b} , \frak{D}_{ \g \d } ] = 4 \bar \m \m \Big(\ve_{\g (\a} M_{\b) \d}+ \ve_{\d (\a} M_{\b) \g}\Big)~,  
\eea
\end{subequations} 
with $\m\neq 0$ being a  complex parameter.
As compared with \eqref{algebra-final}, we have denoted $\cR = \m$.
In particular, of some use during calculations are the following identities,
which can derived from 
the algebra \eqref{AdS11}:
\begin{subequations} 
\label{A1.4}
\bea 
\frak{D}_\a \frak{D}_\b
\!&=&\!\frac{1}{2}\ve_{\a\b}\frak{D}^2-2{\bar \m}\,M_{\a\b}~,
\quad\qquad \,\,\,
\bar {\frak D}_\a \bar {\frak D}_\b
=-\frac{1}{2}\ve_{\a \b}\bar {\frak D}^2+2\m\,{ M}_{\a \b}~,  \label{A1.4a}\\
\frak{D}_\a \frak{D}^2
\!&=&\!4 \bar \m \,\frak{D}^\b M_{\a\b} + 4{\bar \m}\,\frak{D}_\a~,
\quad\qquad
\frak{D}^2 \frak{D}_\a
=-4\bar \m \,\frak{D}^\b M_{\a\b} - 2\bar \m \, \frak{D}_\a~, \label{A1.4b} \\
\bar {\frak D}_\a \bar {\frak D}^2
\!&=&\!4 \m \, \bar {\frak D}^\b { M}_{\a \b}+ 4\m\, \bar {\frak D}_\a~,
\quad\qquad
\bar{\frak D}^2 \bar {\frak D}_\a
=-4 \m \,\bar{\frak D}^\b {M}_{\a \b}-2\m\, \bar {\frak D}_\a~,  \label{A1.4c}\\
\left[\bar {\frak D}^2, \frak{D}_\a \right]
\!&=&\!4\ri \frak{D}_{\a\b} \bar{\frak D}^\b +6 \m\,\frak{D}_\a = 
4\rm i \bar {\frak D}^\b \frak{D}_{\a\b} -6 \m\,\frak{D}_\a~,
 \label{A1.4d} \\
\left[\frak{D}^2, \bar {\frak D}_\a \right]
\!&=&\!-4\ri \frak{D}_{\a\b}\frak{D}^\b +6\bar \m\, \bar {\frak D}_\a = 
-4\rm i \frak{D}^\b \frak{D}_{\a\b} -6 \bar \m\,\bar {\frak D}_\a~,
 \label{A1.4e}
\eea
\end{subequations} 
These relations imply the identity 
\bea
\frak{D}^\a (\bar {\frak D}^2- 6 \mu) \frak{D}_\a = \bar{\frak D}_\a (\frak{D}^2 - 6 \mub) \bar {\frak D}^\a ~,
\label{id11covdev}
\eea
which guarantees the reality of the actions considered 
in later sections.

The covariantly  transverse linear and longitudinal linear superfields
on an arbitrary  supergravity background were described in 
the previous section. In the case of (1,1) AdS superspace, 
such superfields play an important role.
One can define projectors $P^{\perp}_{n}$ and $P^{||}_{n}$
on the spaces of  transverse linear and longitudinal linear superfields, respectively.
The projectors are 
\begin{subequations}
\bea
P^{\perp}_{n}&=& \frac{1}{4 (n+1)\m} (\bar {\frak D}^2+2n\m) ~,\\
P^{||}_{n}&=&- \frac{1}{4 (n+1)\m} (\bar {\frak D}^2-2(n+2)\m ) ~,
\eea
\end{subequations} 
with the properties 
\bea
\big(P^{\perp}_{n}\big)^2 =P^{\perp}_{n} ~, \quad 
\big(P^{||}_{n}\big)^2=P^{||}_{n}~,
\quad P^{\perp}_{n} P^{||}_{n}=P^{||}_{n}P^{\perp}_{n}=0~,
\quad P^{\perp}_{n} +P^{||}_{n} ={\mathbbm 1}~.
\eea

Given a complex tensor superfield $V_{\a(n)} $ with $n \neq 0$, 
it can be represented
as a sum of transverse linear and longitudinal linear multiplets, 
\bea
V_{ \a(n)} = &-& 
\frac{1}{2 \mu (n+2)} \bar{\frak D}^\g \bar{\frak D}_{(\g} V_{ \a_1 \dots  \a_n)} 
- \frac{1}{2 \mu (n+1)} \bar{\frak D}_{(\a_1} \bar{\frak D}^{|\g|} V_{ \a_2 \dots \a_{n} ) \g} 
~ . ~~~
\eea
Choosing $V_{ \a(n)} $ to be  
longitudinal linear ($G_{ \a(n)} $)
or transverse linear ($\G_{ \a(n)} $), the above identity 
gives the relations \eqref{2.19a} and \eqref{2.19b}
for some prepotentials $\J_{\a(n-1)}$ and $\F_{\a(n+1)} $, respectively.

In order to study rigid supersymmetric field theories in (1,1) AdS superspace, a superfield description of the corresponding isometry transformations is required. There exists a universal formalism to determine isometries of curved superspace backgrounds
in diverse dimensions \cite{Ideas}. Real supervector fields $\l^{\cA} E_{\cA} $ on ${\rm AdS}^{(3|1,1)} $ are called Killing supervector fields if 
\bea
\Big{[}\L+\hf l^{ab}M_{ab},\frak{D}_{C}\Big{]}=0~,
\quad
\L := \l^a\frak{D}_a+\l^\a\frak{D}_\a+\bar \l_\a\bar {\frak D}^\a~,
\quad \overline{\l^a}=\l^a~, \label{Kil-11}
\eea 
and $l^{ab}$ corresponds to some local Lorentz parameter.
As demonstrated in \cite{KT-M11},
the master equation \eqref{Kil-11} implies that the parameters $\l^\a$ and $l^{ab}$ 
are uniquely expressed in terms of the vector  $\l^a$,
\bea
\l_\a =\frac{\ri}{6} \bar{\frak D}^\b \l_{\a\b}~,\qquad
l_{\a\b} =2\frak{D}_{(\a}\l_{\b)}~,
\label{2.5}
\eea
and the vector parameter obeys the equation
\bea
\frak{D}_{(\a}\l_{\b\g)}=0 \quad \Longleftrightarrow \quad 
\bar{\frak D}_{(\a}\l_{\b\g)}=0~.
\eea
In comparison with the 3D $\cN=2$ Minkowski superspace,
the specific feature of ${\rm AdS}^{(3|1,1)} $ is that any two of the three parameters 
 $\{ \l_{\ab}, \l_\a, l_{\a\b}\}$ are expressed in terms of the third parameter, 
in particular
\bea
\l_{\a\b} =\frac{\ri}{\m} \bar {\frak D}_{(\a}\l_{\b)}~,\qquad 
\l_\a = \frac{1}{12 \bar \m} \frak{D}^\b l_{\a\b}
~.
\label{2.7-11}
\eea
From \eqref{2.5} and \eqref{2.7-11} we deduce
\bea
\bar {\frak D}^\a\l_\a=
\frak{D}_\a \l^\a=0~.
\label{1,1-SK_1}
\eea
These Killing supervector fields can be shown to generate  the isometry group of ${\rm AdS}^{(3|1,1)}$, which is
${\rm OSp(1|2; {\mathbb{R}})} \times {\rm OSp(1|2; {\mathbb{R}})}$.

In Minkowski superspace ${\mathbb M}^{3|4}$, there are two ways to generate supersymmetric invariants, 
one of which corresponds to the 
integration over the full superspace and the other over its chiral subspace. 
In (1,1) AdS superspace, every chiral integral can always be recast as 
a full superspace integral.
Associated with a scalar superfield $\cL$ is the following 
supersymmetric  invariant
\bea
\int \rd^3x \rd^2 \q  \rd^2 \bar \q
\,\bm E \,{\cal L} &=& 
-\frac{1}{4} \int
\rd^3x \rd^2 \q  
\, \cE\, {(\bar {\frak D}^2 - 4 \m)} {\cal L} ~, \qquad
\bm E^{-1}= {\rm Ber}\, (E_{\cA}{}^{\cM})~,~~~~
\label{3.11-11}
\eea
where 
$\cE$ denotes the chiral integration measure.
Let $\cL_{\rm c}$ be a covariantly chiral scalar Lagrangian, 
$\bar {\frak D}_\a \cL_{\rm c} =0$~. 
It generates a supersymmetric invariant of the form
$
\int \rd^3x \rd^2 \q  \, \cE \,{\cal L}_{\rm c}. 
$
The specific feature
of (1,1) AdS superspace is that the chiral action can equivalently
be written as an integral over the full superspace \cite{KT-M11}
\bea
\int \rd^3x \rd^2 \q  \, \cE \,{\cal L}_{\rm c} 
= \frac{1}{\m} \int \rd^3x \rd^2 \q  \rd^2 \bar \q
\,{\bm E}\, {\cal L}_{\rm c} ~.
\eea
Unlike the flat superspace case, the integral on the right does not vanish in AdS.

Supersymmetric invariant \eqref{3.11-11} can be reduced to component fields by the rule
\cite{KLRST-M}
\bea
\int \rd^3x \rd^2 \q  \rd^2 \bar \q
\,\bm E\,{\cal L} 
=\frac{1}{16}
\int\rd^3x\,e\,
(\frak{D}^2- 16 \bar \m) (\bar {\frak D}^2 - 4 \m) \cL 
\big|~,
\label{comp-ac-1}
\eea
with $e^{-1} := \det (e_a{}^m)$. Here $e_a{}^m$ is the inverse vielbein, which determines 
the torsion-free covariant derivative of AdS space
\bea
\nabla_a=e_a+\hf\o_a{}^{bc} (e)M_{bc}~, \qquad
e_a:=e_a{}^m \pa_m~.
\label{stcd-2}
\eea
In general, the $\q ,\bar \q$-independent component,  $T |_{\q=\bar \q=0}$, 
of a superfield $T(x,\q, \bar \q)$  is denoted by $T|$. To complete the formalism 
of component reduction, we only need the following relation 
\bea
\big(\frak{D}_a T \big) \big| = \nabla_a T|~.
\eea

In what follows, we will work with full superspace integrals only and 
make use of the notation $\rd^{3|4} z:= \rd^3x \rd^2 \q  \rd^2 \bar \q$.

\subsection{(2,0) AdS superspace} \label{ss522}
Let us briefly review the key results concerning (2,0) AdS superspace, ${\rm AdS}^{(3|2,0)}$; see \cite{KT-M11, BKT-M} for the details. There are two ways to describe the geometry of (2,0) AdS superspace, which correspond to making use of either a real or complex basis for the spinor covariant derivatives. Here we first consider the formulation in the complex basis. 

The geometry of ${\rm AdS}^{(3|2,0)}$ is described by covariant derivatives
\bea
\cD_{{\cA}}=(\cD_{{a}}, \cD_{{\a}},\bar \cD^\a)
=E_\cA{}^\cM \frac{\pa}{\pa z^\cM} + \hf\O_{\cA}{}^{cd} M_{cd}
+\ri \F_{{\cA}} J
\label{CovDev}
\eea
obeying the following algebra 
\begin{subequations} \label{deriv20}
\bea
\{\cD_\a,\cD_\b\}&=&0~,\qquad
\{\bar \cD_\a,\bar \cD_\b\}=0~,\\
\{\cD_\a,\cDB_\b\}
&=&
-2 \ri \big( \cD_{\a\b} -2 \cS M_{\a \b} \big)
-4 \ri \ve_{\a\b} \cS J ~,\\
{[}\cD_{a},\cD_\b {]}
&=&(\g_a)_\b{}^\g \cS \cD_{\g}~, \quad
{[}\cD_{a},\cDB_\b{]}
= (\g_a)_\b{}^{\g}\cS \bar \cD_{\g}~, \\
{[}\cD_a,\cD_b]{}
&=&-4 \cS^2 M_{ab} ~.
\eea
\end{subequations}
Here the parameter $\mathcal{S}$ is related to the AdS scalar curvature as $R= -24 \mathcal{S}^2$. 

The covariant derivatives of (2,0) AdS superspace hold various identities, which can be easily derived from 
the algebra \eqref{deriv20}. Some of the useful ones include
\begin{subequations} \label{cov-id}
\bea 
\left[\cD^\a, \bar \cD^2 \right]
&=& 4\ri \cD^{\a\b} \bar \cD_\b + 4\ri \cS  \bar \cD^\a - 8\ri \cS \bar \cD^\a J - 8\ri \cS \bar \cD_\b M^{\a \b} ~, \\
\left[\bar \cD^\a, \cD^2 \right]
&=& -4\ri \cD^{\a\b} \cD_\b - 4\ri \cS \cD^\a - 8\ri \cS \cD^\a J + 8\ri \cS \cD_\b M^{\a \b} ~,
\\
\big[\cD_a, \bar \cD^2\big] &=&0~,  \qquad \big[\cD_a,  \cD^2\big] =0~. \label{A.2c}
\eea
\end{subequations} 
These relations imply 
\bea
\cD^\a \bar \cD^2 \cD_\a = \cDB_\a \cD^2 \cDB^\a ~,
\label{1.44}
\eea
which guarantees the reality of the actions considered 
in the later sections.

In accordance with the general formalism of \cite{Ideas}, the isometries of (2,0) AdS superspace are generated by the Killing supervector fields 
$\z^\cA E_\cA$, which are defined to solve the master equation 
\begin{subequations}\label{Killing}
\bea
\big{[}\z+\hf l^{bc} M_{bc}+\ri \t J
\, ,\cD_\cA\big{]}=0~,
\eea
where 
\bea
\z= \z^\cB \cD_\cB =\z^b\cD_b+\z^\b\cD_\b+\bar \z_\b\bar \cD^\b~,
\qquad \overline{\z^b}=\z^b~,
\eea 
\end{subequations}
and $\t$ and $l^{bc}$ are some real U(1)${}_R$ and Lorentz superfield parameters, respectively.
It follows from eq. \eqref{Killing} that the parameters $\z_\a, \t$  and $l_{\a\b}$ are uniquely expressed in terms of the vector parameter $\z_{\a\b}$ as follows:
\bea
\z_\a= \frac{\ri}{6} \bar \cD^\b \z_{\b\a}~,\quad 
\t =\frac{\ri}{2} \cD^\a\z_\a 
~, \qquad 
l_{\a\b} =2\big( \cD_{(\a } \z_{\b)} -\cS \z_{\a\b} \big) ~.
\eea 
The vector parameter $\z_{\a\b}$ satisfies the equation 
\bea
\cD_{(\a}\z_{\b\g)}=0~.
\eea
This implies the standard Killing equation, 
\bea
\cD_a \z_b + \cD_b \z_a=0~.
\eea
One may also prove the following relations
\bea
 \bar \cD_\a \t
=\frac{\ri}{3}\bar \cD^\b l_{\a\b} = 4\cS \z_\a~,
\qquad 
\bar \cD_\a\z_\b=0~, \qquad 
\cD_{(\a} l_{\b\g)}=0~,
\eea
see \cite{KT-M11} for derivations. 
The Killing supervector fields prove to
generate the supergroup 
$\rm OSp(2|2;{\mathbb R}) \times Sp(2,{\mathbb R})$, 
which is the isometry group of (2,0) AdS superspace. Rigid supersymmetric field theories  in (2,0) AdS superspace are required to be  invariant under the isometry transformations. An infinitesimal isometry transformation acts on a tensor superfield $\bm U$ (with suppressed indices) by the rule
\bea
\d_\z {\bm U} = \big(\z+\hf l^{bc} M_{bc}+\ri \t J
\big) {\bm U}~.
\eea
Associated with a real scalar superfield $\cL$ is the following 
supersymmetric  invariant
\bea
\int \rd^3x \rd^2 \q  \rd^2 \bar \q
\, \bm E\,{ \cL} &=& 
-\frac{1}{4} \int
\rd^3x \rd^2 \q  
\, \cE\, {\bar \cD}^2  {\cL} ~.
\eea

\section{Massless half-integer superspin gauge theories in (1,1) AdS superspace} 
\label{s53}
The results presented in this section were obtained by Daniel Ogburn \cite{HKO}.

The conformal higher-spin gauge superfields $\frak{H}_{\a(n)}$ (see \ref{ss513})
at least for $n=2s$, with $s=1,2,\dots,$
can be used to construct massless actions
 in two of the three $\cN=2$ maximally symmetric backgrounds, which are  
 Minkowski superspace \cite{KO} and (1,1) AdS superspace \cite{HKO}.
Such actions, however, involve
not only ${\mathfrak H}_{\a(n)}$ but also some compensators. 

It is worth pointing out that all massless higher-spin supermultiplets in 3D (1,1) AdS superspace
may be obtained from their counterparts in 4D $\cN=1$ AdS superspace \cite{KS94}
by dimensional reduction. In practice, however, carrying out such a reduction 
proves to be a non-trivial technical task. To explain this, let us consider 
the longitudinal formulation for 
massless superspin-$(s+\hf)$ multiplets, with $s=1,2,\dots$,
 in four and three dimensions.\footnote{The $s=1$ case
corresponds to linearised supergravity.} 
In the 4D $\cN=1$ AdS case \cite{KS94}, 
the massless superspin-$(s+\hf)$ multiplet
is described by a real unconstrained gauge superfield 
$H_{\a_1 \dots \a_s \ad_1 \dots \ad_s}= H_{(\a_1 \dots \a_s )( \ad_1 \dots \ad_s)}$,
 a complex longitudinal linear compensator
$G_{\a_1 \dots \a_{s-1} \ad_1 \dots \ad_{s-1}}
= G_{(\a_1 \dots \a_{s-1} )( \ad_1 \dots \ad_{s-1})}$ and its conjugate.
The dimensional reduction 
of $H_{\a_1 \dots \a_s \ad_1 \dots \ad_s}$ 
leads to a family of real unconstrained symmetric 
superfields $H_{\a_1 \dots \a_{2s}}$, $H_{\a_1 \dots \a_{2s-2}}$, $\cdots$, $H$.
Next, the dimensional reduction 
of $G_{\a_1 \dots \a_{s-1} \ad_1 \dots \ad_{s-1}}$ 
 leads to a family of constrained 3D superfields, 
which include a complex longitudinal linear compensator
$G_{\a_1 \dots \a_{2s-2}}$ and some lower-spin supermultiplets.

As will be shown later, the massless superspin-$(s+\hf)$ multiplet in 3D (1,1) AdS superspace
is described by the gauge superfield $H_{\a_1 \dots \a_{2s}}$, the 
compensator $G_{\a_1 \dots \a_{2s-2}}$ and its conjugate.
The above consideration makes it clear  
that the naive $\rm 4D \to 3D$ dimensional reduction 
leads to the  massless superspin-$(s+\hf)$ multiplet intertwined with 
lower-superspin multiplets. The non-trivial technical task is to disentangle 
the pure superspin-$(s+\hf)$ multiplet from the rest.
This was explicitly done in \cite{KT-M11} for the $s=1$ case,
for which dimensional reduction  leads to two supermultiplets in (1,1) AdS superspace:
a  massless superspin-$\frac{3}{2}$ multiplet 
and a massless vector supermultiplet. 
Instead of carrying out dimensional reduction, 
it proves to be more efficient to recast
the 4D gauge principle of \cite{KS94} in a 3D form 
and use it to construct gauge-invariant actions. 
This is the approach advocated in \cite{KO, HKO}. 

We recall the constructions presented in \cite{KO}. There exist two off-shell formulations for the massless ${\cal N}= 2$ multiplet 
of superspin-$(s+\hf), \,s= 2, 3, \dots$, which describe two propagating massless fields with spin-$(s+1)$ and spin-$(s+ \hf)$ on Minkowski space \cite{KO}. These dually equivalent formulations, known as transverse and longitudinal, differ in the compensators used. 

Let us extend these gauge theories to (1,1) AdS superspace. There exist two formulations which are described in terms of the following dynamical variables
\bsubeq
\bea
\cV^\bot_{(s+\hf )} &=& 
\Big\{ {\mathfrak H}_{\a(2s)}, \G_{\a(2s-2)}, \bar{\G}_{\a(2s-2)} \Big\} ~, \label{541} \\
\cV^{\|}_{(s+\hf)} &=& \Big\{ {\mathfrak H}_{\a(2s)}, G_{\a(2s-2)}, \bar{G}_{\a(2s-2)} \Big\} ~. \label{542}
\eea
\esubeq
Here ${\mathfrak H}_{\a(2s)}$
is an unconstrained real superfield. The complex superfields
$\G_{\a(2s-2)}$ and $G_{\a(2s-2)}$ are transverse linear and longitudinal linear in the sense that they obey the constraints \eqref{2.144-11} and \eqref{2.111}, respectively. 
In accordance with \eqref{2.19-11},
these constraints can be solved in terms of  unconstrained complex prepotentials as follows:
\bsubeq
\bea
 \G_{\a(2s-2)}&=& \bar {\frak D}^\b
{ \Phi}_{(\b \a_1 \dots \a_{2s-2} )} ~, \label{543}\\
 G_{\a(2s-2)} &=& \bar {\frak D}_{(\a_1}
{ \J}_{ \a_2 \dots \a_{2s-2}) } ~.
\label{544}
\eea
\esubeq
These prepotentials are defined modulo gauge transformations of the form 
\bsubeq
\bea
\d_\x \Phi_{\a(2s-1)} 
&=&  \bar {\frak D}^\b 
{ \x}_{(\b \a_1 \dots \a_{2s-1})} ~,
\label{tr-prep-gauge-11}\\
\d_\z \J_{\a(2s-3)} 
&=&  \bar {\frak D}_{(\a_1 }
{ \z}_{\a_2 \dots \a_{2s-3})} ~,
\label{long-prep-gauge-11}
\eea
\esubeq
with the gauge parameters ${\x_{\a(2s)}}$ and ${\z_{\a(2s-4)}}$ being unconstrained complex. 

The dynamical superfields ${\mathfrak H}_{\a(2s)}$ and $ \G_{\a(2s-2)}$
are postulated to be defined modulo gauge transformations of the form 
\bsubeq \label{gaugeLambda-11}
\bea
\d_\l {\mathfrak H}_{\a(2s)}&=& 
\bar {\frak D}_{(\a_1} \l_{\a_2 \dots \a_{2s})}-{\frak D}_{(\a_1}\bar \l_{\a_2 \dots \a_{2s})}
\equiv
g_{\a(2s)}+\bar{g}_{\a(2s)} ~, \label{H-gauge-11} \\ 
\d_\l \Gamma_{\a(2s-2)} &=&
-\frac{1}{4}\bar{\frak D}^{\b} 
\big( {\frak D}^2  +2(2s-1) \bar \m \big)\bar{\l}_{\b\a(2s-2)}
= \frac{s}{2s+1}\bar{\frak D}^{\b}\frak{D}^{\g}\bar g_{(\b \g \a_1 \dots \a_{2s-2})}
~,~~~~~~~~~
\label{gamma-gauge-11} \\
\delta_\l G_{\a(2s-2)} &=& 
-\frac{1}{4}\big( \bar{\frak D}^{2} -4s\m\big)\frak{D}^{\b}\l_{\a(2s-2)\b}
+\ri (s-1)\bar{\frak D}_{(\a_{1}} \frak{D}^{|\b\g|} \l_{\a_2 \dots \a_{2s-2})\b\g} ~. \label{G-gauge-11}
\eea
\esubeq
where the complex gauge parameter $\l_{\a(2s-1)}$ is unconstrained. 
The gauge transformation of ${\mathfrak H}_{\a(2s)}$
coincides with \eqref{2.28-11} for $n=2s$.
From $\d_\l \Gamma_{\a(2s-2)}$, we can read off 
the gauge transformation of the prepotential $\Phi_{\a(2s-1)}$, which is 
\bea
\d_\l \Phi_{\a(2s-1)} = -\frac{1}{4}
\big( {\frak D}^2  +2(2s-1) \bar \m \big)\bar{\l}_{\a(2s-1)}~. \label{545}
\eea

In the transverse formulation, the quadratic action invariant under 
the gauge transformations \eqref{H-gauge-11} and \eqref{gamma-gauge-11} is
\bea
S^{\perp}_{(s+\hf)}
&=& \Big(-\hf \Big)^s \int 
 \rd^{3|4}z \, \bm E
\bigg\{ \frac{1}{8} {\mathfrak H}^{\a(2s)}  \frak{D}^\b ({\bar{ \frak D}}^2- 6\mu) \frak{D}_\b 
{\mathfrak H}_{\a(2s)}
\non \\
&&+ 2s(s-1)\bar \m \m {\mathfrak H}^{\a(2s)} {\mathfrak H}_{\a(2s)}
+ {\mathfrak H}^{\a(2s)}\big(\frak{D}_{\a_1} \bar {\frak D}_{\a_2} \G_{\a_3 \dots \a_{2s}} 
- \bar {\frak D}_ {\a_1} \frak{D}_{\a_2} {\bar \G}_{\a_3 \dots \a_{2s}} \big)
\non \\
&&+ \frac{2s-1}{s} \bar \G^{\a(2s-2)} \G_{\a(2s-2)} 
+ \frac{2s+1}{2s} 
\big(\G^{\a(2s-2)}\G_{\a(2s-2)} + \bar \G^{\a(2s-2)} \bar \G_{\a(2s-2)}\big) \bigg\}
\label{tr-action-half-11}~.~~~~~~~~~~
\eea
In the flat superspace limit, this action 
reduces to the one derived in \cite{KO}.

The $s=1$ choice was excluded from the above consideration, since 
the constraint  \eqref{2.144-11} is not defined for $n=0$.
However, the corollary \eqref{2.15}
of  \eqref{2.144-11} is perfectly consistent for $n=0$ and defines  a covariantly 
transverse linear scalar superfield \eqref{2.18}, 
\bea
(\bar {\frak D}^2 -\m ) \G = 0~.
\label{545-11}
\eea
We therefore postulate  $\G$ and its conjugate $\bar \G$
to be the compensators in the $s=1$ case. 
The gauge transformations \eqref{H-gauge-11} and \eqref{gamma-gauge-11} then become
\begin{subequations}
\bea
\d_\l {\mathfrak H}_{\a \b}&=& 
\bar {\frak D}_{(\a} \l_{\b)}-{\frak{D}}_{(\a}\bar \l_{\b)} ~, 
\\ 
\d_\l \Gamma &=&
-\frac{1}{4}\bar{\frak D}^{\b} 
\big( {\frak D}^2  +2 \bar \m \big)\bar{\l}_{\b}~.
\eea
\end{subequations}
The variation $\d_\l \G$ is compatible with 
the constraint \eqref{545-11}, that is $(\bar {\frak D}^2 -\m ) \d_\l \G=0$.
Finally, choosing $s=1$ in \eqref{tr-action-half-11} gives the linearised 
action for non-minimal (1,1) AdS supergravity, 
which was originally derived in section 9.2 of \cite{KT-M11}.

%
In the longitudinal formulation, the action invariant under the gauge transformations \eqref{H-gauge-11} and \eqref{G-gauge-11} is 
\bea
S^{\|}_{(s+\hf)}
&=& \Big(-\hf \Big)^{s}\int 
\rd^{3|4}z
\, \bm E \bigg\{\frac{1}{8}{\mathfrak H}^{\a(2s)}
\frak{D}^{\b}(\bar{\frak D}^{2}-6\mu)\frak{D}_{\b}
{\mathfrak H}_{\a(2s)} \non \\
&& +2s(s-1)\mu\bar{\mu} {\mathfrak H}^{\a(2s)}{\mathfrak H}_{\a(2s)}
-\frac{1}{16}([\frak{D}_{\b},\bar{\frak D}_{\g}]{\mathfrak H}^{\b \g \a(2s-2)})
[\frak{D}^{\d},\bar{\frak D}^{\r}]{\mathfrak H}_{\d \r \a(2s-2)}
 \non \\
&& +\frac{s}{2}(\frak{D}_{\b \g}{\mathfrak H}^{\b \g \a(2s-2)})
\frak{D}^{\d \r}{\mathfrak H}_{\d \r \a(2s-2)}
\non \\
&&+  \frac{2s-1}{2s} \Big[ \ri
(\frak{D}_{\b \g}{\mathfrak H}^{\b \g \a(2s-2)}) 
\left( G_{\a(2s-2)}-\bar{G}_{\a(2s-2)} \right)
+\frac{1}{s}
\bar{G}^{\a(2s-2)} G_{\a(2s-2)} \Big]
\non \\
&& 
-\frac{2s+1}{4s^{2}}\left(G^{\a(2s-2)}G_{\a(2s-2)}+\bar{G}^{\a(2s-2)} \bar G_{\a(2s-2)}\right)
\bigg\}~.
\label{long-action-half-11}
\eea
As shown in \cite{HKO}, the longitudinal action may be derived from the transverse one by performing a superfield duality transformation. 

In the $s=1$ case, the compensator $G$ becomes covariantly chiral, $\bar {\frak D}_\a G=0$.
Choosing $s=1$ in \eqref{long-action-half-11} gives the linearised 
action for minimal (1,1) AdS supergravity, 
which was originally derived in section 9.1 of \cite{KT-M11}, 
provided we identify $G=3\s$. 
The corresponding gauge transformations are
\begin{subequations}
\bea
\d_\l {\mathfrak H}_{\a \b}&=& 
\bar {\frak D}_{(\a} \l_{\b)}-{\frak{D}}_{(\a}\bar \l_{\b)} ~, \\
\delta_\l G&=& 
-\frac{1}{4}\big( \bar{\frak D}^{2} -4\m\big)\frak{D}^{\b}\l_{\b}~.
\eea
\end{subequations}
It is clear that the variation $\delta_\l G$ is covariantly chiral.


\section{Massless integer superspin gauge theories in (1,1) AdS superspace} 
\label{s54}
The results in subsections \ref{subsection5.1} and \ref{subsect5.2} were obtained by Daniel Ogburn \cite{HKO}. 

When attempting to develop a Lagrangian formulation for a massless multiplet 
of superspin $s$, where $s=1,2,\dots$, a naive expectation is that 
the dynamical variables of such a theory should consist of 
a conformal gauge superfield ${\mathfrak H}_{\a(2s-1)} =\bar {\mathfrak H}_{\a(2s-1)} $,
introduced in subsection \ref{ss513}, in conjunction with some compensator(s).
Instead, our approach in this section will be based on developing
3D $\cN=2$ analogues of the two dually equivalent off-shell
formulations, the so-called longitudinal and transverse  ones, 
for the  massless ${\cal N}=1$ multiplets of 
integer superspin in AdS${}_4$ \cite{KS94}. As the next step, we will construct a generalised longitudinal model,
in a way similar to the one proposed in the 4D $\cN=1$ AdS case in subsection \ref{ss421}.
Such a reformulation naturally leads to the appearance of 
the conformal gauge superfield ${\mathfrak H}_{\a(2s-1)}  $.

\subsection{Longitudinal formulation} \label{subsection5.1}

Given an integer $s \geq 1$, the longitudinal formulation 
for the massless superspin-$s$ multiplet 
is realised in terms of the following dynamical variables:
\bea
\cV^{\|}_{(s)} = \Big\{U_{\a(2s-2)}, G_{\a(2s)}, \bar{G}_{\a(2s)} \Big\} ~.
\label{5.1}
\eea
Here $U_{\a(2s-2)}$ is an unconstrained real superfield, and the complex superfield $G_{\a(2s)}$ is longitudinal linear, eq.  \eqref{2.111}. 
In accordance with \eqref{2.19a}, 
the constraint \eqref{2.111} can be solved in terms of an  unconstrained complex prepotential $\J_{\a(2s-1)}$,
\bea
G_{\a_1 \dots \a_{2s}} := 
\bar {\frak D}_{(\a_1} \J_{\a_2 \dots \a_{2s})}~,
\label{g2.5}
\eea
which is defined modulo gauge transformations of the form 
\be
\d_\z \J_{\a(2s-1)} =  \bar {\frak D}_{(\a_1 }
{ \z}_{\a_2 \dots \a_{2s-1})} ~,
\ee
with the gauge parameter ${\z_{\a(2s-2)}}$ being unconstrained complex.

We postulate the dynamical superfields $ U_{\a(2s-2)}$ and $ G_{\a(2s)}$
 to be  defined modulo gauge transformations of the form 
 \begin{subequations}\label{5.4ab}
\bea
\d_L U_{\a(2s-2)}
&=&\frak{D}^{\b}L_{\b \a_1 \dots \a_{2s-2}}-\bar{\frak D}^{\b}\bar{L}_{\b \a_1 \dots \a_{2s-2}} 
\equiv \bar \g_{\a(2s-2)}+{\g}_{\a(2s-2)}
~, \label{H-gauge-int} \\
\d_L G_{\a(2s)} 
&=&-\frac{1}{2}\bar{\frak D}_{(\a_{1}}\Big(\frak{D}^{2}
-2(2s+1) \bar \m \Big)
L_{\a_2 \dots \a_{2s})}
= \bar{\frak D}_{(\a_{1}} \frak{D}_{\a_{2}}\bar \g_{\a_3 \dots \a_{2s})}
~.
\label{int-long-gauge}
\eea
\end{subequations} 
Here the gauge parameter $L_{\a(2s-1)}$ is an unconstrained complex superfield, 
and ${\g}^{\a(2s-2)}:= \bar {\frak D}_{\b}\bar L^{\b \a(2s-2)}$ is transverse linear. 
From \eqref{int-long-gauge} we read off the gauge transformation law of the prepotential, 
\bea
\d_L \J_{\a(2s-1)} =-\frac{1}{2} \Big(\frak{D}^{2}
-2(2s+1) \bar \m \Big) L_{ \a (2s-1)}
=\frak{D}_{(\a_1} \frak{D}^{|\b |} L_{\a_2 \dots \a_{2s-1}) \b}~. 
\eea

Modulo an overall  normalisation factor, there is a unique quadratic action 
which is invariant under 
the gauge transformations \eqref{5.4ab}. The action is
\bea
S_{(s)}^{\|}
&=& \Big(-\hf\Big)^{s}\int 
\rd^{3|4}z\, \bm E \,
\bigg\{\frac{1}{8}U^{\a(2s-2)}\frak{D}^{\g}(\bar {\frak D}^{2}-6\mu)\frak{D}_{\g}U_{\a(2s-2)}
\non \\
&&+\frac{s}{2s+1}U^{\a(2s-2)}\Big(\frak{D}^{\b} \bar {\frak D}^{\g}
G_{\b \g \a(2s-2) }
-\bar {\frak D}^{\b}{\frak D}^{\g}\bar{G}_{\b \g \a(2s-2)} \Big) \non \\
&&+\frac{s}{2s-1}  \bar{G}^{\a(2s)} G_{\a(2s)}
+ \frac{s}{2(2s+1)}\Big(G^{\a(2s)}G_{\a(2s)}+\bar{G}^{\a(2s)}\bar G_{\a(2s)}\Big) 
\non \\
&&+2s(s+1)\mu\bar{\mu} U^{\a(2s-2)}U_{\a(2s-2)}
\bigg\}~.
\label{long-action-int}
\eea
The special $s=1$ case, which corresponds to the massless gravitino  
multiplet, will be studied in more detail in subsection \ref{subsection5.4}.


\subsection{Transverse formulation} \label{subsect5.2}

The transverse formulation for the massless superspin-$s$ multiplet 
is realised in terms of the following dynamical variables:
\bea
\cV^{\perp}_{(s)} = \Big\{U_{\a(2s-2)}, \G_{\a(2s)}, \bar{\G}_{\a(2s)} \Big\} ~.
\eea
Here $U_{\a(2s-2)}$ is the same as in \eqref{5.1}, 
and the complex superfield $\G_{\a(2s)}$ is transverse linear, eq. \eqref{2.144-11}.
In accordance with \eqref{2.19b},
the constraint on $\G_{\a(2s)} $ is
 solved in terms of an unconstrained prepotential $\Phi_{\a(2s+1)}$,
\bea
 \G_{\a(2s)}= \bar {\frak D}^\b
{ \Phi}_{(\b \a_1 \dots \a_{2s} )} ~,
\label{5.8}
\eea
which is defined modulo gauge transformations of the form 
\bea
\d_\x \Phi_{\a(2s+1)} 
=  \bar {\frak D}^\b 
{ \x}_{(\b \a_1 \dots \a_{2s+1})} ~,
\eea
with the gauge parameter ${\x_{\a(2s+2)}}$ being unconstrained.

The transverse formulation for the massless superspin-$s$ multiplet 
is described by the following action
\bea
S_{(s)}^{\perp}
&=& \Big(-\hf\Big)^{s}\int 
\rd^{3|4}z
\, \bm E \,
\bigg\{\frac{1}{8}U^{\a(2s-2)}\frak{D}^{\g}({\bar {\frak D}}^{2}-6\mu)\frak{D}_{\g}U_{\a(2s-2)}\non \\
&& -\frac{2s-1}{16(2s+1)} \Big( 
8s  {\frak D}^{\a_1 \a_2} U^{\a_3 \dots \a_{2s}} {\frak D}_{(\a_1 \a_2} U_{\a_3 \dots \a_{2s})} \non \\
&& + [\frak{D}^{\a_1}, {\bar {\frak D}}^{\a_2}]U^{\a_3 \dots \a_{2s}} 
[\frak{D}_{(\a_1}, {\bar {\frak D}}_{\a_2}]U_{\a_3 \dots \a_{2s})}\Big) \non \\
&& 
+2s(s+1)\mu\bar{\mu} U^{\a(2s-2)}U_{\a(2s-2)}
-\ri U^{\a_1 \dots \a_{2s-2}} {\frak D}^{\a_{2s-1}  \a_{2s}} 
\big( \G_{\a(2s)} -\bar \G_{\a(2s)} \big)\non \\
&&- \frac{2}{2s-1} \bar \G^{\a(2s)} \G_{\a(2s)} 
+\frac{1}{2s+1} (\G^{\a(2s)} \G_{\a(2s)}+\bar{\G}^{\a(2s)} \bar \G_{\a(2s)})
 \bigg\}~,
\label{tr-action-int}
\eea
which is invariant under the gauge transformation  \eqref{H-gauge-int}
accompanied with
\bea
\d_L \G_{\a(2s)}&=&
-\frac{1}{4}(\bar{\frak{D}}^{2} +4s\mu)\frak{D}_{(\a_{1}}\bar{L}_{\a_2 \dots \a_{2s})}
+  \frac{\ri}{2}  (2s+1)\bar{\frak D}^{\g}\frak{D}_{(\g\a_{1}}\bar{L}_{\a_2 \dots \a_{2s})} \non \\
&=&
\frac{1}{2}\frak{D}_{(\alpha_{1}}\bar{\frak D}_{\alpha_{2}}{\gamma}_{\a_3 \dots \a_{2s})}-\frac{\ri}{2} (2s-1)
 \frak{D}_{(\alpha_{1}\alpha_{2}}{\gamma}_{\a_3 \dots \a_{2s})} 
~,
\label{tr-gauge-int}
\eea
where ${\gamma}_{\a(2s-2)} =-\bar{\frak D}^{\b}\bar{L}_{\b \a_1 \dots \a_{2s-2}} $.


\subsection{Reformulation of the longitudinal theory }\label{ss543} 

Let us take a step further and consider a generalisation of the longitudinal formulation \eqref{long-action-int}. This can be achieved by enlarging the gauge freedom \eqref{5.4ab}, where we choose to work with an \textit{unconstrained} complex gauge parameter $\mathfrak{V}_{\a(2s-2)}$, instead of the transverse linear superfield $\g_{\a(2s-2)}$. As a result, we are required to introduce a new purely gauge superfield, in addition to 
$U_{\a(2s-2)}$, $\J_{\a(2s-1)}$ and $\bar\J_{\a(2s-1)}$.
In such a setting, the  gauge freedom of  $\J_{\a(2s-1)}$ 
coincides with that of a {\it complex} conformal  gauge superfield.

Given a positive integer $s \geq 2$, a massless superspin-$s$ multiplet in ${\rm AdS}^{(3|1,1)}$ can be described using
a complex unconstrained prepotential $\J_{\a(2s-1)}$, a real superfield $U_{\a(2s-2)}$ and a complex superfield $\S_{\a(2s-3) }$ constrained to be transverse linear,  
\bea
\bar {\frak D}^\b \S_{\b \a(2s-4)} =0~.
\label{2.1-11}
\eea
The constraint \eqref{2.1-11} is solved in terms of a complex unconstrained prepotential $Z_{\a(2s-2)}$,
\bea
\S_{\a(2s-3)} = \bar {\frak D}^\b Z_{ (\b \a_1 \dots \a_{2s-3} )} ~,
\label{2.2-11}
\eea
which is defined modulo gauge shift
\bea
\d_\x Z_{\a(2s-2)}=  \bar {\frak D}^\b \x_{ (\b \a_1 \dots \a_{2s-2} )} ~.
\label{2.3-11}
\eea
Here the gauge parameter  $\x_{\a(2s-1)}$ is unconstrained.

The gauge freedom of $\J_{\a(2s-1)} $ is
given by
\begin{subequations} \label{2.4-11}
\bea
 \d_{ {\mathfrak V} ,\z} \J_{\a_1 \dots\a_{2s-1}} 
 &=& {\frak D}_{(\a_1}  {\mathfrak V}_{\a_2 \dots \a_{2s-1})}
+  \bar {\frak D}_{(\a_1} \z_{\a_2 \dots \a_{2s-1} )}  ~ , \label{2.4a-11}
\eea
with unconstrained gauge parameters ${\mathfrak V}_{\a(2s-2)}$ 
and $\z_{\a(2s-2)}$. 
We further postulate the linearised gauge transformations for the superfields $U_{\a(2s-2)}$
and $\S_{\a(2s-3)}$ as follows
\bea
\d_{\mathfrak V} U_{\a(2s-2)}&=& {\mathfrak V}_{\a(2s-2)} +\bar {\mathfrak V}_{\a(2s-2)}
~, \label{2.4b-11}\\
\d_{\mathfrak V} \S_{\a(2s-3) }&=&  \bar {\frak D}^\b \bar {\mathfrak V}_{\b \a(2s-3)}
\quad \Longrightarrow \quad \d_{\mathfrak V} Z_{\a(2s-2)}
=\bar  {\mathfrak V}_{\a(2s-2)}~.~~~~~
\label{2.4c-11}
\eea
\end{subequations}
The longitudinal linear superfield $G_{\a(2s)}$ defined by \eqref{g2.5}
is invariant under the $\z$-transformation \eqref{2.4a-11}. It varies under the $\mathfrak V$-transformation as 
\bea
 \d_{ {\mathfrak V} } G_{\a_1 \dots \a_{2s}} 
 &=&  \bar {\frak D}_{(\a_1} \frak{D}_{\a_2}  {\mathfrak V}_{\a_3 \dots \a_{2s})}~.
\eea

The action 
\bea
S^{\|}_{(s)} &=&
\Big( - \frac{1}{2}\Big)^s  \int 
\rd^{3|4}z\, \bm E
\left\{ \frac{1}{8} U^{ \a (2s-2) }  \frak{D}^\b ({\bar {\frak D}}^2- 6\mu) \frak{D}_\b 
U_{\a (2s-2)} \right. \non \\
&&+ \frac{s}{2s+1}U^{ \a(2s-2) }
\Big( \frak{D}^{\b}  {\bar {\frak D}}^{\g} G_{\b \g \a(2s-2)}
- {\bar {\frak D}}^{\b}  \frak{D}^{\g} 
{\bar G}_{\b \g \a(2s-2) } \Big) \non \\
&&+ 2s(s+1) \bar \mu \mu U^{\a (2s-2)} U_{\a (2s-2)} \non \\
&&+ \frac{s}{2s-1} \bar G^{ \a (2s)} G_{ \a (2s)} 
+ \frac{s}{2(2s+1)}\Big( G^{ \a (2s) } G_{ \a (2s)} 
+ \bar G^{ \a (2s) }  \bar G_{ \a (2s) } 
 \Big) \non \\
 &&+ \hf \frac{s-1}{2s-1}U^{ \a(2s-2) }
\Big( \frak{D}_{\a_1} \bar {\frak D}^2 \bar \S_{\a_2 \dots \a_{2s-2}}
 - {\bar {\frak D}}_{\a_1}  \frak{D}^2 \S_{\a_2 \dots \a_{2s-2} } \Big)  \non \\
&&+\frac{1}{2s-1} \J^{\a(2s-1)} \Big( 
\frak{D}_{\a_1} \bar {\frak D}_{\a_2} -2\ri (s-1) \frak{D}_{\a_1 \a_2} \Big)
\S_{\a_3 \dots \a_{2s-1} }\non  \\
&&+\frac{1}{2s-1} \bar \J^{\a(2s-1)} \Big( 
\bar {\frak D}_{\a_1} \frak{D}_{\a_2} -2\ri (s-1) \frak{D}_{\a_1 \a_2} \Big)
\bar \S_{\a_3 \dots \a_{2s-1} }\non  \\
&&- \mu (s+3) U^{\a (2s-2)} \frak{D}_{\a_1} 
\bar \S_{\a_2 \dots \a_{2s-2}} + \bar \mu (s+3) U^{\a (2s-2)} \bar {\frak D}_{\a_1} 
 \S_{ \a_2 \dots \a_{2s-2}} \non \\
&&+ \frac{s-1}{4(2s-1)} \Big( \S^{\a(2s-3) } \frak{D}^2 \S_{\a(2s-3)} 
- \bar \S^{\a(2s-3) }\bar {\frak D}^2 \bar \S_{\a(2s-3)} \Big) \non \\
&&- \frac{1}{2s-1}\bar \S^{\a(2s-3)} \Big( (2s^2-s+1) \frak{D}^\b \bar {\frak D}_{\a_1 }
+2\ri \frac{ (s-1)(2s-3)}{2s-1} \frak{D}^\b{}_{\a_1} \Big) \S_{\b \a_2 \dots \a_{2s-3}} \non\\
&&+ \mu (s+3) \bar \S^{\a(2s-3)} \bar \S_{\a(2s-3)} + \bar \mu (s+3) \S_{\a(2s-3)}  \S^{\a(2s-3)} \Big\}
~,~~~
\label{action-11}
\eea
possesses the gauge invariance \eqref{2.4-11} and, by construction, \eqref{2.3-11}. The above action is real due to the identity \eqref{1.44}.

Due to the $\mathfrak V$-gauge freedom \eqref{2.4-11}, we are free to make the gauge choice 
\bea
 \S_{\a(2s-3)}=0~,
\label{2.10-11}
\eea
by which we regain the original longitudinal action for the massless superspin-$s$ multiplet \eqref{long-action-int}. The gauge condition \eqref{2.10-11} does not 
fix completely the $\mathfrak V$-gauge freedom. The residual gauge transformations  
are generated by 
\bea
{\mathfrak V}_{\a(2s-2)} = \frak{D}^\b L_{(\b \a_1 \dots \a_{2s-2})}~,
\label{2.11-11}
\eea
with $L_{\a(2s-1)}$ being an unconstrained superfield. With this expression for ${\mathfrak V}_{\a(2s-2)}$, the gauge transformations \eqref{2.4a-11}  and \eqref{2.4b-11} coincide with \eqref{int-long-gauge}.
Thus, the action \eqref{action-11} indeed provides an off-shell formulation for the massless superspin-$s$ multiplet in (1,1) AdS superspace.

The action \eqref{action-11} contains a single term which involves the gauge prepotential $\bar \J_{\a(2s-1)} $ and not the field strength $\bar G_{\a(2s)} $. 
This term can be written as
\bea
 \int  
 \rd^{3|4}z \, \bm E
 \,
 \bar \J^{\a(2s-1)} \Big( 
 \bar {\frak D}_{\a_1} \frak{D}_{\a_2} &-&2\ri (s-1) \frak{D}_{\a_1 \a_2} \Big)
\bar \S_{\a_3 \dots \a_{2s-1} } \non \\
= -\frac{2s}{2s+1}  
\int 
\rd^{3|4}z \, \bm E
\,
 \bar G^{\a(2s)} \Big( \frak{D}_{\a_1} \bar {\frak D}_{\a_2}  
&+&\ri (2s+1) \frak{D}_{\a_1 \a_2} \Big)
\bar Z_{\a_3 \dots \a_{2s} }~.~~~
\label{2.14-11}
\eea
The former makes the gauge symmetry \eqref{2.3-11} manifestly realised, 
while the latter
turns the $\z$-transformation \eqref{2.4a-11} into a manifest symmetry.
If we instead wish to make use of \eqref{2.14-11}, we obtain a different representation 
for the action \eqref{action-11}. It is 
\bea
S^{\|}_{(s)} &=&
\Big( - \frac{1}{2}\Big)^s  \int 
\rd^{3|4}z
\, \bm E
\left\{ \frac{1}{8} U^{ \a (2s-2) }  \frak{D}^\b ({\bar {\frak D}}^2- 6\mu) \frak{D}_\b 
U_{\a (2s-2)} \right. \non \\
&&+ \frac{s}{2s+1}U^{ \a(2s-2) }
\Big( \frak{D}^{\b}  {\bar {\frak D}}^{\g} G_{\b \g \a(2s-2)}
- {\bar {\frak D}}^{\b}  \frak{D}^{\g} 
{\bar G}_{\b \g \a(2s-2) } \Big) \non \\
&&+ 2s(s+1) \bar \mu \mu U^{\a (2s-2)} U_{\a (2s-2)} \non \\
&&+ \frac{s}{2s-1} \bar G^{ \a (2s)} G_{ \a (2s)} 
+ \frac{s}{2(2s+1)}\Big( G^{ \a (2s) } G_{ \a (2s)} 
+ \bar G^{ \a (2s) }  \bar G_{ \a (2s) } 
 \Big) \non \\
 &&+ \hf \frac{s-1}{2s-1}U^{ \a(2s-2) }
\Big( \frak{D}_{\a_1} \bar {\frak D}^2 \bar \S_{\a_2 \dots \a_{2s-2}}
 - {\bar {\frak D}}_{\a_1}  \frak{D}^2 \S_{\a_2 \dots \a_{2s-2} } \Big)  \non \\
&&+\frac{2s}{(2s-1)(2s+1)} G^{\a(2s)} \Big( \bar {\frak D}_{\a_1} \frak{D}_{\a_2} +\ri (2s+1) \frak{D}_{\a_1 \a_2} \Big)
Z_{\a_3 \dots \a_{2s} }\non  \\
&&-\frac{2s}{(2s-1)(2s+1)} \bar G^{\a(2s)} \Big( \frak{D}_{\a_1} \bar {\frak D}_{\a_2} +\ri (2s+1) \frak{D}_{\a_1 \a_2} \Big)
\bar Z_{\a_3 \dots \a_{2s} }\non  \\
&&- \mu (s+3) U^{\a (2s-2)} \frak{D}_{\a_1} 
\bar \S_{\a_2 \dots \a_{2s-2}} + \bar \mu (s+3) U^{\a (2s-2)} \bar {\frak D}_{\a_1} 
 \S_{ \a_2 \dots \a_{2s-2}} \non \\
&&+ \frac{s-1}{4(2s-1)} \Big( \S^{\a(2s-3) } \frak{D}^2 \S_{\a(2s-3)} 
- \bar \S^{\a(2s-3) }\bar {\frak D}^2 \bar \S_{\a(2s-3)} \Big) \non \\
&&- \frac{1}{2s-1}\bar \S^{\a(2s-3)} \Big( (2s^2-s+1) \frak{D}^\b \bar {\frak D}_{\a_1 }
+2\ri \frac{ (s-1)(2s-3)}{2s-1} \frak{D}^\b{}_{\a_1} \Big) \S_{\b \a_2 \dots \a_{2s-3}} \non\\
&&+ \mu (s+3) \bar \S^{\a(2s-3)} \bar \S_{\a(2s-3)} + \bar \mu (s+3) \S_{\a(2s-3)}  \S^{\a(2s-3)} \Big\}
~.~~~
\label{action2-11}
\eea

It is worth discussing the 
structure of the dynamical variable $\J_{\a(2s-1)}$. 
This superfield is unconstrained complex, and its gauge transformation law
is given by eq. \eqref{2.4a-11}. Comparing \eqref{2.4a-11} 
with the gauge transformation law \eqref{2.28-11} with  $n =2s-1$, 
which corresponds to  the conformal gauge superfield ${\mathfrak H}_{\a(2s-1)} $, 
we see that $\J_{\a(2s-1)}$ may be interpreted as a complex 
conformal gauge superfield.  

\subsection{Massless gravitino multiplet}\label{subsection5.4}

The massless gravitino multiplet, which corresponds to the $s=1$ case, 
was excluded from our consideration of the previous subsection. 
Here we will fill the gap.

The (generalised) longitudinal formulation for the gravitino 
multiplet is described by the action 
\bea
S^{\|}_{\rm GM} &=& -\hf \int \rd^{3|4}z
\, \bm E\,
\bigg\{ \frac{1}{8} U  \frak{D}^\b ({\bar {\frak D}}^2- 6\mu) \frak{D}_\b U
+ \frac{1}{3}U\big( \frak{D}^{\a}  {\bar {\frak D}}^{\b} G_{\a \b}
- {\bar {\frak D}}^{\a}  \frak{D}^{\b} {\bar G}_{\a \b } \big)  \non \\
&& + \bar G^{\a\b} G_{\a\b} 
+\frac 16 \big( G^{\a\b} G_{\a\b} +\bar G^{\a\b} \bar G_{\a\b} \big) \non \\
&&+|\m|^2 \Big(2U - \frac{\F}{\m} -\frac{\bar \F}{\bar \m} \Big)^2
+2\Big( \frac{\F}{\m} +\frac{\bar \F}{\bar \m}\Big)
\Big( \m \frak{D}^\a \J_\a + \bar \m \bar {\frak D}_\a \bar \J^\a \Big) 
\bigg\}~,
\label{5.21}
\eea
where $\F$ is a covariantly chiral scalar superfield, $\bar {\frak D}_\a \F=0$, and
\bea
G_{\a\b} = \bar {\frak D}_{(\a} \J_{\b)} ~, \qquad 
\bar G_{\a\b} = -\frak{D}_{(\a} \bar \J_{\b)} ~.
\eea
  This action is invariant under gauge transformations of the form 
\begin{subequations}\label{5.23}
\bea
\d U&=& {\mathfrak V} +\bar {\mathfrak V}~, \\
\d \J_\a &=& =  \frak{D}_\a  {\mathfrak V}+ \bar {\frak D}_\a \z~, 
 \\
\d \F &=& -\frac 14 (\bar {\frak D}^2 -4\m) \bar  {\mathfrak V}~,
\eea
\end{subequations}
where the gauge parameters $\mathfrak V$ and $\z$ are unconstrained complex
superfields. 
  
The gauge $\mathfrak V$-freedom \eqref{5.23} allows us to
 impose the condition $\F=0$. In this gauge 
the action \eqref{5.21}  turns into  \eqref{long-action-int} with $s=1$, and the residual 
gauge $\mathfrak V$-freedom is described by 
${\mathfrak V} = {\frak D}^\b L_\b$, where the spinor
gauge parameter $L_\a$ is unconstrained complex.
  
The action \eqref{5.21} involves the chiral scalar $\F$ and its conjugate only in the combination $(\vf + \bar \vf)$, where $\vf = \F /\m$. This means that the model 
\eqref{5.21} possesses a dual formulation realised in terms of a real linear superfield
subject to the constraint
\eqref{2.22}.



\subsection{${\cN=2} \rightarrow {\cN=1}$ superspace reduction}
Every supersymmetric field theory in (1,1) AdS superspace ${\rm AdS}^{3|(1,1)}$ may be reformulated in terms of superfields on $\cN=1$ AdS superspace.\footnote{In the case of $\cN=1$ AdS supersymmetry, both notations 
$(1,0)$ and $\cN=1$ are used in the literature. We will also use the notation $\rm AdS^{3|2}$ for ${\cN}=1$ AdS superspace.} Let us briefly discuss how to perform such a reduction. 

First, it proves to be advantageous to switch to the real basis for the (1,1) AdS spinor covariant derivatives. Following \cite{KLT-M12}, we can introduce a real basis for the spinor covariant derivatives
which is obtained by replacing the complex operators ${\frak D}_\a$ and $\bar {\frak D}_\a$ with 
$\de_\a^I$, where $I ={\1}, {\2}$,
 defined by  
\bea
 \frak{D}_\a=\frac{\re^{\ri\vf}}{\sqrt{2}}(\de_\a^{\1}-\ri\de_\a^{\2})~,\qquad
  \bar {\frak D}_\a=-\frac{\re^{-\ri\vf}}{\sqrt{2}}(\de_\a^{\1}+\ri\de_\a^{\2})~,
 \eea 
where we have represented  $\mu=-\,\ri\,\re^{2\ri\vf} |\m|$.
The new covariant derivatives  can be shown to obey the algebra:
\bsubeq
\bea
\{\nabla_\a^{\1},\de_\b^{\1}\}&=&
2\ri\de_{\a\b}
-4\ri |\m| M_{\a\b}
~, \qquad
\{\de_\a^{\2},\de_\b^{\2}\}=
2\ri\de_{\a\b}
+4\ri |\m| M_{\a\b}
~, \\
\{\de_\a^{\1},\de_\b^{\2}\}&=&0
~,~~~~~~~
\label{1_1-alg-AdS-2-1b}
\\
{[}\de_{a},\de_\b^{\1}{]}
&=&
|\m|(\g_a)_\b{}^\g\de_{\g}^{\1}
~, \qquad
{[}\de_{a},\de_\b^{\2}{]}
=
-|\m|(\g_a)_\b{}^\g\de_{\g}^{\2}
~, \\
{[}\de_{a},\de_b{]} &=&-4 |\m|^2 M_{ab}
~.~~~~
\label{1_1-alg-AdS-2}
\eea
\esubeq
The graded commutation relations for the operators $\de_a$ and $\de_\a^{\1}$ form a closed algebra. Indeed, they are isomorphic to those defining the $\cN=1$ AdS superspace, see \cite{KLT-M12}
for the details.
These properties mean that (1,0) AdS superspace is naturally  embedded in (1,1) AdS  superspace  as a subspace. The Grassmann variables
  $\q^\m_I =(\q^\m_{\1}, \q^\m_{\2} )$ may be chosen in such a way that (1,0) AdS corresponds to the surface defined by $\q^\m_{\2} =0$. It is thus possible to carry out a consistent (1,1) $\to$ (1,0) AdS superspace reduction for all the higher-spin supersymmetric gauge theories constructed in sections \ref{s53} and \ref{s54}. 
 Implementation of this program will be 
 described elsewhere.

For concreteness, let us consider the ${\cN=2} \rightarrow {\cN=1}$ superspace 
reduction of the longitudinal model for massless superspin-$s$ multiplet \eqref{long-action-int}.
Here our analysis is restricted to the flat superspace case for simplicity. 

In order to be consistent with the previous work \cite{KT}, 
in which the ${\cN=2} \rightarrow {\cN=1}$ superspace reduction 
of the massless superspin-$(s+ \hf)$ models of \cite{KO} 
was studied,
we denote by 
${\mathbb D}_\a$ and $\bar {\mathbb D}_\a$ the spinor covariant 
derivatives
of $\cN=2$ Minkowski superspace ${\mathbb M}^{3|4}$.
They obey the 
anti-commutation relations
\bea
\{{\mathbb D}_\a, \bar {\mathbb D}_\b\}=-2\ri\, \pa_{\a\b}~,\qquad
\{{\mathbb D}_\a,{\mathbb D}_\b\}=\{ \bar {\mathbb D}_\a, \bar {\mathbb D}_\b\}=0~.
\label{N=2acd}
\eea
In order to carry out the $\cN=2 \to \cN=1$ superspace reduction, 
it is useful to introduce 
real Grassmann coordinates  $\q^\a_I$ for ${\mathbb M}^{3|4}$, 
where $I =\1, \2$. We define these coordinates 
by choosing the corresponding spinor covariant derivatives
$D^I_\a$ as in \cite{KPT-MvU}:
\bea
&&
{\mathbb D}_\a=\frac{1}{ \sqrt{2}}(D_\a^{\1}-\ri D_\a^{\2})~,\qquad
\bar {\mathbb D}_\a=-\frac{1}{ \sqrt{2}}(D_\a^{\1}+\ri D_\a^{\2})~.~~~
\label{N1-deriv}
\eea
From \eqref{N=2acd} we deduce
\bea
\big\{ D^I_\a , D^J_\b \big\} = 2{\rm i}\, \d^{IJ}  (\g^m)_{\a\b}\,\pa_m~, 
\qquad I ,J=\1, \2~.
\eea

Given an $\cN=2$ superfield $U(x, \q_I)$, we define its $\cN=1$ bar-projection
\bea
U|:= U(x, \q_I)|_{\q_{\2} =0}~,
\eea
which
is a superfield on $\cN=1$ Minkowski superspace  
${\mathbb M}^{3|2}$ parametrised
by real Cartesian coordinates $z^A= (x^a, \q^\a)$, where $\q^\a:=\q^\a_{\1}$.
The spinor covariant derivative of $\cN=1$ Minkowski superspace
$D_\a := D_\a^{\1}$
obeys the anti-commutation relation
\bea
\big\{ D_\a , D_\b \big\} = 2{\rm i}\,   (\g^m)_{\a\b}\,\pa_m~.
\eea
Finally, the ${\cN=2} \rightarrow {\cN=1}$ superspace reduction 
of the $\cN=2$ supersymmetric action is carried out using the rule
 \cite{KT}
\bea
S =\int \rd^{3|4}z \, L_{(\cN=2)}
= \int \rd^{3|2} z \, L_{(\cN=1)}~, \qquad 
L_{(\cN=1)} := -\frac{\ri}{4} (D^{\2})^2 L_{(\cN=2)}\Big|~.
\label{B.6-11}
\eea

Given an integer $s \geq 1$, the longitudinal formulation
for the massless superspin-$s$ multiplet 
is realised in terms of the following dynamical variables:
\bea
\cV^{\|}_{(s)} = \Big\{\mathbb{U}_{\a(2s-2)}, \mathbb{G}_{\a(2s)}, \mathbb{\bar{G}}_{\a(2s)} \Big\} ~.
\eea
Here $\mathbb{U}_{\a(2s-2)}$ is an unconstrained real superfield, and the complex superfield $\mathbb{G}_{\a(2s)}$ is longitudinal linear,
\be
\mathbb{\bar{D}}_{(\a_1} \mathbb{G}_{\a_2 \dots \a_{2s+1})} =0~.
\ee
The dynamical superfields are defined modulo gauge transformations of the form 
\begin{subequations} \label{int-gauge-flat}
\bea
\d \mathbb{U}_{\a(2s-2)}
&=& \bar \g_{\a(2s-2)}+{\g}_{\a(2s-2)}
~, \label{U-gauge-flat} \\
\d \mathbb{G}_{\a(2s)} 
&=& \mathbb{\bar{D}}_{(\a_{1}} \mathbb{D}_{\a_{2}}\bar \g_{\a_3 \dots \a_{2s})}
~,
\eea
\end{subequations} 
where the  gauge parameter $\g_{\a(2s-2)}$ is an arbitrary transverse linear superfield,
\be
\mathbb{{\bar D}}^{\b} \g_{\b \a_1 \dots \a_{2s-3}} =0~. \label{ee1-11}
\ee
The gauge-invariant action is 
\bea
S_{(s)}^{\|}
&=& \Big(-\hf\Big)^{s}\int 
\rd^{3|4}z \,
\bigg\{\frac{1}{8}\mathbb{U}^{\a(2s-2)}\mathbb{D}^{\g}{\mathbb{\bar D}}^{2}\mathbb{D}_{\g} \mathbb{U}_{\a(2s-2)}
\non \\
&&+\frac{s}{2s+1} \mathbb{U}^{\a(2s-2)}\Big(\mathbb{D}^{\b} \mathbb{{\bar D}}^{\g}
\mathbb{G}_{\b \g \a(2s-2) }
-\mathbb{\bar D}^{\b}\mathbb{D}^{\g} \mathbb{\bar{G}}_{\b \g \a(2s-2)} \Big) \non \\
&&+\frac{s}{2s-1}  \mathbb{\bar{G}}^{\a(2s)} \mathbb{G}_{\a(2s)}
+ \frac{s}{2(2s+1)}\Big(\mathbb{G}^{\a(2s)} \mathbb{G}_{\a(2s)}+ \mathbb{\bar{G}}^{\a(2s)} \mathbb{\bar G}_{\a(2s)}\Big) 
\bigg\}~.
\label{long-action-flat}
\eea

Making use of the representation \eqref{N1-deriv}, the transverse linear constraint \eqref{ee1-11} takes the form
\be 
D^{\underline{2} \, \b} \g_{\b \a_1 \dots \a_{2s-3}} 
= \ri D^{\underline{1} \b} \g_{\b \a_1 \dots \a_{2s-3}}~.
\ee
It follows that $\g_{\a(2s-2)}$ has 
two independent $\theta_{\underline{2}}$-components, which are:
\bea
\g_{\a(2s-2)} |, \qquad D^{\underline{2}}_{(\a_1} \g_{\a_2 \dots \a_{2s-1})}|~.
\eea
The gauge transformation of $\mathbb{U}_{\a(2s-2)}$, eq. \eqref{int-gauge-flat}, 
allows us to impose two conditions
\bea
\mathbb{U}_{\a(2s-2)} | =0~, \qquad D^{\underline{2}}_{(\a_1} \mathbb{U}_{\a_2 \dots \a_{2s-1})} | =0~.
\label{q1-111}
\eea
In this gauge we define the following unconstrained real ${\cN=1}$ superfields:
\begin{subequations} \label{q2-11}
\bea
U_{\a(2s-3)} &:=& \frac{\ri}{s} D^{\underline{2} \b} \mathbb{U}_{\b \a(2s-3)} | ~, \label{U1} \\
U_{\a(2s-2)} &:=& -\frac{\ri}{4s}(D^{\underline{2}})^2 \mathbb{U}_{\a(2s-2)}|~. \label{U2}
\eea
\end{subequations}
The residual gauge freedom, which preserves the gauge conditions \eqref{q1-111}, 
is described by unconstrained real ${\cN=1}$ superfield 
parameters $\z_{\a(2s-2)}$ and $\l_{\a(2s-1)}$ defined by 
\begin{subequations}
\bea
\g_{\a(2s-2)} | &=& \frac{\ri}{2} \z_{\a(2s-2)}~, \qquad {\bar \z}_{\a(2s-2)} = \z_{\a(2s-2)}~, \label{q3.a-11} \\
D^{\underline{2}}_{(\a_1} \g_{\a_2 \dots \a_{2s-1})} | &=& \hf \l_{\a(2s-1)} ~, \qquad {\bar \l}_{\a(2s-1)} = \l_{\a(2s-1)}~. \label{q3.b-11}
\eea
\end{subequations}
The gauge transformation laws of the superfields \eqref{q2-11} are
\begin{subequations}
\bea
\d U_{\a(2s-3)} &=& -\frac{\ri}{s} D^{\b} \z_{\b \a(2s-3)}~, \label{q4.a-11}\\
\d U_{\a(2s-2)} &=&  \frac{1}{2s} D^{\b} \l_{\b \a(2s-2)} \label{q4.b-11}~.
\eea
\end{subequations}

We now turn to reducing $\mathbb{G}_{\a(2s)}$ to ${\cN=1}$ superspace. 
From the point of view of ${\cN=1}$ supersymmetry, 
$\mathbb{G}_{\a(2s)}$ is equivalent to two unconstrained complex  superfields, 
which we define as follows: 
\begin{subequations} \label{G-comp}
\bea
\mathbb{G}_{\a(2s)}| &=& -\hf(G_{\a(2s)} + \ri H_{\a(2s)})~, \\
\ri D^{\underline{2} \,\b} \mathbb{G}_{\b \a(2s-1)} | &=& \Phi_{\a(2s-1)} + \ri \Psi_{\a(2s-1)}~.
\eea
\end{subequations}
Making use of the gauge transformation \eqref{int-gauge-flat} gives
\begin{subequations} \label{G-gauge2}
\bea
\d \mathbb{G}_{\a(2s)}
&=& -\ri \pa_{(\a_1 \a_2} {\bar \g}_{\a_3 \dots \a_{2s})} + \ri D^{\underline{1}}_{(\a_1} D^{\underline{2}}_{\a_2} {\bar \g}_{\a_3 \dots \a_{2s})}
~, \label{U-gauge-flat2} \\
\ri D^{\underline{2} \,\b} \d \mathbb{G}_{\b \a(2s-1)}
&=& \ri \Big\{-\ri \frac{2s-1}{2s} \pa^{\b}\,_{\a_1} D^{\underline{2}}_{(\b} {\bar \g}_{\a_2 \dots \a_{2s-1})} \non \\
\qquad &+& \frac{s-1}{s} \pa_{(\a_1 \a_2} D^{\b} {\bar \g}_{\a_3 \dots \a_{2s-1}) \b} -2 D^{\b} \pa_{\b(\a_1} {\bar \g}_{\a_2 \dots \a_{2s-1})}\non \\
\qquad &+& \frac{2s+1}{4s} D^2 D^{\underline{2}}_{(\a_1} {\bar \g}_{\a_2 \dots \a_{2s-1})} \Big\}
~.
\eea
\end{subequations} 
At this stage one should recall that upon imposing the ${\cN=1}$ supersymmetric gauge conditions \eqref{q1} the residual gauge freedom is described by the gauge parameters \eqref{q3.a-11} and \eqref{q3.b-11}. From \eqref{G-gauge2} we read off the gauge transformations of the ${\cN=1}$ complex superfields \eqref{G-comp}
\begin{subequations} \label{N1-gauge}
\bea
\d \mathbb{G}_{\a(2s)} |
&=& -\hf \Big\{\pa_{(\a_1 \a_2} \z_{\a_3 \dots \a_{2s})} +  \ri D_{(\a_1} \l_{\a_2 \dots \a_{2s})}\Big\}
~, \\
\ri D^{\underline{2} \,\b} \d \mathbb{G}_{\b \a(2s-1)} |
&=& - \frac{2s-1}{4s} \pa^{\b}\,_{(\a_1} {\l}_{\a_2 \dots \a_{2s-1})\b} 
- \ri \frac{2s+1}{8s} D^2 \l_{\a(2s-1)} \non \\
\qquad &+& \frac{s-1}{2s} \pa_{(\a_1 \a_2} D^{\b} {\z}_{\a_3 \dots \a_{2s-1}) \b} - D^{\b} \pa_{\b(\a_1} {\z}_{\a_2 \dots \a_{2s-1})}
~.
\eea
\end{subequations} 

In the ${\cN=1}$ supersymmetric gauge \eqref{q1-111}, 
$\mathbb{U}_{\a(2s-2)}$ is described by two unconstrained real superfields $U_{\a(2s-3)}$ and $U_{\a(2s-2)}$ defined according to \eqref{q2-11}, and their gauge transformation laws are given by eqs. \eqref{q4.a-11} and \eqref{q4.b-11}, respectively. It follows from the gauge transformations \eqref{q4.a-11}, \eqref{q4.b-11} and \eqref{N1-gauge} that in fact we are dealing with two different gauge theories. One of them is formulated in terms of the  unconstrained real gauge superfields
\bea
\{G_{\a(2s)}, U_{\a(2s-3)}, \J_{\a(2s-1)}\}
\label{B.21}
\eea
which are defined modulo gauge transformations of the form
\begin{subequations} \label{type1-gauge}
\bea
\d G_{\a(2s)} 
&=& \pa_{(\a_1 \a_2} \z_{\a_3 \dots \a_{2s})} ~, \label{B.22a}\\
\d U_{\a(2s-3)} &=& -\frac{\ri}{s} D^{\b} \z_{\b \a(2s-3)}~, \label{B.22b}\\
\d \J_{\a(2s-1)} 
&=& -\ri \frac{s-1}{2s} \pa_{(\a_1 \a_2} D^{\b} {\z}_{\a_3 \dots \a_{2s-1}) \b} +\ri D^{\b} \pa_{\b(\a_1} {\z}_{\a_2 \dots \a_{2s-1})}
~,
\eea
\end{subequations} 
where the gauge parameter $\z_{\a(2s-2)}$ is unconstrained real. 
The other theory is described by the gauge superfields 
\bea
\{H_{\a(2s)}, U_{\a(2s-2)}, \Phi_{\a(2s-1)}\}
\label{B.23}
\eea
with the following gauge freedom 
\begin{subequations} \label{type2-gauge}
\bea
\d H_{\a(2s)}
&=&  D_{(\a_1} \l_{\a_2 \dots \a_{2s})}~, \label{B.24a} \\
\d U_{\a(2s-2)} &=&  \frac{1}{2s}D^\b \l_{\b \a(2s-2)} ~, \label{B.24b} \\
\d \Phi_{\a(2s-1)}
&=& - \frac{1}{8s} \Big\{(4s-2) \pa^{\b}\,_{(\a_1} {\l}_{\a_2 \dots \a_{2s-1})\b} 
+ \ri (2s+1) D^2 \l_{\a(2s-1)}\Big\}
~.
\eea
\end{subequations}

Applying the reduction rule \eqref{B.6-11}
to the action \eqref{long-action-flat} gives two decoupled $\cN=1$ supersymmetric
actions, which are described  in terms of the dynamical variables
\eqref{B.21} and \eqref{B.23}, respectively. 
In the former case, the superfield $\J_{\a(2s-1)}$ is  auxiliary.
Integrating it out, we arrive at the following action:
\bea
S
&=& -\Big(-\hf\Big)^{s} \,\frac{s^2(s-1)}{2s-1} \frac{\ri}{2}\int 
\rd^{3|2}z \,
\bigg\{ 
 \frac{1}{2s} G^{\a(2s)} D^2 G_{\a(2s)}
 \non \\
&& 
-
\frac{\ri}{s-1} G^{\a(2s-1) \b} \pa_{\b}\, ^{\g} G_{\a(2s-1) \g}
-2\ri U^{\a(2s-3)} \pa^{\b \g} D^{\d} G_{\b \g \d \a(2s-3)}
 \non \\
&&+2 \,U^{\a(2s-3)} \Box U_{\a(2s-3)} 
+  \frac{(2s-3)(s-2)}{2s-1} \, \pa_{\d \l} U^{\d \l \a(2s-5)} \pa^{\b \g} U_{\b \g \a(2s-5)} \non \\
&&- \hf \,\frac{2s-3}{2s-1} D_\b U^{\a(2s-4)\b} D^2 D^\g U_{\g \a(2s-4)} 
 \bigg\}~.
\label{type1-action}
\eea
This action is invariant under  the gauge transformations
\eqref{B.22a} and \eqref{B.22b}.

In the latter case, the superfield $\Phi_{\a(2s-1)}$ is  auxiliary. 
Integrating it out, we obtain the following gauge-invariant action: 
\bea
S
&=& \Big(-\hf\Big)^{s} \frac{s }{2s-1} \ri \int 
\rd^{3|2}z \,
\bigg\{ 
\frac{1}{2} H^{\a(2s)} D^2 H_{\a(2s)} 
+\ri H^{\a(2s-1) \b} \pa_{\b} \,^{\g} H_{\a(2s-1) \g} \non \\
&&+ 2 \ri (2s-1)U^{\a(2s-2)} \pa^{\b \g} H_{\b \g \a(2s-2)}
+ (2s-1 ) U^{\a(2s-2)} D^2 U_{\a(2s-2)}
\non \\
&& 
+ 2 (2s-1)(s-1) D_{\b} U^{\b \a(2s-3)} D^{\g} U_{\g \a(2s-3)} 
\bigg\}~.
\label{type2-action}
\eea
This action is invariant under  the gauge transformations
\eqref{B.24a} and \eqref{B.24b}. Modulo an overall normalisation factor, 
\eqref{type2-action} coincides with 
the off-shell $\cN=1$ supersymmetric action 
for massless superspin-$s$ multiplet \cite{KT}
in the form given in \cite{KP1}.

The action \eqref{type1-action} defines a new $\cN=1$ supersymmetric 
higher-spin theory which did not appear in the analysis of \cite{KT}.
It may be shown that at the component level it reduces, upon imposing 
a Wess-Zumino gauge and eliminating the auxiliary fields, 
to a sum of two massless actions. One of them is the bosonic Fronsdal-type  
spin-$s$ model and the other is the fermionic Fang-Fronsdal-type 
spin-$(s+\hf)$ model. 

  

\section{Higher-spin (1,1) AdS supercurrents} \label{s55}

Inspired by the analysis of Dumitrescu and Seiberg \cite{DS},
the most general supercurrent multiplets for theories with 
(1,1) AdS or (2,0) AdS supersymmetry were introduced in \cite{KT-M11},
with the (1,1) AdS case being 
a natural extension of the 4D $\cN=1 $ AdS supercurrents classified in 
\cite{BK12,BK11}. Here we will formulate {\it higher-spin} supercurrent multiplets in (1,1) AdS superspace by making use of the off-shell massless supersymmetric higher-spin theories constructed in the previous two sections. Our analysis will be mostly analogous to that in the 4D case.

\subsection{Non-conformal supercurrents: Half-integer superspin} \label{ss541}
The two formulations for the massless half-integer superspin which were described in section \ref{s53} lead to different higher-spin supercurrent multiplet. Following similar derivations as in subsections \ref{ss321} and \ref{ss432}, one may show that the most general half-integer superspin current multiplet is described by the conservation equation
\bsubeq \label{half11}
\bea
\bar{\frak D}^{\b}J_{\b \a(2s-1)} &=& -\hf \big( \frak{D}_{(\a_1} \bar{\frak D}_{\a_2} -2 \ri (s-1) \frak{D}_{(\a_1 \a_2} \big) T_{\a_3 \dots \a_{2s-1})} \non\\
&&+ \frac{1}{4} \big(\bar{\frak D}^2 + 2 \mu (2s-1) \big)\mathbb{F}_{\a(2s-1)}~.~~
\eea
Here the higher-spin supercurrent $J_{\a(2s)}$ is a real superfield. The trace multiplets $T_{\a(2s-3)}$ and $\mathbb{F}_{\a(2s-1)}$ are complex superfields constrained by
\bea
\bar{\frak D}^{\b}T_{\b \a(2s-4)} &=& 0~, \label{trace-lon11}\\
{\frak D}_{(\a_1}\mathbb{F}_{\a_2 \dots \a_{2s})} &=& 0~,
\eea
\esubeq
and therefore $T_{\a(2s-3)}$ is a transverse linear superfield, while $\bar{\mathbb{F}}_{\a(2s-1)}$ is longitudinal linear. 
The multiplet with $\mathbb{F}_{\a(2s-1)} = 0$ corresponds to the longitudinal formulation for massless superspin-$(s+\hf)$ multiplet \eqref{long-action-half-11}. The case $T_{\a(2s-3)}=0$ is associated with the transverse formulation \eqref{tr-action-half-11}. In this way, we have 3D counterparts of the 4D  half-integer superspin current multiplets given by \eqref{7.3} and \eqref{TSupercurrent}.

We can also construct a well-defined improvement transformation which converts the longitudinal higher-spin supercurrent to the transverse one, thus showing that they are indeed equivalent. 
The most general higher-spin supercurrent \eqref{half11} can be modified by an improvement transformation
\begin{subequations} \label{614}
\bea
{J}_{\a(2s)} &\longrightarrow & J_{\a(2s)}
+\frac{s}{2} \big[ \frak{D}_{(\a_1}, \bar{ \frak D}_{\a_2} \big]
U_{\a_3 \dots \a_{2s} ) }
+ s \frak{D}_{(\a_1 \a_2 } V_{\a_3 \dots \a_{2s})}~, ~~~\\
T_{\a(2s-3)}  &\longrightarrow & T_{\a(2s-3)} - \bar {\frak D}^{\b} \Big( U_{\b\a(2s-3)} + \ri V_{\b \a(2s-3)} \Big)~, ~~~\\
{\mathbb F}_{\a(2s-1)} & \longrightarrow & 
{\mathbb F}_{\a(2s-1)} + \frak{D}_{(\a_1} \Big( 2s \,U_{\a_2 \dots \a_{2s-1)}}
- \ri V_{\a_2 \dots \a_{2s-1})}\Big)~,~~~
\eea
\end{subequations}
with $U_{\a(2s-2)}$ and $V_{\a(2s-2)}$ well-defined operators. 

The transverse linearity constraint \eqref{trace-lon11} can always be solved in the (1,1) AdS geometry as
\bea
T_{\a(2s-3)} = \bar{\frak D}^{\b}\Big( U_{\b \a(2s-3)}+ \ri V_{\b \a(2s-3)}\Big)~,
\eea
for well-defined real tensor operators $U_{\a(2s-2)}$ and $V_{\a(2s-2)}$~.
This property means that we can always set $T_{\a(2s-3)}$ to zero by applying a certain improvement transformation \eqref{614}.
The above analysis shows that the longitudinal and transverse supercurrents are equivalent.
The situation proves to be analogous in the integer superspin case, 
for which we will formulate in the next subsection a higher-spin supercurrent
associated with  the new gauge formulation \eqref{action2-11}. Therefore, it suffices to work with one of them, say, 
the longitudinal supercurrent multiplet $(J_{\a(2s)}, T_{\a(2s-3)})$, which obeys the conservation equation
\bea
\bar {\frak D}^\b J_{\b \a(2s-1)} 
= - \hf \Big( \frak{D}_{(\a_1} \bar {\frak D}_{\a_2}
-2\ri (s-1) \frak{D}_{ (\a_1 \a_2 } \Big)  T_{\a_3\dots \a_{2s-1})}~.
\label{lonsc11}
\eea
For completeness, we also give the conjugate equation
\bea
\frak{D}^\b J_{\b \a(2s-1)} 
 = \hf \Big( \bar {\frak D}_{(\a_1}  \frak{D}_{\a_2}
-2\ri (s-1) \frak{D}_{ (\a_1 \a_2 } \Big)  \bar T_{\a_3 \dots \a_{2s-1})} ~. \label{lonscc11}
\eea
%

Before we proceed to the construction of higher-spin supercurrents for (1,1) AdS supersymmetric field theories, let us first recall our condensed notation in complete analogy with the four-dimensional analysis. We introduce auxiliary real variables  
$\z^\a \in {\mathbb R}^2$ and associate with any tensor superfield $U_{\a(m)}$ the following index-free  field
\bea
U_{(m)} (\z):= \z^{\a_1} \dots \z^{\a_m} U_{\a_1 \dots \a_m}~, \label{eee1}
\eea
which is a homogeneous polynomial of degree $m$ in $\z^\a$.
Furthermore, we make use of the bosonic variables $\z^\a$ and the corresponding 
partial derivatives $\pa/\pa \z^\a$ to convert the spinor and vector covariant derivatives into index-free operators.  
In the case of (1,1) AdS superspace, 
we introduce operators that increase the degree 
of homogeneity in $\z^\a$, 
\begin{subequations}
\bea
{\frak{D}}_{(1)} &:=& \z^\a \frak{D}_\a~, \qquad 
\bar{\frak D}_{(1)} :=  \z^\a \bar {\frak D}_\a~,\\
{\frak{D}}_{(2)} &:=& \ri \z^\a \z^\b \frak{D}_{\a\b} = -\hf \{ \frak{D}_{(1)}, \bar{\frak D}_{(1)} \}~.
\eea
\end{subequations}
We also introduce two operators that decrease the degree 
of homogeneity in the variable $\z^\a$, specifically
\bea
\frak{D}_{(-1)} &:=& \frak{D}^\a \frac{\pa}{\pa \z^\a}~, \qquad
\bar {\frak D}_{(-1)} := \bar {\frak D}^\a \frac{\pa}{\pa  \z^\a}~.
\eea

The transverse linear condition \eqref{trace-lon11} and its conjugate can be written as
\bsubeq
\bea
\bar{\frak D}_{(-1)} T_{(2s-3)} &=& 0~, \label{trace-lon11a} \\
\frak{D}_{(-1)} \bar{T}_{(2s-3)}&=&0~.\label{trace-lon11b}
\eea
\esubeq
The conservation equations \eqref{lonsc11} and \eqref{lonscc11} turn into
\begin{subequations}
\bea
\frac{1}{2s}\bar {\frak D}_{(-1)} J_{(2s)} &=& \hf A_{(2)} T_{(2s-3)}~, \label{lonsc11a}\\
\frac{1}{2s}\frak{D}_{(-1)} J_{(2s)} &=& \hf \bar A_{(2)} \bar T_{(2s-3)}~. \label{lonsc11b}
\eea
\end{subequations}
where 
\bea
A_{(2)} := -\frak{D}_{(1)} \bar {\frak D}_{(1)} + 2(s-1) \frak{D}_{(2)} ~, \quad
\bar A_{(2)} := \bar {\frak D}_{(1)}  \frak{D}_{(1)} -2(s-1) \frak{D}_{(2)} ~. 
\eea
Since 
$(\bar {\frak D}_{(-1)}) ^2 J_{(2s)} =0$,
the conservation equation \eqref{lonsc11a} is consistent provided
\bea
\bar {\frak D}_{(-1)}  A_{(2)} T_{(2s-3)}=0~.
\eea
This is indeed true, as a consequence of the transverse linear condition
\eqref{trace-lon11a}. 

\subsubsection{Models for a chiral superfield} \label{ss5511}

We now give several examples of higher-spin supercurrents introduced above by 
studying rigid supersymmetric field theories in (1,1) AdS superspace. 

Our first example is the superconformal theory of a single chiral scalar superfield 
\bea
S = \int 
\rd^{3|4}z
\, \bm E \, \bar \F \F ~,
\label{chiral-11}
\eea
where  $\F$ is covariantly chiral, $\bar {\frak D}_\a \F =0$.
The corresponding conformal higher-spin supercurrent is given by 
\bea
J_{(2s)} &=& \sum_{k=0}^s (-1)^k
\left\{ \hf \binom{2s}{2k+1} 
{\frak{D}}^k_{(2)} \bar {\frak D}_{(1)} \bar \F \,\,
{\frak{D}}^{s-k-1}_{(2)} \frak{D}_{(1)} \F  
+ \binom{2s}{2k} 
{\frak{D}}^k_{(2)} \bar \F \,\, {\frak{D}}^{s-k}_{(2)} \F \right\}~,~~~~~~~~
\label{7.15-11}
\eea
which is a minimal extension of the conserved supercurrent constructed in flat ${\cal N}=2$ Minkowski superspace \cite{NSU}.
It may be checked that for $s > 0$, the real higher-spin supercurrent $J_{(2s)}$ satisfies the conservation equation
\bea
\frak{D}_{(-1)} J_{(2s)} = 0 \quad \Longleftrightarrow \quad 
\bar {\frak D}_{(-1)} J_{(2s)} = 0 ~,~
\label{7.9-11}
\eea
by virtue of the massless equations of motion, $(\frak{D}^2-4\bar \m)\, \F = 0$~. 

Let us now add the mass term to~\eqref{chiral-11} and consider the following action
\bea
S = \int 
\rd^{3|4}z
\, \bm E\, \bar \F \F
+\Big\{ \hf \int 
\rd^{3|4}z
\, \bm E \, \frac{m}{\m}\F^2 +{\rm c.c.} \Big\}~,
\label{chiralm11}
\eea
with $m$ a complex mass parameter. The equations of motion are
\bea \label{5517}
-\frac{1}{4} (\frak{D}^2-4\bar\m) \F  +\bar m \bar \F =0~, \qquad
-\frac{1}{4} (\bar {\frak D}^2-4\m) \bar \F +m \F =0~.
\eea
After some lengthy calculations (see \cite{HKO} for the derivation), the equations of motion imply that on-shell the higher-spin supercurrent multiplet takes the form 
\bsubeq
\bea
&&J_{(2s)} = \sum_{k=0}^s (-1)^k
\left\{ \hf \binom{2s}{2k+1} 
{\frak{D}}^k_{(2)} \bar {\frak D}_{(1)} \bar \F \,\,
{\frak{D}}^{s-k-1}_{(2)} \frak{D}_{(1)} \F  
+ \binom{2s}{2k} 
{\frak{D}}^k_{(2)} \bar \F \,\, {\frak{D}}^{s-k}_{(2)} \F \right\}~,~~~~~~~~~~\\
&&\bar T_{(2s-3)} = \bar m \sum_{k=0}^{s-2} c_k 
{\frak D}^k_{(2)} \bar \F\,
{\frak D}^{s-k-2}_{(2)}
 \bar {\frak D}_{(1)} \bar \F ~,
\eea
with the coefficients $c_k$ given by
\bea
c_k &=& (-1)^{s+k-1} \frac{(2s+1)(s-k-1)}{2s(s-1)}
\sum_{l=0}^k \frac{1}{s-l} \binom{2s}{2l+1} \left\{ 1+(-1)^s \frac{2l+1}{2s-2l+1} \right\}  ~,~~~~~~~~~~~~  \\
&& \qquad \qquad \qquad  k=0,1,\dots s-2~. \non
\eea
\esubeq
This is the (1,1) AdS analogue of the non-conformal supercurrents presented in \ref{subsection5.2}.
Indeed, the same selection rules also emerge since one can verify that the conservation equation \eqref{lonsc11b} and the transverse linearity constraint \eqref{trace-lon11b} are identically satisfied only for the \textit{odd} values of $s, s= 2n+1,$ with $n = 1,2, \dots$. In this sense our (1,1) AdS higher-spin supercurrents are very similar to the 4D ${\cN}=1$ Minkowski and AdS cases studied in subsections \ref{ss3211} and \ref{subsection5.2}, respectively.


\subsubsection{Superconformal model with $N$ chiral superfields}

Another interesting example is a generalisation of the superconformal model \eqref{chiral-11} to the case of $N$ covariantly chiral scalar superfields $\F^i$, $i=1,\dots N$,
\bea
S = \int 
\rd^{3|4}z
\, \bm E \,{\bar \F}^i \F^i ~,
\qquad \bar {\frak D}_\a \F^i = 0 ~.~
\label{Nchiral}
\eea
This model is characterised by two different types of conformal supercurrents, which we denote by
\bea
J^+_{(2s)} &=& S^{ij} \sum_{k=0}^s (-1)^k
\left\{ \hf \binom{2s}{2k+1} 
{\frak D}^k_{(2)} \bar {\frak D}_{(1)} \bar \F^i \,\,
{\frak D}^{s-k-1}_{(2)} {\frak D}_{(1)} \F^j  
\right. \non \\ 
&& \left.
 \qquad \qquad
+ \binom{2s}{2k} 
{\frak D}^k_{(2)} \bar \F^i \,\, {\frak D}^{s-k}_{(2)} \F^j \right\}~, \qquad S^{ij}= S^{ji} 
\label{J-sym}
\eea
and
\bea
J^-_{(2s)} &=& \ri \, A^{ij} \sum_{k=0}^s (-1)^k
\left\{ \hf \binom{2s}{2k+1} 
{\frak D}^k_{(2)} \bar {\frak D}_{(1)} \bar \F^i \,\,
{\frak D}^{s-k-1}_{(2)} \frak{D}_{(1)} \F^j  
\right. \non \\ 
&& \left.
 \qquad \qquad
+ \binom{2s}{2k} 
{\frak D}^k_{(2)} \bar \F^i \,\, {\frak D}^{s-k}_{(2)} \F^j \right\}~, \qquad A^{ij}= -A^{ji} 
\label{J-anti}
\eea
Here $S$ and $A$ are arbitrary real symmetric and antisymmetric constant matrices, respectively. We have put an overall factor  $\sqrt{-1} $ in eq.~\eqref{J-anti} in order to make $J^-_{(2s)}$ real. 
The currents \eqref{J-sym} and \eqref{J-anti} obey the conservation equation
\bea
\frak{D}_{(-1)} J^\pm_{(2s)} = 0 \quad \Longleftrightarrow \quad 
\bar {\frak D}_{(-1)} J^\pm_{(2s)} = 0 ~.~
\eea

The above results can be recast in terms of the matrix conformal supercurrent
$J_{(2s)} =\big(J^{ij}_{(2s)} \big)$ with components
\bea
J^{ij}_{(2s)} &:=& \sum_{k=0}^s (-1)^k
\left\{ \hf \binom{2s}{2k+1} 
{\frak D}^k_{(2)} \bar {\frak D}_{(1)} \bar \F^i \,\,
{\frak D}^{s-k-1}_{(2)} \frak{D}_{(1)} \F^j 
\right. \non \\ 
&& \left.
 \qquad \qquad
+ \binom{2s}{2k} 
{\frak D}^k_{(2)} \bar \F^i \,\, {\frak D}^{s-k}_{(2)} \F^j \right\}~, 
\label{520}
\eea
which is  Hermitian, $J_{(2s)}^\dagger = J_{(2s)}$. 
The chiral action  \eqref{Nchiral}
possesses rigid ${\rm U}(N)$ symmetry acting on the chiral column-vector $\F = (\F^i$) 
by $\F \to g \F$, with $g \in {\rm U}(N)$. This implies that 
the supercurrent \eqref{520} transforms
as $J_{(2s)} \to gJ_{(2s)} g^{-1}$.


\subsection{Non-conformal supercurrents: Integer superspin}

Let us now consider the new gauge formulation \eqref{action-11}, 
or  equivalently \eqref{action2-11}, for the integer superspin-$s$ multiplet to derive the 3D analogue of the non-conformal higher-spin supercurrents formulated in \ref{subsection4.4}.

As usual, we first couple the prepotentials 
$U_{ \a (2s-2) } $, $Z_{ \a (2s-2) }$ and $\Psi_{ \a (2s-1) } $ to some external sources through the action
\bea
S^{(s)}_{\rm source} &=& \int 
\rd^{3|4}z
\, \bm E\, \Big\{ 
\Psi^{ \a (2s-1)  } J_{ \a (2s-1) }
-\bar \Psi^{ \a (2s-1) } \bar J_{ \a (2s-1) }
+U^{ \a (2s-2)  } S_{ \a (2s-2)  } \non \\
&&\qquad \qquad ~~
+ Z^{ \a (2s-2) } T_{ \a (2s-2) } 
+ \bar Z^{ \a (2s-2)  } \bar T_{ \a (2s-2) }
 \Big\}~.
\eea
The action $S^{(s)}_{\rm source}$ should be invariant under the $\z$-transformation 
\eqref{2.4a-11}, which demands the source  $J_{ \a (2s-1) }$ to be transverse linear,
\bea
\bar {\frak D}^\b J_{\b \a(2s-2)} =0 \quad \Longleftrightarrow \quad
\frak{D}^\b \bar J_{\b \a(2s-2)} =0 ~.
\label{4.2a-11}
\eea
Next, the action $S^{(s)}_{\rm source}$ should also preserve the $\xi$-gauge freedom  \eqref{2.3-11}. This requires  $T_{\a(2s-2)}$ to be longitudinal linear
\bea
\bar {\frak D}_{(\a_1} T_{\a_2 \dots \a_{2s-1})} =0
 \quad \Longleftrightarrow \quad
 \frak{D}_{(\a_1} \bar T_{\a_2 \dots \a_{2s-1})} =0~.
\label{4.2b-11}
\eea
Finally, imposing the invariance of $S^{(s)}_{\rm source}$ under the 
$\mathfrak V$-transformation 
\eqref{2.4-11} leads to the following conservation equation
\begin{subequations} \label{3.4-11}
\bea
- \frak{D}^\b J_{\b \a(2s-2)} 
+S_{\a(2s-2)} + \bar T_{\a(2s-2) } =0
\label{4.2c-11}
\eea
as well as its conjugate
\bea
 \bar {\frak D}^\b \bar J_{ \b \a(2s-2)} 
+S_{\a(2s-2)} + T_{\a(2s-2)} =0~. 
\label{3.4b-11}
\eea
\end{subequations}

Taking the sum of \eqref{4.2c-11} and \eqref{3.4b-11}
leads to
\bea
 \frak{D}^\b J_{\b \a(2s-2)} 
+ \bar {\frak D}^\b \bar J_{\b \a(2s-2)}
+ T_{\a(2s-2)}-\bar T_{\a(2s-2)} =0~. 
\label{4.3-11}
\eea
As a consequence of \eqref{4.2b-11}, the conservation equation \eqref{4.3-11} 
implies
\bea
 \frak{D}_{(\a_1} \left\{\frak{D}^{|\b|} J_{\a_2 \dots \a_{2s-1}) \b} 
+ \bar {\frak D}^\b \bar J_{ \a_2 \dots \a_{2s-1}) \b}\right\}
+\frak{D}_{(\a_1} T_{\a_2 \dots \a_{2s-1})} =0~. 
\label{4.4-11}
\eea

Employing the condensed notation, 
the transverse linear condition \eqref{4.2a-11} turns into 
\bea
\bar {\frak D}_{(-1)} J_{(2s-1)} &=& 0~,  \label{4.5a-11}
\eea
while the longitudinal linear condition \eqref{4.2b-11} takes the form
\bea
\bar {\frak D}_{(1)} T_{(2s-2)} &=& 0~. \label{4.5b-11}
\eea
The conservation equation \eqref{4.2c-11} becomes
\bea
-\frac{1}{(2s-1)} \frak{D}_{(-1)} J_{(2s-1)} + S_{(2s-2)} + \bar T_{(2s-2)} = 0
\label{4.6-11}
\eea
and \eqref{4.4-11} takes the form
\bea
\frac{1}{(2s-1)} \frak{D}_{(1)} \left\{\frak{D}_{(-1)} J_{(2s-1)} + \bar {\frak{D}}_{(-1)} \bar J_{(2s-1)}\right\}
+\frak{D}_{(1)} T_{(2s-2)} =0~. 
\label{4.7-11}
\eea

%

As an example, let us go back to the massive chiral multiplet model \eqref{chiralm11}
\bea
S = \int 
\rd^{3|4}z
\,E \, \bar \F \F
+\Big\{ \hf\int 
\rd^{3|4}z
\,E \,  \frac{M}{\m} \F^2  +{\rm c.c.} \Big\}~,
\label{hyper2}
\eea
where the mass parameter $M$ is now real.\footnote{This is analogous to the massive hypermultiplet model considered in \eqref{hyper1}, where it is always possible to make the mass parameter real by changing of variables.}

In the massless case, $M=0$, this model is characterised by a fermionic supercurrent $J_{\a(2s-1)}$, which only exists for even values of $s$. In condensed notation, it has the form
\bea
J_{(2s-1)} &=& 2 \sum_{k=0}^{s-1} (-1)^k  \binom{2s-1}{2k+1} 
{\frak D}^k_{(2)}
 \frak{D}_{(1)} \F \,\,
{\frak D}^{s-k-1}_{(2)}
 \F  
\label{4.8-11}
\eea
The above is the (1,1) AdS counterpart of the integer supercurrent \eqref{6.14}.
%
One may check that for $s > 1$, the conservation equations   
\bea
\frak{D}_{(-1)} J_{(2s-1)} = 0, \qquad
\bar {\frak D}_{(-1)} J_{(2s-1)} = 0 
\label{4.9-11}
\eea
hold on-shell. 

In the massive case, we need to solve a more general conservation equation given by \eqref{4.7-11}. After some calculations, one may show that the on-shell conditions \eqref{5517} imply
\bsubeq
\bea
\bar {\frak D}_{(-1)} J_{(2s-1)} &=& 0~,\\
\frak{D}_{(-1)} J_{(2s-1)} &=& 8Ms \sum_{k=0}^{s-1} (-1)^{k+1} \binom{2s-1}{2k} \non \\
&\times & \Big\{ {\frak D}^k_{(2)} \F \,{\frak D}^{s-k-1}_{(2)} \bar \F  \,\,
 + \frac{k}{2k+1} 
{\frak D}^{k-1}_{(2)} \,\bar {\frak D}_{(1)} \bar \F \,\,{\frak D}^{s-k-1}_{(2)} {\frak D}_{(1)} \F \Big\}
~.~ \label{4.10-11}
\eea
\esubeq
The latter allows us to deduce the explicit form of the trace multiplet $T_{(2s-2)}$, which is a longitudinal linear superfield \eqref{4.5b-11} and obeys \eqref{4.7-11}, as a consequence of the conservation equation \eqref{4.6-11}. 
This guides us to choose an ansatz of the form
\bea
T_{(2s-2)} &=& 
\sum_{k=0}^{s-1} c_k \,{\frak D}^k_{(2)} \F \,\, {\frak D}^{s-k-1}_{(2)} \bar \F  \non \\
&&+ \sum_{k=1}^{s-1} d_k \,{\frak D}^{k-1}_{(2)} \frak{D}_{(1)} \F \,\, {\frak D}^{s-k-1}_{(2)} \bar {\frak D}_{(1)} \bar \F  ~.
\label{T4.11-11}
\eea
Condition \eqref{4.7-11} implies that the coefficients must be related by
\begin{subequations} 
\bea
c_0 = 0~, \qquad c_k = 2d_k ~. 
\eea
For $k=1,2, \dots s-2$, the following recurrence relations are obtained by the requirement \eqref{4.6-11}:
\bea 
d_k + d_{k+1} &=& -\frac{8Ms}{2s-1} (-1)^{k+1} \binom{2s-1}{2k} 
 \frac{4ks+3s-1-2s^2}{(2k+1)(2k+3)} ~.
\eea
It also follows from \eqref{4.7-11} that
\bea
d_1 =-\frac{8}{3} Ms(s-1)~, \qquad d_{s-1} &=& -\frac{8}{2s-1} Ms(s-1)~.
\eea
\end{subequations}
The above conditions lead to a simple expression for $d_k$:
\bea
d_k &=& \frac{8Ms}{2s-1} \frac{k}{2k+1} (-1)^{k} \binom{2s-1}{2k} ~,
\eea
\label{qqq2-11}
where $ k=1,2,\dots s-1$ and the parameter $s$ is even 
for $J_{(2s-1)}$ to be non-zero.


\section{Higher-spin supercurrents for chiral matter in (2,0) AdS superspace} \label{s62}

We now turn to describing the off-shell constructions of higher-spin gauge supermultiplets in (2,0) AdS superspace \cite{HK18}, which prove to be less trivial. As pointed out in the introduction, the massless 3D constructions of \cite{HKO,KO}, 
were largely modelled on the 4D results of \cite{KS94,KPS}.
With respect to 3D (2,0) AdS supersymmetry, unfortunately 
there is no 4D intuition to guide us,
and new ideas are required in order to construct higher-spin gauge supermultiplets.
The approach employed in \cite{HK18} was based on an observation 
that has often been used in the past 
to formulate off-shell supergravity multiplets \cite{BdeRdeW,SW,SohniusW2,SohniusW3,HL,Howe5Dsugra}.
The idea is to make use of a higher-spin extension of the supercurrent. 
Specifically, for a simple supersymmetric model in (2,0) AdS superspace 
we identify a multiplet of conserved higher-spin currents. 
In general, the multiplet of currents is always off-shell. 
Using the constructed higher-spin supercurrent, we may identify  a corresponding off-shell
supermultiplet of higher-spin fields. 

We begin with some simple models for a chiral scalar supermultiplet
in $(2,0)$ AdS superspace and try to derive the corresponding higher-spin supercurrent multiplet. 

\subsection{Massless models}

Let us first consider a massless model. Its action
\bea
S = \int \rd^3x \rd^2 \q  \rd^2 \bar \q
\,\bm E\, \bar \F \F ~, \qquad \bar \cD_\a \F =0
\label{chiral20}
\eea
is invariant under the isometry transformations of (2,0) AdS superspace
for any U(1)${}_R$ charge $r$ of the chiral superfield, 
\bea
{J} \F = - r \F ~, \qquad r ={\rm const} ~.
\eea
The action is superconformal provided  $r =\hf$.

Let us first consider the superconformal case, $r =\hf$.  The analysis
given in subsection \ref{ss5511} implies that  the theory possesses a
real, bosonic supercurrent $\mathbb{J}_{(2s)} = \bar{ \mathbb J}_{(2s)}$, for any positive integer $s$, 
which obeys the conservation equation 
\bea
\cD_{(-1)} \mathbb{J}_{(2s)} &=& 0~.
\label{61}
\eea
This supercurrent proves to have the same form as in the (1,1) AdS case, given by \eqref{7.15-11}.
Specifically, the higher-spin supercurrent is given by
\bea
{\mathbb J}_{(2s)} &=& \sum_{k=0}^s (-1)^k
\left\{ \hf \binom{2s}{2k+1} 
{\cD}^k_{(2)} \bar \cD_{(1)} \bar \F \,\,
{\cD}^{s-k-1}_{(2)} \cD_{(1)} \F  
+ \binom{2s}{2k} 
{\cD}^k_{(2)} \bar \F \,\, {\cD}^{s-k}_{(2)} \F \right\}~.~~~~~
\label{611}
\eea
Making use of
the massless equations of motion, $\cD^2 \F = 0$, one may check that
\eqref{611} does obey the conservation equation \eqref{61}. In the flat superspace limit, 
the supercurrent \eqref{611} reduces to the one constructed in \cite{NSU}.

Now we turn to the non-superconformal case, $r\neq \hf$.
Direct calculations give
\begin{subequations} \label{62}
\bea
\cD_{(-1)} {\mathbb J}_{(2s)} &=&  {\cD}_{(1)} {\mathbb T}_{(2s-2)}~,
\label{68a}
\eea
where we have denoted 
\bea
{\mathbb T}_{(2s-2)}&=& 2\ri (1-2r){\cS}(2s+1)(s+1) \sum_{k=0}^{s-1}\frac{1}{2s-2k+1} (-1)^{k} \binom{2s}{2k+1}
\non \\ 
&& 
\times {\cD}^k_{(2)} \bar \F \,\,{\cD}^{s-k-1}_{(2)} \F ~.
\eea
The trace multiplet ${\mathbb T}_{(2s-2)}$ is covariantly  linear,
\be
\bar {\cD}^2 {\mathbb T}_{(2s-2)} =0 ~, \qquad {\cD}^2 {\mathbb T}_{(2s-2)} =0 ~,
\label{68c}
\ee
as a consequence of the equations of motion and the identity \eqref{A.2c}.
It is seen that 
${\mathbb T}_{(2s-2)}$ has nonzero real and imaginary parts,  
\bea
{\mathbb T}_{(2s-2)} = {\mathbb Y}_{(2s-2)} -\ri {\mathbb Z}_{(2s-2)}~,
\qquad \bar {\mathbb Y}_{(2s-2)} ={\mathbb Y}_{(2s-2)}~, \qquad
\bar {\mathbb Z}_{(2s-2)}={\mathbb Z}_{(2s-2)}~,
\eea
\end{subequations}
except for the  $s=1$  case which is characterised by ${\mathbb Y}=0$.
For $s=1$ the above results agree with \cite{KT-M11}.

The above results can be used to derive higher-spin supercurrents
in a non-minimal scalar supermultiplet model described by the action
\bea
S = -\int \rd^3x \rd^2 \q  \rd^2 \bar \q
\,\bm E\, \bar \G \G ~, \qquad \bar \cD^2  \G =0~,
\label{non-minimal}
\eea
with $\G$ being a complex linear superfield. 
The non-minimal theory \eqref{non-minimal} proves to be dual to \eqref{chiral20}
provided the U(1)${}_R$ weight of $\G$ is opposite to that of $\F$, 
\bea
{J} \G =r \G ~.
\eea
Replacing $\F \to \bar \G$ and $\bar \F \to \G$ in \eqref{62} gives the higher-spin 
supercurrents in the non-minimal theory \eqref{non-minimal}, 
which is similar to the 4D case \cite{KKvU,BHK}.

Let us also mention that in deriving eq. \eqref{62}, one may find the following identities useful. 
We start with the obvious relations
\begin{subequations}
\bea
\frac{\pa}{\pa \z^\a} {\cD}_{(2)} &=&2\ri { \z}^\b {\cD}_{\a \b}~, \\
\frac{\pa}{\pa \z^\a} {\cD}^k_{(2)} &=& 
\sum_{n=1}^k\,{\cD}^{n-1}_{(2)} \,\,  2\ri \, {\z}^\b {\cD}_{\a \b}\,\, {\cD}^{k-n}_{(2)} ~, \qquad k>1
~.\label{63}
\eea
\end{subequations}
To simplify eq.~\eqref{63}, we may push ${\z}^\b{\cD}_{\a \b}$, say,  to the left 
provided that we take into account its commutator with ${\cD}_{(2)}$:
\bea
[{ \z}^\b {\cD}_{\a \b}\,, {\cD}_{(2)}] = -4\ri \,{\cS}^2 \z_\a  { \z}^\b{\z}^\g {M}_{\b \g}~.
\label{64}
\eea
Associated with the Lorentz generators are the operators
\bea
{M}_{(2)} &:=& {\z}^\a {\z}^\b {M}_{\a \b}~,
\eea
where ${M}_{(2)}$ appears in the right-hand side of \eqref{64}.
This operator annihilates every superfield $U_{(m)}(\z) $ of the form 
\eqref{eee1},
\bea
{M}_{(2)} U_{(m)} =0~.
\eea
From the above consideration, it follows that
\begin{subequations}
\bea
[{\z}^\b {\cD}_{\a \b}\,, {\cD}^k_{(2)}]\, U_{(m)} &=& 0 ~, \\
\Big(\frac{\pa}{\pa \z^\a} {\cD}^k_{(2)}\Big)U_{(m)} &=& 2\ri k\, {\z}^\b {\cD}_{\a \b}\, {\cD}^{k-1}_{(2)}U_{(m)}~.
\eea
\end{subequations}
We also state some other properties which we often use throughout our calculations
\begin{subequations}
\bea
{\cD}^2_{(1)} &=& 0 ~,\\
\big[ {\cD}_{(1)}\,, {\cD}_{(2)} \big] 
&=& 
\big[ \bar \cD_{(1)}\,, \cD_{(2)} \big] = 0~,\\
\big[ \cD^\a, \cD_{(2)} \big] &=& 2\ri \, {\cS}\, \z^\a {\cD}_{(1)} ~,\\
\big[\cD^\a, \cD^k_{(2)}\big] &=& 2\ri \, {\cS} \,k \,\z^\a \cD^{k-1}_{(2)} {\cD}_{(1)}~,\\
\big[\cD^\a, \z^\b \cD_{\a \b}\big] &=& 3 {\cS} \cD_{(1)}~.
\eea
\end{subequations}

\subsection{Massive model}

We consider the addition of a mass term to the functional \eqref{chiral20} 
\bea
S = \int \rd^3x \rd^2 \q  \rd^2 \bar \q \, \bm E\, \bar \F \F
+\Big\{ \frac{m}{2} \int \rd^3x \rd^2 \q \, \cE \, \F^2 +{\rm c.c.} \Big\}~,
\label{chiral-massive20}
\eea
with $m$ a complex mass parameter. 
In the $m\neq 0$ case,   the U(1)${}_R$ weight of $\F$ 
is uniquely fixed to be $r=1$, in order for the action to be $R$-invariant.

Making use of the massive equations of motion
\bea
-\frac{1}{4} \cD^2 \F  +\bar m \bar \F =0, \qquad
-\frac{1}{4} \bar \cD^2 \bar \F +m \F =0,
\eea
we obtain 
\bea
\cD_{(-1)}{\mathbb J}_{(2s)} &=& -2\ri{\cS}(2s+1)(s+1) \cD_{(1)}
\sum_{k=0}^{s-1}\frac{1}{2s-2k+1} (-1)^{k} \binom{2s}{2k+1}
\non \\ 
&& 
\times {\cD}^k_{(2)} \bar \F \,\,{\cD}^{s-k-1}_{(2)}  \F 
\non \\
&& + \bar m \,(-1)^s (2s+1)\sum_{k=0}^{s-1}\left\{1+ (-1)^s \frac{2k+1}{2s-2k+1}\right\} (-1)^{k} \binom{2s}{2k+1}
\non \\ 
&& 
\times {\cD}^k_{(2)} \bar \F \,\,{\cD}^{s-k-1}_{(2)} \bar \cD_{(1)} \bar \F  ~,
\label{51}
\eea
where ${\mathbb J}_{(2s)}$ is defined by \eqref{611}.
We observe that \eqref{51} can also be written in the form 
\bea
\cD_{(-1)}{\mathbb J}_{(2s)} &=& \hf (-1)^s \,\cD_{(-1)} \sum_{k=0}^{s-1} (-1)^{k} \binom{2s}{2k+1} \cD^{k}_{(2)} \cD_{(1)} \F \,\, \cD^{s-k-1}_{(2)} \bar \cD_{(1)} \bar \F 
\non \\ 
&&-\hf \cD_{(1)} \sum_{k=0}^{s-1} (2k+1)(-1)^{k} \binom{2s}{2k+1}  \cD^{k}_{(2)} \cD^{\a} \F \,\, \cD^{s-k-1}_{(2)} \bar \cD_{\a} \bar \F 
\non \\
&&+ 2\ri \cS \, \cD_{(1)} \sum_{k=0}^{s-1} \left[(2k+1) + (-1)^{s-1} s(2s-2k-1) \right] 
\non \\
&&\qquad \times (-1)^{k} \binom{2s}{2k+1} \cD^{k}_{(2)}  \F \,\, \cD^{s-k-1}_{(2)}  \bar \F 
\non \\
&&+ \ri [1+ (-1)^s ] \sum_{k=0}^{s-1} (2k+1) (-1)^{k} \binom{2s}{2k+1} 
\non \\
&& \qquad \times \cD^{k}_{(2)} \cD^{\a} \F \,\, \cD^{s-k-1}_{(2)} \z^{\b} \cD_{\a \b} \bar \F ~.
\label{52}
\eea
Thus, for all odd values of $s$, 
\begin{subequations}
\bea
s= 2n+1~, \qquad n =0, 1, \dots~, \label{313a}
\eea
we end up with the conservation equation 
\bea
\cD_{(-1)}  \hat{{\mathbb J}}_{(2s)} = \cD_{(1)} \hat{\mathbb T}_{(2s-2)}
\label{313b}
\eea
where we have denoted 
\bea
\hat{{\mathbb J}}_{(2s)} &=& {\mathbb J}_{(2s)}
- \hf \sum_{k=0}^s (-1)^k
 \binom{2s}{2k+1} 
{\cD}^k_{(2)} \bar \cD_{(1)} \bar \F \,\,
{\cD}^{s-k-1}_{(2)} \cD_{(1)} \F  
~,\\
\hat{\mathbb T}_{(2s-2)} &=& -\hf \sum_{k=0}^{s-1} (2k+1)(-1)^{k} \binom{2s}{2k+1} 
 \cD^{k}_{(2)} \cD^{\a} \F \,\, \cD^{s-k-1}_{(2)} \bar \cD_{\a} \bar \F 
\non \\
&&+ 2\ri \cS \,  \sum_{k=0}^{s-1} \left[(1-s)(2k+1)+ 2s^2 \right] (-1)^{k} \binom{2s}{2k+1} 
\cD^{k}_{(2)}  \F \,\, \cD^{s-k-1}_{(2)}  \bar \F
~.~~~~~~~~
\eea
The trace multiplet $\hat{\mathbb T}_{(2s-2)}$ is covariantly linear, 
\bea
\bar \cD^2 \hat{\mathbb T}_{(2s-2)} = 0 ~,\qquad \cD^2 \hat{\mathbb T}_{(2s-2)} =0~.
\label{313e}
\eea
\end{subequations}
The conservation equation defined by eqs.~\eqref{313b} and \eqref{313e} 
coincides with that defined by eqs. \eqref{68a} and \eqref{68c}.

The above analysis demonstrates that in the massive case, the higher-spin supercurrent $ \hat{{\mathbb J}}_{(2s)} $ exists only for the odd values of $s$. This conclusion is again analogous to our previous results 
in 4D and (1,1) AdS superspace. As demonstrated in the construction of AdS higher-spin supercurrents (see \ref{ss444}), 
the even values of $s$ are also allowed provided there are several massive chiral superfields
in the theory. This analysis may be extended to the (2,0) AdS case. 


\section{Massless higher-spin gauge theories in (2,0) AdS superspace}
\label{s57} 

The explicit  structure of the higher-spin supercurrent multiplet defined by eqs.~\eqref{68a}
and \eqref{68c} 
allows us to develop two off-shell formulations for 
a massless multiplet of half-integer superspin-$(s+\hf)$, with $s = 2,3, \dots$~. We will call them type II and type III series\footnote{Type I series will be referred to as the longitudinal formulation 
for the gauge massless half-integer superspin multiplets in (1,1) AdS superspace
\eqref{long-action-half-11} and Minkowski superspace \cite{KO}.
The type I series and its dual are naturally related to 
the off-shell formulations for massless higher-spin $\cN=1$ 
supermultiplets in four dimensions \cite{KPS,KS,KS94}.
The type II and type III series have no four-dimensional counterpart. }  
to comply with the terminology introduced in \cite{KT-M11}
for the minimal off-shell formulations for $\cN=2$ supergravity ($s=1$).


\subsection{Type II series}

Given a positive integer $s \geq 2$, 
we propose to describe a massless multiplet of  superspin-$(s+\hf)$ 
in terms of two unconstrained real superfields
\be
\cV^{(\rm II)}_{(s+\hf )} = 
\Big\{ {\mathfrak H}_{\a(2s)}, \mathfrak{L}_{\a(2s-2)} \Big\} ~. \label{21}
\ee
Here ${\mathfrak H}_{\a(2s)} = {\mathfrak H}_{(\a_1 \dots \a_{2s})}$ 
and ${\mathfrak L}_{\a(2s-2)} = {\mathfrak L}_{(\a_1 \dots \a_{2s-2})}$
are symmetric in their spinor indices.

We postulate gauge transformations for the dynamical superfields:
\begin{subequations} \label{lambda-gauge20}
\bea
\d_\l {\mathfrak H}_{\a(2s)}&=& 
{\bar \cD}_{(\a_1} \l_{\a_2 \dots \a_{2s})}-{\cD}_{(\a_1}\bar \l_{\a_2 \dots \a_{2s})}
\equiv
g_{\a(2s)}+\bar{g}_{\a(2s)} ~,  
\label{H-gauge20} \\ 
\d_\l {\mathfrak L}_{\a(2s-2)} &=& -\frac{\ri}{2}
\big( \bar \cD^{\b} \l_{\b \a(2s-2)}+ \cD^{\b} \bar \l_{\b \a(2s-2)} \big)~,
\label{L-gauge20}
\eea
\end{subequations}
where the gauge parameter $\l_{\a(2s-1)}$ is unconstrained complex.
Eq. \eqref{H-gauge20} implies that the complex gauge parameter $g_{\a(2s)}$
is a covariantly  longitudinal linear superfield,
\bea
g_{\a(2s)}:= {\bar \cD}_{(\a_1} \l_{\a_2 \dots \a_{2s})}~, \qquad
{{\bar \cD}}_{(\a_1} g_{\a_2 \dots \a_{2s+1})} =0~. \label{gparam20}
\eea
The gauge transformation of ${\mathfrak H}_{\a(2s)}$, eq. \eqref{H-gauge20}, corresponds to the conformal 
superspin-$(s+\hf)$ gauge prepotential reviewed in subsection \ref{ss513}. It is natural to interpret 
${\mathfrak L}_{\a(2s-2)} $ as a compensating multiplet.
In order for $\d_\l {\mathfrak H}_{\a(2s)}$ and $\d_\l {\mathfrak L}_{\a(2s-2)}$
to be real,  $\l_{\a(2s-1)}$ must be
charged under the $R$-symmetry group U(1)${}_{R}$:
\bea
J \l_{\a(2s-1)} = \l_{\a(2s-1)}~, \qquad J \bar \l_{\a(2s-1)}= -\bar \l_{\a(2s-1)}~.
\eea

In addition to \eqref{L-gauge20}, the compensator ${\mathfrak L}_{\a(2s-2)} $ also possesses its own 
gauge freedom
\bea
\d_\x {\mathfrak L}_{\a(2s-2)} 
=  { \x}_{\a(2s-2)}+ \bar { \x}_{\a(2s-2)} ~, \qquad  \bar \cD_{\b} \x_{\a(2s-2)}=0~,
\label{prep-gauge20}
\eea
with the gauge parameter ${\x_{\a(2s-2)}}$ being covariantly chiral, 
but otherwise arbitrary. 
It should be pointed out that in (1,1) AdS superspace covariantly chiral superfields
exist only in the scalar case, since the constraint $\bar \cD_\b\J_{\a(n)}=0$ is inconsistent 
for $n>0$. Therefore, the gauge transformation law \eqref{prep-gauge20}
is specific for the (2,0) AdS supersymmetry.

Associated with ${\mathfrak L}_{\a(2s-2)} $
is the real field strength 
\bea
 \mathbb{L}_{\a(2s-2)} = \ri \cD^{\b} \bar \cD_{\b} {\mathfrak L}_{\a(2s-2)} ~,
 \qquad \mathbb{L}_{\a(2s-2)}= \bar{\mathbb{L}}_{\a(2s-2)}~,
\label{22}
\eea
which is a covariantly linear superfield, 
\bea
{\cD}^2 \mathbb{L}_{\a(2s-2)}=0 \quad \Longleftrightarrow \quad 
\bar \cD^2 \mathbb{L}_{\a(2s-2)}=0~.
\label{RLconst}
\eea
It is inert under the gauge transformation \eqref{prep-gauge20}, $\d_\x  \mathbb{L}_{\a(2s-2)} =0$.
From \eqref{L-gauge20}
we can read off the $\l$-gauge transformation of the field strength 
\bea
\d_\l \mathbb{L}_{\a(2s-2)}&=&
\frac{1}{4}
\big( \cD^{\b} {\bar \cD}^2 \l_{\b \a(2s-2)}- \bar \cD^{\b} {\cD}^2 \bar \l_{\b \a(2s-2)}\big)
\non \\
\qquad &=& -\frac{s}{2s+1} \cD^{\b} {\bar \cD}^{\g}\big(g_{\b \g \a(2s-2)} + \bar g_{\b \g \a(2s-2)}\big)
-\frac{2 \ri s}{2s+1} \cD^{\b \g} \bar g_{\b \g \a(2s-2)} ~.
\label{bbL-gauge20}
\eea
The reason why we express the gauge transformations of $\mathfrak{H}_{\a(2s)}$ and $\mathbb{L}_{\a(2s-2)}$ in terms of the constrained superfield $g_{\a(2s)}$ is that such representation will be useful to carry out the $(2,0) \rightarrow (1,0)$ AdS reduction in chapter \ref{ch6}.

Modulo an overall  normalisation factor, there is a unique quadratic action 
which is invariant under 
the gauge transformations \eqref{lambda-gauge20}. It is given by 
\bea
S^{(\rm II)}_{(s+\hf)}[{\mathfrak H}_{\a(2s)} ,{\mathfrak L}_{\a(2s-2)} ]
&=& \Big(-\hf \Big)^{s} 
\int 
\rd^3x \rd^2 \q  \rd^2 \bar \q
\, \bm E \bigg\{\frac{1}{8}{\mathfrak H}^{\a(2s)}
\cD^{\b}\bar{\cD}^{2} \cD_{\b}
{\mathfrak H}_{\a(2s)} \non \\
&&-\frac{s}{8}([\cD_{\b},\bar{\cD}_{\g}]{\mathfrak H}^{\b \g \a(2s-2)})
[\cD^{\d},\bar{\cD}^{\r}]{\mathfrak H}_{\d \r \a(2s-2)}
 \non \\
&& +\frac{s}{2}(\cD_{\b \g}{\mathfrak H}^{\b \g \a(2s-2)})
\cD^{\d \r}{\mathfrak H}_{\d \r \a(2s-2)}+ 2\ri s \,{\cS} {\mathfrak H}^{\a(2s)} {\cD}^\b {\bar \cD}_{\b} {\mathfrak H}_{\a(2s)}
\non \\
&&-  \frac{2s-1}{2} \Big( \mathbb{L}^{\a(2s-2)} [\cD^{\b}, \bar \cD^{\g}] {\mathfrak H}_{\b \g \a(2s-2)}
+ 2
\mathbb{L}^{\a(2s-2)} \mathbb{L}_{\a(2s-2)} \Big) \non\\
&& 
-\frac{(s-1)(2s-1)}{4s} \Big(  \cD_{\b} \mathfrak{L}^{\b \a(2s-3)} \bar \cD^2 \cD^{\g} \mathfrak{L}_{\g \a(2s-3)} +{\rm c.c.} \Big)
\non \\
&&-4(2s-1) \cS \mathfrak{L}^{\a(2s-2)} \mathbb{L}_{\a(2s-2)}
\bigg\}~.
\label{action20-t2}
\eea
By construction, the action is also invariant under \eqref{prep-gauge20}. This action differs from the massless half-integer superspin actions in (1,1) AdS superspace, \eqref{tr-action-half-11} and \eqref{long-action-half-11}, due to the presence of a Chern-Simons-type term.

Setting $s=1$ in \eqref{action20-t2} gives the linearised 
action for (2,0) AdS supergravity, which was originally derived 
in section 10.1 of \cite{KT-M11}. Ref. \cite{KT-M11} made use of the curvature
parameter $\r$, which is related to our $\cS$ as $\r =4\cS$. 
It should be remarked that the structure $\cD_{\b} \mathfrak{L}^{\b \a(2s-3)} \bar \cD^2 \cD^{\g} \mathfrak{L}_{\g \a(2s-3)}$ in  \eqref{action20-t2} is not defined for $s=1$. However, this term contains an overall  numerical factor $(s-1)$
and therefore it does not contribute for $s=1$.


\subsection{Type III series}

Our second model for the massless multiplet of superspin-$(s+\hf)$
is realised in terms of dynamical variables 
that are completely similar to \eqref{21}, 
\be
\cV^{(\rm III)}_{(s+\hf )} = 
\Big\{ {\mathfrak H}_{\a(2s)}, \mathfrak{V}_{\a(2s-2)} \Big\} ~.
\ee
Here ${\mathfrak H}_{\a(2s)}$
and ${\mathfrak V}_{\a(2s-2)}$
are unconstrained real tensor superfields. 

The dynamical superfields are defined modulo gauge transformations of the form
\begin{subequations}\label{410}
\bea
\d_\l {\mathfrak H}_{\a(2s)}&=& 
{\bar \cD}_{(\a_1} \l_{\a_2 \dots \a_{2s})}-{\cD}_{(\a_1}\bar \l_{\a_2 \dots \a_{2s})}
=g_{\a(2s)}+\bar{g}_{\a(2s)} ~,  \\ 
\d_\l {\mathfrak V}_{\a(2s-2)} &=& \frac{1}{2s}
\big( \bar \cD^{\b} \l_{\b \a(2s-2)}- \cD^{\b} \bar \l_{\b \a(2s-2)} \big)~,
\label{V-gauge20}
\eea
\end{subequations}
where the gauge parameter $\l_{\a(2s-1)}$ is unconstrained complex, and the longitudinal linear parameter $g_{\a(2s)}$ is defined as in \eqref{gparam20}. As in the type II case, ${\mathfrak H}_{\a(2s)}$ is the superconformal 
gauge multiplet, while ${\mathfrak V}_{\a(2s-2)} $ is a compensating multiplet.
The only difference from the type II case occurs in the gauge transformation law for the compensator ${\mathfrak V}_{\a(2s-2)}$. 

The compensator ${\mathfrak V}_{\a(2s-2)} $ is required to have its own 
gauge freedom of the form 
\bea
\d_\x {\mathfrak V}_{\a(2s-2)} 
=  { \x}_{\a(2s-2)}+ \bar { \x}_{\a(2s-2)} ~, \qquad  \bar \cD_{\b} \x_{\a(2s-2)}=0~,
\label{prep-gauge2}
\eea
with the gauge parameter ${\x_{\a(2s-2)}}$ being covariantly chiral, 
but otherwise arbitrary. 

Associated with $ \mathfrak{V}_{\a(2s-2)}$ is the real field strength
\bea
\mathbb{V}_{\a(2s-2)} = \ri \cD^{\b} \bar \cD_{\b} {\mathfrak V}_{\a(2s-2)} ~,
 \qquad \mathbb{V}_{\a(2s-2)}= \bar{\mathbb{V}}_{\a(2s-2)}~, \label{5713}
\eea
which is inert under \eqref{prep-gauge2}, $\d_\x  \mathbb{V}_{\a(2s-2)} =0$. It is not difficult to see that 
$\mathbb{V}_{\a(2s-2)}$ is covariantly linear, 
\bea
{\cD}^2 \mathbb{V}_{\a(2s-2)}=0 \qquad \Longleftrightarrow \quad  \bar {\cD}^2 \mathbb{V}_{\a(2s-2)}=0~. \label{RLV}
\eea
It varies under the $\l$-gauge transformation \eqref{410} as 
\bea
\d_\l \mathbb{V}_{\a(2s-2)}&=&
\frac{\ri}{4s}
\big( \cD^{\b} {\bar \cD}^2 \l_{\b \a(2s-2)}+ \bar \cD^{\b} {\cD}^2 \bar \l_{\b \a(2s-2)}\big) ~.~~~
\non \\
\qquad &=& -\frac{\ri}{2s+1} \cD^{\b} {\bar \cD}^{\g}\big(g_{\b \g \a(2s-2)} - \bar g_{\b \g \a(2s-2)}\big)
-\frac{2}{2s+1} \cD^{\b \g} \bar g_{\b \g \a(2s-2)} ~.~~~~
\label{bbV-gauge20}
\eea

Modulo normalisation, there exists a unique action being invariant under the gauge transformations \eqref{410} and \eqref{prep-gauge2}. It is given by 
\bea
S^{(\rm III)}_{(s+\hf)}
&=& \Big(-\hf \Big)^{s} 
\int 
\rd^3x \rd^2 \q  \rd^2 \bar \q
\, \bm E \bigg\{\frac{1}{8}{\mathfrak H}^{\a(2s)}
\cD^{\b}\bar{\cD}^{2} \cD_{\b}
{\mathfrak H}_{\a(2s)} \non \\
&&-\frac{1}{16}([\cD_{\b},\bar{\cD}_{\g}]{\mathfrak H}^{\b \g \a(2s-2)})
[\cD^{\d},\bar{\cD}^{\r}]{\mathfrak H}_{\d \r \a(2s-2)}
 \non \\
&& +\frac{1}{4}(\cD_{\b \g}{\mathfrak H}^{\b \g \a(2s-2)})
\cD^{\d \r}{\mathfrak H}_{\d \r \a(2s-2)}+ \ri  \,{\cS} {\mathfrak H}^{\a(2s)} {\cD}^\b {\bar \cD}_{\b} {\mathfrak H}_{\a(2s)}
\non \\
&&-  \frac{2s-1}{2} \Big( \mathbb{V}^{\a(2s-2)} \cD^{\b \g} {\mathfrak H}_{\b \g \a(2s-2)}
+ \frac{1}{2}
\mathbb{V}^{\a(2s-2)} \mathbb{V}_{\a(2s-2)} \Big) \non\\
&&+2s(2s-1) \cS \mathfrak{V}^{\a(2s-2)} \mathbb{V}_{\a(2s-2)}
\non \\
&& 
+\frac{1}{8}(s-1)(2s-1)\Big(
\cD_{\b} \mathfrak{V}^{\b \a(2s-3)} \bar \cD^2 \cD^{\g} \mathfrak{V}_{\g \a(2s-3)}
+{\rm c.c.} \Big)
\bigg\}~.
\label{action2-t3}
\eea
Although the structure 
$\cD_{\b} \mathfrak{V}^{\b \a(2s-3)} \bar \cD^2 \cD^{\g} \mathfrak{V}_{\g \a(2s-3)}$
in \eqref{action2-t3} is not defined for $s=1$, it comes with the factor $(s-1)$ 
and drops out from \eqref{action2-t3} for the $s=1$ case.
In this case the action 
coincides with the type III supergravity 
action\footnote{Type III supergravity is known only at the 
linearised level. In the super-Poincar\'e case, it is a 3D analogue of the 
massless superspin-3/2 multiplet proposed in \cite{BGLP}.} in (2,0) AdS superspace, which was originally derived 
in section 10.2 of \cite{KT-M11}.


\section{Summary and discussion} \label{s58}
Let us summarise the main results obtained thus far. Sections \ref{s53} and \ref{s54} are devoted to the superfield descriptions of off-shell massless higher-spin gauge theories in (1,1) AdS superspace, which are essentially analogous to their 4D ${\cN}=1$ AdS counterparts. A useful application includes the possibility to derive off-shell massless higher-spin ${\cN}=1$ supermultiplets by performing ${\cN}=2 \to {\cN}=1$ superspace reduction. As an example, we carried out reduction of the longitudinal theory for the massless superspin-$s$ multiplet \eqref{long-action-int} in the super-Poincar\'e limit and ended up with
a new model for massless ${\cN}=1$ higher-spin supermultiplet that was not described in \cite{KP1, KT}. In section \ref{s55}, the off-shell gauge formulations enabled us to derive consistent higher-spin supercurrent multiplets with (1,1) AdS supersymmetry. By studying models for chiral scalar superfields, we presented explicit expressions of such supercurrents.

With regards to (2,0) AdS supersymmetry, we employed a ``bottom-up" approach. The starting point was some simple dynamical systems 
in (2,0) AdS superspace, \textit{i.e} models for a free chiral scalar superfield. In such models, we deduced that the corresponding multiplet of higher-spin currents 
is described by the conservation equations
\begin{subequations} \label{1.1}
\bea
\cD^\b \mathbb{J}_{\b \a_1 \dots \a_{2s-1}} = \cD_{(\a_1} {\mathbb T}_{\a_2 \dots \a_{2s-1})}~, 
\qquad \bar \cD^\b \mathbb{J}_{\b \a_1 \dots \a_{2s-1}} 
= \bar \cD_{(\a_1} \bar {\mathbb T}_{\a_2 \dots \a_{2s-1})}~,
\eea
with the real superfield $\mathbb{J}_{\a(2s)}$ denotes the higher-spin supercurrent, and  ${\mathbb T}_{\a(2s-2)} $  
the corresponding trace supermultiplet constrained to be 
covariantly linear
\bea
\bar \cD^2 {\mathbb T}_{\a(2s-2)}=0~, 
\qquad \cD^2 {\mathbb T}_{\a(2s-2)}=0~.\label{1.1b}
\eea
In general, the trace supermultiplet is complex, 
\bea
{\mathbb T}_{\a(2s-2)} = {\mathbb Y}_{\a(2s-2)} -\ri {\mathbb Z}_{\a(2s-2)}~, 
\quad 
{\rm Im} \,{\mathbb Y}_{\a(2s-2)}=0~,
\quad 
{\rm Im} \,{\mathbb Z}_{\a(2s-2)}=0~.
\eea
\end{subequations} 
In the $s=1$ case, the above conservation equation coincides with that 
for the (2,0) AdS supercurrent \cite{KT-M11}.

We did not carry out a systematic analysis 
(similar to that given by  Dumitrescu and Seiberg \cite{DS} 
for ordinary supercurrents in Minkowski space)
of the higher-spin supercurrent \eqref{1.1}.
However, the formal consistency of \eqref{1.1} follows from 
the structure of the massless superspin-$(s+\hf)$ gauge theories constructed 
in section \ref{s57}. For instance, within the framework of
the type II formulation, let us couple the prepotentials 
${\mathfrak H}_{ \a (2s)} $ and ${\mathfrak L}_{ \a (2s-2)} $
to external sources
\bea
S^{(s+\hf)}_{\rm source}= \int \rd^3x \rd^2 \q  \rd^2 \bar \q \, \bm E\, \Big\{ 
{\mathfrak H}^{ \a (2s)} {\mathbb J}_{ \a (2s)}
-2 {\mathfrak L}^{ \a (2s-2)} {\mathbb Z}_{ \a (2s-2)}
 \Big\}~.
\eea
Requiring $S^{(s+\hf)}_{\rm source}$ 
to be invariant under the gauge transformations 
\eqref{prep-gauge20} tells us that the real supermultiplet ${\mathbb Z}_{\a(2s-2)}$
is covariantly linear, 
\bea
\bar \cD^2 {\mathbb Z}_{\a(2s-2)}=0~. 
\eea
If we also require $S^{(s+\hf)}_{\rm source}$ to be invariant under the gauge transformations
\eqref{lambda-gauge20}, we obtain the conservation equation
\bea
\bar \cD^\b {\mathbb J}_{\b \a_1 \dots \a_{2s-1}} 
= \ri \bar \cD_{(\a_1}  {\mathbb Z}_{\a_2 \dots \a_{2s-1})}~.
\eea
Additionally, taking the type III formulation into account leads to the  
general conservation equation 
\bea\label{ce1}
\bar \cD^{\b} {\mathbb J}_{\b \a_1 \dots \a_{2s-1}}  = \bar \cD_{(\a_1} \big( \mathbb{Y}_{\a_2 \dots \a_{2s-1})} + \ri \mathbb{Z}_{\a_2 \dots \a_{2s-1})} \big)~,
\eea
where the real trace supermultiplets $\mathbb{Y}_{\a(2s-2)} $ and $\mathbb{Z}_{\a(2s-2)} $
are covariantly linear.
The off-shell construction of a massless multiplet of $\textit{integer}$ superspin with (2,0) AdS supersymmetry would definitely deserve further study.

An improvement transformation exists for the higher-spin  supercurrent multiplet \eqref{1.1}.
Let us introduce
\begin{subequations}\label{55}
\bea
\widetilde{{\mathbb J}}_{\a(2s)}&:=& {\mathbb J}_{\a(2s)}+ [\cD_{(\a_1}, \bar \cD_{\a_2}] \mathbb{S}_{\a_3 \dots \a_{2s})} 
+ 2 \cD_{(\a_1 \a_2} \mathbb{R}_{\a_3 \dots \a_{2s})}~, \\
\widetilde{\mathbb{Y}}_{\a(2s-2)}&:=& 
\mathbb{Y}_{\a(2s-2)}- {\ri} \cD^{\g} \bar \cD_{\g} \mathbb{R}_{\a(2s-2)} 
+ 4(s+1) \cS \mathbb{R}_{\a(2s-2)}
\non
\\
\qquad \qquad &&+ \frac{2}{s}(s-1)\cD^{\b}\,_{(\a_1}\mathbb{R}_{\a_2 \dots \a_{2s-2}) \b} ~, \\
\widetilde{\mathbb{Z}}_{\a(2s-2)}&:=& \mathbb{Z}_{\a(2s-2)} - \ri \frac{s+1}{s} 
\cD^{\g} \bar \cD_{\g} \mathbb{S}_{\a(2s-2)} - 4 (s+1)\cS \mathbb{S}_{\a(2s-2)}
\non \\
\qquad \qquad &&- \frac{2}{s}(s-1)\cD^{\b}\,_{(\a_1}\mathbb{S}_{\a_2 \dots \a_{2s-2}) \b} ~,
\eea
\end{subequations}
with $\mathbb{S}_{\a(2s-2)}$ and $\mathbb{R}_{\a(2s-2)}$ real linear superfields.
One may check that $\widetilde{{\mathbb J}}_{\a(2s)}, \widetilde{\mathbb{Y}}_{\a(2s-2)}$ 
and $\widetilde{\mathbb{Z}}_{\a(2s-2)}$ obey the conservation equation and 
constraints described by 
\eqref{1.1}.
In the $s=1$ case, we reproduce the result given in section 10.4 of \cite{KT-M11}. 

As a final remark, there is one special feature of the supergravity case, $s=1$, 
for which the supercurrent conservation equation takes the form \cite{KT-M11}
\bea
\bar \cD^{\b} {\mathbb J}_{\b \a} = \bar \cD_{\a} \big( \mathbb{Y} + \ri \mathbb{Z} \big)~,
\label{56}
\eea
with the real trace supermultiplets $\mathbb{Y}$ and $ \mathbb{Z} $ being covariantly 
linear. Building on the thorough analysis of  \cite{DS},
it was pointed out in \cite{KT-M11} that there exists a well-defined improvement 
transformation that results with $\mathbb{Y} =0$. For all the supersymmetric 
field theories in (2,0) AdS superspace considered in \cite{KT-M11}, 
the supercurrent is characterised by the condition $\mathbb{Y} =0$.
Actually, this condition is easy to explain.
The point is that every 3D $\cN=2$ supersymmetric field theory with U(1) $R$-symmetry
may be coupled to the (2,0) AdS supergravity, which implies $\mathbb{Y} =0$ upon 
freezing the supergravity multiplet to its maximally supersymmetric 
(2,0) AdS background. There is another way to explain why $\mathbb Y$
may always be improved to zero. For simplicity, let us consider the case of $\cN=2$
Poincar\'e supersymmetry, with $D_\a$ and $\bar D_\a$ being the flat-superspace 
covariant derivatives. In Minkowski superspace eq. \eqref{56} implies 
$\pa^{\a\b} {\mathbb J}_{\a\b} = \ri D^\a \bar D_\a {\mathbb Y}$, and therefore 
$\mathbb{Y}=\ri D^{\a} \bar D_{\a} \mathbb{R}$, for some real linear superfield 
$\mathbb R$. If we now apply the flat-superspace version of \eqref{55}
with $\mathbb S=0$, we will end up with ${\mathbb Y}=0$.
However, in the higher-spin case it no longer seems possible 
to improve  the trace supermultiplet $\mathbb{Y}_{\a(2s-2)} $ to vanish, 
as our analysis in section \ref{s62} indicates.



\chapter{Field theories with (2,0) AdS supersymmetry in ${\cN}=1$ \\
AdS superspace} \label{ch6}
In the preceding chapter, it was pointed out that 3D ${\cN}$-extended AdS supergravity exists in several incarnations and they are known as $(p,q)$ AdS supergravity theories. Various aspects of  ${\cN}=2$ supersymmetric higher-spin gauge theories in 3D anti-de Sitter space, ${\rm AdS}_3$~, have also been elaborated in some detail.


This chapter has two main objectives. The first is to present a formalism which was developed in \cite{HK19} to reduce every field theory with (2,0) AdS supersymmetry to ${\cN}=1$ AdS superspace. This formalism is then applied  to carry out the (2,0) $\to$ (1,0) AdS reduction of the two off-shell massless higher-spin supermultiplets constructed in section \ref{s57}.
Our motivation came from certain theoretical arguments which suggest the existence of more general off-shell massless higher-spin 
$\cN=1$ supermultiplets in AdS${}_3$ than those described in \cite{KP1}. 
The second objective is to study $\cN=1$ supermultiplets of conserved higher-spin currents in AdS${}_3$, which were derived for the first time in \cite{HK19}.\footnote{It should be pointed out that the superconformal multiplets
of conserved currents  in Minkowski superspace \cite{NSU} can readily be lifted to AdS${}_3$.}

\section{(2,0) $\to$ (1,0) AdS superspace reduction} \label{section72}

The aim of this section is to elaborate on the details of procedure for reducing field theories  in  (2,0) AdS superspace to ${\cal N}=1$ AdS superspace. 
Explicit examples of such a reduction are given by considering supersymmetric 
nonlinear $\s$-models. 


\subsection{Geometry of (2,0) AdS superspace: Real basis}

In section \ref{ss522} the geometry of (2,0) AdS superspace was described in terms of the complex basis for the spinor covariant derivatives, eq. \eqref{deriv20}. It proves to be more convenient to switch to a real basis in order 
to carry out reduction to $\cN=1$ AdS superspace ${\rm AdS}^{3|2}$.
Following \cite{KLT-M12}, such a basis is introduced by replacing the complex operators $\cD_\a$ and $\bar \cD_\a$ with 
${\bm \de}_\a^I= ( {\bm \de}_\a^\1, {\bm \de}_\a^\2 )$ defined as follows:
 \bea
& \cD_\a=\frac{1}{\sqrt{2}}({\bm \nabla}_\a^{\1}-\ri {\bm \nabla}_\a^{\2})~,~~~
  \bar \cD_\a=-\frac{1}{\sqrt{2}}({\bm \nabla}_\a^{\1}+\ri {\bm \nabla}_\a^{\2})~.
  \label{N1-deriv20}
 \eea
In a similar way, we introduce real coordinates, $z^\cM = (x^m, \q^\m_I)$, 
to parametrise (2,0) AdS superspace. 
Defining ${\bm \nabla}_a =\cD_a$, the algebra of  (2,0) AdS covariant derivatives 
\eqref{deriv20} turns into\footnote{The antisymmetric tensors $\ve^{IJ}$ and $\ve_{IJ} $
are normalised as $\ve^{\1\2} = \ve_{\1\2} =1$.}
\bsubeq \label{2_0-alg-AdS}
\bea
&\{ {\bm \nabla}_\a^I, {\bm \nabla}_\b^J\}=
2\ri\d^{IJ} {\bm \nabla}_{\a\b}
-4\ri \d^{IJ} \cS M_{\a\b}
+4\ve_{\a\b}\ve^{IJ} \cS J
~,
\label{2_0-alg-AdS-1}
\\
&{[} {\bm \nabla}_{a}, {\bm \nabla}_\b^J{]}
=
\cS (\g_a)_\b{}^\g {\bm \nabla}_{\g}^J
~, \qquad 
{[} {\bm \nabla}_{a}, {\bm \nabla}_b{]}
=
-4\cS^2 M_{ab}
~,
\label{2_0-alg-AdS-2}
\eea
\esubeq
The action of the ${\rm U(1)}_R$ generator on the spinor covariant derivatives 
is given by
\bea
{[}J, {\bm \nabla}_\a^I{]} = -\ri \ve_{I J}  {\bm \nabla}_{\a}^J~.
\eea

As may be seen from \eqref{2_0-alg-AdS},
the graded commutation relations for the operators $ {\bm \de}_a$ and 
$ {\bm \de}_\a^{\1}$
have the following properties: 
\begin{enumerate} 
\item These (anti-)commutation relations do not involve $ {\bm \de}_\a^{\2}$,
\bsubeq \label{2.144}
\bea
&\{  {\bm \nabla}_\a^\1,  {\bm \nabla}_\b^\1\}=
2\ri  {\bm \nabla}_{\a\b}
-4\ri  \cS M_{\a\b}
~,
\\
&{[}  {\bm \nabla}_{a},  {\bm  \nabla}_\b^\1{]}
=
\cS (\g_a)_\b{}^\g  {\bm \nabla}_{\g}^\1
~,\qquad 
{[}  {\bm \nabla}_{a},  {\bm \nabla}_b{]}
=
-4\cS^2 M_{ab}
~.
\eea
\esubeq
\item
 Relations \eqref{2.144} are isomorphic
to the algebra of the covariant derivatives of 
${\rm AdS}^{3|2}$, see \cite{KLT-M12} for the details.
\end{enumerate}
We thus see that ${\rm AdS}^{3|2}$ is naturally  embedded in
(2,0) AdS  superspace  as a subspace. The real Grassmann variables of (2,0) AdS superspace, 
$\q^\m_I =(\q^\m_{\1}, \q^\m_{\2} )$, may be chosen in such a way that ${\rm AdS}^{3|2}$ corresponds to the surface defined by $\q^\m_{\2} =0$.
We also note that no U$(1)_R$ curvature  
is present in the algebra of $\cN=1$ AdS covariant derivatives.
These properties make possible a consistent  $(2,0)  \to (1,0)$
AdS superspace reduction.

Now we will recast  the fundamental properties of  
the (2,0) AdS Killing supervector fields in the real representation  \eqref{N1-deriv20}.
The isometries of  (2,0) AdS superspace are described in terms of 
those first-order operators 
\begin{subequations}\label{2.155}
\bea
\z := \z^{\cB}  {\bm  \de}_{\cB} = \z^b  {\bm \de}_b+\z^\b_J  {\bm \de}_\b^J~, 
\qquad J= \1,\2~,
\label{realvect}
\eea
which solve the equation
\bea
\big{[}\z + \hf l^{bc} M_{bc} + \ri \t J ,  {\bm \de}_\cA\big{]}=0~, \label{Killing-real}
\eea
\end{subequations}
for some real parameters $\t$ and $l^{ab} = -l^{ba}$. 
Equation \eqref{Killing-real} is equivalent to
\begin{subequations} \label{K-eq-real1}
\bea
 {\bm \de}_{\a}^I \z_{\b}^J &=& -\ve_{\a \b} \ve^{IJ} \t + \cS \d^{IJ} \z_{\a \b}+ \hf \d^{IJ} l_{\a \b}~, \label{K-eq-real11}\\
 {\bm \de}_{\a}^I \z_{b} &=& 2 \ri \, \z^{\b I} (\g_b)_{\a \b}~,\\
 {\bm \de}_{\a}^I \t &=& -4 \ri\, \cS \ve^{IJ} \z_{\a J}~,\\
 {\bm \de}_{\a}^I l_{\b \g} &=& 8 \ri\, \cS \ve_{\a (\b} \z_{\g)}^I~, \label{K-eq-real14}
\eea
\end{subequations}
and
\begin{subequations} 
\bea
 {\bm \de}_{a} \z_{b} &=& l_{ab}= -l_{ba}~, \label{keq} \\
 {\bm \de}_{a} \z^{\b}_I &=& -\cS \z^{\a}_I (\g_a)_{\a}^{\,\b}~, \label{ks-eq} \\
 {\bm \de}_{a} \t &=& 0~,\\
 {\bm \de}_{a} l^{bc} &=& 4 \cS^2(\d_a^b \z^c - \d_a^c \z^b)~.
\eea
\end{subequations}
Some nontrivial implications of the above equations 
which will be important for our subsequent consideration are: 
\begin{subequations} \label{K-eq-real2}
\bea
 {\bm \de}^I_{(\a} \z_{\b \g)} &=&0 ~, \qquad   {\bm \de}^I_{(\a} l_{\b \g)} = 0~, \\
 {\bm \de}^{I}_{(\a} \z^{J}_{\b)} &=& 2 \cS \, \d^{IJ} \z_{\a \b}~,
\qquad
 {\bm \de}^{\g (I}\z^{J)}_{\g}=0~,\\
 \z^{I \a} &=& \frac{\ri}{6}  {\bm \de}^I_{\b} \z^{\a \b} 
 = \frac{\ri}{12 \cS}  {\bm \de}^I_{\b} l^{\a \b} 
= -\frac{\ri}{4 \cS} \ve^{IJ}  {\bm \de}_J^{\a}\, \t  ~,\\
\t &=& -\frac{1}{4} \ve_{IJ}  {\bm \de}^{\g I} \z^J_{\g}~.
\eea 
\end{subequations}
Equation \eqref{keq} implies that $\z_a$ is a Killing vector field,
\bea
 {\bm \de}_a \z_b +  {\bm \de}_b \z_a =0~,
\eea
while \eqref{ks-eq} is a Killing spinor equation. The real parameter $\t$ 
is constrained by 
\bea
( {\bm \de}^{\2})^2 \t = (  {\bm \de}^\1)^2 \t = 8\ri \cS \t~, \quad
{\bm \de}_a \t =0~.
\eea


\subsection{Reduction from (2,0) to (1,0) AdS superspace}

Given a tensor superfield ${\bm U}(x, \q_I)$ on (2,0) AdS superspace, 
 its $\cN=1$ projection (or bar-projection) is defined by 
\bea
{\bm U}|:= {\bm U}(x, \q_I)|_{\q_{\2} =0}
\eea
in a {\it special coordinate system} to be specified below.
By definition, ${\bm U}|$ depends on 
the real  coordinates $z^M= (x^m, \q^\m)$, with $\q^\m:=\q^\m_{\1}$,  
which will be used to parametrise $\cN=1$ AdS superspace ${\rm AdS}^{3|2}$. 
For the (2,0) AdS covariant derivative
\bea
{\bm \nabla}_{\cA}=({\bm \nabla}_a , {\bm \nabla}_\a^I) 
=
E_\cA{}^\cM \frac{\pa}{\pa z^\cM}
+\hf\O_{\cA}{}^{bc} M_{bc}
+\ri \F_{{\cA}} J~,
\eea
its bar-projection is defined as
\bea
{\bm \nabla}_{\cA}|=E_\cA{}^\cM | \frac{\pa}{\pa z^\cM}
+\hf\O_{\cA}{}^{bc} | M_{bc}
+\ri \F_{{\cA}} |J~.
\eea

We use the freedom to perform general coordinate, local Lorentz and  U$(1)_R$ 
transformations  to choose the following  gauge condition
\bea
{\bm \nabla}_a |=\nabla_a~, \qquad {\bm \nabla}^\1_\a|=\nabla_\a~, 
\label{224}
\eea
where 
\bea
\nabla_A = (\nabla_a , \nabla_\a ) = E_A{}^M \frac{\pa}{\pa z^M}
+\hf\o_{A}{}^{bc} M_{bc}
\eea
denotes the set of covariant derivatives for ${\rm AdS}^{3|2}$, which obey 
 the following graded commutation relations: 
\bsubeq 
\bea
&\{  { \nabla}_\a,  { \nabla}_\b \}=
2\ri  { \nabla}_{\a\b}
-4\ri  \cS M_{\a\b}
~,
\\
&{[}  { \nabla}_{a},  {  \nabla}_\b{]}
=
\cS (\g_a)_\b{}^\g  { \nabla}_{\g}
~,\qquad 
{[}  { \nabla}_{a},  { \nabla}_b{]}
=
-4\cS^2 M_{ab}
~.
\eea
\esubeq
In such a coordinate system,
the operator ${\bm \nabla}_\a^\1|$   contains no 
partial derivative with respect to $\q_\2$. As a consequence, 
 $\big({\bm \nabla}^\1_{{\a}_1} \cdots {\bm \nabla}^\1_{{\a}_k} {\bm U} \big)\big|
= \nabla_{{\a}_1} \cdots  \nabla_{{\a}_k} {\bm U}|$, for any positive integer $k$,  
where $\bm U$ is a tensor superfield on (2,0) AdS superspace.
Let us study how the $\cN=1$ descendants of $\bm U$ defined by
$U_{\a_1 \dots \a_k}:=
\big({\bm \nabla}^\2_{{\a}_1} \cdots {\bm \nabla}^\2_{{\a}_k} {\bm U} \big)\big|$
 transform under the (2,0) AdS isometries, 
 with $k$ a non-negative integer.

We introduce 
the $\cN=1$ projection of the (2,0) AdS Killing supervector field \eqref{2.155}
\bea
\z| = \x^b \de_b + \x^{\b} \de_{\b}+ \e^{\b} {\bm \de}_{\b}^{\2}|~,\qquad 
\x^b := \z^b|~, ~~\x^{\b}:= \z_{\1}^{\b}|~, ~~\e^{\b} :=  \z_{\2}^{\b}|~.
\eea
We also introduce the $\cN=1$ projections of the  
Lorentz and U$(1)_R$ parameters in \eqref{2.155}:
\bea
\l^{bc} := l^{bc}|~,~ \qquad \e := \t|~.
\eea
It follows from \eqref{2.155} that the $\cN=1$ parameters $\x^B = (\x^b, \x^\b)$ 
and $\l^{bc}$ obey the equation
\bea
\big{[}\x + \hf \l^{bc} M_{bc},\de_A\big{]}=0~, \qquad 
\x =\x^B \nabla_B=  \x^b \de_b + \x^{\b} \de_{\b}~,
\eea
which tells us that $\x^B$ is a Killing supervector field of $\cN=1$ AdS superspace
 \cite{KLT-M12}. This equation is equivalent to 
\bsubeq
\bea
 \de_{(\a} \x_{\b \g)}&=&0~, \qquad \de_{\b}\x^{\b \a} =- 6 \ri \x^{\a} ~, \\
\de_{\a}\x_{\b} &=& \hf \l_{\a \b}+ \cS \x_{\a \b} ~,\\
\de_{(\a} \l_{\b \g)}&=&0~, 
 \qquad \de_{\b}\l^{\b \a} =- 12 \ri \cS \x^{\a} ~. 
 \eea
\esubeq
These relations automatically follow from the (2,0) AdS Killing equations, 
eqs.~\eqref{K-eq-real11}~--~\eqref{K-eq-real14}, upon $\cN=1$ projection. 
Thus $(\x^a, \x^{\a}, \l^{ab})$ parametrise the infinitesimal isometries of 
${\rm AdS}^{3|2}$ \cite{KLT-M12} (see also\cite{KP1}). 

The remaining parameters $\e^{\a}$ and $\e$ generate the second supersymmetry and ${\rm U(1)}_R$ transformations, respectively.
Using the Killing equations \eqref{K-eq-real2}, it can be shown that they satisfy the following properties 
\bsubeq \label{u1-c}
\bea
\e_{\a}= \frac{\ri}{4\cS} \de_{\a} \e ~, \qquad
\e&=& -\hf \de^{\a} \e_{\a}~,  \label{u1-c-a} \\
(\ri \de^2 + 8 \cS)\e =0~, \qquad \de_a \e &=& 0~. 
\label{u1-c-b}
\eea
\esubeq
These imply that the only independent components of $\e$ are $\e|_{\theta=0}$ and $\de_{\a}\e|_{\theta=0}$. They correspond to the  ${\rm U(1)}_R$ and
second supersymmetry transformations, respectively.

Given a matter tensor superfield $\bm U$, its (2,0) AdS transformation law
\bea
\d_\z {\bm U} = \big(\z  +\hf l^{bc} M_{bc} +\ri \t J  \big){\bm U}
\label{isometry20-real}
\eea
turns into
\bsubeq\label{isometry1-real}
\bea
\d_\z {\bm U} | &=& \d_\x {\bm U} |  +\d_\e {\bm U} | ~, \\
\d_\x {\bm U} | &=&
\Big(\x^b \de_b + \x^{\b} \de_{\b}+ \hf \l^{bc} M_{bc}\Big)
{\bm U}| ~, \\
\d_\e {\bm U} | &=& 
 \e^{\b} ({\bm \de}_{\b}^{\2} \bm U)| +\ri \e J\,{\bm U}| ~.
\eea
\esubeq
It follows from \eqref{2.155} and \eqref{isometry1-real}
that every $\cN=1$ descendant $U_{\a_1 \dots \a_k}:=
\big( {\bm \nabla}^\2_{{\a}_1} \cdots {\bm \nabla}^\2_{{\a}_k} {\bm U} \big)\big|$
is a tensor superfield on ${\rm AdS}^{3|2}$,
\bea
\d_\x U_{\a_1 \dots \a_k}  = \Big(\x^b \de_b + \x^{\b} \de_{\b}+ \hf \l^{bc} M_{bc}\Big)  U_{\a_1 \dots \a_k}~.
\eea
For the $\e$-transformation we get 
\bea
\d_\e U_{\a_1 \dots \a_k} &=& \e^{\b} 
\big({\bm \de}_{\b}^{\2} {\bm \nabla}^\2_{{\a}_1} \cdots {\bm \nabla}^\2_{{\a}_k} {\bm U} \big)
\big|
+ \ri \e \big( J {\bm \nabla}^\2_{{\a}_1} \cdots {\bm \nabla}^\2_{{\a}_k} {\bm U} \big)\big| \label{u1}
\\ 
&=& \e^\b U_{\b \a_1 \dots \a_k} -\e \sum_{l=1}^{k}
{\bm \nabla}^\2_{{\a}_1} \cdots 
{\bm \nabla}^\2_{\a_{l-1}}  {\bm \nabla}^\1_{\a_{l}} {\bm \nabla}^\2_{\a_{l+1}} 
\cdots {\bm \nabla}^\2_{{\a}_k} {\bm U} \big)\big|
+  \ri  q\e U_{\a_1 \dots \a_k}~,~~\non
\eea
 where $q$ is the U$(1)_R$ charge of $\bm U$ defined  by  
 $J {\bm U} =q {\bm U}$.
 In the second term on the right, we have to push ${\bm \nabla}_{\a_l}^\1$
 to the far left through the $(l-1)$ factors of ${\bm \nabla}^\2$'s
 by making use of the relation
$ \{ {\bm \nabla}_\a^\1, {\bm \nabla}_\b^\2\}= 4\ve_{\a\b}\cS J$ 
and taking into account the relation 
\bea
\big({\bm \nabla}^\1_{\a_{l}} {\bm \nabla}^\2_{{\a}_1} \cdots 
{\bm \nabla}^\2_{\a_{l-1}}  {\bm \nabla}^\2_{\a_{l+1}} 
\cdots {\bm \nabla}^\2_{{\a}_k} {\bm U} \big)\big|
= \nabla_{\a_l} U_{\a_1 \dots \a_{l-1}\a_{l+1} \dots \a_k} ~.
\eea
As the next step, the U$(1)_R$ generator $J$ should be pushed to the right until it hits
$\bm U$ producing on the way insertions of ${\bm \nabla}^\1$. Then the procedure should be repeated. As a result, the variation $ \d_\e U_{\a_1 \dots \a_k} $
is expressed in terms of the superfields $ U_{\a_1 \dots \a_{k+1}},~
 U_{\a_1 \dots \a_k} ,  \cdots U_{\a_1} , U$.

So far we have been completely general and discussed infinitely many descendants 
$U_{\a_1 \dots \a_k}$ of $\bm U$.  However only a few  of them
are functionally independent. Indeed, eq. \eqref{2_0-alg-AdS-1}
tells us  that 
\bea
&\{  {\bm \nabla}_\a^\2,  {\bm \nabla}_\b^\2\}=
2\ri  {\bm \nabla}_{\a\b}
-4\ri  \cS M_{\a\b}
~,
\eea
and thus every  $U_{\a_1 \dots \a_k}$ for $k>2$ can be expressed in terms of 
$U$, $U_\a$ and $U_{\a_1\a_2}$. Therefore, it suffices to consider $k \leq 2$.
 
Let us give two examples of matter superfields 
on (2,0) AdS superspace. We first consider a covariantly chiral scalar superfield $\bm \f,~ \bar \cD_{\a}\bm {\f} =0$, with an arbitrary U$(1)_R$ charge $q$ defined by $J \bm \f = q \bm \f$. It transforms under the (2,0) AdS isometries as
\bea
\d_{\z} \bm \f = (\z +  \ri q \t ) \bm \f~.
\eea
When expressed in the real basis \eqref{N1-deriv20}, the chirality constraint on $\bm \f$ means
\bea
\bm \de^{\2}_{\a} \bm \f = \ri \bm \de^{\1}_{\a} \bm \f~,
\eea
As a result, there is only one independent ${\cal N}=1$ superfield upon reduction,
\bea
\vf := \bm \f|~.
\eea  
We then get the following relations
\bsubeq
\bea
\bm \de^{\2}_{\a} \bm \f| &=& \ri \de_{\a} \vf~, \\
( \bm \de^{\2})^2 \bm \f| &=& - \de^2 \vf - 8 \ri q \cS  \vf~.
\eea
\esubeq
The $\e$-transformation \eqref{u1} is given by 
\bea
\d_{\e} \vf = \ri \e^{\b} \de_{\b}\vf + \ri q \e  \vf~.
\eea

Our second example is a real linear superfield ${\mathbb L}= \bar{\mathbb L}~, \bar \cD^2 {\mathbb L} =0$~. The real linearity constraint relates the ${\cal N}=1$ descendants of $\mathbb{L}$ as follows:
\begin{subequations}
\bea
(\bm{\de}^{\2})^2 \mathbb{L} &=& ( \bm{\de}^{\1})^2 \mathbb{L}~,\\
\bm \de^{\1 \b} \bm \de^{\2}_{\b} \mathbb{L} &=&0~.
\label{qq1}
\eea
\end{subequations}
Thus, ${\mathbb L}$ is equivalent to two independent, real $\cN=1$ superfields:
\bea
X := {\mathbb L} |~, \qquad
W_{\a} :=\ri \bm \nabla_\a^{\2} {\mathbb L} |~.
\eea
Here $X$ is unconstrained, while $W_\a$ obeys 
the constraint \eqref{qq1}
\bea
\nabla^\a W_\a =0~,
\eea
which means that $W_\a$ is the field strength of an $\cN=1$ vector multiplet. 
Since ${\mathbb L}$ is neutral under the $R$-symmetry group U$(1)_R$, $J\, {\mathbb L} =0$, the second supersymmetry and U$(1)_R$ transformation laws of the ${\cal N}=1$ descendants of $\mathbb L$ are as follows:
\bsubeq
\bea
\d_{\e} X &=& \d_{\e} \mathbb{L}| = \e^{\b} (\bm \de^{\2}_{\b} \mathbb{L})| = -\ri \e^{\b} W_{\b}~,\\
\d_{\e} W_{\a} &=& \ri\, \big(\bm{ \de}^{\2}_{\a}  \d_{\e} \mathbb{L}\big)| = \ri \e^{\b}\big(\bm{ \de}^{\2}_{\b} \bm{ \de}^{\2}_{\a} \mathbb{L}\big)|- \e [J, \bm{ \de}^{\2}_{\a}] \mathbb{L}| \non\\
&=& -\e^{\b} \de_{\a \b} X - \frac{\ri}{2} \e_{\a} \de^2 X- \ri \e \de_{\a} X~.
\eea
\esubeq
 
 
\subsection{The (2,0) AdS supersymmetric actions in ${\rm AdS}^{3|2}$} \label{ss2.3}

Every rigid supersymmetric field theory in (2,0) AdS superspace may be reduced to ${\cal N}=1$ AdS superspace. Here we provide the key technical details of the reduction.

In accordance with \cite{KLT-M11,BKT-M,KT-M11,KLRST-M},
there are two ways of constructing supersymmetric actions in (2,0) AdS superspace:
(i) either by integrating a real scalar 
$\cL$ over the full (2,0) AdS superspace,\footnote{The component 
inverse vierbein is defined as usual, 
$e_a{}^m (x) = E_a{}^m |_{\q=0}$, with $e^{-1}=\det(e_a{}^m)$.}
\begin{align}\label{realac}
S&= \int \rd^3x \rd^2 \q \rd^2 \bar \q \, 
{\bm E}\, \cL
= \frac{1}{16} \int \rd^3x\, e\, 
	 \cD^2 \bar \cD^2  \cL   \Big|_{\q=0}
	 = \frac{1}{16} \int \rd^3x\, e\, 
	\bar  \cD^2  \cD^2  \cL   \Big|_{\q=0}
 \\	
	&= \int \rd^3x\, e\, \Big(
	\frac{1}{16} \cD^\alpha \bar \cD^2 \cD_\alpha 
	+ \ri \cS  \bar \cD^\alpha \cD_\alpha 
	\Big)    \cL\Big|_{\q=0} 
	= \int \rd^3x\, e\, \Big(
	\frac{1}{16} \bar \cD_\alpha  \cD^2 \bar \cD^\alpha 
	+ \ri \cS  \cD^\alpha  \bar\cD_\alpha
	\Big)    \cL   \Big|_{\q=0}
	~,
\non
\end{align}
with ${\bm E}^{-1}= {\rm Ber} (E_\cA{}^\cM)$;
or (ii) by integrating a covariantly chiral scalar $\cL_{\rm c}$
over the chiral subspace of the (2,0) AdS superspace,
\begin{align}\label{chiralac}
S_{\rm c} =\int \rd^3x\, \rd^2\q\, \cE\, \cL_{\rm c}
	= -\frac{1}{4} \int \rd^3x\, e\, \cD^2 \cL_{\rm c}\Big|_{\q=0}~, \qquad
\bar\cD^\alpha \cL_{\rm c} =0~,
\end{align}
with $\cE$ being the chiral density.
The superfield Lagrangians $\cL$ and $\cL_c$ are 
neutral and charged, respectively with respect to the group ${\rm U}(1)_R$: 
\begin{align}
J \cL = 0~, \qquad J \cL_{\rm c} = -2 \cL_{\rm c}~.
\end{align}
The two types of supersymmetric actions are related to each other
by the rule
\bea
\int \rd^3x \rd^2 \q  \rd^2 \bar \q
\,{\bm E}\,\cL
= \int \rd^3x \rd^2 \q \, \cE  \cL_{\rm c}~, \qquad \cL_{\rm c}:= -\frac{1}{4} \bar {\cD}^2 \cL~.
\label{rc}
\eea

Instead of reducing the above actions to components,
in this paper we need their reduction 
to $\cN=1$ AdS superspace. 
We remind the reader that the supersymmetric action in ${\rm AdS}^{3|2}$
makes use of a real scalar Lagrangian $L$. The superspace and component forms
of the action are:
\bea
S=  \int \rd^{3|2} z \, E\,L 
=  \frac{1}{4} \int \rd^3 x \, e\,\big(\ri \de^2 +8 \cS \big) L \Big|_{\q=0}~.
\label{252}
\eea
For the action \eqref{realac} we get
\bea
S = \int \rd^3x \rd^2 \q \rd^2 \bar \q \, {\bm E}\, \cL 
= -\frac{\ri}{4}\int 
\rd^{3|2} z
\, E\, (\bm \nabla^{\2})^2 \cL \Big|~,
\label{r-action}
\eea
with $E^{-1} =\mathrm{Ber}(E_A\,^M)$. 
The chiral action \eqref{chiralac} reduces to 
(1,0) AdS
as follows:
\bea
S_{\rm c} = \int \rd^3x \rd^2 \q \, \cE  \cL_{\rm c} = 2 \ri \int 
\rd^{3|2}z\, E \,  \cL_{\rm c} \Big|~.
\label{c-action}
\eea

Making use of the (2,0) AdS transformation law 
$\d \cL = \z \cL$, 
$\d \cL_{\rm c} = (\z - 2 \ri \t) \cL_{\rm c}$, and the Killing equation \eqref{Killing-real}, 
it can be checked explicitly that the ${\cN}=1$ action defined by the right-hand side of \eqref{r-action}, or \eqref{c-action} are invariant under the (2,0) AdS isometry 
transformations. 


\subsection{Supersymmetric nonlinear sigma models}

To illustrate the $(2,0)  \to (1,0)$ AdS superspace reduction described above,  
here we discuss two interesting examples.

Our first example is a general nonlinear $\s$-model with (2,0) AdS supersymmetry \cite{BKT-M,KT-M11}. It is described by the action
\bea
S = \int \rd^3x \,\rd^2 \q  \,\rd^2 \bar \q
\, {\bm E} \,K(\f^{i} , {\bar \f}^{\bar j}) + \bigg\{ \int 
\rd^3x \rd^2 \q\, \cE \,W(\f^{i})+ \rm{c.c} \bigg\}~, \quad \bar \cD_{\a} \f^{i} =0~,~~~~
\label{2.34}
\eea
where $K(\f^i, \bar \f^{\bar j})$ is the K\"ahler potential of a K\"ahler manifold 
and $W(\f^i)$ is a superpotential. 
The U$(1)_R$ generator is realised on the dynamical superfields $\f^i $ and 
$\bar \f^{\bar i}$ as 
\bea
\ri J = {\mathfrak J}^i(\f) \pa_i + \bar{\mathfrak J}^{\bar i} (\bar \f) \pa_{\bar i} ~,
\eea
where ${\mathfrak J}^i (\f)$ is a holomorphic Killing vector field such that
\bea
{\mathfrak J}^i(\f) \pa_i K = -\frac{\ri}{2} {\mathfrak D}(\f, \bar \f) ~, \qquad 
  \bar {\mathfrak D} ={\mathfrak D}~,
   \label{2.36a}
\eea
for some  Killing potential  ${\mathfrak D}(\f, \bar \f)$. The superpotential 
has to obey the condition 
\bea
{\mathfrak J}^i(\f) \pa_i W =- 2\ri W
\eea
in order for the action \eqref{2.34} to be invariant under the (2,0) AdS isometry transformations
\bea
\d \f^i = (\z +  \ri \t J)\f^i~.
\eea

In the real representation \eqref{N1-deriv20}, the chirality condition on $\f^i$ turns into
\bea
\bm \de^{\2}_{\a} \f^i = \ri \bm \de^{\1}_{\a} \f^i~.
\eea
It follows that upon ${\cal N}=1$ reduction, $\f^i$ leads to  just one superfield, 
\bea
\vf^i := \f^i|~.
\eea
In particular, we have the following relations
\bsubeq
\bea
\bm \de^{\2}_{\a} \f^i| &=& \ri \de_{\a} \vf^i~, \\
(\bm \de^{\2})^2 \f^i| &=& - \de^2 \vf^i -8 \cS {\mathfrak J}^i (\vf )~.
\eea
\esubeq
Using the reduction rules \eqref{r-action} and \eqref{c-action}, we obtain
\bea
S &=& \int \rd^{3|2} z \,E \,
\bigg\{ -\ri K_{i \bar j} (\vf, \bar \vf)
\de^{\a} \vf^{i}\de_{\a} \bar \vf^{\bar j}
 + \cS {\mathfrak D}(\vf, \bar \vf) 
+ \Big(  2\ri W(\vf) +{\rm c.c.} \Big)
\bigg\}~, ~~~
\label{N1-sigma}
\eea
where we have made use of the standard notation
\bea
K_{i_1 \cdots i_p \bar j_1 \cdots \bar j_q }:=
\frac{\pa^{p+q} K(\vf , \bar \vf)}{\pa \vf^{i_1}\cdots \pa \vf^{i_p} \pa \bar \vf^{\bar j_1} \cdots \pa \bar \vf^{\bar j_q}}~.
\eea
The action \eqref{N1-sigma} is manifestly ${\cN}=1$ supersymmetric. One may explicitly check that it is also invariant under the second supersymmetry and $R$-symmetry transformations generated by a real scalar  parameter $\e$ subject to the constraints \eqref{u1-c}, which are:
\bea
\d_{\e} \vf^i = \ri \e^{\a} \de_{\a}\vf^i 
+ \e \,{\mathfrak J}^i (\vf )~.
\eea

The family of supersymmetric $\s$-models \eqref{2.34} 
includes a special subclass which is specified by the two conditions:
(ii)   all  $\f$'s are neutral, $J \f^i=0$;  and (ii) no superpotential is present,  $W(\f)=0$. 
In this case no restriction on the K\"ahler potential is imposed by eq. \eqref{2.36a}, 
and the action \eqref{2.34} is invariant under arbitrary K\"ahler transformations 
\bea
K   \rightarrow K + \L + \bar \L, 
\eea
with  $\L(\f^i)$ a holomorphic function. The corresponding action in $\cN=1$ AdS superspace
is obtained from \eqref{N1-sigma} by setting ${\mathfrak D}(\vf, \bar \vf)=0$ and $W(\vf)=0$, 
and thus the action is manifestly K\"ahler invariant. 

Let us also consider a supersymmetric nonlinear $\s$-model formulated in terms 
of several Abelian vector multiplets with action \cite{KT-M11}
\bea
S = - 2 \int \rd^3x \,\rd^2 \q  \,\rd^2 \bar \q
\, {\bm E} \, F({\mathbb L}^i) ~, \qquad \bar \cD^2 {\mathbb L}^i =0~, \quad 
\bar{\mathbb L}^i ={\mathbb L}^i~,
\label{2.46}
\eea
where  $F(x^i)$ is a real analytic function of several variables, which is defined 
modulo linear inhomogeneous shifts
\bea
F(x) \to F(x) + b_i x^i + c~, 
\eea
with real parameters $b_i$ and $c$.
The real linear scalar ${\mathbb L}^i$ is the field strength of a vector multiplet.
Upon reduction to $\cN=1$ AdS superspace,  ${\mathbb L}^i$ 
generates two different $\cN=1$ superfields:
\bea
X^i := {\mathbb L}^i |~, \qquad
W_\a^i :=\ri \bm \nabla_\a^{\2} {\mathbb L}^i |~.
\eea
Here the real scalar $X^i$ is unconstrained, while the real spinor $W_\a^i$ obeys 
the constraint
\bea
\nabla^\a W_\a^i =0~,
\eea
which means that $W_\a^i$ is the field strength of an $\cN=1$ vector multiplet. 
Reducing the action \eqref{2.46} to $\cN=1$ AdS superspace gives
\bea
S = - \frac{\ri}{2}  \int \rd^{3|2} z\, E \, g_{ij}(X) \Big\{ 
\nabla^\a X^i \nabla_\a X^j 
+W^{\a i} W_\a^j \Big\}~,
\label{2.49}
\eea
where we have introduced the target-space metric
\bea
g_{ij} (X) = \frac{\pa^2  F(X) }{\pa X^i \pa X^j} ~.
\label{2.71}
\eea
The vector multiplets in \eqref{2.49} can be dualised into scalar ones, which gives
\bea
S_{\rm dual} = - \frac{\ri}{2}  \int \rd^{3|2} z\, E \, \Big\{ 
g_{ij} (X) 
\nabla^\a X^i \nabla_\a X^j 
+ g^{ij} (X) \nabla^\a Y_i \nabla_\a Y_j \Big\}~,
\eea
with $g^{ij} (X)$ being the inverse metric. 


\section{Massless higher-spin models: Type II series} \label{s72}

In accordance with section \ref{s57}, there exist two off-shell formulations for a massless multiplet
of half-integer superspin-$(s+\hf)$ 
in (2,0) AdS superspace, with $s=2, 3, \dots, $
which are called the type II and type III series. 
In this section we describe the $(2,0) \to (1,0)$ AdS superspace reduction 
of the type II theory. 
The reduction of the type III theory will be given in section \ref{s63}.

\subsection{Reduction of the gauge prepotentials to ${\rm AdS}^{3|2}$}

Let us turn to reducing the gauge prepotentials \eqref{21} 
to $\cN=1$ AdS superspace.\footnote{In the super-Poincar\'e case,
the $\cN=2 \to \cN=1$ reduction of ${\mathfrak H}_{\a(2s)}$ 
 has been carried out in \cite{KT}.} 
 Our first task is to work out such a reduction for 
 the superconformal gauge multiplet ${\mathfrak H}_{\a(2s)}$.
In the real representation \eqref{N1-deriv20}, the longitudinal linear constraint \eqref{gparam20} takes the form
\be 
{\bm \nabla}^{\underline{2}}_{(\a_1} g_{\a_2 \dots \a_{2s+1})} 
= \ri {\bm \nabla}^{\underline{1}} _{(\a_1}g_{\a_2 \dots \a_{2s+1})}~.
\ee
It follows that $g_{\a(2s)}$ has 
two independent $\theta_{\underline{2}}$-components, which are
\bea
g_{\a(2s)} |~, \qquad {\bm \nabla}^{\underline{2}\, \b}g_{\a(2s-1)\b}|~.
\eea
The gauge transformation of ${\mathfrak H}_{\a(2s)}$, eq. \eqref{H-gauge20}, allows us to choose two gauge conditions
\bea
{\mathfrak H}_{\a(2s)}| =0~, \qquad \
{\bm \nabla}^{\underline{2}\, \b}{\mathfrak H}_{\a(2s-1)\b}| =0~.
\label{q1}
\eea
In this gauge we stay with the following unconstrained real ${\cN=1}$ superfields:
\begin{subequations} \label{q2}
\bea
H_{\a(2s+1)} &:=& {\ri} {\bm \nabla}^{\underline{2}}_{(\a_1} {\mathfrak H}_{\a_2 \dots \a_{2s+1})} | ~, \label{H1} \\
H_{\a(2s)} &:=& \frac{\ri}{4}({\bm \nabla}^{\underline{2}})^2 {\mathfrak H}_{\a(2s)}|~. \label{H2}
\eea
\end{subequations}
There exists a residual gauge freedom which preserves the gauge conditions \eqref{q1}. It is described by unconstrained real ${\cN=1}$ superfields $\z_{\a(2s)}$ and $\z_{\a(2s-1)}$ defined by 
\begin{subequations}
\bea
g_{\a(2s)} | &=& -\frac{\ri}{2} \z_{\a(2s)}~,\,\,  \qquad \qquad {\bar \z}_{\a(2s)} = \z_{\a(2s)}~, \label{q3.a} \\
\bm \nabla^{\underline{2}\, \b} g_{\a(2s-1) \b} | &=& \frac{2s+1}{2s} \z_{\a(2s-1)} ~, \qquad {\bar \z}_{\a(2s-1)} = \z_{\a(2s-1)}~. \label{q3.b}
\eea
\end{subequations}
The gauge transformation laws of the superfields \eqref{q2} are given by
\begin{subequations} \label{q4.a-b}
\bea
\d H_{\a(2s+1)} &=& {\ri} \nabla_{(\a_1} \z_{\a_2 \dots \a_{2s+1})}~, \label{q4.a}\\
\d H_{\a(2s)} &=&  \nabla_{(\a_1}\z_{\a_2 \dots \a_{2s})} \label{q4.b}~.
\eea
\end{subequations}

Our next step is  to reduce 
the compensator ${\mathfrak L}_{\a(2s-2)}$ to ${\cN=1}$ AdS superspace. 
Making use of the representation \eqref{N1-deriv20}, we observe that the chirality condition \eqref{prep-gauge20} reads
\bea
{\bm \nabla}^{\underline{2}}_{\b} \x_{\a(2s-2)}= \ri {\bm \nabla}^{\underline{1}}_{\b} \x_{\a(2s-2)}~. \label{l1}
\eea
The gauge transformation \eqref{prep-gauge20} allows us to impose a gauge condition
\bea
{\mathfrak L}_{\a(2s-2)}| =0~. \label{l2}
\eea
Thus, upon reduction to ${\cal N}=1$ superspace, we have the following real superfields
\begin{subequations}
\bea
\J_{\b; \, \a(2s-2)} &:=& \ri {\bm \nabla}^{\underline{2}}_{\b} {\mathfrak L}_{\a(2s-2)}|~, \label{l3} \\
L_{\a(2s-2)} &:=& \frac{\ri}{4}({\bm \nabla}^{\underline{2}})^2 {\mathfrak L}_{\a(2s-2)}|~. \label{l4}
\eea
\end{subequations}
Here $\J_{\b; \, \a(2s-2)}$ is a reducible superfield which belongs
to the representation 
 $ {\bf 2}\otimes (\bf{2s- 1}) $ of ${\rm SL} (2,{\mathbb R})$,  
$\J_{\b; \, \a_1 \dots \a_{2s-2}} = \J_{\b; \, (\a_1 \dots \a_{2s-2})}$. 
The condition \eqref{l2} is preserved by the residual gauge freedom generated by a real unconstrained ${\cal N}=1$ superfield $\eta_{\a(2s-2)}$ defined by
\bea
\x_{\a(2s-2)}| = -\frac{\ri}{2} \eta_{\a(2s-2)}~, \qquad \bar \eta_{\a(2s-2)} = \eta_{\a(2s-2)}~. \label{l5}
\eea
We may now determine how the $\eta$-transformation acts on the superfields 
\eqref{l3} and \eqref{l4}. We obtain
\begin{subequations}
\bea
\d_{\eta} \J_{\b; \, \a(2s-2)} &=& \ri \nabla_{\b} \eta_{\a(2s-2)}~, \label{l6} \\
\d_{\eta} L_{\a(2s-2)} &=& 0~,
\eea
\end{subequations}
where we have used the chirality constraint \eqref{l1} and the expression \eqref{l5} for the residual gauge transformation. 

Next, we analyse the $\l$-gauge transformation and reduce the ${\cal N}=2$ field strength $\mathbb{L}_{\a(2s-2)}$ to ${\rm AdS}^{3|2}$. In the real basis for the covariant derivatives, the real linearity constraint \eqref{RLconst} is equivalent to two constraints:
\begin{subequations}
\bea
({\bm \nabla}^{\underline{2}})^2 \mathbb{L}_{\a(2s-2)} 
&=& ({\bm \nabla}^{\underline{1}})^2 \mathbb{L}_{\a(2s-2)}~, \\
{\bm \nabla}^{\underline{1}\, \b} {\bm \nabla}^{\underline{2}}_{\b} \mathbb{L}_{\a(2s-2)} &=& 0~. \label{BI}
\eea
\end{subequations}
These constraints imply that the resulting ${\cal N}=1$ components of $\mathbb{L}_{\a(2s-2)}$ are given by
\bea
\mathbb{L}_{\a(2s-2)}|~, \qquad \quad \ri 
{\bm \nabla}^{\underline{2}}_{\b} \mathbb{L}_{\a(2s-2)}|~,
\eea
of which the former is unconstrained and the latter is a constrained ${\cN=1}$ 
superfield that proves to be  a gauge-invariant field strength, as we shall see below.
The relation between $\mathbb{L}_{\a(2s-2)}$ and the prepotential ${\mathfrak L}_{\a(2s-2)}$ is given by \eqref{22}, which can be expressed as
\bea
\mathbb{L}_{\a(2s-2)} = -\frac{\ri}{2}\Big\{ ({\bm \nabla}^{\underline{1}})^2 
+ ( {\bm \nabla}^{\underline{2}})^2\Big\} {\mathfrak L}_{\a(2s-2)}~. 
\label{bbl1}
\eea
We now compute the bar-projection of \eqref{bbl1} in the gauge \eqref{l2} and make use of the definition \eqref{l4} to obtain
\bea
\mathbb{L}_{\a(2s-2)}| = -2 L_{\a(2s-2)}~.
\eea
Making use of \eqref{bbl1} and \eqref{l3}, the bar-projection of 
$\ri {\bm \nabla}^{\underline{2}}_{\b} \mathbb{L}_{\a(2s-2)}$ leads to the ${\cal N}=1$ field strength
\bea
 \cW_{\b;\, \a(2s-2)}&:=& \ri {\bm \nabla}^{\underline{2}}_{\b} \mathbb{L}_{\a(2s-2)}|
 =
  -\ri \Big( \nabla^{\g} \nabla_{\b} - 4\ri \cS \delta^{\g}_{\,\b}\Big){\Psi}_{\g; \,\a(2s-2)} ~.
 \label{eqW-20}
\eea
Here $\cW_{\b;\, \a(2s-2)}$ is a real superfield, $\cW_{\b;\, \a(2s-2)}= {\bar \cW}_{\b;\, \a(2s-2)}$, and is a descendant of the real unconstrained prepotential $\J_{\b; \, \a(2s-2)}$ defined modulo gauge transformation \eqref{l6}. The field strength proves to be gauge invariant under \eqref{l6}. It also obeys 
\bea
\nabla^{\b}\cW_{\b;\, \a(2s-2)}=0~,
\eea
as a consequence of \eqref{BI} and the identity \eqref{AA2}. 
Let us express the gauge transformation of $\mathbb{L}_{\a(2s-2)}$, eq.~\eqref{bbL-gauge20} in terms of the real basis for the covariant derivatives,
\bea
\d  \mathbb{L}_{\a(2s-2)} &=& \frac{\ri s}{2s+1} \Big\{ {\bm \nabla}^{\1 \b} {\bm \nabla}^{\underline{2}\g}\Big( g_{\b \g \a(2s-2)} + {\bar g}_{\b \g \a(2s-2)}\Big)
\non \\
&+& {\bm \nabla}^{\b \g}\Big( g_{\b \g \a(2s-2)} - {\bar g}_{\b \g \a(2s-2)}\Big) \Big\}~, 
\eea
In a similar way, one should also rewrite 
${\bm \nabla}^{\2}_{\b} \,\d \mathbb{L}_{\a(2s-2)}$ in the real basis.
This allows us to derive the gauge transformations 
 for $L_{\a(2s-2)}$ and $\cW_{\b;\, \a(2s-2)}$
\begin{subequations}
\bea
\d L_{\a(2s-2)} &=& -\frac{s}{2(2s+1)} \nabla^{\b \g} \z_{\b \g \a(2s-2)}~, \label{N1-Lgauge} \\
\d \cW_{\b; \, \a(2s-2)} &=& \ri \big( \nabla^{\g} \nabla_{\b} -4 \ri \cS \d^{\g}_{\,\b} \big) \z_{\g \a(2s-2)}~. \label{eqWgauge}
\eea
We can then read off the transformation law for the prepotential $\J_{\b; \, \a(2s-2)}$
\bea
\d \J_{\b; \,\a(2s-2)} &=& -\z_{\b \a(2s-2)}+ \ri \nabla_{\b} \eta_{\a(2s-2)}~, 
\label{N1-psigauge}
\eea
\end{subequations}
where we have also taken into account the $\eta$-gauge freedom \eqref{l6}.

Applying the ${\cal N}=1$ reduction rule \eqref{r-action} to the type II action \eqref{action20-t2} and using the commutation relation
\bea
{[}(\bm \de^{\1})^2 (\bm \de^{\1})^2 -4 \ri \cS (\bm \de^{\1})^2, \bm \nabla_{\a}^{\2} {]} &=& 16 \cS \nabla_{\a \b} \bm \nabla^{\2 \b} -16 \cS^2 \bm \nabla_{\a}^{\2} \non\\
&-&32 \cS^2 \bm \nabla^{\2 \b} M_{\a \b} -32 \ri \cS^2 \bm \nabla^{\1}_{\a} J~,
\eea 
we find that \eqref{action20-t2} becomes a sum of two actions,
\bea
S^{(\rm II)}_{(s+\hf)}[{\mathfrak H}_{\a(2s)} ,{\mathfrak L}_{\a(2s-2)} ] 
= S^{\parallel}_{(s+\hf)}[H_{\a(2s+1)} ,L_{\a(2s-2)} ] 
+ S^{\perp}_{(s)}[H_{\a(2s)} ,{\Psi}_{\b; \,\a(2s-2)} ]~.
\eea
Explicit expressions for these ${\cal N}=1$ actions will be given in the next subsection. 


\subsection{Massless higher-spin $\cN=1$ supermultiplets in AdS${}_3$} \label{ss32}

The gauge transformations 
 \eqref{q4.a}, \eqref{q4.b},
  \eqref{N1-Lgauge} and \eqref{N1-psigauge} tell us that in fact 
  we are dealing with two different $\cN=1$ supersymmetric 
  higher-spin gauge theories. 
  
   Given a positive integer $n>0$,
  we say that a supersymmetric gauge theory describes a 
  multiplet of superspin $n/2$ if it is formulated in terms of 
  a superconformal gauge prepotential $H_{\a(n)}$ and possibly a compensating 
  multiplet.  The gauge freedom of the real tensor superfield $H_{\a(n)}$ is 
  \bea 
 \d_\z H_{\a(n)}  = \ri^n (-1)^{ \left \lfloor{n/2}\right \rfloor }
 \nabla_{(\a_1} \z_{\a_2 \dots \a_n)}~,
 \label{3.299-20}
 \eea
 with the gauge parameter $\z_{\a(n-1)}$ being real but otherwise 
 unconstrained. 
  
  
\subsubsection{Longitudinal
formulation for massless superspin-$(s+\hf)$ multiplet} 
 
One of the two $\cN=1$ theories  provides an off-shell formulation 
for the massless superspin-$(s+\hf)$ multiplet. It is formulated in terms of the real unconstrained gauge superfields
\bea
\cV^{\parallel}_{(s+\hf )} = \Big\{H_{\a(2s+1)},\, L_{\a(2s-2)} \Big\} ~, 
\label{hf}
\eea
which are defined modulo gauge transformations 
\begin{subequations} \label{hf-gauge}
\bea
\d H_{\a(2s+1)} &=& \ri \nabla_{(\a_1} \z_{\a_2 \dots \a_{2s+1})}~,\\
\d L_{\a(2s-2)} &=& -\frac{s}{2(2s+1)} \nabla^{\b \g} \z_{\b \g \a(2s-2)}~,
\eea
\end{subequations}
where the parameter $\z_{\a(2s)}$ is unconstrained real.
The  gauge-invariant  action is
\bea
\lefteqn{S^{\parallel}_{(s+\hf)}[H_{\a(2s+1)} ,L_{\a(2s-2)} ]
= \Big(-\hf \Big)^{s} 
\int \rd^{3|2} z
\, E\, \bigg\{-\frac{\ri}{2} H^{\a(2s+1)} {\mathbb{Q}} H_{\a(2s+1)} }
\non \\
&&\qquad -\frac{\ri}{8} \nabla_{\b} H^{\b \a(2s)} \nabla^2 \nabla^{\g}H_{\g \a(2s)}+\frac{\ri s}{4}{\nabla}_{\b \g}H^{\b \g \a(2s-1)} {\nabla}^{\rho \d}H_{\rho \d \a(2s-1)}
\non \\
&& \qquad + (2s-1) L^{\a(2s-2)}\nabla^{\b \g} \nabla^{\d} H_{\b \g \d \a(2s-2)}
\non \\
&& \qquad + 2 (2s-1)\Big( L^{\a(2s-2)} (\ri \nabla^2 - 4 \cS) L_{\a(2s-2)}
- \frac{\ri}{s}(s-1) \nabla_{\b} L^{\b \a(2s-3)} \nabla^{\g}L_{\g \a(2s-3)}\Big)
\non \\
&& \qquad + \cS \Big(s \,\de_{\b}H^{\b \a(2s)} \de^{\g} H_{\g \a(2s)}+ \hf (2s+1)H^{\a(2s+1)}( \nabla^2-4 \ri \cS)H_{\a(2s+1)} \Big)
\bigg\}~,~~~
\label{action-t2-half}
\eea
where ${\mathbb{Q}}$ is the quadratic Casimir operator of the 3D ${\cal N}=1$ AdS supergroup, see eq.~\eqref{casimir}.
The action \eqref{action-t2-half} coincides with the off-shell ${\cal N}=1$ supersymmetric action for massless half-integer superspin in ${\rm AdS}^{3|2}$ in the form given in \cite{KP1}.
Its flat-superspace limit was presented earlier in \cite{KT}.
In what follows, we will refer to the above theory as the longitudinal
formulation for the massless superspin-$(s+\hf)$ multiplet. 

The structure $\nabla_{\b} L^{\b \a(2s-3)} \nabla^{\g}L_{\g \a(2s-3)}$ in 
\eqref{action-t2-half} is not defined for $s=1$. However it comes with the factor 
$(s-1)$ and drops out from \eqref{action-t2-half} for $s=1$. The resulting action 
\bea
S^{\parallel}_{(\frac 32)} [H_{\a(3)} ,L ]
&=& -\hf \int \rd^{3|2} z
\, E\, \bigg\{-\frac{\ri}{2} H^{\a(3)} {\mathbb{Q}} H_{\a(3)} 
-\frac{\ri}{8} \nabla_{\b} H^{\b \a(2)} \nabla^2 \nabla^{\g}H_{\g \a(2)}
\non \\
&&
+\frac{\ri }{4}{\nabla}_{\b \g}H^{\b \g \a} {\nabla}^{\rho \d}H_{\rho \d \a}
+  L \nabla^{\b \g} \nabla^{\d} H_{\b \g \d }
 + 2  L \big( \ri \nabla^2  -4\cS\big) L
 \\
&& 
+ \cS \Big( \de_{\b}H^{\b \a(2)} \de^{\g} H_{\g \a(2)}
+ \frac 32  H^{\a(3)}\big( \nabla^2 -4\ri \cS\big) H_{\a(3)} 
 \Big) \bigg\}
\non
\eea
is the linearised action for $\cN=1$ AdS supergravity. 
In the flat superspace limit, the action is equivalent to the one given in \cite{GGRS}.

  
\subsubsection{Transverse formulation for massless superspin-$s$ multiplet}

The other $\cN=1$ theory provides a formulation 
  for the massless superspin-$s$ multiplet. 
 It is described by the unconstrained real  superfields 
\bea
\cV^{\perp}_{(s)} = \Big\{H_{\a(2s)}, \J_{\b; \, \a(2s-2)} \Big\} ~,
\label{t2new}
\eea
which are defined modulo gauge transformations of the form
\begin{subequations} \label{t2new-gauge}
\bea
\d H_{\a(2s)} &=& \nabla_{(\a_1} \z_{\a_2 \dots \a_{2s})}~,\\
\d \J_{\b; \,\a(2s-2)} &=& -\z_{\b \a(2s-2)}+ \ri \nabla_{\b} \eta_{\a(2s-2)}~,
\eea
\end{subequations}
where the gauge parameters $\z_{\a(2s-1)}$ and $\eta_{\a(2s-2)}$ 
are unconstrained real.
The gauge-invariant action is given by
\begin{subequations} \label{action-t2-new}
\bea
\lefteqn{S^{\perp}_{(s)}[H_{\a(2s)} ,{\Psi}_{\b; \,\a(2s-2)} ]
= \Big(-\hf \Big)^{s} 
\int \rd^{3|2}z\,
 E\, \bigg\{\frac{1}{2} H^{\a(2s)} (\ri \nabla^2 +8 s \cS) H_{\a(2s)}}
\non \\
&& \qquad \qquad - \ri s \nabla_{\b} H^{\b \a(2s-1)} \nabla^{\g}H_{\g \a(2s-1)} 
-(2s-1) \cW^{\b ;\,\a(2s-2)} \nabla^{\g} H_{\g \b \a(2s-2)}
\non \\
&& \qquad \qquad -\frac{\ri}{2} (2s-1)\Big(\cW^{\b ;\, \a(2s-2)} \cW_{\b ;\, \a(2s-2)}+\frac{s-1}{s} \cW_{\b;}\,^{\b \a(2s-3)} \cW^{\g ;}\,_{\g \a(2s-3)} \Big) 
\non\\
&& \qquad \qquad
 -2 \ri (2s-1) \cS \Psi^{\b ;\, \a(2s-2)} \cW_{\b ; \, \a(2s-2)}
\bigg\}~,
\label{action-t2-new-a}
\eea
where $\cW_{\b; \, \a(2s-2)}$ denotes the field strength
\bea
 \cW_{\b;\, \a(2s-2)} =
  -\ri \Big( \nabla^{\g} \nabla_{\b} - 4\ri \cS \delta^{\g}_{\,\b}\Big){\Psi}_{\g; \,\a(2s-2)} ~,
  \qquad \nabla^\b  \cW_{\b;\, \a(2s-2)}=0~.
\eea
\end{subequations}
The action \eqref{action-t2-new} defines a new ${\cal N}=1$ supersymmetric higher-spin theory which was not present in \cite{KP1,HKO, KT}
even in the super-Poincar\'e case.

The structure $\cW_{\b;}\,^{\b \a(2s-3)} \cW^{\g ;}\,_{\g \a(2s-3)} $
in \eqref{action-t2-new-a} is not defined for $s=1$. 
However it comes with the factor $(s-1)$ and drops out from 
\eqref{action-t2-new-a} for $s=1$.
The resulting gauge-invariant action 
\bea
S^{\perp}_{(1)}[H_{\a(2)} ,{\Psi}_{\b} ]
&=& -\hf \int \rd^{3|2}z\, E\, \bigg\{\frac{1}{2} H^{\a(2)} (\ri \nabla^2 + 8 \cS) H_{\a(2)}
- \ri  \nabla_{\b} H^{\b \a } \nabla^{\g}H_{\g \a} 
\non \\
&& 
\qquad - \cW^{\b} \nabla^{\g} H_{\g \b } -\frac{\ri}{2} \cW^{\b } \cW_{\b}
 -2 \ri  \cS \Psi^{\b } \cW_{\b }
\bigg\}
\label{337}
\eea
provides an off-shell realisation for a massless gravitino multiplet in AdS${}_3$.
In the flat-superspace limit, this model reduces to the one described in 
\cite{KT}.

In the $s>1$ case, the gauge freedom of the prepotential $\J_{\b ;\, \a(2s-2)}$ \eqref{t2new-gauge} allows us to impose a gauge condition 
\bea
\J_{(\a_1;\, \a_2 \dots \a_{2s-1})} =0 \quad \Longleftrightarrow \quad 
\J_{\b ;\, \a(2s-2)} =  \sum_{k=1}^{2s-2}\ve_{\b \a_k} \vf_{\a_1 \dots \hat{\a}_k \dots \a_{2s-2}}~,
\label{3.355}
\eea
for some field  $\vf_{\a(2s-3)}$.
Since we gauge away the symmetric part of $\J_{\b ;\, \a(2s-2)}$, the two gauge parameters $\z_{\a(2s-1)}$ and $\eta_{\a(2s-2)}$ are related. The theory is now realised in terms of the following dynamical variables
\bea
\Big\{H_{\a(2s)},~\vf_{\a(2s-3)}\Big\}~,
\eea
with the gauge freedom
\begin{subequations}
\bea
\d H_{\a(2s)} &=& -\nabla_{(\a_1 \a_2} \eta_{\a_3 \dots \a_{2s})}~,\\
\d \vf_{\a(2s-3)}&=& \ri \nabla^{\b} \eta_{\b \a(2s-3)}~.
\eea
\end{subequations}
It follows that in the flat superspace limit, $\cS=0$, 
and in the gauge \eqref{3.355},
the action \eqref{action-t2-new} reduces to \eqref{type1-action}. The component structure of this model will be discussed in appendix \ref{AppendixBB-1}.


\section{Massless higher-spin models: Type III series} \label{s63}

In this section we carry out  the ${\cN=1}$ AdS superspace reduction of the type III theory \eqref{action2-t3} following the procedure employed in section  \ref{s72}. 

\subsection{Reduction of the gauge prepotentials to ${\rm AdS}^{3|2}$}

The reduction of the superconformal gauge multiplet ${\mathfrak H}_{\a(2s)}$ to ${\rm AdS}^{3|2}$ has been carried out in the previous section. We saw that in the gauge \eqref{q1}, ${\mathfrak H}_{\a(2s)}$ is described by the two unconstrained real superfields $H_{\a(2s+1)}$ and $H_{\a(2s)}$ defined according to \eqref{q2}, with their gauge transformation laws given by eqs. \eqref{q4.a} and \eqref{q4.b}, respectively. Now it remains to reduce the prepotential ${\mathfrak V}_{\a(2s-2)}$ to ${\cN=1}$ AdS superspace, following the same approach as outlined in the type II series. 
The gauge transformation \eqref{prep-gauge2} allows us to choose a gauge condition
\bea
{\mathfrak V}_{\a(2s-2)}| =0~. \label{v2}
\eea
The compensator ${\mathfrak V}_{\a(2s-2)}$ is then equivalent to the following real ${\cal N}=1$ superfields, which we define as follows:
\begin{subequations}
\bea
\U_{\b; \, \a(2s-2)} &:=& \ri \bm \nabla^{\underline{2}}_{\b} {\mathfrak V}_{\a(2s-2)}|~, \label{v3} \\
V_{\a(2s-2)} &:=& \frac{\ri}{4}(\bm \nabla^{\underline{2}})^2 {\mathfrak V}_{\a(2s-2)}|~. \label{v4}
\eea
\end{subequations}
The residual gauge freedom, which preserves the gauge condition \eqref{v2} is described by a real unconstrained ${\cal N}=1$ superfield $\eta_{\a(2s-2)}$ defined by
\bea
\x_{\a(2s-2)}| = -\frac{\ri}{2} \eta_{\a(2s-2)}~, \qquad \bar \eta_{\a(2s-2)} = \eta_{\a(2s-2)}~. \label{v5}
\eea
As a result, we may determine how \eqref{v3} and \eqref{v4} vary under $\eta$-transformation
\begin{subequations}
\bea
\d_{\eta} \U_{\b; \, \a(2s-2)} &=& \ri \nabla_{\b} \eta_{\a(2s-2)}~, \label{v6}\\
\d_{\eta} V_{\a(2s-2)} &=&0~.
\eea
\end{subequations}

Next, we analyse the $\l$-gauge transformation and reduce the field strength $\mathbb{V}_{\a(2s-2)}$ to ${\rm AdS}^{3|2}$. 
In the real basis for the covariant derivatives, the real linearity constraint \eqref{RLV} turns into:
\begin{subequations}
\bea
(\bm \nabla^{\underline{2}})^2 \mathbb{V}_{\a(2s-2)} &=& (\bm \nabla^{\underline{1}})^2 \mathbb{V}_{\a(2s-2)}~, \\
\bm \nabla^{\underline{1}\, \b} \bm \nabla^{\underline{2}}_{\b} \mathbb{V}_{\a(2s-2)} &=& 0~.
\eea
\end{subequations}
This tells us that $\mathbb{V}_{\a(2s-2)}$ is equivalent to two real ${\cN=1}$ superfields
\bea
\mathbb{V}_{\a(2s-2)}|~, \qquad \quad \ri \bm \nabla^{\underline{2}}_{\b} \mathbb{V}_{\a(2s-2)}|~.
\eea
The relation between the field strength $\mathbb{V}_{\a(2s-2)}$ and the prepotential ${\mathfrak V}_{\a(2s-2)}$ is given by \eqref{5713}, which can be expressed as
\bea
\mathbb{V}_{\a(2s-2)} = -\frac{\ri}{2}\Big\{ (\bm \nabla^{\underline{1}})^2 + (\bm \nabla^{\underline{2}})^2\Big\} {\mathfrak V}_{\a(2s-2)}~. 
\label{bbv1}
\eea
We now compute the bar-projection of \eqref{bbv1} in the gauge \eqref{v2} and make use of the definition \eqref{v4} to obtain
\bea
\mathbb{V}_{\a(2s-2)}| = -2 V_{\a(2s-2)}~.
\eea
The bar-projection of $\ri \bm \nabla^{\underline{2}}_{\b} \mathbb{V}_{\a(2s-2)}$ leads to the ${\cal N}=1$ field-strength
\bea
 \O_{\b;\, \a(2s-2)}&:=& \ri \bm \nabla^{\underline{2}}_{\b} \mathbb{V}_{\a(2s-2)}|
 \non \\
 \qquad &=& -\ri \Big( \nabla^{\g} \nabla_{\b} - 4\ri \cS \delta_{\b}\,^{\g}\Big){\U}_{\g; \,\a(2s-2)} ~,
 \label{eqO}
\eea
which is a real superfield, $\O_{\b;\, \a(2s-2)}= \bar \O_{\b;\, \a(2s-2)}$, and is a descendant of the real unconstrained prepotential $\U_{\b; \, \a(2s-2)}$ defined modulo gauge transformation \eqref{v6}. One may check that the field strength is invariant under \eqref{v6} and obeys the condition 
\bea
\nabla^{\b}\O_{\b;\, \a(2s-2)}=0~.
\eea
Let us express the gauge transformation of $\mathbb{V}_{\a(2s-2)}$, eq.~\eqref{bbV-gauge20} in terms of the real basis for the covariant derivatives. This leads to 
\bea
\d  \mathbb{V}_{\a(2s-2)} &=& -\frac{1}{2s+1} \Big\{ \bm \nabla^{\1 \b} \bm \nabla^{\underline{2}\g}\Big( g_{\b \g \a(2s-2)} - {\bar g}_{\b \g \a(2s-2)}\Big)
\non \\
&+& \bm \nabla^{\b \g}\Big( g_{\b \g \a(2s-2)} + {\bar g}_{\b \g \a(2s-2)}\Big) \Big\}~. 
\eea
One should also express its corollary $\bm \nabla^{\underline{2}}_{\b} \d \mathbb{V}_{\a(2s-2)}$ in the real basis for the covariant derivatives. We determine the gauge transformations law for $V_{\a(2s-2)}$ and $\O_{\b;\, \a(2s-2)}$ to be
\begin{subequations}
\bea
\d V_{\a(2s-2)} &=& \frac{1}{2s} \nabla^{\b} \z_{\b \a(2s-2)}~, \label{N1-Vgauge}\\
\d \O_{\b; \, \a(2s-2)} &=& \frac{1}{2s+1}\big( \nabla^{\g} \nabla_{\b} \nabla^{\d} -4 \ri \cS \nabla^{\d} \d_{\b} \,^{\g} \big) \z_{\d \g \a(2s-2)}~.\label{N1-Ogauge}
\eea
\end{subequations}
From \eqref{N1-Ogauge} we read off the transformation law for the prepotential $\U_{\b;\, \a(2s-2)}$:
\bea
\d \U_{\b; \,\a(2s-2)} &=& \frac{\ri}{2s+1}\Big(\nabla^{\g}\z_{\g \b \a(2s-2)}+ (2s+1) \nabla_{\b} \eta_{\a(2s-2)}\Big)~, \label{N1-Ugauge}
\eea
where we have also taken into account the $\eta$-gauge freedom \eqref{v6}.

Performing ${\cal N}=1$ reduction to the original type III action \eqref{action2-t3}, we arrive at two decoupled ${\cal N}=1$ actions 
\bea
S^{(\rm III)}_{(s+\hf)}[{\mathfrak H}_{\a(2s)} ,{\mathfrak V}_{\a(2s-2)} ] 
= S^{\perp}_{(s+\hf)}[H_{\a(2s+1)} ,{\U}_{\b; \,\a(2s-2)} ] 
+ S^{\parallel}_{(s)}[H_{\a(2s)} ,{V}_{\a(2s-2)} ]~.~~~~
\eea
We will present the exact form of these actions in the next subsection. 


\subsection{Massless higher-spin ${\cal N}=1$ supermultiplets in AdS${}_3$} \label{ss42}

Upon reduction to ${\cal N}=1$ superspace, the type III theory leads to 
two ${\cal N}=1$ supersymmetric gauge theories. 


\subsubsection{Longitudinal formulation for massless superspin-$s$ multiplet}

One of the two $\cN=1$ theories provides an off-shell realisation for 
massless superspin-$s$ multiplet described in terms of the real unconstrained superfields
\bea
\cV^{\parallel}_{(s)}= \Big\{H_{\a(2s)}, V_{\a(2s-2)} \Big\} ~,
\label{int}
\eea
which are defined modulo gauge transformations of the form
\begin{subequations} \label{int-gauge}
\bea
\d H_{\a(2s)} &=& \nabla_{(\a_1} \z_{\a_2 \dots \a_{2s})}~, \\
\d V_{\a(2s-2)} &=& \frac{1}{2s} \nabla^{\b} \z_{\b \a(2s-2)}~,
\eea
\end{subequations}
where the gauge parameter $\z_{\a(2s-1)}$ is unconstrained real.
The gauge-invariant action is given by
\bea\label{action-t3}
\lefteqn{S^{\parallel}_{(s)}[H_{\a(2s)} ,V_{\a(2s-2)} ]
= \Big(-\hf \Big)^{s} 
\int 
\rd^{3|2}z
\, E \, \bigg\{
\frac{1}{2} H^{\a(2s)} \big(\ri \de^2  +4 \cS\big)H_{\a(2s)} }
\non \\
&& \qquad 
- \frac{\ri}{2}\de_{\b}H^{\b \a(2s-1)} \de^{\g}H_{\g \a(2s-1)}
-(2s-1) V^{\a(2s-2)} \nabla^{\b \g} H_{\b \g \a(2s-2)}
 \\
&& \qquad +(2s-1)\Big(\hf V^{\a(2s-2)} \big( \ri \nabla^2 +8s\cS\big)V_{\a(2s-2)}
+\ri (s-1) \nabla_{\b}V^{\b \a(2s-3)} \nabla^{\g}V_{\g \a(2s-3)} \Big) \bigg\}
~.
\non
\eea
Modulo an overall normalisation factor, \eqref{action-t3} coincides with the off-shell ${\cal N}=1$ supersymmetric action for massless superspin-$s$ multiplet in the form given in \cite{KP1}. In the flat superspace limit it reduces to the action derived in 
\cite{KT}.

Although the structure $\nabla_{\b}V^{\b \a(2s-3)} \nabla^{\g}V_{\g \a(2s-3)} $ 
in \eqref{action-t3} is not defined for $s=1$, 
it comes with the factor $(s-1)$  and thus drops out from \eqref{action-t3}
for $s=1$. The resulting gauge-invariant action 
\bea
S^{\parallel}_{(1)}[H_{\a(2)} ,V ]
&=& -\hf 
\int 
\rd^{3|2}z
\, E \, \bigg\{
\frac{1}{2} H^{\a(2)} \big( \ri \de^2 +4\cS\big) H_{\a(2)} 
- \frac{\ri}{2}\de_{\b}H^{\b \a} \de^{\g}H_{\g \a}
\non \\
&& \qquad  
- V \nabla^{\b \g} H_{\b \g}
+\frac{1}{2} V \big( \ri \nabla^2  +8\cS\big) V
\bigg\}
\label{425}
\eea
describes an off-shell massless gravitino multiplet in AdS${}_3$. 
In the flat superspace limit, it reduces to the gravitino multiplet model described in \cite{Siegel} (see also \cite{KT}).


\subsubsection{Transverse formulation for massless superspin-$(s+\hf)$ multiplet}

The other theory provides an off-shell formulation for massless  
superspin-$(s+\hf) $ multiplet. It is described by  the unconstrained superfields
\bea
\cV^{\perp}_{(s+\hf )}= \Big\{H_{\a(2s+1)}, \U_{\b; \, \a(2s-2)} \Big\} ~, 
\label{t3new}
\eea
which are defined modulo gauge transformations of the form 
\begin{subequations} \label{t3new-gauge}
\bea
\d H_{\a(2s+1)} &=& \ri \nabla_{(\a_1} \z_{\a_2 \dots \a_{2s+1})}~,\\
\d \U_{\b; \,\a(2s-2)} &=& \frac{\ri}{2s+1}\big( \nabla^{\g} \z_{\g \b \a(2s-2)}+ (2s+1) \nabla_{\b} \eta_{\a(2s-2)}\big)~.
\eea
\end{subequations}

The gauge-invariant  action is 
\begin{subequations}
\label{action-t3-new-complete}
\bea
&&S^{\perp}_{(s+\hf)}[{H}_{\a(2s+1)} ,\U_{\b; \,\a(2s-2)} ]
= \Big(-\hf \Big)^{s} 
\int \rd^{3|2}z\, E \,\bigg\{-\frac{\ri}{2} H^{\a(2s+1)} {\mathbb{Q}} H_{\a(2s+1)}
\non \\
&& \qquad \qquad -\frac{\ri}{8} \nabla_{\b} H^{\b \a(2s)} \nabla^2 \nabla^{\g}H_{\g \a(2s)}+\frac{\ri}{8}{\nabla}_{\b \g}H^{\b \g \a(2s-1)} {\nabla}^{\rho \d}H_{\rho \d \a(2s-1)}
\non \\
&& \qquad \qquad 
-\frac{\ri}{4}(2s-1) \O^{\b; \,\a(2s-2)} \nabla^{\g \d}H_{\g \d \b \a(2s-2)}
\non\\
&& \qquad \qquad -\frac{\ri}{8}(2s-1)\Big(\O^{\b ;\, \a(2s-2)} \O_{\b ;\, \a(2s-2)}
-2(s-1)\O_{\b;}\,^{\b \a(2s-3)} \O^{\g ;}\,_{\g \a(2s-3)}  \Big) 
\non \\
&& \qquad \qquad 
+ \cS \Big( H^{\a(2s+1)} \big( \nabla^2 - 4\ri \cS\big) H_{\a(2s+1)} 
+ \hf \ \de_{\b}H^{\b \a(2s)} \de^{\g}H_{\g \a(2s)}
\Big)
\non \\
&& \qquad \qquad + \ri s (2s-1) \cS \,\U^{\b ;\, \a(2s-2)} \O_{\b ; \, \a(2s-2)}
\bigg\}~,
\label{action-t3-new}
\eea
where $\O_{\b; \a(2s-2)}$ denotes the real  field strength
\bea
 \O_{\b;\, \a(2s-2)}
  = -\ri \Big( \nabla^{\g} \nabla_{\b} - 4\ri \cS \delta_{\b}\,^{\g}\Big){\U}_{\g; \,\a(2s-2)} ~,
 \qquad \nabla^\b  \O_{\b;\, \a(2s-2)}=0~.
\eea
\end{subequations}
This action defines a new ${\cal N}=1$ supersymmetric higher-spin theory which did not appear in \cite{HKO,KT,KP1}.

The structure $\O_{\b;}\,^{\b \a(2s-3)} \O^{\g ;}\,_{\g \a(2s-3)} $
in \eqref{action-t3-new} is not defined for $s=1$. However it comes 
with the factor $(s-1)$ and hence drops out from \eqref{action-t3-new} 
for $s=1$. The resulting gauge-invariant action 
\bea
S^{\perp}_{(\frac 32)}[{H}_{\a(3)} ,\U_{\b} ]
&=& -\hf \int \rd^{3|2}z
\, E \,\bigg\{-\frac{\ri}{2} H^{\a(3)} {\mathbb{Q}} H_{\a(3)} 
  -\frac{\ri}{8} \nabla_{\b} H^{\b \a(2)} \nabla^2 \nabla^{\g}H_{\g \a(2)}
\non \\
&&
 +\frac{\ri}{8}{\nabla}_{\b \g}H^{\b \g \a} {\nabla}^{\rho \d}H_{\rho \d \a}
-\frac{\ri}{4} \O^{\b} \nabla^{\g \d}H_{\g \d \b }
\non \\
&&
+ \cS \Big( H^{\a(3)} \big( \nabla^2 -4\ri \cS\big) H_{\a(3)} 
+ \hf  \de_{\b}H^{\b \a(2)} \de^{\g}H_{\g \a(2)}
\Big)
\non \\
&& 
 -\frac{\ri}{8} \O^{\b  } \O_{\b }
+ \ri  \cS \,\U^{\b } \O_{\b }\bigg\}
\eea
provides an off-shell formulation for a linearised supergravity 
multiplet in AdS${}_3$. In the flat superspace limit, it reduces to the linearised
supergravity model proposed in \cite{KT}.


\section{Analysis of the results} \label{s64}

Let $s>0$ be a positive integer. For each superspin value, 
integer $(s)$ or half-integer $(s+\hf)$, we have constructed two off-shell 
formulations which have been called longitudinal and transverse. 
Now we have to explain this terminology.

Consider a field theory in ${\cN}=1$ AdS superspace that is described in terms of a real tensor 
superfield $V_{\a(n)}$.  We assume the action to have the form
\bea
S^{\parallel} [ V_{\a(n)} ]= \int \rd^{3|2}z \, E\, \cL \big(\ri^{n+1}  \nabla_\b V_{\a(n)} \big)~.
\label{long5.1}
\eea 
It is natural to call $\nabla_\b V_{\a(n)} $ a longitudinal superfield, 
by analogy with a longitudinal vector field.
This theory possesses a dual formulation that is obtained by introducing 
a first-order action 
\bea
S_{\text{first-order}} = \int \rd^{3|2}z \, E\, \Big\{ \cL \big( \S_{\b; \, \a(n)} \big)
+ \ri^{n+1} \cW^{\b; \, \a(n) }  \S_{\b; \,\a(n)} \Big\}~,
\label{first-order}
\eea
where $\S_{\b;\a(n)} $ is unconstrained and the Lagrange multiplier is 
\bea
 \cW_{\b; \,\a(n)} =
  \ri^{n+1} \Big( \nabla^{\g} \nabla_{\b} - 4\ri \cS \delta^{\g}_{\,\b}\Big){\Psi}_{\g; \,\a(n)} ~,
  \qquad \nabla^\b \cW_{\b; \,\a(n)} =0~,
  \label{W5.3}
 \eea
for some unconstrained prepotential ${\Psi}_{\g; \,\a(n)} $. 
Varying \eqref{first-order} with respect to ${\Psi}_{\g; \,\a(n)} $ gives
\bea
\nabla^\b \nabla_\g \S_{\b; \,\a(n)} - 4\ri \cS \S_{\g; \,\a(n)} =0 
\quad \implies \quad \S_{\b; \,\a(n)} = \ri^{n+1} \nabla_\b V_{\a(n)}~,
\eea
and then $S_{\text{first-order}} $ reduces to the original action \eqref{long5.1}.
On the other hand, we may start from $S_{\text{first-order}}$ 
and integrate $\S_{\b;\a(n)} $ out.  This will lead to a dual action of the form 
\bea
S^{\perp} [ {\Psi}_{\g; \,\a(n)} ]= \int \rd^{3|2}z \, E\,  
\cL_{\rm dual} \big(  \cW_{\b; \, \a(n) } \big)~.
\label{tran5.5}
\eea
This is a gauge theory since the action is invariant under 
gauge transformations
\bea
\d {\Psi}_{\g; \,\a(n)} = \ri^{n+1} \nabla_\g \eta_{\a(n)} ~.
\eea
The gauge-invariant field strength $\cW_{\b; \, \a(n) } $ can be called a transverse superfield, due to the constraint \eqref{W5.3} it obeys. 
It is natural to call the dual formulations \eqref{long5.1} and \eqref{tran5.5} 
as longitudinal and transverse, respectively.

Now,  let us consider the transverse and longitudinal  formulations 
for the massless superspin-$s$ models, which 
are given by eqs. \eqref{action-t2-new} and \eqref{action-t3}, respectively. 
These actions depend parametrically on $\cS$, the curvature of AdS superspace.
We denote by $S^{\perp}_{(s)}[H_{\a(2s)} ,{\Psi}_{\b; \,\a(2s-2)}]_{\rm FS}$
and $S^{\parallel}_{(s)}[H_{\a(2s)} ,V_{\a(2s-2)} ]_{\rm FS}$
these actions in the limit $\cS=0$, which corresponds to a flat-superspace.
The dynamical systems  $S^{\perp}_{(s)}[H_{\a(2s)} ,{\Psi}_{\b; \,\a(2s-2)}]_{\rm FS}$
and $S^{\parallel}_{(s)}[H_{\a(2s)} ,V_{\a(2s-2)} ]_{\rm FS}$ prove to 
be related to each other by  the Legendre transformation described above.
Thus $S^{\perp}_{(s)}[H_{\a(2s)} ,{\Psi}_{\b; \,\a(2s-2)}]_{\rm FS}$
and $S^{\parallel}_{(s)}[H_{\a(2s)} ,V_{\a(2s-2)} ]_{\rm FS}$ are dual formulations of the same theory. This duality does not survive if $\cS$ is non-vanishing.

The same feature characterises the longitudinal and transverse  formulations 
for the massless superspin-$(s+\hf)$ multiplet, which 
are described by the actions \eqref{action-t2-half} and \eqref{action-t3-new-complete}, respectively. The flat-superspace counterparts of these higher-spin models, which we denote by
$S^{\parallel}_{(s+\hf)}[H_{\a(2s+1)} ,L_{\a(2s-2)} ]_{\rm FS}$ and 
$S^{\perp}_{(s+\hf)}[{H}_{\a(2s+1)} ,\U_{\b; \,\a(2s-2)} ]_{\rm FS}$,
are dual to each other. However, this duality does not survive 
if we turn on a non-vanishing AdS curvature.

\section{Non-conformal higher-spin supercurrents} \label{s65}
In the previous sections, we have shown that there exist
two different off-shell formulations for the massless higher-spin ${\cal N}=1$ supermultiplets. Massless half-integer superspin theory can be realised in terms of the dynamical variables \eqref{hf} and \eqref{t3new}, while the models \eqref{t2new} and \eqref{int} define massless multiplet of integer superspin $s$, with $s >1$. These models lead to different ${\cal N}=1$ higher-spin supercurrent multiplets. Our aim in this section is to describe the general structure of ${\cal N}=1$ supercurrent multiplets in AdS. 


\subsection{${\cal N}=1$ supercurrents: Half-integer superspin case}

Our half-integer supermultiplet in the longitudinal formulation \eqref{hf} can be coupled to external sources
\bea
S^{(s+\hf)}_{\rm source}= \int \rd^{3|2}z \, E\, \Big\{ 
{\ri} {H}^{ \a (2s+1)} J_{ \a (2s+1)}
+ 4{L}^{ \a (2s-2)} {S}_{ \a (2s-2)}
 \Big\}~.
\eea
The condition that the above action is invariant under the gauge transformations \eqref{hf-gauge} gives the conservation equation
\bea
\de^{\b}J_{\b \a(2s)} = -\frac{2s}{(2s+1)} \de_{(\a_1 \a_2} S_{\a_3 \cdots \a_{2s})}~.
\label{ce-hf11}
\eea

For the transverse theory \eqref{t3new} described by the prepotentials $\{ H_{\a(2s+1)},\U_{\b;\, \a(2s-2)}\}$, we construct an action functional of the form
\bea
S^{(s+\hf)}_{\rm source}= \int \rd^{3|2} z
\, E\, \Big\{ 
{\ri} {H}^{ \a (2s+1)} J_{ \a (2s+1)}
+ 2 \ri s\, {\U}^{\b; \, \a (2s-2)} {U}_{ \b; \, \a (2s-2)}
 \Big\}~.
\eea
Requiring that the action is invariant under the gauge transformations \eqref{t3new-gauge} leads to
\bea \label{ce-hf12}
\de^{\b}J_{\b \a(2s)} = \frac{2s}{2s+1} \de_{(\a_1} U_{\a_2 \cdots \a_{2s})}~, \quad 
\de^{\b} U_{\b; \, \a(2s-2)} = 0~. 
\eea
From the above consideration, it follows that the most general conservation equation in the half-integer superspin case takes the form
\bsubeq \label{ce-hf13}
\bea
\de^{\b}J_{\b \a(2s)} &=& \frac{2s}{2s+1} \bigg(\de_{(\a_1} U_{\a_2 \cdots \a_{2s})} 
- \de_{(\a_1 \a_2} S_{\a_3 \cdots \a_{2s})} \bigg)~, \label{ce-hf13a} \\
\de^{\b} U_{\b; \, \a(2s-2)} &=& 0~.\label{ce-hf13b}
\eea
\esubeq


\subsection{${\cal N}=1$ supercurrents: Integer superspin case}
In complete analogy with the half-integer superspin case, we couple the prepotentials \eqref{int} in terms of which the integer superspin-$s$ is described, to external sources 
\bea
S^{(s)}_{\rm source}= \int \rd^{3|2}z \, E\, \Big\{ 
 {H}^{ \a (2s)} J_{ \a (2s)}
+ 2s \,{V}^{ \a (2s-2)} {R}_{ \a (2s-2)}
 \Big\}~.
\eea
For such an action to be invariant under the gauge freedom \eqref{int-gauge}, the sources must be conserved 
\bea
\de^{\b}J_{\b \a(2s-1)} = \de_{(\a_1} R_{\a_2 \cdots \a_{2s-1})}~.
\label{ce-int11}
\eea

Next, we turn to the transverse formulation \eqref{t2new} characterised by the prepotentials $\{ H_{\a(2s)},\J_{\b;\, \a(2s-2)}\}$ and construct an action functional
\bea
S^{(s)}_{\rm source}= \int \rd^{3|2}z
\, E\, \Big\{ 
{H}^{ \a (2s)} J_{ \a (2s)}
+ \ri {\J}^{\b; \, \a (2s-2)} {T}_{ \b; \, \a (2s-2)}
 \Big\}~.
\eea
Demanding that the action be invariant under the gauge transformations \eqref{t2new-gauge}, we derive the following conditions
\bea
\de^{\b}J_{\b \a(2s-1)} = \ri\, T_{\a(2s-1)}~, \qquad \de^{\b} T_{\b; \, \a(2s-2)} = 0~. \label{ce-int12}
\eea
From the above consideration, the most general conservation equation for the multiplet of currents in the integer superspin case is given by
\bsubeq \label{ce-int13}
\bea
\de^{\b}J_{\b \a(2s-1)} &=&  \de_{(\a_1} R_{\a_2 \cdots \a_{2s-1})}+ 
\ri T_{\a(2s-1)}~,\\
\de^{\b} T_{\b; \, \a(2s-2)} &=& 0~. 
\eea
\esubeq


\subsection{From ${\cal N}=2$ supercurrents to $\cN=1$ supercurrents}

In the previous chapter (see section \ref{s62}), we formulated the general conservation equation for the ${\cal N}=2$ higher-spin supercurrent multiplets in (2,0) AdS superspace, which takes the form
\bea
\label{ce11}
\bar \cD^{\b} \mathbb{J}_{\b \a(2s-1)} = \bar \cD_{(\a_1} \big( \mathbb{Y}_{\a_2 \dots \a_{2s-1})} + \ri \mathbb{Z}_{\a_2 \dots \a_{2s-1})} \big)~.
\eea
Here $\mathbb{J}_{\a(2s)}$ denotes the higher-spin supercurrent, while the trace supermultiplets $\mathbb{Y}_{\a(2s-2)} $ and $\mathbb{Z}_{\a(2s-2)} $
are both real and covariantly linear superfields,
\bea
\bar{ \mathbb Y}_{\a(2s-2)} 
- \mathbb{Y}_{\a(2s-2)} = \bar {\mathbb Z}_{\a(2s-2)} - \mathbb{Z}_{\a(2s-2)} = 0~, \quad \bar \cD^2 \mathbb{Y}_{\a(2s-2)} = \bar \cD^2 \mathbb{Z}_{\a(2s-2)}= 0~.~~~~~~ 
\label{ce-tr}
\eea 
The explicit form of this multiplet of currents was presented by considering simple ${\cal N}=2$ supersymmetric models for a chiral scalar superfield. 
Unlike in 4D ${\cal N}=1$ supergravity where every supersymmetric matter theory can be coupled to only one of the off-shell supergravity formulations (either old-minimal or new-minimal), here in the (2,0) AdS case our trace multiplets require both type II and type III compensators to couple to. 

The general conservation equation \eqref{ce11} naturally gives rise to the ${\cal N}=1$ higher-spin supercurrent multiplets discussed in the previous subsection. One may show that in the real basis, \eqref{ce11} turns into:
\bsubeq \label{ce2}
\bea
{\bm \de^{\1 \b}} \mathbb{J}_{\b \a(2s-1)} &=& \bm \de^{\1}_{(\a_1} \mathbb{Y}_{\a_2 \cdots \a_{2s-1})} - \bm \de^{\2}_{(\a_1} \mathbb{Z}_{\a_2 \cdots \a_{2s-1})}~, \\
\bm \de^{\2 \b} \mathbb{J}_{\b \a(2s-1)} &=& \bm \de^{\1}_{(\a_1} \mathbb{Z}_{\a_2 \cdots \a_{2s-1})} + \bm \de^{\2}_{(\a_1} \mathbb{Y}_{\a_2 \cdots \a_{2s-1})}~,
\eea
\esubeq
The real linearity constraints on the trace supermultiplets, eq.\,\eqref{ce-tr}, are equivalent to
\begin{subequations} \label{ce3}
\bea
(\bm \nabla^{\underline{2}})^2 \mathbb{Y}_{\a(2s-2)} &=& (\bm \nabla^{\underline{1}})^2 \mathbb{Y}_{\a(2s-2)}~, \qquad 
\bm \nabla^{\underline{1}\, \b} \bm \nabla^{\underline{2}}_{\b} \mathbb{Y}_{\a(2s-2)} = 0~, \\
(\bm \nabla^{\underline{2}})^2 \mathbb{Z}_{\a(2s-2)} &=& (\bm \nabla^{\underline{1}})^2 \mathbb{Z}_{\a(2s-2)}~, \qquad 
\bm \nabla^{\underline{1}\, \b} \bm \nabla^{\underline{2}}_{\b} \mathbb{Z}_{\a(2s-2)} = 0~.
\eea
\end{subequations}

It follows from \eqref{ce2} and \eqref{ce3} that $\mathbb{J}_{\a(2s)}$ contains two independent real ${\cal N}=1$ supermultiplets:
\bsubeq \label{J}
\bea
J_{\a(2s)} &:=& \mathbb{J}_{\a(2s)}|~, \\
J_{\a(2s+1)} &:=& \ri \bm \de^{\2}_{(\a_1} \mathbb{J}_{\a_2 \cdots \a_{2s+1})}|~,
\eea
\esubeq
while the independent real ${\cal N}=1$ components of $\mathbb{Y}_{\a(2s-2)}$ and $\mathbb{Z}_{\a(2s-2)}$ are defined by
\bsubeq \label{trace}
\bea
R_{\a(2s-2)} &:=& \mathbb{Y}_{\a(2s-2)}|~, \qquad U_{\b;\, \a(2s-2)}:= \ri \bm \de^{\2}_{\b} \mathbb{Y}_{\a(2s-2)}|~, \\
S_{\a(2s-2)} &:=& \mathbb{Z}_{\a(2s-2)}|~, \qquad T_{\b;\, \a(2s-2)}:= \ri \bm \de^{\2}_{\b} \mathbb{Z}_{\a(2s-2)}|~.
\eea
\esubeq
Making use of \eqref{ce3}, one may readily show that 
\bsubeq
\bea
\de^{\b} U_{\b; \, \a(2s-2)} = 0~, \label{ce-hf1} \\
\de^{\b} T_{\b; \, \a(2s-2)} = 0~. \label{ce-int1} 
\eea
\esubeq
On the other hand, eq.~\eqref{ce2} implies that the ${\cal N}=1$ superfields obey the following conditions
\bsubeq
\bea
&&\de^{\b}J_{\b \a(2s)} = \frac{2s}{2s+1} \Big(\de_{(\a_1} U_{\a_2 \cdots \a_{2s})} 
- \de_{(\a_1 \a_2} S_{\a_3 \cdots \a_{2s})} \Big)~, \label{ce-hf2}\\
&&\de^{\b}J_{\b \a(2s-1)} =  \de_{(\a_1} R_{\a_2 \cdots \a_{2s-1})}+ 
\ri T_{\a(2s-1)}~.  \label{ce-int2}
\eea
\esubeq
Indeed, the right-hand side of eq.~\eqref{ce-hf2} coincides with \eqref{ce-hf13a}. Therefore, eqs.~\eqref{ce-hf1} and \eqref{ce-hf2} define the ${\cal N}=1$ higher-spin current multiplets associated with the massless half-integer superspin formulations \eqref{hf} and \eqref{t3new}. In a similar way, it can be observed that eqs.~\eqref{ce-int1} and \eqref{ce-int2} correspond to the ${\cal N}=1$ higher-spin supercurrents for the two integer superspin models  \eqref{t2new} and \eqref{int}.


\section{Examples of ${\cal N}=1$ higher-spin supercurrents} \label{s66}

In this section we give an explicit realisation of the ${\cal N}=1$ multiplet of higher-spin 
supercurrent introduced earlier.

We recall the action \eqref{chiral20} for a massless chiral scalar in (2,0) AdS superspace
\bea 
S = \int \rd^3x \,\rd^2 \q  \,\rd^2 \bar \q
\, {\bm E} \,\bar \F \F ~, \qquad \bar \cD_{\a} \F =0~.
\label{ch-action}
\eea
The chiral superfield is charged under  the $R$-symmetry group $\rm U(1)_R$,
\bea
J \F = -r \F~, \qquad r ={\rm const} ~.
\eea
This action is a special case of the supersymmetric nonlinear sigma model studied in subsection \eqref{ss2.3} with a vanishing superpotential, $W(\F)=0$~. 
Making use of \eqref{N1-sigma}, the reduction of the action 
\eqref{ch-action} to $\cN=1$ AdS superspace is given by 
\bea
S &=& \int \rd^{3|2}z\,E \,
\Big\{-\ri \de^{\a} \bar \vf \de_{\a}  \vf + 4 r \cS\, \bar \vf \vf \Big\}~, 
\eea
where we have denoted $\vf:= \F|$~. This action is manifestly ${\cal N}=1$ supersymmetric. It also possesses hidden second supersymmetry and $\rm U(1)_R$ invariance. These transformations are
\bea
\d_{\e} \vf = \ri \e^{\a} \de_{\a}\vf -{\ri}\e \, r \vf ~, \qquad
\d_{\e} \bar \vf = \ri \e^{\a} \de_{\a} \bar \vf 
+{\ri}\e \,r {\bar \vf}~,
\eea
where $\e^\a$ is given in terms or $\e$ according to \eqref{u1-c-a},
and the real parameter $\e$ is constrained by \eqref{u1-c-b}.
It can be seen that $\vf$ and $\bar \vf$ obey the  equations of motion
\bea 
(\ri \de^2 +4 r \cS)\vf = 0~, \qquad (\ri \de^2 +4 r \cS)\bar \vf = 0~.
\label{eom}
\eea

Making use of our condensed notation, we may define the operators associated with the real spinor (2,0) AdS covariant derivatives 
$ {\bm \de}^{I}_\a$
\bea
{\bm \de}^{I}_{(1)} &:=& \z^\a \bm \de^{I}_\a , \qquad 
{\bm \de}_{(2)} := \ri \z^\a \z^\b \bm \de_{\a\b}~, \\
{\bm \de}^{I}_{(-1)} &:=& \bm \de^{I \a} \frac{\pa}{\pa \z^\a}~.
\eea
Analogous operators are introduced in the case of $\cN=1$ AdS superspace.
They are
\bea
{\de}_{(1)} &:=& \z^\a \de_\a , \qquad 
 \de_{(2)} := \ri \z^\a \z^\b  \de_{\a\b}~, \\
{\de}_{(-1)} &:=& \de^{\a} \frac{\pa}{\pa \z^\a}~.
\eea

It was shown in section \ref{s62} that by using the massless equation of motion, $\cD^2 \F=0$, the ${\cal N}=2$ higher-spin supercurrent multiplet associated with the theory \eqref{ch-action} is described by the conservation equation
\bsubeq 
\bea
\cD_{(-1)} \mathbb{J}_{(2s)} &=&  {\cD}_{(1)} {\mathbb T}_{(2s-2)}~.
\eea
Here the real supercurrent $ \mathbb{J}_{(2s)} = \bar {\mathbb{J}}_{(2s)}$ is given by
\bea
\mathbb{J}_{(2s)} &=& \sum_{k=0}^s (-1)^k
\left\{ \hf \binom{2s}{2k+1} 
{\cD}^k_{(2)} \bar \cD_{(1)} \bar \F \,\,
{\cD}^{s-k-1}_{(2)} \cD_{(1)} \F  
+ \binom{2s}{2k} 
{\cD}^k_{(2)} \bar \F \,\, {\cD}^{s-k}_{(2)} \F \right\}~,~~~~~~~~~~~
\eea
while the trace multiplet $\mathbb{T}_{(2s-2)}$ has the form
\bea
{\mathbb T}_{(2s-2)}&=& 2\ri {\cS}(1-2r)(2s+1)(s+1) \sum_{k=0}^{s-1}\frac{1}{2s-2k+1} (-1)^{k} \binom{2s}{2k+1}
\non \\ 
&& 
\times {\cD}^k_{(2)} \bar \F \,\,{\cD}^{s-k-1}_{(2)} \F ~.
\label{7.6a}
\eea
One may check that  ${\mathbb T}_{(2s-2)}$ is covariantly linear,
\bea
\bar \cD^2 \mathbb{T}_{(2s-2)} =0~, \qquad \cD^2 \mathbb{T}_{(2s-2)} =0~.
\eea
\esubeq
As is seen from \eqref{7.6a}, ${\mathbb T}_{(2s-2)}$ vanishes for  $r = \hf $, 
in which case $\F$ is an $\cN=2$  superconformal multiplet.

The complex trace multiplet $\mathbb{T}_{(2s-2)}$  
may be split into its real and imaginary parts:
\bsubeq
\bea
\mathbb{T}_{(2s-2)} = \mathbb{Y}_{(2s-2)}- \ri \mathbb{Z}_{(2s-2)}~,
\eea
with 
\bea
\mathbb{Y}_{(2s-2)} &=& 2\ri {\cS}(1-2r)(2s+1)(s+1) \sum_{k=0}^{s-1}\frac{2k-s+1}{(2k+3)(2s-2k+1)} 
\non \\ 
&& 
\times (-1)^{k} \binom{2s}{2k+1} {\cD}^k_{(2)} \bar \F \,\,{\cD}^{s-k-1}_{(2)} \F ~,\\
\mathbb{Z}_{(2s-2)} &=& -2 {\cS}(1-2r)(2s+1)(s+1)(s+2) \sum_{k=0}^{s-1}\frac{1}{(2k+3)(2s-2k+1)}
\non \\ 
&& 
\times (-1)^{k} \binom{2s}{2k+1} {\cD}^k_{(2)} \bar \F \,\,{\cD}^{s-k-1}_{(2)} \F ~.
\eea
\esubeq

In accordance with \eqref{J}, the supercurrent $\mathbb{J}_{(2s)}$ reduces to two different multiplets upon projection to ${\cal N}=1$ superspace:
\bsubeq 
\bea
J_{(2s)} &:=& \mathbb{J}_{(2s)}\big| \non\\
&=& \sum_{k=0}^s (-1)^{k+1}
\bigg\{ \binom{2s}{2k+1} 
{\de}^k_{(2)} \de_{(1)} \bar \vf \,\,
{\de}^{s-k-1}_{(2)} \de_{(1)} \vf  \non\\
&&- \binom{2s}{2k} 
{\de}^k_{(2)} \bar \vf \,\, {\de}^{s-k}_{(2)} \vf \bigg\}~, \\
J_{(2s+1)} &:=& \ri \bm \de^{\2}_{(1)}\mathbb{J}_{(2s)} \big| =
-\frac{1}{\sqrt{2}}\big(\cD_{(1)} + \bar \cD_{(1)} \big) \mathbb{J}_{(2s)}\big|~,\non\\
&=& (2s+1)\sum_{k=0}^s
\frac{1}{2s-2k+1} (-1)^{k+1} \binom{2s}{2k} \bigg\{
{\de}^k_{(2)} \bar \vf \,\, {\de}^{s-k}_{(2)} \de_{(1)} \vf \non\\
&&+ (-1)^{s-1} {\de}^k_{(2)}\vf \,\,
{\de}^{s-k}_{(2)} \de_{(1)} \bar \vf \bigg\}~,
\eea
\esubeq
of which the former corresponds to the integer superspin current and the latter half-integer superspin current. 

In the case of half-integer superspin, the conservation equation \eqref{ce-hf13} is satisfied provided we impose \eqref{eom}:
\bsubeq
\bea
\de_{(-1)} J_{(2s+1)} = \frac{2s}{2s+1} \bigg( \de_{(1)} U_{(2s-1)}+ \ri \de_{(2)} S_{(2s-2)}\bigg)~,\quad
\de^{\b} U_{\b; \,(2s-2)}= 0~,
\eea
with 
\bea
S_{(2s-2)} &:=& \mathbb{Z}_{(2s-2)} \big| \non\\
&=& -2 {\cS}(1-2r)(2s+1)(s+1)(s+2) \sum_{k=0}^{s-1}\frac{1}{(2k+3)(2s-2k+1)}
\non \\ 
&& 
\times (-1)^{k} \binom{2s}{2k+1} {\de}^k_{(2)} \bar \vf \,\,{\de}^{s-k-1}_{(2)} \vf ~,\\
U_{\b; \,(2s-2)} &:=& -\frac{1}{\sqrt{2}}\big(\cD_{\b} + \bar \cD_{\b} \big) \mathbb{Y}_{(2s-2)}\big|~,\non\\
&=& -2\ri {\cS}(1-2r)(2s+1)(s+1) \sum_{k=0}^{s-1}\frac{2k-s+1}{(2k+3)(2s-2k+1)} 
(-1)^{k} \binom{2s}{2k+1} \non\\
&& \times \bigg\{{\de}^k_{(2)} \bar \vf \,\,{\de}^{s-k-1}_{(2)} \de_{\b} \vf + (-1)^{s+1} {\de}^k_{(2)} \vf \,\,{\de}^{s-k-1}_{(2)} \de_{\b} \bar \vf \non\\
&&\qquad + 2\ri \cS (s-k-1) \z_{\b} \bigg({\de}^k_{(2)} \bar \vf \,\,{\de}^{s-k-2}_{(2)} \de_{(1)} \vf \non\\
&&\qquad+ (-1)^{s+1} {\de}^k_{(2)} \vf \,\,{\de}^{s-k-2}_{(2)} \de_{(1)} \bar \vf \bigg) \bigg\}~.
\eea
\esubeq

It may also be verified that the ${\cal N}=1$ supercurrent multiplet for integer superspin obeys the conditions \eqref{ce-int13} on-shell:
\bsubeq
\bea
\de_{(-1)} J_{(2s)} = \de_{(1)} R_{(2s-2)}+ \ri T_{(2s-1)}~, \quad \de^{\b}T_{\b; \,(2s-2)}=0
\eea
with
\bea
R_{(2s-2)} &:=& \mathbb{Y}_{(2s-2)} \big| \non\\
&&= 2\ri {\cS}(1-2r)(2s+1)(s+1)\sum_{k=0}^{s-1}\frac{2k-s+1}{(2k+3)(2s-2k+1)} 
\non \\ 
&& 
\times (-1)^{k} \binom{2s}{2k+1} {\de}^k_{(2)} \bar \vf \,\,{\de}^{s-k-1}_{(2)} \vf ~,\\
T_{\b; \,(2s-2)} &:=& -\frac{1}{\sqrt{2}}\big(\cD_{\b} + \bar \cD_{\b} \big) \mathbb{Y}_{(2s-2)}\big|~,\non\\
&=& 2 {\cS}(1-2r)(2s+1)(s+1)(s+2) \sum_{k=0}^{s-1}\frac{1}{(2k+3)(2s-2k+1)} 
(-1)^{k} \binom{2s}{2k+1} \non\\
&& \times \bigg\{{\de}^k_{(2)} \bar \vf \,\,{\de}^{s-k-1}_{(2)} \de_{\b} \vf + (-1)^{s} {\de}^k_{(2)} \vf \,\,{\de}^{s-k-1}_{(2)} \de_{\b} \bar \vf \non\\
&&\qquad + 2\ri \cS (s-k-1) \z_{\b} \bigg({\de}^k_{(2)} \bar \vf \,\,{\de}^{s-k-2}_{(2)} \de_{(1)} \vf \non\\
&&\qquad+ (-1)^{s} {\de}^k_{(2)} \vf \,\,{\de}^{s-k-2}_{(2)} \de_{(1)} \bar \vf \bigg) \bigg\}~.
\eea
\esubeq

The above technique can also be used to construct ${\cal N}=1$ higher-spin supercurrents for the Abelian vector multiplets model described by the action \eqref{2.46}. We will not elaborate on such a construction here.


\section{Applications and open problems}

Let us briefly summarise the results obtained in this chapter. In section \ref{section72}, a formalism to reduce every field theory with (2,0) AdS supersymmetry to ${\cal N}=1$ AdS superspace was developed. As nontrivial examples, we considered supersymmetric nonlinear sigma models formulated in terms of ${\cN}=2$ chiral and linear supermultiplets. 
In sections \ref{s72} and \ref{s63}, we applied the reduction technique and presented the ${\cN}=1$ superfield descriptions of the off-shell massless higher-spin supermultiplets with (2,0) AdS supersymmetry, which were constructed in chapter \ref{ch5}.
For each superspin value $\hat{s}$, integer ($\hat{s} = s$) or half-integer $ (\hat{s}=s+\hf) $, with $s= 1, 2, \dots$, the reduction produced two off-shell gauge formulations (called longitudinal and transverse) for a massless ${\cal N}=1$ superspin-$\hat{s}$ multiplet in AdS$_{3}$. The transverse formulations are new gauge theories.
In section \ref{s64}, we proved that for each superspin value the longitudinal and transverse theories are dually equivalent only in the flat superspace limit. 
In section \ref{s65}  we formulated the non-conformal higher-spin ${\cal N}=1$ supercurrent in ${\rm AdS}_3$. 
In section \ref{s66} we provided the explicit examples of these supercurrents
in simple models of a chiral scalar superfield. 

There are several interesting applications of the results presented in this chapter. 
In particular, the massless higher-spin $\cN=1$ supermultiplets in AdS${}_3$, 
which were derived in sections \ref{s72} and \ref{s63}, can be used 
to construct new topologically massive higher-spin off-shell supermultiplets in AdS${}_3$
by extending the approaches advocated in \cite{KO,KT,KP1}. Such a massive 
supermultiplet is described by  a gauge-invariant action being the sum of 
massless and  superconformal higher-spin actions.\footnote{ We will not review such a construction in this thesis, see \cite{HK19} for details.} This procedure follows the philosophy of topologically massive theories
 \cite{Siegel,Schonfeld,DJT1,DJT2}.

We now present two off-shell formulations for the massive $\cN=1$ gravitino supermultiplet  in  AdS${}_3$  and analyse the corresponding 
equations of motion.\footnote{The
construction of the models \eqref{massL} and \eqref{massT}
is similar to those used to derive the off-shell formulations for massive superspin-1
and superspin-3/2 multiplets in four dimensions 
\cite{OS2,BL1,BL2,AB1,AB2,BGLP1,BGLP2,GSS,BGKPmass,Gates:2005su,GKT-M}.}
The massive extension of the longitudinal theory \eqref{425} is described by the action
\bea
S^{||}_{(1), \,\m}
&=& -\frac{1}{2}
\int 
\rd^{3|2}z
\, E \,\bigg\{
\frac{\ri}{2} H^{\a \b} {\de}^2 H_{\a \b}
-\frac{\ri}{2} \de_{\b}H^{\a \b} \nabla^{\g} H_{\g \a}  - V \nabla^{\a \b} H_{\a \b}
\non \\
&&
+\frac{\ri}{2} V \nabla^2 V + (\m + 2 \cS) H^{\a \b} H_{\a \b} -2( \m -2 \cS) V^2
\bigg\}~,
\label{massL}
\eea
with $\m$ a real mass parameter. 
The massive gravitino action is thus constructed from the massless one by adding mass-like terms. In the limit $\m\to 0$, the action reduces to \eqref{425}.
The equations of motion for the dynamical superfields $H^{\a \b}$ and $V$ are
\bsubeq
\bea
&& \qquad 2 \de^{\g}\,_{(\a}H_{\b) \g} -\ri \de^2 H_{\a \b}-2 \de_{\a \b}V - 4 \m H_{\a \b} = 0~, \label{m1}\\
&& \qquad \de^{\a \b} H_{\a \b} = \big(\ri \de^2 + 8\cS -4\m \big)V~.
\label{m2}
\eea
\esubeq
Multiplying \eqref{m1} by $\de^{\a \b}$ and noting that $[\de_{\a \b}, \de^2]=0$ yields
\bea
-\ri \de^2 \de^{\a \b}H_{\a \b}+ 4 \Box V -4 \m \de^{\a \b} H_{\a \b}=0~. \label{m3}
\eea
Substituting \eqref{m2} into \eqref{m3} leads to
\bea
V=0~.
\eea
Now that $V=0$ on-shell, eq.~\eqref{m2} turns into
\bea
\de^{\a \b}H_{\a \b}=0~,
\eea
while \eqref{m1} can equivalently be written as
\bea
-\ri \de^{\g} \de_{\a}H_{\b \g}-(2 \m + 4 \cS) H_{\a \b}=0~. \label{m4}
\eea
Making use of the identity \eqref{AA2},
it immediately follows from \eqref{m4} that
\bea
\de^{\a} H_{\a \b}=0~,
\eea
and then \eqref{m4} is equivalent to
\bea
-\frac{\ri}{2}\de^2 H_{\a \b} = (\m + 2\cS) H_{\a \b}~.
\eea
Therefore, we have demonstrated that the model \eqref{massL} leads to the following conditions on the mass shell:
\bsubeq \label{8.15}
\bea
V &=& 0~,
\\
\de^{\a}H_{\a \b}&=&0 \quad \implies \quad \de^{\a \b}H_{\a \b}=0~, \\
-\frac{\ri}{2}\de^2 H_{\a \b} &=& (\m +2\cS ) H_{\a \b}~.
\eea
\esubeq
Such conditions are required to describe an irreducible on-shell massive gravitino multiplet in 3D ${\cal N}=1$ AdS superspace \cite{KNT-M}. 

In the transverse formulation \eqref{337}, the action for a massive gravitino multiplet 
is given by
\bea
S^{\perp}_{(1),\, \m}
&=&
- \hf \int 
\rd^{3|2}z 
\, E \, \bigg\{\frac{\ri}{2} H^{\a \b} {\de}^2 H_{\a \b}
-{\ri} \de_{\b}H^{\a \b} \nabla^{\g} H_{\g \a} 
-H^{\a \b} {\de}_{\a} \cW_{\b}
- \frac{\ri}{2} \cW^{\a}\cW_{\a} \non\\
&&+(\m + 4\cS) H^{\a \b}H_{\a \b} 
- \ri (\mu + 2\cS) \Big(\J^{\a}\cW_{\a} 
+2 \m \J^\a\J_\a \Big)
\bigg\}~.
\label{massT}
\eea
In the limit $\m\to 0$, the action reduces to \eqref{337}.
One may check that the equations of motion for this model imply that 
\bsubeq \label{8.17}
\bea
\J_{\a} &=& 0~,
\\
\de^{\a}H_{\a \b}&=&0 \quad \implies \quad \de^{\a \b}H_{\a \b}=0~, \\
-\frac{\ri}{2}\de^2 H_{\a \b} &=& (\m +4\cS ) H_{\a \b}~.
\eea
\esubeq

The actions \eqref{massL} and \eqref{massT} can be made into gauge-invariant ones using the Stueckelberg construction. 

In the Minkowski superspace limit, the massive models \eqref{massL} and \eqref{massT} 
lead to the identical equations of motion described in terms of $H_{\a\b}$:
\bea
D^{\a}H_{\a \b}=0 ~,\qquad 
-\frac{\ri}{2}D^2 H_{\a \b} &=& \m H_{\a \b}~.
\eea
In the AdS case, the equations \eqref{8.15} and \eqref{8.17} lead to equivalent 
dynamics modulo a redefinition of $\m$.
It is an interesting open problem to understand whether there exists a duality 
 transformation relating these models. 

It should be pointed out that there also exists  an on-shell construction 
of gauge-invariant Lagrangian formulations for massive higher-spin supermultiplets in 3D Minkowski and AdS spaces, which were  developed in \cite{BSZ3,BSZ4}.
It is obtained by combining the massive bosonic 
and fermionic higher-spin actions \cite{BSZ1,BSZ2}, and therefore 
this construction is intrinsically on-shell.
The formulations given in \cite{BSZ1,BSZ2,BSZ3,BSZ4}
are based on the gauge-invariant approaches 
to  the dynamics of massive higher-spin fields, which were advocated by Zinoviev \cite{Zinoviev} and Metsaev \cite{Metsaev}.
It is an interesting open problem to understand whether there exists an off-shell 
uplift of these models.



\chapter{Conclusion} \label{ch7}

Over the course of this thesis, we have presented various non-conformal higher-spin supercurrents and their associated off-shell massless higher-spin supermultiplets in three and four spacetime dimensions. All of our analyses were performed using the superspace approach, which is an efficient means of formulating supercurrent multiplets and off-shell supersymmetric theories. 
In four dimensions, a major part of this work was devoted to the explicit construction of conserved higher-spin currents with ${\cN}=1$ Poincar\'e and AdS supersymmetry. In three dimensions, we studied both the ${\cN}=1$ and ${\cN}=2$ AdS cases. 
A number of avenues which could be explored for future studies have been highlighted at the end of each chapter.
Here we summarise the key results of this thesis and discuss their implications.

Higher-spin supercurrents in 4D ${\cN}=1$ Minkowski and AdS superspaces \cite{HK1, HK2, BHK} were studied in chapter \ref{ch3} and \ref{ch4}, respectively. The main ingredients in deriving the supercurrents are the known off-shell massless higher-spin supermultiplets \cite{KPS,KS, KS94} and their gauge symmetries. We also formulated the higher-spin supercurrent multiplet associated with the new off-shell model for the massless {\it integer} superspin. Having understood the structure of the current multiplets and their improvement transformations, we obtained closed-form expressions of conserved supercurrents for various supersymmetric theories in AdS. For instance, a model with $N$ massive chiral scalar superfields with an arbitrary mass matrix, and the free theories of tensor and complex linear superfields. For the latter cases, we employed the complex linear-chiral and the minimal scalar-tensor dualities. 
The structure of the conserved supercurrents is determined by the type (integer or half-integer) and value (even or odd) of the superspin, as well as the mass matrix. This has been summarised in section \ref{s46}.
%

A natural extension of the analysis presented in chapters \ref{ch3} and \ref{ch4} would be to construct 4D ${\cN}=2$ higher-spin supercurrents, by making use of the known off-shell gauge supermultiplets \cite{GKS1}.


In regards to the off-shell massless higher-spin ${\cN}=1$ supermultiplets in 4D Minkowski and AdS backgrounds, we also developed a new off-shell formulation for the massless \textit{integer} superspin multiplet. It was shown that the gauge-invariant action generalises that of the longitudinal theory. It is described in terms of the complex superconformal higher-spin prepotential $\J_{\a(s) \ad(s-1)}$, in conjunction with two compensating superfields. Making use of the superfield Legendre transformation, we constructed its dual action and demonstrated that it reduces to the transverse formulation. 

Chapter \ref{ch5} discussed ${\cN}=2$ supersymmetric higher-spin gauge theories in $\rm AdS_3$ based on the results of \cite{HKO, HK18}. Along the same lines, we generalised the 4D gauge principles used in chapter \ref{ch4} to construct off-shell linearised actions for massless higher-spin supermultiplets around the (1,1) AdS background. In addition, we derived the corresponding consistent supercurrents and gave their explicit expressions for models of chiral superfields. 
Within the framework of (2,0) AdS supersymmetry, the problem of constructing off-shell massless higher-spin gauge supermultiplets has not been fully resolved. In section \ref{s62}, we identified a multiplet of conserved higher-spin currents, which allowed us to construct two off-shell actions for the massless \textit{half-integer} superspin multiplet. In each of the formulations, the corresponding gauge-invariant action contains a higher-spin extension of a Chern-Simons term. In the limit of $s=1$, these actions reduce to the linearised actions for (2,0) AdS supergravity \cite{KT-M11}. 
It remains an open problem to construct an off-shell formulation for a massless \textit{integer} superspin multiplet. For completeness, it would be useful to study the component actions of \eqref{action20-t2} and \eqref{action2-t3} in order to understand their actual differences with the (1,1) AdS actions. 

In chapter \ref{ch6} we derived four series of off-shell massless higher-spin ${\cN}=1$ supermultiplets in $\rm AdS_3$, two of which were new supersymmetric gauge theories. This was accomplished via the $(2,0) \to (1,0)$ AdS superspace reduction procedure \cite{HK19}. Further analysis showed that these off-shell models are related by a superfield Legendre transformation in the flat superspace limit, but the duality is not lifted to the AdS case. The massless ${\cN}=1$ supersymmetric higher-spin actions in $\rm AdS_3$ were used to formulate (i) conserved ${\cN}=1$ higher-spin supercurrents; and (ii) two new off-shell massive ${\cN}=1$ gravitino supermultiplets in $\rm AdS_3$. Additionally, we elaborated on the component structure of the two new ${\cN}=1$ supersymmetric higher-spin models \eqref{action-t2-new} and \eqref{action-t3-new-complete} in flat superspace (see appendix \ref{AppendixBB}). Whilst it was shown that the action \eqref{action-t2-new} reduces to the 3D (Fang-)Fronsdal actions upon elimination of the auxiliary fields, an interesting feature appeared in the analysis of \eqref{action-t3-new-complete}.
At the component level, the corresponding multiplet is a 3D counterpart
of the so-called (reducible) higher-spin triplet systems. In $\rm AdS_D$ an action for higher-spin triplets was constructed in \cite{Sagnotti:2003qa} and \cite{Sorokin:2008tf,Agugliaro:2016ngl}, for the bosonic and fermionic cases, respectively. This demonstrates that our superfield construction provides a manifestly supersymmetric generalisation of these systems. 

All supersymmetric higher-spin models presented in this thesis are linearised actions for higher-spin multiplets. We believe that they can be used to construct interacting theories, which 
will allow us to make contact with the 3D gauge theories developed by Vasiliev and collaborators \cite{PV1, PV2, PSV}. As a next step towards complete superfield formulation for higher-spin supergravity, an important problem is to go beyond the linearised approximation, {\it i.e.}\,finding the relevant deformations of superfield higher-spin actions, gauge transformations and their corresponding supercurrents. 

It would be of particular interest to examine the off-shell structure of supersymmetric higher-spin multiplets and their associated conserved supercurrents in 3D with ${\cN}>2$ supersymmetry. It is also expected that new techniques are required. For example, in order to study the ${\cN}=3$ case in ${\rm AdS}_3$, one may apply the projective-superspace formalism developed in \cite{KLT-M12}. 

All off-shell higher-spin $\cN=2$ supermultiplets in AdS${}_3$ presented in chapter \ref{ch5}
are reducible gauge theories (in the terminology of the Batalin-Vilkovisky 
quantisation \cite{BV}), similar to the massless higher-spin supermultiplets 
in AdS${}_4$ \cite{KS94}. The Lagrangian quantisation of such theories is nontrivial, as demonstrated in \cite{BKS} in the 4D case. All off-shell higher-spin $\cN=1$ supermultiplets 
in AdS${}_3$, which were constructed in chapter \ref{ch6}, are irreducible gauge theories. They can be quantised using the Faddeev-Popov procedure \cite{FP}
as in the non-supersymmetric case, see e.g. \cite{FH,GGS}. 

As a final comment, all off-shell supersymmetric massless higher-spin models presented in this thesis (both in three and four dimensions) share a common feature. After we reformulated the massless integer superspin theories, one obtains a universal picture in which every gauge-invariant action is now realised in terms of two dynamical variables: a superconformal gauge prepotential and an appropriate set of compensating superfield(s).

 



\appendix

\chapter{Notation and conventions} \label{AppA}
In this appendix we collect important definitions and identities that have been used throughout the thesis. For a more rigorous presentation, the reader is referred to \cite{Ideas}, which our 4D notation and conventions mainly follow. Below are the types of indices that we use:
\begin{itemize}
\item
Lower case letters from the beginning (middle) of the Latin alphabet,\\ {\it i.e.} $ a, b, \dots (m, n, \dots)$ correspond to flat (curved) spacetime indices.
\item
Lower case letters from the beginning of the Greek alphabet, {\it i.e.} $\a, \b, \dots$ denote indices for two-component Weyl spinors.
\item
Likewise, upper case letters from the beginning and middle of the Latin alphabet denote flat and curved superspace coordinates respectively.
\end{itemize} 
\section{4D spinor and tensor identities}
We use the mostly positive convention for the Minkowski metric:
\bea
\eta_{a b} := \mbox{diag}(-1,1,1,1)~,
\eea
in order to raise and lower spacetime indices of tangent space tensors, $V_{a} = \eta_{ab} V^b~, V^a = \eta^{ab}V_{b}~,$ with $a, b = 0,1,2,3$. On the other hand, the indices of curved spacetime tensors can be raised or lowered using the curved metric $g_{mn}$, 
\bea
g_{mn} = e_m{}^a e_n{}^{b} \eta_{a b},
\eea
where $e_m{}^{a}$ is the vierbein. The inverse vierbein $e_a{}^{m}$ is introduced by $e_a{}^{m} e_m{}^{b} = \d_a{}^{b}$ and $e_m{}^{a} e_a{}^{n} = \d_m{}^{n}$. 

The (brackets) parentheses denote (anti-)symmetrisation of tensor or spinor indices, which include a normalisation factor, for instance
\bea
V_{[a_1 a_2 \dots a_n]} := \frac{1}{n!}\sum_{\pi \in S_n} \mbox{sgn}(\pi) V_{a_{\pi(1)} \dots a_{\pi(n)}}~, \quad V_{(\a_1 \dots \a_n )} := \frac{1}{n!} \sum_{\pi \in S_n} \mbox{sgn}(\pi) V_{\a_{\pi(1)} \dots \a_{\pi(n)}}~,~~~~~~
\eea
with $S_n$ being the symmetric group of $n$ elements. 

The totally antisymmetric Levi-Civita tensor, $\ve_{abcd} \equiv \ve_{[abcd]}$, is normalised such that
\bea
\ve^{0123} = -\ve_{0123} = 1~.
\eea
A product of Levi-Civita tensors can be written as
\bea
\ve^{abcd} \ve_{a' b' c' d'} = -4!  \d^{a}_{[a'}\d^{b}_{b'}\d^{c}_{c'}\d^{d}_{d']}~.
\eea

Central to the description of 4D ${\cN}=1$ supersymmetry is the formalism of two-component Weyl spinors. These are representations of ${\rm SL}(2, \mathbb{C})$, which is the covering group of the restricted Lorentz group ${\rm SO}_{0}(3,1)$. Specifically, an object $\psi_{\a}\, (\a = 1,2$) which transforms under the fundamental representation of $\rm SL(2, \mathbb{C})$,
\bea
\psi_{\a}' = N_{\a}{}^{\b}\psi_{\b}~, \qquad N_{\a}{}^{\b} \in \rm SL(2, \mathbb{C})
\eea
is called a left-handed Weyl spinor. 
This is denoted by $(\hf, 0)$ and is known as the left-handed spinor representation of the Lorentz group. 
On the other hand, a right-handed Weyl spinor $\bar \psi_{\ad} \,(\ad = \dot{1}, \dot{2})$ transforms in the conjugate representation
\bea
\bar \psi_{\ad}' = \bar N_{\ad}{}^{\bd}\bar \psi_{\bd}~, \qquad (N_{\a}{}^{\b})^* = \bar N_{\ad}{}^{\bd}~.
\eea
This is denoted by $(0, \hf)$ and is called the right-handed spinor representation of the Lorentz group. 


The undotted and dotted indices of two-component spinors may be raised and lowered with the help of $\ve$ tensors:
\bea
\psi_{\a} = \ve_{\a \b} \psi^{\b}~, \quad
\psi^{\a} = \ve^{\a \b} \psi_{\b}~, \quad  \quad \bar \chi_{\ad} = \ve_{\ad \bd} \bar \chi^{\bd}~, \quad \bar \chi^{\ad} = \ve^{\ad \bd} \bar \chi_{\bd}~.
\eea
The antisymmetric tensors,  $\ve_{\a\b}=-\ve_{\b\a}$ and $\ve_{\ad\bd}=-\ve_{\bd\ad}$ are invariant under $\rm SL(2, \mathbb{C})$. They are defined by
\bea
\ve^{\a \b} \ve_{\b \g} = \d^{\a}{}_{\g}~,\,\,\,\, \ve^{\ad \bd} \ve_{\bd \gd} = \d^{\ad}{}_{\gd}~,\,\, \,\, \ve^{12} = \ve_{21} = 1~, \,\,\,\, \ve^{{\dot 1}{\dot 2}}=\ve_{{\dot 2}{\dot 1}}=1~.
\eea
We will adopt the following rules for contraction of spinor indices:
\bea
\psi \chi := \psi^{\a}\chi_{\a} = \chi \psi~, \quad \bar \psi \bar \chi := \bar \psi_{\ad} \bar \chi^{\ad} = \bar \chi \bar \psi~, 
\eea
and $\psi^2 = \psi \psi~, \bar \psi^2 = \bar \psi \bar \psi$. Here spinor conjugation is understood as Hermitian conjugation,
\bea
(\psi \chi)^* = (\psi^{\a} \chi_{\a})^* = (\chi_{\a})^* (\psi^\a)^* = \bar\chi_{\ad} \bar \psi^{\ad}.
\eea

We define the sigma matrices $\sigma_{a}:= (\sigma_a)_{\a\ad}$ as
\bea
(\sigma_{a}) := (\mathbbm{1}, \vec{\sigma})~, \quad 
(\tilde{\sigma}_a):= (\mathbbm{1}, -\vec{\sigma})~,
\eea
{\it i.e.}
\be
\s_{0}=
\left(\begin{array}{cc}1 & 0 \\0 & 1\end{array}\right),\quad\!
\s_{1}=
\left(\begin{array}{cc}0 & 1 \\1 & 0\end{array}\right),\quad\!
\s_{2}=
\left(\begin{array}{cc}0 & -{\rm i} \\{\rm i} & 0\end{array}\right),\quad\!
\s_{3}=
\left(\begin{array}{cc}1 & 0 \\0 & -1\end{array}\right)~.
\ee
The tilded sigma matrices with raised indices are denoted by
\bea
(\tilde{\s}_a)^{\ad \a} = \ve^{\ad \bd} \ve^{\a \b} (\s_a)_{\b \bd}~. 
\eea
The sigma matrices satisfy some useful properties, for instance
\bea
(\s_{a}{\tilde \s}_{b}+\s_{b}{\tilde \s}_{a})_\a{}^\b
=-2\,\eta_{ab}\,\d_\a{}^\b~,\qquad&&\qquad\quad\!
{\rm Tr}(\s_{a}{\tilde \s}_{b})
=-2\,\eta_{ab}~,\\
({\tilde \s}_{a}\s_{b}+{\tilde \s}_{b}\s_{a})^\ad{}_\bd
=-2\,\eta_{ab}\,\d^\ad{}_\bd~,
\qquad&&\quad
(\s^{a})_{\a\ad}({\tilde \s}_{a})^{\bd\b}
=-2\,\d_\a^\b\,\d^\bd_\ad~. \non
\eea
We can introduce the antisymmetric traceless matrices
\bsubeq
\bea
(\s_{ab})_\a{}^{\b} = -\frac{1}{4}(\s_a \tilde{\s}_b - \s_b \tilde{\s}_a)_{\a}{}^{\b}~,\\
(\tilde{\s}_{ab})^{\ad}{}_{\bd} = -\frac{1}{4}(\tilde{\s}_a {\s}_b - \tilde{\s}_b {\s}_a)^{\ad}{}_{\bd}~,
\eea
\esubeq
which are 
(anti) self-dual,
\be
\frac{1}{2} \,\ve_{abcd} \, \s^{cd} = - {\rm i} \,\s_{ab}~,
\quad\qquad
\frac{1}{2} \,\ve_{abcd} \, \tilde{\s}^{cd} = {\rm i} \,\tilde{\s}_{ab}~.
\ee
They also satisfy the Lorentz algebra
\be \label{eq:lorentz-algebra}
[\s_{ab},\s_{cd}]=\eta_{ad}\s_{bc}-\eta_{ac}\s_{bd}
+\eta_{bc}\s_{ad}-\eta_{bd}\s_{ac}~.
\ee
 
Given a vector $V_a$, one can convert the vector index to a pair of spinor indices using the $\sigma$-matrices. The rules are as follows
\bea
V_{\a \bd} = (\sigma^a)_{\a \bd}V_{a}~, \qquad V_{a} = -\hf (\tilde{\sigma}_a)^{\bd \a} V_{\a \bd}~.
\eea
As an example, let us consider a real and antisymmetric rank-2 tensor, $F_{ab} = -F_{ba}$. The decomposition is
\bsubeq
\bea
F_{\a \ad, \b \bd}= (\sigma^a)_{\a \ad}(\sigma^b)_{\b \bd} F_{ab} = 2 \ve_{\a\b}\bar F_{\ad \bd} + 2 \ve_{\ad \bd}F_{\a \b}~.
\eea
Here we have defined
\bea
F_{\a \b}= \hf (\sigma^{ab})_{\a \b}F_{a b}~, \qquad \bar F_{\ad \bd} = -\hf (\tilde{\sigma}^{ab})_{\ad \bd} F_{ab}~.
\eea
\esubeq 
In particular, this applies to the Lorentz generators, $M_{ab}=-M_{ba} \Leftrightarrow (M_{\a\b},{\bar M}_{\ad\bd})$, which satisfy the same algebra \eqref{eq:lorentz-algebra} as the $\s_{ab}$ matrices. They act on arbitrary spinors as follows:
\bea
M_{\a\b}(\j_{\g})=
\frac{1}{2}(\ve_{\g\a}\j_{\b}+\ve_{\g\b}\j_{\a})
~,\quad&&\qquad M_{\a\b}({\bar \j}_{\gd})=0~,\\
{\bar M}_{\ad\bd}({\bar \j}_{\gd})=
\frac{1}{2}(\ve_{\gd\ad}{\bar \j}_{\bd}+\ve_{\gd\bd}{\bar \j}_{\a})
~,\quad&&\qquad {\bar M}_{\ad\bd}(\j_{\g})=0~.\non
\eea

Let $D_{A} = (\pa_a, D_\a, \bar D^{\ad})$ be the set of covariant derivatives of ${\cN}=1$ Minkowski superspace.
The spinor covariant derivatives, $D_{\a}$ and $\bar D_{\ad}$, are related by complex conjugation, which works as follows
\be
\overline{D_\a V} =(-1)^{\e(V)}{\bar D}_\ad \bar{V}~,
\quad\qquad
\overline{D^2 V} ={\bar D}^2\bar  V~. \label{cc-4d}
\ee
Here $\bar V$ is the complex conjugate of $V$. The Grassmann parity of $V$ is denoted by $\e(V)$ {\it i.e.} $\e(V)=0$ for a bosonic superfield and $\e(V)=1$, if $V$ is fermionic. We also note that $D^2 = D^{\a}D_{\a}$ and $\bar D^2 = \bar D_{\ad} \bar D^{\ad}$. It is important to keep in mind the following rules when doing calculations with the covariant derivatives:
\bea
D_{A}(UV) &=& D_{A}(U) V + (-1)^{\e(U)\e(D_A)} U D_{A}(V)~,\non\\
\e(D_{A} V) &=& \e(D_A) + \e(V)\,\,\,\,
\rm{(mod\, 2)}
\eea
for arbitrary superfields $U$ and $V$.

\section{3D notation and AdS identities} \label{AppA2}
We summarise our 3D notation and conventions following
\cite{KLT-M11, KPT-MvU}. The 3D Minkowski metric is
$\eta_{ab}=\mbox{diag}(-1,1,1)$.
The spinor indices are raised and lowered by the rule
\bea
\psi^{\a}=\ve^{\a\b}\psi_\b~, \qquad \psi_{\a}=\ve_{\a\b}\psi^\b~.
\label{sp}
\eea
Here the antisymmetric $\rm SL(2,{\mathbb R})$ invariant tensors $\ve_{\a \b}= -\ve_{\b \a}$ and $\ve^{\a \b}= -\ve^{\b \a}$ are normalised as $\ve_{12} = -1~, \ve^{12}=1$~.

We make use of real Dirac $\g$-matrices,  $\g_a := \big( (\g_a)_\a{}^\b \big)$ defined by
\bea
(\g_a)_\a{}^\b := \ve^{\b \g} (\g_a)_{\a \g} = (-\ri \s_2, \s_3, \s_1)~.
\eea
They obey the algebra
\be
\gamma_a \gamma_b=\eta_{ab}{\mathbbm{1}} + \varepsilon_{abc}
\gamma^c~,
\ee
where the Levi-Civita tensor is normalised as
$\varepsilon^{012}=-\varepsilon_{012}=1$. 
Some useful relations involving $\g$-matrices are 
\bsubeq
\bea
(\gamma^a)_{\alpha\beta}(\gamma_a)^{\rho\sigma}
&=&-(\delta_\alpha^\rho\delta_\beta^\sigma
+\delta_\alpha^\sigma\delta_\beta^\rho)~, \\
\ve_{abc}(\g^b)_{\a\b}(\g^c)_{\g\d}&=&
\ve_{\g(\a}(\g_a)_{\b)\d}
+\ve_{\d(\a}(\g_a)_{\b)\g}
~,
\\
\tr[\g_a\g_b\g_{c}\g_d]&=&
2\eta_{ab}\eta_{cd}
-2\eta_{ac}\eta_{db}
+2\eta_{ad}\eta_{bc}
~.
\eea
\esubeq

Given a three-vector $x_a$,
it  can be equivalently described by a symmetric second-rank spinor $x_{\a\b}$
defined as
\bea
x_{\a\b}:=(\g^a)_{\a\b}x_a=x_{\b\a}~,\qquad
x_a=-\hf(\g_a)^{\a\b}x_{\a\b}~.
\eea
In the 3D case,  an
antisymmetric tensor $F_{ab}=-F_{ba}$ is Hodge-dual to a three-vector $F_a$, 
specifically
\bea
F_a=\hf\ve_{abc}F^{bc}~,\qquad
F_{ab}=-\ve_{abc}F^c~.
\label{hodge-1}
\eea
Then, the symmetric spinor $F_{\a\b} =F_{\b\a}$, which is associated with $F_a$, can 
equivalently be defined in terms of  $F_{ab}$: 
\bea
F_{\a\b}:=(\g^a)_{\a\b}F_a=\hf(\g^a)_{\a\b}\ve_{abc}F^{bc}
~.
\label{hodge-2}
\eea
These three algebraic objects, $F_a$, $F_{ab}$ and $F_{\a \b}$, 
are in one-to-one correspondence to each other, 
$F_a \leftrightarrow F_{ab} \leftrightarrow F_{\a\b}$.
The corresponding inner products are related to each other as follows:
\bea
-F^aG_a=
\hf F^{ab}G_{ab}=\hf F^{\a\b}G_{\a\b}
~.
\eea

The Lorentz generators with two vector indices ($M_{ab} =-M_{ba}$),  one vector index ($M_a$)
and two spinor indices ($M_{\a\b} =M_{\b\a}$) are related to each other by the rules:
\bea
M_{ab} = -\ve_{abc}M^c~, \,\,\, M_a=\hf \ve_{abc}M^{bc}~, \,\,\, M_{\a\b}=(\g^a)_{\a\b}M_a~, \,\,\, M_{a}= -\hf (\g_a)^{\a \b} M_{\a \b}~.~~~~~
\eea
These generators 
act on a vector $V_c$ 
and a spinor $\J_\g$ 
as follows:
\bea
M_{ab}V_c=2\eta_{c[a}V_{b]}~, ~~~~~~
M_{\a\b}\J_{\g}
=\ve_{\g(\a}\J_{\b)}~.
\label{generators}
\eea

We collect some useful properties for ${\cal N}=1$ AdS covariant derivatives, which we denote by 
$\nabla_{A} = \left( \nabla_a, \nabla_{\a} \right)$.
We first note the unusual complex conjugation property of the spinor covariant derivative, which can be compared with the 4D case, see \eqref{cc-4d}. Given an arbitrary superfield $V$ and its complex conjugate $\bar V $, it holds that 
\bea
\overline{\nabla_{\a} V} = - (-1)^{\e(V)} \nabla_{\a} \bar V~,
\eea
where $\e(V)$ denotes the Grassmann parity of $V$~.

Making use of the (anti)-commutation relation \eqref{2_0-alg-AdS-1} and \eqref{2_0-alg-AdS-2}, we obtain the following identities
\begin{subequations}
\bea
\nabla_{\a} \nabla_{\b} &=& \hf \ve_{\a \b} \nabla^2 + \ri \nabla_{\a \b} - 2 \ri \cS M_{\a \b}~, \label{AA1} \\
\nabla^{\b} \nabla_{\a} \nabla_{\b} &=& 4\ri \cS \nabla_{\a}~, \label{AA2} \\
\nabla^2 \nabla_{\a} &=& -\nabla_{\a} \nabla^2 + 4 \ri \cS \nabla_{\a} = 2 \ri \nabla_{\a \b} \nabla^{\b} +2 \ri \cS \nabla_{\a} - 4 \ri \cS \nabla^{\b} M_{\a \b}~, \label{AA3} \\
-\frac{1}{4} \nabla^2 \nabla^2 &=& \Box -2 \ri \cS \nabla^2 + 2\cS \nabla^{\a \b} M_{\a \b} -2 \cS^2 M^{\a \b} M_{\a \b}~, \label{AA4}
\eea
\end{subequations}
where $\nabla^2 = \nabla^{\a} \nabla_{\a}$ and $\Box = \nabla^{a} \nabla_{a} = -\hf \nabla^{\a \b} \nabla_{\a \b}$~. An important corollary of \eqref{AA1} and \eqref{AA3} is 
\bea
{[} \nabla_{\a} \nabla_{\b}, \nabla^2 {]} =0 
\quad \implies \quad {[} \nabla_{\a \b}, \nabla^2 {]}=0~.
\eea
The left-hand side of \eqref{AA4} can be expressed in terms of the quadratic Casimir operator of the 3D ${\cal N}=1$ AdS supergroup \cite{KP1}:
\bea
\mathbb{Q} = -\frac{1}{4} \nabla^2 \nabla^2 + \ri \cS \nabla^2~, \qquad {[} \mathbb{Q}, \nabla_A {]} =0~.
\label{casimir}
\eea

\chapter{Conserved higher-spin currents in four dimensions}
\label{AppendixC}


In appendix \ref{AppendixC-scalar} we consider the construction of conserved higher-spin currents in free scalar field theory in flat space. Similar analysis for free fermions will be done in the next section \ref{AppendixD-spinor}. This material has been drawn from \cite{BHK}.

\section{Free real scalars}
\label{AppendixC-scalar}

Given an integer $s\geq 2$, 
the massless spin-$s$ field  \cite{Fronsdal} is described   by
real potentials
$h_{\a(s) \ad(s)}$ and $h_{\a(s-2) \ad(s-2)}$
with the gauge freedom\footnote{We follow the description of Fronsdal's
theory \cite{Fronsdal} given in section 
6.9 of \cite{Ideas}.}
\begin{subequations} \label{gauge1}
\bea
 \d h_{\a_1 \dots \a_{s} \ad_1 \dots \ad_{s} } 
&=& \pa_{(\a_1 (\ad_1} \l_{\a_2\dots \a_{s}) \ad_2 \dots \ad_{s})}~, \\
\d h_{\a_1 \dots \a_{s-2} \ad_1 \dots \ad_{s-2} } 
&=& \frac{s-1}{s^2} \pa^{\b \bd} \l_{\b \a_1\dots \a_{s-2} \bd \ad_1 \dots \ad_{s-2}}~,
\eea
\end{subequations}
for an arbitrary real gauge parameter 
$\l_{\a(s-1) \ad(s-1)}$. 
The field $h_{\a(s) \ad(s)}$ may be interpreted as a conformal spin-$s$ field \cite{FT,FL}.

To construct non-conformal higher-spin currents, we couple $h_{\a(s) \ad(s)}$ and $h_{\a(s-2) \ad(s-2)}$ to external sources
\bea
S^{(s)}_{\rm source} &=& \int \rd^4x \, \Big\{ 
h^{ \a (s) \ad (s) } j_{ \a (s) \ad (s) }
+h^{ \a (s-2) \ad (s-2) } t_{ \a (s-2) \ad (s-2) } \Big\}~.
\label{source}
\eea
Requiring that $S^{(s)}_{\rm source}$ be invariant under the $\l$-transformation 
in \eqref{gauge1} gives the conservation equation
\bea
{\pa}^{\b \bd} j_{\b \a_1\dots \a_{s-1} \bd \ad_1 \dots \ad_{s-1}} 
+ \frac{s-1}{s^2} \pa_{(\a_1 (\ad_1} t_{\a_2 \dots \a_{s-1}) \ad_2 \dots \ad_{s-1})}= 0 ~.~
\label{cons-eq1}
\eea
Our derivation of \eqref{cons-eq1} is analogous to that given in \cite{Anselmi}.

Let us introduce the following operators 
\begin{subequations} \label{notation}
\bea
{\pa}_{(1,1)} &:=& 2\ri \z^\a \bar \z^\ad \pa_{\a\ad}~, \\
{\pa}_{(-1,-1)} &:=& 2\ri {\pa}^{\a \ad} \frac{\pa}{\pa \z^\a} \frac{\pa}{\pa \bar \z^\ad}~.
\eea
\end{subequations}
The conservation equation \eqref{cons-eq1} then becomes
\bea
{\pa}_{(-1,-1)} j_{(s,s)} + (s-1) {\pa}_{(1,1)} t_{(s-2,s-2)} = 0
\label{cons-eq2}
\eea
Note that both $j_{(s,s)}$ and $t_{(s-2,s-2)}$ are real.

Let us now consider the model for $N$ massless real scalar fields $\f^i$, with $i=1,\dots N$,
in Minkowski space
\bea
S = - \hf  \int \rd^4x \, \pa_\m \f^i \pa^\m \f^i ~,
\label{Nreal}
\eea
which admits conserved higher spin currents of the form
\bea
j_{(s,s)} = {\rm i}^s \,C^{ij}\sum_{k=0}^s (-1)^k
\binom{s}{k} \binom{s}{k} 
{\pa}^k_{(1,1)} \f^i \,\,
{\pa}^{s-k}_{(1,1)} \f^j ~,
\label{C.1}
\eea
where $C^{ij}$ is a constant matrix. It can be shown that $j_{(s,s)}=0$ if $s$ is odd and $C^{ij}$ is symmetric. Similarly, $j_{(s,s)}=0$ if $s$ is even and $C^{ij}$ is antisymmetric. Thus, we have to consider two separate cases: the case of even $s$ with symmetric $C$ and, the case of odd $s$ with antisymmetric $C$. Using the massless equation of motion $\Box \f^i =0 ~,$ one may show that $j_{(s,s)}$ satisfies the conservation equation
\bea
{\pa_{(-1,-1)}}j_{(s,s)}=0 ~.
\eea

We now turn to the massive model
\bea
S = -\hf \int \rd^4x \, \Big\{  \pa_\m \f^i \pa^\m \f^i+(M^2)^{ij} \f^i \f^j \Big\}~,
\label{Nrealm}
\eea
where $M =(M^{ij})$ is a real, symmetric $N\times N$ mass matrix. In the massive theory, the conservation equation is described by \eqref{cons-eq2} and so we first need to compute $\pa_{(-1,-1)} j_{(s,s)}$ using the massive equations of motion
\bea
\Box \f^i  -(M^2)^{ij}\f^j=0~.
\eea
For symmetric $C$, we obtain 
\bea
{\pa}_{(-1,-1)} j_{(s,s)} &=&- 8(s+1)^2 (C M^2)^{ij} \sum_{k=0}^{s-1} (-1)^{k} \binom{s}{k} \binom{s}{k} 
\non\\
&&
\times \frac{(s-k)^2}{(k+1)(k+2)}
 {\pa}^{k}_{(1,1)} \,{\f}^j \,\,{\pa}^{s-k-1}_{(1,1)}{\f}^i~.
\label{C.4}
\eea
If $C^{ij}$ is antisymmetric, we get
\bea
{\pa}_{(-1,-1)} j_{(s,s)} &=& 8\ri(s+1)^2 (C M^2)^{ij} \sum_{k=0}^{s-1} (-1)^{k} \binom{s}{k} \binom{s}{k} 
\non\\
&&
\times \frac{(s-k)^2}{(k+1)(k+2)}
 {\pa}^{k}_{(1,1)} \,{\f}^j \,\,{\pa}^{s-k-1}_{(1,1)}{\f}^i ~.
\label{C.4.anti}
\eea
Thus, in the massive real scalars there are four cases to consider:
\begin{enumerate}
\item Both $C$ and $CM^2$ are symmetric $\Longleftrightarrow [C,M^2]=0, \,\, s$ even.
\item $C$ is symmetric; $CM^2$ is antisymmetric $\Longleftrightarrow \{C,M^2\}=0, \,\,s$ even.
\item $C$ is antisymmetric; $CM^2$ is symmetric $\Longleftrightarrow \{C,M^2\}=0, \,\,s$ odd.
\item Both $C$ and $CM^2$ are antisymmetric $\Longleftrightarrow [C,M^2]=0, \,\,s$ odd.
\end{enumerate}

\textbf{Case 1:} Eq. \eqref{C.4} is equivalent to 
\bea
{\pa}_{(-1,-1)} j_{(s,s)} &=& -4(s+1)^2 (C M^2)^{ij} \sum_{k=0}^{s-1} (-1)^{k} \binom{s}{k} \binom{s}{k} (s-k)
\non \\ 
&& 
\times \left\{\frac{s-k}{(k+1)(k+2)}+(-1)^{s-1} \frac{1}{s-k+1}\right\} 
 {\pa}^{k}_{(1,1)} {\f}^j \,\,{\pa}^{s-k-1}_{(1,1)} {\f}^i ~.
\label{C.4a}
\eea
We look for $t_{(s-2, s-2)}$ such that (i) it is real; and (ii) it satisfies the conservation equation \eqref{cons-eq2}. We consider a general ansatz
\bea
t_{(s-2, s-2)} = -(C M^2)^{ij}\sum_{k=0}^{s-2} d_k \,
{\pa}^k_{(1,1)} \f^j\,
{\pa}^{s-k-2}_{(1,1)} \f^i ~.
\label{C.5}
\eea
For $k = 1,2,...s-2$, condition (ii) gives
\begin{subequations}\label{C.6}
\begin{align}
d_{k-1} + d_k &= -4\frac{(s+1)^2}{s-1} (-1)^k\binom{s}{k} \binom{s}{k} (s-k)\label{C.6a}
\non \\
& \qquad \qquad \times \left\{ \frac{s-k}{(k+1)(k+2)} + (-1)^{s-1} \frac{1}{s-k+1} \right\} ~.
\end{align}
Condition (ii) also implies that 
\begin{align}
d_{s-2} + d_0 &= -4s (s+1)(s+2)~,\label{C.6b}
\end{align}
\end{subequations}
Equations \eqref{C.6} lead to the following expression for $d_k,\, k=1,2,\dots s-2$ 
\begin{subequations}\label{C.7}
\begin{align}
d_k &= (-1)^k d_0 - \frac{4(s+1)^2}{s-1}
\sum_{l=1}^k (-1)^k \binom{s}{l} \binom {s}{l}(s-l)\left\{ \frac{s-l}{(l+1)(l+2)}- \frac{1}{s-l+1} \right\}  ~, 
\end{align}
\begin{align}
d_0 &= d_{s-2} = -2s(s+1)(s+2)~.
\end{align}
\end{subequations}
One can check that the equations \eqref{C.6a}--\eqref{C.6b} are identically 
satisfied if $s$ is even. 

\textbf{Case 2:} If we take $C M^2$ to be antisymmetric, a similar analysis shows that no solution for $t_{(s-2,s-2)}$ exists for even $s$.

\textbf{Case 3:} Now we consider the case where $C$ is antisymmetric and $CM^2$ symmetric. Again, similar consideration shows that no solution for $t_{(s-2,s-2)}$ exists for odd $s$.

\textbf{Case 4:} Eq. \eqref{C.4.anti} is equivalent to
\bea
{\pa}_{(-1,-1)} j_{(s,s)} &=& 4\ri(s+1)^2 (C M^2)^{ij} \sum_{k=0}^{s-1} (-1)^{k} \binom{s}{k} \binom{s}{k} (s-k)
\non \\ 
&& 
\times \left\{\frac{s-k}{(k+1)(k+2)}- \frac{1}{s-k+1}\right\} 
 {\pa}^{k}_{(1,1)} {\f}^j \,\,{\pa}^{s-k-1}_{(1,1)} {\f}^i ~.
\eea
We consider a general ansatz
\bea
t_{(s-2, s-2)} = -\ri (C M^2)^{ij}\sum_{k=0}^{s-2} d_k \,
{\pa}^k_{(1,1)} \f^j\,
{\pa}^{s-k-2}_{(1,1)} \f^i ~.
\label{C.5.anti}
\eea
Imposing (i) and (ii) and keeping in mind that $s$ is odd, we obtain the following conditions for $d_k$:
\begin{subequations}\label{C.6.anti}
\begin{align}
d_{k-1} + d_k &= 4\frac{(s+1)^2}{s-1} (-1)^k\binom{s}{k} \binom{s}{k} (s-k)\label{C.6a.anti}
\non \\
& \qquad \qquad \times \left\{ \frac{s-k}{(k+1)(k+2)} - \frac{1}{s-k+1} \right\} ~.
\end{align}
Condition (ii) also implies that 
\begin{align}
d_{s-2} - d_0 &= -4s (s+1)(s+2)~,\label{C.6b.anti}
\end{align}
\end{subequations}
Equations \eqref{C.6.anti} lead to the following expression for $d_k,\, k=1,2,\dots s-2$ 
\begin{subequations}\label{C.7.anti}
\begin{align}
d_k &= (-1)^k d_0 + \frac{4(s+1)^2}{s-1}
\sum_{l=1}^k (-1)^k \binom{s}{l} \binom {s}{l}\left\{ \frac{(s-l)^2}{(l+1)(l+2)}- \frac{s-l}{s-l+1} \right\}  ~, 
\end{align}
\begin{align}
d_0 &=-d_{s-2} = 2s(s+1)(s+2)~.
\end{align}
\end{subequations}
One can check that the equations \eqref{C.6a.anti}--\eqref{C.6b.anti} are identically 
satisfied if $s$ is odd.



\section{Free Majorana fermions}
\label{AppendixD-spinor}


Let us now consider $N$ free massless Majorana fermions
\bea
S = -\ri \,\int \rd^4x \,\, {\j}^{\a i} {\pa}_{\a \ad} {\bar \j}^{\ad i}~,
\label{Nfermions}
\eea
with the equation of motion 
\bea
{\pa}_{\a \ad} {\bar \j}^{\ad i} \Longrightarrow \Box {\bar \j}_\ad^i =0 ~, \qquad i=1,\dots N ~.~
\label{eom1}
\eea
We can construct the following higher spin currents
\bea
j_{(s,s)} &=& C^{ij}\sum_{k=0}^{s-1} (-1)^k
\binom{s}{k} \binom{s}{k+1}\, 
{\pa}^k_{(1,1)}
 \z^\a \psi^{i}_\a \,\,
{\pa}^{s-k-1}_{(1,1)}
{\bar \z}^\ad {\bar \psi}^{j}_\ad ~, \qquad C^{ij}= C^{ji} ~,~
\label{D.1} \\
j_{(s,s)} &=& \ri\, C^{ij}\sum_{k=0}^{s-1} (-1)^k
\binom{s}{k} \binom{s}{k+1}\, 
{\pa}^k_{(1,1)}
 \z^\a \psi^{i}_\a \,\,
{\pa}^{s-k-1}_{(1,1)}
{\bar \z}^\ad {\bar \psi}^{j}_\ad ~, \qquad C^{ij}= -C^{ji} ~,~
\label{D.2}
\eea
where we put an extra $\ri$ in eq. \eqref{D.2} since $j_{(s,s)}$ has to be real. Using the equation of motion \eqref{eom}, it can be shown that the currents \eqref{D.1}, \eqref{D.2} are conserved
\bea
{\pa}_{(-1,-1)} j_{(s,s)} = 0~.~
\label{D.3}
\eea

We now look at the massive model
\bea
S = -\int \rd^4x \,\, \Big\{\ri {\j}^{\a i} {\pa}_{\a \ad} {\bar \j}^{\ad i}
+\Big(\hf M^{ij} {\j}^{\a i} \j_\a^j 
+ \hf \bar M^{ij} {\bar \j}_\ad^ i {\bar \j}^{\ad j}\Big)\Big\}~,
\label{fermions-massive}
\eea
where $M^{ij}$ is a constant symmetric $N\times N$ mass matrix.
To construct the conserved currents, we compute $\pa_{(-1,-1)} j_{(s,s)}$ using the massive equations of motion ($i=1,\dots,N$)
\begin{subequations}
\bea
\ri {\pa}_{\a \ad} {\bar \j}^{\ad i} + M^{ij}\j_\a^j &=&0 
\quad\Longrightarrow  \quad \Box {\bar \j}_\ad^i = (M \bar M)^{ij} \bar \j_\ad^j~,\\
-\ri {\pa}_{\a \ad} {\j}^{\a i} + \bar M^{ij} \bar \j_\ad^j &=&0 
\quad \Longrightarrow \quad \Box {\j}_\a^i = (\bar M M)^{ij} \j_\a^j~.
\eea
\end{subequations}
If $C^{ij}$ is a real symmetric matrix, we find
\bea 
{\pa}_{(-1,-1)} j_{(s,s)} &=& -2(s+1)\sum_{k=0}^{s-1}\frac{k+1}{s-k+1} (-1)^k \binom{s}{k} \binom{s}{k+1} 
\non \\
&& 
\times \left\{(CM)^{ij} {\pa}^{k}_{(1,1)} {\j}^{\a i} \,\,{\pa}^{s-k-1}_{(1,1)} {\j}_\a^j +(-1)^s (C\bar M)^{ij} {\pa}^{k}_{(1,1)} {\bar \j}_\ad^i \,\,{\pa}^{s-k-1}_{(1,1)} {\bar \j}^{\ad j}\right\}
\non \\ 
&&+ 4(s+1)(s+2)\sum_{k=1}^{s-1} k (-1)^k \binom{s}{k} \binom{s}{k+1}
\non \\
&&\times \left\{ \frac{1}{k+2} (M \bar M C)^{ij} -\frac{k+1}{(s-k+2)(s-k+1)} (C M \bar M)^{ij} \right\} 
\non \\
&&\times {\pa}^{k-1}_{(1,1)} \z^\a {\j}_\a^ i \,\,{\pa}^{s-k-1}_{(1,1)} \bar \z^\ad {\bar \j}_\ad^j ~.~
\label{D.4}
\eea 
If $C^{ij}$ is antisymmetric, we have
\bea 
{\pa}_{(-1,-1)} j_{(s,s)} &=& -2\ri(s+1)\sum_{k=0}^{s-1}\frac{k+1}{s-k+1} (-1)^k \binom{s}{k} \binom{s}{k+1} 
\non \\
&& 
\times \left\{(CM)^{ij} {\pa}^{k}_{(1,1)} {\j}^{\a i} \,\,{\pa}^{s-k-1}_{(1,1)} {\j}_\a^j +(-1)^{s-1}(C\bar M)^{ij} {\pa}^{k}_{(1,1)} {\bar \j}_\ad^i \,\,{\pa}^{s-k-1}_{(1,1)} {\bar \j}^{\ad j}\right\}
\non \\ 
&&+ 4\ri(s+1)(s+2)\sum_{k=1}^{s-1} k (-1)^k \binom{s}{k} \binom{s}{k+1}
\non \\
&&\times \left\{ \frac{1}{k+2} (M \bar M C)^{ij} -\frac{k+1}{(s-k+2)(s-k+1)} (C M \bar M)^{ij} \right\} 
\non \\
&&\times {\pa}^{k-1}_{(1,1)} \z^\a {\j}_\a^ i \,\,{\pa}^{s-k-1}_{(1,1)} \bar \z^\ad {\bar \j}_\ad^j ~.
\label{D.4a}
\eea

There are four cases to consider:
\begin{enumerate}
\item $C, CM, CM \bar M$ are symmetric $\Longleftrightarrow [C,M]=[C,\bar M]=0, [M,\bar M]=0$.
\item $C, C M \bar M$ symmetric; $C M$ antisymmetric  $\Longleftrightarrow \{C,M\}=\{C,\bar M\}=0, [M,\bar M]=0$.
\item $C, CM \bar M$ antisymmetric; $CM$ symmetric $\Longleftrightarrow \{C,M\}=\{C,\bar M\}=0, [M,\bar M]=0$.
\item $C, CM, CM \bar M$ are antisymmetric $\Longleftrightarrow [C,M]=[C,\bar M]=0, [M,\bar M]=0$.
\end{enumerate}

\textbf{Case 1:} Eq. \eqref{D.4} becomes
\bea
{\pa}_{(-1,-1)} j_{(s,s)} &=& -(s+1)\sum_{k=0}^{s-1}(-1)^k \binom{s}{k} \binom{s}{k+1} 
\non \\
&&\times \left\{\frac{k+1}{s-k+1} +(-1)^{s-1} \frac{s-k}{k+2}\right\}(CM)^{ij}\,\, {\pa}^{k}_{(1,1)} {\j}^{\a i} \,\,{\pa}^{s-k-1}_{(1,1)} {\j}_\a^j 
\non \\
&&+ (-1)^{s-1}(s+1)\sum_{k=0}^{s-1}(-1)^k \binom{s}{k} \binom{s}{k+1} 
\non \\
&& 
\times \left\{\frac{k+1}{s-k+1} +(-1)^{s-1} \frac{s-k}{k+2}\right\}(C\bar M)^{ij}\,\, {\pa}^{k}_{(1,1)} {\bar \j}_{\ad}^ i \,\,{\pa}^{s-k-1}_{(1,1)} {\bar \j}^{\ad j} 
\non \\ 
&&+ 4(s+1)(s+2)\sum_{k=1}^{s-1} k (-1)^k \binom{s}{k} \binom{s}{k+1}
\non \\
&& \times \left\{ \frac{1}{k+2} -\frac{k+1}{(s-k+2)(s-k+1)}\right\}
\non \\
&&\times (C M \bar M)^{ij} \,\, {\pa}^{k-1}_{(1,1)} \z^\a {\j}_\a^ i \,\,{\pa}^{s-k-1}_{(1,1)} \bar \z^\ad {\bar \j}_\ad^j ~.
\label{D.4new}
\eea
We look for $t_{(s-2, s-2)}$ such that (i) it is real; and (ii) it satisfies the conservation equation \eqref{cons-eq2}:
\bea
{\pa}_{(-1,-1)} j_{(s,s)} = -(s-1) {\pa}_{(1,1)} t_{(s-2, s-2)}~. 
\eea
Consider a general ansatz
\bea
t_{(s-2, s-2)} &=& (C M)^{ij}\,\sum_{k=0}^{s-2} c_k \,
{\pa}^k_{(1,1)} \j^{\a i}\,\,{\pa}^{s-k-2}_{(1,1)} \j^j_\a
\non \\
&&+(-1)^s(C \bar M)^{ij}\,\sum_{k=0}^{s-2} c_k \,
{\pa}^k_{(1,1)} {\bar \j}^i_\ad \,\,{\pa}^{s-k-2}_{(1,1)} {\bar \j}^{\ad j}
\non \\
&&+ (C M \bar M)^{ij} \, \sum_{k=1}^{s-2} g_k \,{\pa}^{k-1}_{(1,1)} \z^\a {\j}_\a^ i \,\,{\pa}^{s-k-1}_{(1,1)} \bar \z^\ad {\bar \j}_\ad^j  ~.
\label{D.5}
\eea
For $k = 1,2,...s-2$, condition (i) gives
\begin{subequations}\label{D.6}
\begin{align}
g_k &= (-1)^{s-1} g_{s-1-k}~, \label{D.6a}
\end{align}
while condition (ii) gives
\begin{align}
c_{k-1} + c_k &= \frac{s+1}{s-1} (-1)^k\binom{s}{k} \binom{s}{k+1} \left\{\frac{k+1}{s-k+1}+(-1)^{s-1} \frac{s-k}{k+2} \right\}~,\label{D.6b}
\end{align}
\begin{align}
g_{k-1} + g_k &= -4\frac{(s+1)(s+2)}{s-1} (-1)^k\binom{s}{k} \binom{s}{k+1} k\left\{\frac{1}{k+2}- \frac{k+1}{(s-k+2)(s-k+1)} \right\}~.\label{D.6c}
\end{align}
Condition (ii) also implies that 
\begin{align}
c_{s-2} + c_0 &= \frac{1}{s-1} \left\{2s + (-1)^{s-1} s^2(s+1)\right\}~,\label{D.6d}
\end{align}
\begin{align}
g_1 &= \frac{2s(s-2)}{3} (s^2+5s+6)~,\label{D.6e}
\end{align}
\begin{align}
g_{s-2} &= (-1)^{s-1} \frac{2s(s-2)}{3} (s^2+5s+6)~~.\label{D.6f}
\end{align}
\end{subequations}
The above conditions lead to the following expressions for $c_k$ and $g_k$ ($k=1,2,\dots s-2$) 
\begin{subequations}\label{D.7}
\begin{align}
c_k &= (-1)^k c_0 + \frac{s+1}{s-1} 
\sum_{l=1}^k (-1)^k \binom{s}{l} \binom {s}{l+1}\left\{ \frac{l+1}{s-l+1}+ (-1)^{s-1} \frac{s-l}{l+2} \right\}  ~, \label{D.7a}
\end{align}
\begin{align}
g_k &= 4(-1)^k \frac{(s+1)(s+2)}{s-1} 
\sum_{l=1}^k \binom{s}{l} \binom {s}{l+1}\left\{ \frac{l(l+1)}{(s-l+1)(s-l+2)}- \frac{l}{l+2} \right\}  ~. \label{D.7b}
\end{align}
If the parameter $s$ is even, \eqref{D.7a} gives 
\begin{align}
c_{s-2} &= c_0 = -\hf s(s+2)\label{D.7c}
\end{align}
\end{subequations}
and \eqref{D.6a}-\eqref{D.6f} are identically satisfied. However, when $s$ is odd, there appears an inconsistency: 
the right-hand side of \eqref{D.6d} is positive, while the left-hand side 
is negative, $c_{s-2} + c_0 < 0$. Therefore, our solution \eqref{D.7} is only consistent for $s=2n, n=1,2,\dots$. 

\textbf{Case 2:} If $C M$ is antisymmetric while $C M \bar M$ symmetric, eq.~\eqref{D.4} is slightly modified
\bea
{\pa}_{(-1,-1)} j_{(s,s)} &=& -(s+1)\sum_{k=0}^{s-1}(-1)^k \binom{s}{k} \binom{s}{k+1} 
\non \\
&&\times \left\{\frac{k+1}{s-k+1} +(-1)^{s} \frac{s-k}{k+2}\right\}(CM)^{ij}\,\, {\pa}^{k}_{(1,1)} {\j}^{\a i} \,\,{\pa}^{s-k-1}_{(1,1)} {\j}_\a^j 
\non \\
&&+ (-1)^{s-1}(s+1)\sum_{k=0}^{s-1}(-1)^k \binom{s}{k} \binom{s}{k+1} 
\non \\
&& 
\times \left\{\frac{k+1}{s-k+1} +(-1)^{s} \frac{s-k}{k+2}\right\}(C\bar M)^{ij}\,\, {\pa}^{k}_{(1,1)} {\bar \j}_{\ad}^ i \,\,{\pa}^{s-k-1}_{(1,1)} {\bar \j}^{\ad j} 
\non \\ 
&&+ 4(s+1)(s+2)\sum_{k=1}^{s-1} k (-1)^k \binom{s}{k} \binom{s}{k+1}
\non \\
&& \times \left\{ \frac{1}{k+2} -\frac{k+1}{(s-k+2)(s-k+1)}\right\}
\non \\
&&\times (C M \bar M)^{ij} \,\, {\pa}^{k-1}_{(1,1)} \z^\a {\j}_\a^ i \,\,{\pa}^{s-k-1}_{(1,1)} \bar \z^\ad {\bar \j}_\ad^j ~.
\label{D.4b}
\eea
Starting with a general ansatz
\bea
t_{(s-2, s-2)} &=& (C M)^{ij}\,\sum_{k=0}^{s-2} d_k \,
{\pa}^k_{(1,1)} \j^{\a i}\,\,{\pa}^{s-k-2}_{(1,1)} \j^j_\a
\non \\
&&+(-1)^s(C \bar M)^{ij}\,\sum_{k=0}^{s-2} d_k \,
{\pa}^k_{(1,1)} {\bar \j}^i_\ad \,\,{\pa}^{s-k-2}_{(1,1)} {\bar \j}^{\ad j}
\non \\
&&+ (C M \bar M)^{ij} \, \sum_{k=1}^{s-2} g_k \,{\pa}^{k-1}_{(1,1)} \z^\a {\j}_\a^ i \,\,{\pa}^{s-k-1}_{(1,1)} \bar \z^\ad {\bar \j}_\ad^j  
\label{D.8}
\eea
and imposing conditions (i) and (ii) yield 
\begin{subequations}\label{D.9}
\begin{align}
g_k &= (-1)^{s-1} g_{s-1-k}~, \label{D.9a}
\end{align}
\begin{align}
d_{k-1} + d_k &= \frac{s+1}{s-1} (-1)^k\binom{s}{k} \binom{s}{k+1} \left\{\frac{k+1}{s-k+1}-(-1)^{s-1} \frac{s-k}{k+2} \right\}~,\label{D.9b}
\end{align}
\begin{align}
g_{k-1} + g_k &= -4\frac{(s+1)(s+2)}{s-1} (-1)^k\binom{s}{k} \binom{s}{k+1} k\left\{\frac{1}{k+2}- \frac{k+1}{(s-k+2)(s-k+1)} \right\}~,\label{D.9c}
\end{align}
\begin{align}
d_{0} - d_{s-2} &= \frac{1}{s-1} \left\{2s + (-1)^{s} s^2(s+1)\right\}~,\label{D.9d}
\end{align}
\begin{align}
g_1 &= \frac{2s(s-2)}{3} (s^2+5s+6)~,\label{D.9e}
\end{align}
\begin{align}
g_{s-2} &= (-1)^{s-1} \frac{2s(s-2)}{3} (s^2+5s+6)~~.\label{D.9f}
\end{align}
\end{subequations}
As a result, the coefficients $d_k$ and $g_k$ are given by ($k=1,2,\dots s-2$) 
\begin{subequations}\label{D.10}
\begin{align}
d_k &= (-1)^k d_0 + \frac{s+1}{s-1} 
\sum_{l=1}^k (-1)^k \binom{s}{l} \binom {s}{l+1}\left\{ \frac{l+1}{s-l+1}- (-1)^{s-1} \frac{s-l}{l+2} \right\}  ~, \label{D.10a}
\end{align}
\begin{align}
g_k &= 4(-1)^k \frac{(s+1)(s+2)}{s-1} 
\sum_{l=1}^k \binom{s}{l} \binom {s}{l+1}\left\{ \frac{l(l+1)}{(s-l+1)(s-l+2)}- \frac{l}{l+2} \right\}  ~. \label{D.10b}
\end{align}
When the parameter $s$ is odd, \eqref{D.10a} gives 
\begin{align} \label{D.10c}
d_{s-2} &= -d_0 = \hf s(s+2)
\end{align}
\end{subequations}
and \eqref{D.9a}-\eqref{D.9f} are identically satisfied. However, when $s$ is even, there appears an inconsistency: 
the right-hand side of \eqref{D.9d} is positive, while the left-hand side 
is negative, $d_{0} - d_{s-2} < 0$. Therefore, our solution \eqref{D.10} is only consistent for $s=2n+1, n=1,2,\dots$. 

Finally, we consider $C^{ij}=-C^{ji}$ with the corresponding $j_{(s,s)}$ given by \eqref{D.2}. Similar considerations show that in \textbf{Case 3}, the non-conformal currents exist only if $s$ is even. The trace $t_{(s-2,s-2)}$ is given by \eqref{D.5} with the coefficients $c_k$ and $g_k$ given by
\begin{subequations}
\begin{align}
c_k &= \ri(-1)^k c_0 + \ri \frac{s+1}{s-1} 
\sum_{l=1}^k (-1)^k \binom{s}{l} \binom {s}{l+1}\left\{ \frac{l+1}{s-l+1}+ (-1)^{s-1} \frac{s-l}{l+2} \right\}  ~, \label{ck.a}
\end{align}
\begin{align}
g_k &=4 \ri  \, (-1)^k \frac{(s+1)(s+2)}{s-1} 
\sum_{l=1}^k \binom{s}{l} \binom {s}{l+1}\left\{ \frac{l(l+1)}{(s-l+1)(s-l+2)}- \frac{l}{l+2} \right\}  ~. \label{gk.a}
\end{align}
\end{subequations}
In \textbf{Case 4}, the non-conformal currents exist only for odd values of $s$. The trace $t_{(s-2,s-2)}$ is given by \eqref{D.8} with the coefficients $d_k$ and $g_k$ given by
\begin{subequations}
\begin{align}
d_k &= \ri (-1)^k d_0 + \ri \frac{s+1}{s-1} 
\sum_{l=1}^k (-1)^k \binom{s}{l} \binom {s}{l+1}\left\{ \frac{l+1}{s-l+1}- (-1)^{s-1} \frac{s-l}{l+2} \right\}  ~, \label{dk.b}
\end{align}
\begin{align}
g_k &= 4\ri \, (-1)^k \frac{(s+1)(s+2)}{s-1} 
\sum_{l=1}^k \binom{s}{l} \binom {s}{l+1}\left\{ \frac{l(l+1)}{(s-l+1)(s-l+2)}- \frac{l}{l+2} \right\}  ~. \label{gk.b}
\end{align}
\end{subequations}
We observe that the coefficients $c_k$ and $g_k$ in eq.~\eqref{ck.a} and \eqref{gk.a}, respectively differ from similar coefficients in \eqref{D.7a} and \eqref{D.7b} by a factor of $\ri$. Hence, for even $s$ we may define a more general supercurrent
\bea
j_{(s,s)} = C^{ij}\sum_{k=0}^{s-1} (-1)^k
\binom{s}{k} \binom{s}{k+1}\, 
{\pa}^k_{(1,1)}
 \z^\a \psi^{i}_\a \,\,
{\pa}^{s-k-1}_{(1,1)}
{\bar \z}^\ad {\bar \psi}^{j}_\ad ~,
\label{j.even}
\eea
where $C^{ij}$ is a generic matrix which can be split into the symmetric and antisymmetric parts: $C^{ij} =S^{ij}+\ri A^{ij}$. Here both $S$ and $A$ are real and we put an $\ri$ in front of $A$ because $j_{(s,s)} $ must be real. From the above consideration it then follows that the corresponding more general solution for $t_{(s-2, s-2)}$ reads
\bea
t_{(s-2, s-2)} &=& (C M)^{ij}\,\sum_{k=0}^{s-2} c_k \,
{\pa}^k_{(1,1)} \j^{\a i}\,\,{\pa}^{s-k-2}_{(1,1)} \j^j_\a
\non \\
&&+(-1)^s(\bar C \bar M)^{ij}\,\sum_{k=0}^{s-2} c_k \,
{\pa}^k_{(1,1)} {\bar \j}^i_\ad \,\,{\pa}^{s-k-2}_{(1,1)} {\bar \j}^{\ad j}
\non \\
&&+ (C M \bar M)^{ij} \, \sum_{k=1}^{s-2} g_k \,{\pa}^{k-1}_{(1,1)} \z^\a {\j}_\a^ i \,\,{\pa}^{s-k-1}_{(1,1)} \bar \z^\ad {\bar \j}_\ad^j  ~,
\label{t.even}
\eea
where $[S, M]=[S, \bar M]=0$, $\{A, M\}=\{A, \bar M\}=0$ and $[M, \bar M]=0$. The coefficients $c_k$ and $g_k$ are given by eqs.~\eqref{D.7a} and \eqref{D.7b}, respectively.
Similarly, the coefficients $d_k$ and $g_k$ in~\eqref{dk.b} and \eqref{gk.b} differ from similar coefficients in~\eqref{D.10a} and \eqref{D.10b} by a factor of $\ri$. This means that for odd $s$ we can define a more general supercurrent~\eqref{j.even}, where $C^{ij}$ is a generic matrix which we can split as before into the symmetric and antisymmetric parts,  $C^{ij} = S^{ij} + \ri A^{ij}$. 
From the above consideration it then follows that the corresponding more general solution for $t_{(s-2, s-1)}$ reads
\bea
t_{(s-2, s-1)} &=& (C M)^{ij}\,\sum_{k=0}^{s-2} d_k \,
{\pa}^k_{(1,1)} \j^{\a i}\,\,{\pa}^{s-k-2}_{(1,1)} \j^j_\a
\non \\
&&+(-1)^s(\bar C \bar M)^{ij}\,\sum_{k=0}^{s-2} d_k \,
{\pa}^k_{(1,1)} {\bar \j}^i_\ad \,\,{\pa}^{s-k-2}_{(1,1)} {\bar \j}^{\ad j}
\non \\
&&+ (C M \bar M)^{ij} \, \sum_{k=1}^{s-2} g_k \,{\pa}^{k-1}_{(1,1)} \z^\a {\j}_\a^ i \,\,{\pa}^{s-k-1}_{(1,1)} \bar \z^\ad {\bar \j}_\ad^j  ~,
\label{t.odd}
\eea
where $\{S, M\}=\{S, \bar M\}=0$, $[A, M]=[A, \bar M]=0$ and $[M, \bar M]=0$. The coefficients $d_k$ and $g_k$ are given by eqs.~\eqref{D.10a} and \eqref{D.10b}, respectively. 


\chapter{Component analysis of ${\cal N}=1$ higher-spin actions in three dimensions} 
\label{AppendixBB}
In this appendix we discuss the component structure of the two new off-shell ${\cal N}=1$ supersymmetric higher-spin theories in three dimensions: the transverse massless superspin-$s$ multiplet \eqref{action-t2-new}, and the transverse massless superspin-$(s+\hf)$ multiplet \eqref{action-t3-new-complete}. The longitudinal actions \eqref{action-t2-half} and \eqref{action-t3} can be reduced to components in a similar fashion. For simplicity we will carry out our analysis in flat Minkowski superspace. 
This material has been drawn from \cite{HK19}.


\section{Massless superspin-$s$ action} \label{AppendixBB-1}
In accordance with \eqref{252}, the component form of an ${\cal N}=1$ supersymmetric action is computed by the rule
\bea
S= \int \rd^{3|2}z
 \, L = \frac{\ri}{4} \int \rd^3 x \, D^2 L \Big|_{\q=0}~, \qquad L = \bar{L}~. 
\label{comp}
\eea

Let us first work out the component structure of the massless integer superspin model \eqref{action-t2-new}. In the flat superspace limit, the transverse action \eqref{action-t2-new} takes the form
\bea
\lefteqn{S^{\perp}_{(s)}[H_{\a(2s)} ,{\Psi}_{\b; \,\a(2s-2)} ]
= \Big(-\hf \Big)^{s} 
\int 
\rd^{3|2}z\,
 \bigg\{\frac{\ri}{2} H^{\a(2s)} D^2 H_{\a(2s)}}
\non \\
&&\qquad - \ri s D_{\b} H^{\b \a(2s-1)} D^{\g}H_{\g \a(2s-1)}  -(2s-1) \cW^{\b \a(2s-2)} D^{\g} H_{\g \b \a(2s-2)} 
\non \\
&& \qquad -\frac{\ri}{2} (2s-1)\Big(\cW^{\b ;\, \a(2s-2)} \cW_{\b ;\, \a(2s-2)}+\frac{s-1}{s} \cW_{\b;}\,^{\b \a(2s-3)} \cW^{\g ;}\,_{\g \a(2s-3)} \Big) 
\bigg\}~.~~~~~
\eea

As described in \eqref{3.355}, it is possible to choose a gauge condition $\J_{(\a_1;\, \a_2 \cdots \a_{2s-1})}=0$~, such that the above action turns into 
\bea
\lefteqn{S^{\perp}_{(s)}[H_{\a(2s)} ,{\Psi}_{\b; \,\a(2s-2)} ]
= \Big(-\hf \Big)^{s} 
\int 
\rd^{3|2}z\,
 \bigg\{\frac{\ri}{2} H^{\a(2s)} D^2 H_{\a(2s)}}
\non \\
&& \qquad \quad - \ri s D_{\b} H^{\b \a(2s-1)} D^{\g}H_{\g \a(2s-1)} -2(s-1) \vf^{\a(2s-3)} \pa^{\b \g} D^{\d} H_{\b \g \d \a(2s-3)}
\non \\
&&\qquad \quad -\frac{2\ri}{s}(s-1) \vf^{\a(2s-3)} \Box \vf_{\a(2s-3)} -\frac{\ri (s-1)(s-2)(2s-3)}{s(2s-1)} \pa_{\d \l}\vf^{\d \l \a(2s-5)} \pa^{\b \g} \vf_{\b \g \a(2s-5)} \non \\
&&\qquad \quad + \frac{\ri(s-1)(2s-3)}{2s(2s-1)} D_{\b} \vf^{\b \a(2s-4)} D^2 D^{\g} \vf_{\g \a(2s-4)}
\bigg\}~.
\label{BB.1}
\eea
It is invariant under the following gauge transformations
\begin{subequations}
\bea
\d H_{\a(2s)} &=& -\pa_{(\a_1 \a_2} \eta_{\a_3 \dots \a_{2s})}~,\\
\d \vf_{\a(2s-3)}&=& \ri D^{\b} \eta_{\b \a(2s-3)}~,
\eea
\label{BB.2}
\end{subequations}
where the gauge parameter $\eta_{\a(2s-2)}$ is a real unconstrained superfield. 

The gauge freedom \eqref{BB.2} can be used to impose a Wess-Zumino gauge
\bea
\vf_{\a(2s-3)}|_{\q=0}=0~, \qquad D_{(\a_1} \vf_{\a_2 \cdots \a_{2s-2})}|_{\q=0} =0~. 
\label{BB.3}
\eea
In order to preserve these gauge conditions, the residual gauge freedom has to be constrained by
\bea
D^{\b}\eta_{\b \a(2s-3)}|_{\q=0}=0~, \qquad D^2 \eta_{\a(2s-2)}|_{\q=0} = 2\ri \, \pa^{\b}\,_{(\a_1} \eta_{\a_2 \cdots \a_{2s-2}) \b}|_{\q=0}~.
\eea
These imply that there remain two independent, real components of $\eta_{\a(2s-2)}$:
\bea
\xi_{\a(2s-2)}:= \eta_{\a(2s-2)}|_{\q=0}~, \qquad \l_{\a(2s-1)} := \ri D_{(\a_1} \eta_{\a_2 \cdots \a_{2s-1})}|_{\q=0}~.
\eea
In the gauge \eqref{B.3}, the independent component fields of $\vf_{\a(2s-3)}$ can be chosen as 
\bea
y_{\a(2s-4)}:= -\frac{2s-2}{2s-1} D^{\b} \vf_{\b \a_1 \cdots \a_{2s-4}}|_{\q=0}~, \qquad y_{\a(2s-3)} := \frac{\ri}{2} D^2 \vf_{\a(2s-3)}|_{\q=0}~.
\eea
We define the component fields of $H_{\a(2s)}$ as
\bea
h_{\a(2s)} &:=& -H_{\a(2s)}|_{\q=0}~,\\
h_{\a(2s+1)} &:=& \ri \frac{s}{2s+1} D_{(\a_1} H_{\a_2 \cdots \a_{2s+1})}|_{\q=0}~, \qquad 
y_{\a(2s-1)} := \ri D^{\b} H_{\b \a_1 \cdots \a_{2s-1}}|_{\q=0}~,~~\\
F_{\a(2s)} &:=& \frac{\ri}{4} D^2 H_{\a(2s)} |_{\q=0}~.
\eea

Applying the reduction rule \eqref{comp} to the ${\cal N}=1$ action \eqref{BB.1}, we find that it splits into bosonic and fermionic parts:
\bea
S^{\perp}_{(s)}[H_{\a(2s)} ,{\Psi}_{\b; \,\a(2s-2)} ] = S_{\rm bos} + S_{\rm ferm}~.
\eea
The bosonic action is given by
\bea
S_{\rm bos}
&=& \Big(-\hf \Big)^{s}
\int 
\rd^3x \,
 \bigg\{ 2(1-s) F^{\a(2s)} F_{\a(2s)} + 2s F^{\a(2s-1) \b} \pa_{\b}\,^{\g}  h_{\a(2s-1)\g}
\non\\
&&-\hf (s-1) h^{\a(2s)} \Box h_{\a(2s)}-\frac{(2s-1)(2s-3)}{2s(s-1)}y^{\a(2s-4)} \Box y_{\a(2s-4)} 
\non\\
&&- \frac{(2s-1)(2s-3)}{4(s-1)} y^{\a(2s-4)}\pa^{\b\g}\pa^{\d \l}h_{\b \g \d \l \a(2s-4)}
\non\\
&&- \frac{(s-2)(2s-1)(2s-3)(2s-5)}{16s(s-1)^2} \pa_{\d \l} y^{\d \l \a(2s-6)} \pa^{\b \g} y_{\b \g \a(2s-6)}
\bigg\}~.
\eea
Integrating out the auxiliary field $F_{\a(2s)}$ leads to 
\bea
S_{\rm bos}
&=& \Big(-\hf \Big)^{s} \,\frac{2s-1}{2s-2}
\int 
\rd^3x \,
 \bigg\{h^{\a(2s)} \Box h_{\a(2s)}-\frac{s}{2}\pa_{\d \l} h^{\d \l \a(2s-2)} \pa^{\b \g} h_{\b \g \a(2s-2)}
\non\\ 
&&-\frac{2s-3}{2s} \Big[ s y^{\a(2s-4)}\pa^{\b\g}\pa^{\d \l}h_{\b \g \d \l \a(2s-4)}+ 2 y^{\a(2s-4)} \Box y_{\a(2s-4)} 
\non\\
&&+ \frac{(s-2)(2s-5)}{4(s-1)} \pa_{\d \l} y^{\d \l \a(2s-6)} \pa^{\b \g} y_{\b \g \a(2s-6)} \Big]
\bigg\}~.
\label{sbos}
\eea
This action is invariant under the gauge transformations
\bea
\d_{\x} h_{\a(2s)} &=& \pa_{(\a_1 \a_2} \x_{\a_3 \cdots \a_{2s})}~,\\
\d_{\x} y_{\a(2s-4)} &=& \frac{2s-2}{2s-1} \pa^{\b\g} \x_{\b \g \a_1 \cdots \a_{2s-4}}~.
\eea
The gauge transformations for the fields $h_{\a(2s)}$ and $y_{\a(2s-4)}$ can be easily read off from the gauge transformations of the superfields $H_{\a(2s)}$ and $\vf_{\a(2s-3)}$~, respectively. 
Modulo an overall normalisation factor, \eqref{sbos} corresponds to the massless Fronsdal spin-$s$ action $S^{(2s)}_F$ described in \cite{KP1}.

The fermionic sector of the component action is described by the real dynamical fields $h_{\a(2s+1)}$, $y_{\a(2s-1)}$, $y_{\a(2s-3)}$~, defined modulo gauge transformations of the form
\bea
\d_{\l} h_{\a(2s+1)} &=& \pa_{(\a_1 \a_2} \l_{\a_3 \cdots \a_{2s+1})}~,\\
\d_{\l} y_{\a(2s-1)} &=&\frac{1}{2s+1} \pa^{\b}\,_{(\a_1} \l_{\a_2 \cdots \a_{2s-1})\b}~, \\
\d_{\l} y_{\a(2s-3)} &=& \pa^{\b\g} \l_{\b \g \a_1 \cdots \a_{2s-3}}~.
\eea
The gauge-invariant action is
\bea
S_{\rm ferm}
&=& \Big(-\hf \Big)^{s} \,\frac{\ri}{2}
\int 
\rd^3x \,
 \bigg\{h^{\a(2s)\b} \pa_{\b }\,^{\g} h_{\a(2s)\g}+ 2(2s-1) y^{\a(2s-1)} \pa^{\b \g } h_{\b \g \a(2s-1)}
\non\\ 
&&+ 4(2s-1) y^{\a(2s-2) \b}\pa_{\b }\,^{\g} y_{\a(2s-2)\g} + \frac{2}{s}(2s+1)(s-1) y^{\a(2s-3)} \pa^{\b \g } y_{\b \g \a(2s-3)}
\non\\
&&- \frac{(s-1)(2s-3)}{s(2s-1)} y^{\a(2s-4)\b} \pa_{\b }\,^{\g} y_{\a(2s-4)\g}
\bigg\}~.
\label{ff-action}
\eea
It may be shown that $S_{\rm ferm}$ coincides with the Fang-Fronsdal spin-$(s+\hf)$ action, $S^{(2s+1)}_{FF}$ \cite{KP1}. 

We have thus proved that at the component level and upon elimination of the auxiliary field, the transverse theory \eqref{B.1} is equivalent to a sum of two massless models: the bosonic Fronsdal spin-$s$ model and the fermionic Fang-Fronsdal spin-$(s+\hf)$ model. 

\section{Massless superspin-$(s+\hf)$ action}
We now elaborate on the component structure of the massless half-integer superspin model in the transverse formulation \eqref{action-t3-new}. The theory is described in terms of the real unconstrained prepotentials $H_{\a(2s+1)}$ and $\U_{\b; \,\a(2s-2)}$. 
In Minkowski superspace, the action \eqref{action-t3-new} simplifies into
\bea
\lefteqn{S^{\perp}_{(s+\hf)}[{H}_{(2s+1)} ,\U_{\b; \,\a(2s-2)} ]
= \Big(-\hf \Big)^{s} 
\int 
\rd^{3|2}z\,
\bigg\{-\frac{\ri}{2} H^{\a(2s+1)} {\Box} H_{\a(2s+1)} } 
\non \\
&&\qquad -\frac{\ri}{8} D_{\b} H^{\b \a(2s)} D^2 D^{\g}H_{\g \a(2s)}+\frac{\ri}{8}{\pa}_{\b \g}H^{\b \g \a(2s-1)} {\pa}^{\rho \d}H_{\rho \d \a(2s-1)}
\non \\
&&\qquad -\frac{\ri}{4} (2s-1) \O^{\b; \, \a(2s-2)} \pa^{\g \d}H_{\b \g \d \a(2s-2)}
\non \\
&&\qquad  -\frac{\ri}{8}(2s-1)\Big(\O^{\b ;\, \a(2s-2)} \O_{\b ;\, \a(2s-2)}
-2(s-1)\O_{\b;}\,^{\b \a(2s-3)} \O^{\g ;}\,_{\g \a(2s-3)}  \Big) 
\bigg\}~,
\label{CC.1}
\eea
with the following gauge symmetry
\begin{subequations}
\bea
\d H_{\a(2s+1)} &=& \ri D_{(\a_1} \z_{\a_2 \dots \a_{2s+1})}~,\\
\d \U_{\b;\, \a(2s-2)}&=& \frac{\ri}{2s+1}\left( D^{\g} \z_{\g \b \a(2s-2)}+ (2s+1) D_{\b} \eta_{\a(2s-2)} \right)~.
\eea \label{C1}
\end{subequations}
The action \eqref{CC.1} involves the real field strength $\O_{\b;\, \a(2s-2)}$
\bea
 \O_{\b; \a(2s-2)} = -\ri D^{\g} D_{\b}\U_{\g;\a(2s-2)}~, \qquad D^{\b}\O_{\b;\, \a(2s-2)} =0~.
\eea

The gauge transformations \eqref{C1} allow us to impose a Wess-Zumino gauge on the prepotentials:
\bea
H_{\a(2s+1)}|_{\q=0}=0~, \,\, D^{\b}H_{\b \a_1 \cdots \a_{2s}}|_{\q=0}=0~, \,\, \U_{\b; \, \a(2s-2)}|_{\q=0}=0~, \,\, D^{\b} \U_{\b; \, \a(2s-2)}|_{\q=0}=0~.~~~~~~~~
\label{C.2}
\eea
The residual gauge symmetry preserving the conditions \eqref{C.2} is characterised by
\bsubeq
\bea
D_{(\a_1} \z_{\a_2 \cdots \a_{2s+1})}|_{\q=0}&=&0~, \qquad 
D^2 \z_{\a(2s)}|_{\q=0} = -\frac{2\ri s}{s+1} \pa^{\b}\,_{(\a_1} \z_{\a_2 \cdots \a_{2s}) \b}|_{\q=0}~, \\
D_{\b} \eta_{\a(2s-2)}|_{\q=0} &=& D_{(\b} \eta_{\a(2s-2))}|_{\q=0} = -\frac{1}{2s+1} D^{\g} \z_{\g \b \a(2s-2)}|_{\q=0}~, \\
D^2 \eta_{\a(2s-2)}|_{\q=0} &=&-\frac{\ri}{2s+1} \pa^{\b \g} \z_{\b \g \a(2s-2)}|_{\q=0}~.
\eea
\esubeq
As a result, there are three independent, real gauge parameters at the component level, which we define as
\bea
\xi_{\a(2s)}:= \z_{\a(2s)}|_{\q=0}~, \quad \l_{\a(2s-1)} := -\ri \frac{s}{2s+1} D^{\b} \z_{\b \a(2s-1)}|_{\q=0}~, \quad \rho_{\a(2s-2)}:= \eta_{\a(2s-2)}|_{\q=0}~.~~~~~~
\eea
Let us now represent the prepotential $\U_{\b; \, \a(2s-2)}$ in terms of its irreducible components,
\bea
\U_{\b; \,\a(2s-2)} = Y_{\b \a_1 \dots \a_{2s-2}} + \sum_{k=1}^{2s-2}\ve_{\b \a_k} Z_{\a_1 \dots \hat{\a}_k \dots \a_{2s-2}}~,
\eea
where we have introduced the two irreducible components of $\U_{\b; \, \a(2s-2)}$ by the rule
\bea
Y_{\b \a_1 \cdots \a_{2s-2}}:= \U_{(\b; \, \a_1 \cdots \a_{2s-2})}~, \qquad Z_{\a_1 \dots \a_{2s-3}} := \frac{1}{2s-1} \U^{\b;}\,_{\b \a_1 \dots \a_{2s-3}}~. 
\eea 
The next step is to determine the remaining independent component fields of $H_{\a(2s+1)}$ and $\U_{\b; \, \a(2s-2)}$ in the Wess-Zumino gauge \eqref{C.2}.

In the bosonic sector, we have the following set of fields:
\bsubeq
\bea
h_{\a(2s+2)} &:=& -D_{(\a_1}H_{\a_2 \cdots \a_{2s+2})}|_{\q=0}~,\\
y_{\a(2s)} &:=&  D_{(\a_1} Y_{\a_2 \cdots \a_{2s})}|_{\q=0}~,\\
z_{\a(2s-2)} &:=& -\frac{1}{s}(2s-1) D_{(\a_1} Z_{\a_2 \cdots \a_{2s-2})}|_{\q=0}~,\\
z_{\a(2s-4)}&:=& -(2s-1)D^{\b}Z_{\b \a(2s-4)}|_{\q=0}~.
\eea
\esubeq
Reduction of the action \eqref{CC.1} to components leads to the following bosonic action:
\bea
{S}_{\rm bos}
&=& \Big(-\hf \Big)^{s} 
\int 
\rd^3x \,
 \bigg\{-\frac{1}{4}h^{\a(2s+2)} \Box h_{\a(2s+2)}+\frac{3}{16}\pa_{\d \l} h^{\d \l \a(2s)} \pa^{\b \g} h_{\b \g \a(2s)}
\non\\ 
&&+\frac{1}{4}(2s-1)\pa_{\d \l} h^{\d \l \a(2s)} \pa^{\b}\,_{(\a_1} y_{\a_2 \cdots \a_{2s}) \b}
- \frac{1}{4} (2s-1)(s-1) z^{\a(2s-2)} \pa^{\b \g} \pa^{\d \l} h_{\b \g \d \l \a(2s-2)}
\non \\
&&- \frac{1}{4}(2s-1) y^{\a(2s)} \Box y_{\a(2s)}- \frac{1}{8}(s-2)(2s-1)\pa_{\d \l} y^{\d \l \a(2s-2)} \pa^{\b \g} y_{\b \g \a(2s-2)}
\non\\
&&-(s-1)(2s-1) z^{\a(2s)} \Box z_{\a(2s)}
\non\\
&&- \frac{1}{4}(s-1)(s+2)(2s-1)(2s-3)\pa_{\d \l} z^{\d \l \a(2s-4)} \pa^{\b \g} z_{\b \g \a(2s-4)}
\non\\
&&+(s-1)(2s-1)\pa_{\b \g} y^{\b \g \a(2s-2)} \pa^{\d}\,_{(\a_1} z_{\a_2 \cdots \a_{2s-2})\d}
\non \\
&&-\frac{s}{4} \frac{2s-3}{(s-1)(2s-1)}(4s^2-12s+11) z^{\a(2s-4)} \Box z_{\a(2s-4)}
\non\\
&&+\frac{3s}{8(s-1)(2s-1)}(s-2)(2s-3)(2s-5)\pa_{\d \l} z^{\d \l \a(2s-6)} \pa^{\b \g} z_{\b \g \a(2s-6)}
\non\\
&&+ \frac{1}{4}(s+1) (2s-3) z^{\a(2s-4)} \pa^{\b \g} \pa^{\d \l} y_{\b \g \d \l \a(2s-4)}
\non \\
&&+\hf (s-2)(2s+1)(2s-3)\pa_{\b \g} z^{\b \g \a(2s-4)} \pa^{\d}\,_{(\a_1} z_{\a_2 \cdots \a_{2s-4})\d}
\bigg\}~,
\eea
which proves to be invariant under gauge transformations of the form
\bsubeq
\bea
\d_{\x} h_{\a(2s+2)} &=& \pa_{(\a_1 \a_2} \x_{\a_3 \cdots \a_{2s+2})}~, \label{bg1a}\\
\d_{\x, \rho} y_{\a(2s)} &=&-\frac{1}{s+1} \pa^{\b}\,_{(\a_1} \x_{\a_2 \cdots \a_{2s})\b}- \pa_{(\a_1 \a_2} \rho_{\a_3 \cdots \a_{2s})}~, \label{bg2a}\\
\d_{\x, \rho} z_{\a(2s-2)} &=& \frac{1}{2s(2s+1)} \pa^{\b \g} \x_{\b \g \a(2s-2)} + \frac{1}{s} \pa^{\b}\,_{(\a_1} \rho_{\a_2 \cdots \a_{2s-2})\b}~, \label{bg3a} \\
\d_{\rho} z_{\a(2s-4)} &=& \pa^{\b \g} \rho_{\b \g \a(2s-4)}~. \label{bg4a}
\eea
\esubeq

Let us consider the fermionic sector. We find that the independent fermionic fields are:
\bsubeq
\bea
h_{\a(2s+1)} &:=& \frac{\ri}{4} D^2 H_{\a(2s+1)} |_{\q=0}~, \\
y_{\a(2s-1)} &:=& \frac{\ri}{8} D^2 Y_{\a(2s-1)}|_{\q=0}~,\\
y_{\a(2s-3)} &:=& \frac{\ri}{2}s(2s-1) D^2 Z_{\a(2s-3)}|_{\q=0}~,
\eea
\esubeq
and their gauge transformation laws are given by
\bsubeq
\bea
\d_{\l} h_{\a(2s+1)} &=& \pa_{(\a_1 \a_2} \l_{\a_3 \cdots \a_{2s+2})}~, \label{fg1a}\\ 
\d_{\l} y_{\a(2s-1)} &=& \frac{1}{2s+1} \pa^{\b}\,_{(\a_1} \l_{\a_2 \cdots \a_{2s-1})\b}~, \label{fg2a}\\
\d_{\l} y_{\a(2s-3)} &=& \pa^{\b \g } \l_{\b \g \a(2s-3)}~. \label{fg3a}
\eea
\esubeq
The above fermionic fields correspond to the dynamical variables of the Fang-Fronsdal spin-$(s + \hf)$ model. As follows from \eqref{fg1a}, \eqref{fg2a} and \eqref{fg3a}, their gauge freedom is equivalent to that of the
massless spin-$(s + \hf)$ gauge field. 
Indeed, direct calculations of the component action give the standard massless gauge-invariant spin-$(s+\hf)$ action $S^{(2s+1)}_{FF}$.

The component structure of the obtained supermultiplets is a three-dimensional counterpart of so-called (reducible) higher-spin triplet systems. 
In AdS${}_D$ an action for bosonic higher-spin triplets was constructed in
\cite{Sagnotti:2003qa}
and for fermionic triplets in \cite{Sorokin:2008tf,Agugliaro:2016ngl}.
 Our superfield construction  provides a manifestly off-shell supersymmetric generalisation of these systems.
It might be of interest to extend it to AdS${}_4$.


\begin{footnotesize}

\end{footnotesize}


\begin{thebibliography}{66}


  
  \bibitem{HK1} 
  J.~Hutomo and S.~M.~Kuzenko,
  ``Non-conformal higher spin supercurrents,''
  Phys.\ Lett.\ B {\bf 778}, 242 (2018)
   [arXiv:1710.10837 [hep-th]].
 


\bibitem{HK2} 
  J.~Hutomo and S.~M.~Kuzenko,
  ``The massless integer superspin multiplets revisited,''
  JHEP {\bf 1802}, 137 (2018)
  [arXiv:1711.11364 [hep-th]]. 
  
  \bibitem{BHK} 
E.~I.~Buchbinder, J.~Hutomo and S.~M.~Kuzenko,
``Higher spin supercurrents in anti-de Sitter space,''
JHEP {\bf 1809}, 027 (2018)
[arXiv:1805.08055 [hep-th]].  


  \bibitem{HKO} 
  J.~Hutomo, S.~M.~Kuzenko and D.~Ogburn,
  ``${\cal N}=2$ supersymmetric higher spin gauge theories and current multiplets in three dimensions,''
  Phys.\ Rev.\ D {\bf 98}, no. 12, 125004 (2018)
  [arXiv:1807.09098 [hep-th]].
  
 \bibitem{HK18} 
  J.~Hutomo and S.~M.~Kuzenko,
  ``Higher spin supermultiplets in three dimensions: (2,0) AdS supersymmetry,''
  Phys.\ Lett.\ B {\bf 787}, 175 (2018)
  [arXiv:1809.00802 [hep-th]].
  
\bibitem{HK19} 
  J.~Hutomo and S.~M.~Kuzenko,
  ``Field theories with (2,0) AdS supersymmetry in ${\cal N}=1$ AdS superspace,''
  Phys.\ Rev.\ D {\bf 100}, no. 4, 045010 (2019)
  [arXiv:1905.05050 [hep-th]].
  






\bibitem{Wein1}
S.~Weinberg, 
{\it The Quantum Theory of Fields}, Vol. 1:
{\it Foundations.} 
Cambridge University Press (1995).  



\bibitem{Wein2}
S.~Weinberg, 
{\it The Quantum Theory of Fields}, Vol. 2:
{\it Modern Applications.} 
Cambridge University Press (1996).  



\bibitem{GL} 
  Y.~A.~Golfand and E.~P.~Likhtman,
  ``Extension of the algebra of Poincare group generators and violation of p invariance,''
  JETP Lett.\  {\bf 13}, 323 (1971)
  [Pisma Zh.\ Eksp.\ Teor.\ Fiz.\  {\bf 13}, 452 (1971)].
  
\bibitem{VA} 
  D.~V.~Volkov and V.~P.~Akulov,
  ``Possible universal neutrino interaction,''
  JETP Lett.\  {\bf 16}, 438 (1972)
  [Pisma Zh.\ Eksp.\ Teor.\ Fiz.\  {\bf 16}, 621 (1972)];
  ``Is the neutrino a Goldstone particle?''
  Phys.\ Lett.\  {\bf 46B}, 109 (1973).
  

  
\bibitem{WZ1} 
  J.~Wess and B.~Zumino,
  ``Supergauge transformations in four-dimensions,''
  Nucl.\ Phys.\ B {\bf 70}, 39 (1974).
  ``A Lagrangian model invariant under supergauge transformations,''
  Phys.\ Lett.\  {\bf 49B}, 52 (1974). 
  

\bibitem{Dine}
M.~Dine,
{\it Supersymmetry and String Theory: Beyond the Standard Model}, Cambridge University Press (2016).
  
\bibitem{VSor} 
  D.~V.~Volkov and V.~A.~Soroka,
  ``Higgs effect for Goldstone particles with spin 1/2,''
  JETP Lett.\  {\bf 18}, 312 (1973)
  [Pisma Zh.\ Eksp.\ Teor.\ Fiz.\  {\bf 18}, 529 (1973)].

\bibitem{FvNF}
  D.~Z.~Freedman, P.~van Nieuwenhuizen and S.~Ferrara,
  ``Progress toward a theory of supergravity,''
  Phys.\ Rev.\ D {\bf 13}, 3214 (1976).
  
\bibitem{DZ-sugra} 
  S.~Deser and B.~Zumino,
  ``Consistent supergravity,''
  Phys.\ Lett.\  {\bf 62B}, 335 (1976).

\bibitem{FSrev} 
  S.~Ferrara and A.~Sagnotti,
  ``Supergravity at 40: reflections and perspectives,''
  Riv.\ Nuovo Cim.\  {\bf 40}, no. 6, 279 (2017)
  [J.\ Phys.\ Conf.\ Ser.\  {\bf 873}, no. 1, 012014 (2017)]
  [arXiv:1702.00743 [hep-th]].
  

  

  
  
\bibitem{Wein3}
S.~Weinberg, 
{\it The Quantum Theory of Fields}, Volume 3:
{\it Supersymmetry.} 
Cambridge University Press (2005).
  
  \bibitem{GSWbook} 
  M.~B.~Green, J.~H.~Schwarz and E.~Witten,
  {\it Superstring Theory}, vols 1 \& 2,
  Cambridge University Press (1987) 
  (Cambridge Monographs On Mathematical Physics).  
  
  \bibitem{Pol} 
  J.~Polchinski,
  {\it String Theory}, vols. 1 \& 2, Cambridge University Press, Cambridge, 1998.  
  
\bibitem{Vas04} 
  M.~A.~Vasiliev,
  ``Higher spin gauge theories in various dimensions,''
  Fortsch.\ Phys.\  {\bf 52}, 702 (2004)
  [hep-th/0401177].
  
  \bibitem{Dirac36} 
  P.~A.~M.~Dirac,
  ``Relativistic wave equations,''
  Proc.\ Roy.\ Soc.\ Lond.\ A {\bf 155}, 447 (1936).
  
\bibitem{FPauli} 
  M.~Fierz and W.~Pauli,
  ``On relativistic wave equations for particles of arbitrary spin in an electromagnetic field,''
  Proc.\ Roy.\ Soc.\ Lond.\ A {\bf 173}, 211 (1939).

\bibitem{RS41} 
  W.~Rarita and J.~Schwinger,
  ``On a theory of particles with half integral spin,''
  Phys.\ Rev.\  {\bf 60}, 61 (1941).
  
\bibitem{Wigner} 
  E.~P.~Wigner,
  ``On unitary representations of the inhomogeneous Lorentz group,''
  Annals Math.\  {\bf 40}, 149 (1939)
  [Nucl.\ Phys.\ Proc.\ Suppl.\  {\bf 6}, 9 (1989)].

  
\bibitem{Sorokin} 
  D.~Sorokin,
  ``Introduction to the classical theory of higher spins,''
  AIP Conf.\ Proc.\  {\bf 767}, no. 1, 172 (2005)
  [hep-th/0405069].
  
\bibitem{FTrev} 
  A.~Fotopoulos and M.~Tsulaia,
 ``Gauge invariant Lagrangians for free and interacting higher spin fields. A review of the BRST formulation,''
  Int.\ J.\ Mod.\ Phys.\ A {\bf 24}, 1 (2009)
  [arXiv:0805.1346 [hep-th]].
  

\bibitem{BBS} 
  X.~Bekaert, N.~Boulanger and P.~Sundell,
  ``How higher-spin gravity surpasses the spin two barrier: no-go theorems versus yes-go examples,''
  Rev.\ Mod.\ Phys.\  {\bf 84}, 987 (2012)
  [arXiv:1007.0435 [hep-th]].
  
\bibitem{BCIV} 
  X.~Bekaert, S.~Cnockaert, C.~Iazeolla and M.~A.~Vasiliev,
  ``Nonlinear higher spin theories in various dimensions,''
  hep-th/0503128.


  
 

  
  
\bibitem{SH1} 
  L.~P.~S.~Singh and C.~R.~Hagen,
  ``Lagrangian formulation for arbitrary spin. 1. The boson case,''
  Phys.\ Rev.\ D {\bf 9}, 898 (1974).
  
\bibitem{SH2} 
  L.~P.~S.~Singh and C.~R.~Hagen,
  ``Lagrangian formulation for arbitrary spin. 2. The fermion case,''
  Phys.\ Rev.\ D {\bf 9}, 910 (1974).
   
 \bibitem{Fronsdal}
  C.~Fronsdal,
  ``Massless fields with integer spin,''
  Phys.\ Rev.\  D {\bf 18} (1978) 3624.
  
\bibitem{FronsdalAdS} 
  C.~Fronsdal,
  ``Singletons and massless, integral spin fields on de Sitter space (elementary particles in a curved space,''
  Phys.\ Rev.\ D {\bf 20}, 848 (1979).

\bibitem{FF}
  J.~Fang and C.~Fronsdal,
  ``Massless fields with half integral spin,''
  Phys.\ Rev.\  D {\bf 18} (1978)  3630.
  

\bibitem{FFAdS} 
  J.~Fang and C.~Fronsdal,
  ``Massless, half integer spin fields in de Sitter space,''
  Phys.\ Rev.\ D {\bf 22}, 1361 (1980).
    
\bibitem{Ideas} 
I.~L. Buchbinder and S.~M. Kuzenko, {\it Ideas and Methods of Supersymmetry and
Supergravity, Or a Walk Through Superspace},
 IOP, Bristol, 1995 (Revised Edition 1998).
 
 \bibitem{CM} 
  S.~R.~Coleman and J.~Mandula,
  ``All possible symmetries of the $S$-matrix,''
  Phys.\ Rev.\  {\bf 159}, 1251 (1967).
%

\bibitem{WT1} 
  S.~Weinberg,
  ``Photons and gravitons in  $S$-matrix theory: derivation of charge conservation and equality of gravitational and inertial mass,''
  Phys.\ Rev.\  {\bf 135}, B1049 (1964). 
  
\bibitem{WW} 
  S.~Weinberg and E.~Witten,
  ``Limits on massless particles,''
  Phys.\ Lett.\  {\bf 96B}, 59 (1980).   
 
\bibitem{BBB83} 
  A.~K.~H.~Bengtsson, I.~Bengtsson and L.~Brink,
  ``Cubic interaction terms for arbitrarily extended supermultiplets,''
  Nucl.\ Phys.\ B {\bf 227}, 41 (1983).
  
\bibitem{BBvD2} 
  F.~A.~Berends, G.~J.~H.~Burgers and H.~Van Dam,
  ``On spin three self-interactions,''
  Z.\ Phys.\ C {\bf 24}, 247 (1984); 
  ``On the theoretical problems in constructing interactions involving higher spin massless particles,''
  Nucl.\ Phys.\ B {\bf 260}, 295 (1985).
  
  
\bibitem{Metsaev93}
   R.~R.~Metsaev,
  ``Generating function for cubic interaction vertices of higher spin fields in any dimension,''
  Mod.\ Phys.\ Lett.\ A {\bf 8}, 2413 (1993).
 
\bibitem{FV87} 
  E.~S.~Fradkin and M.~A.~Vasiliev,
  ``Cubic interaction in extended theories of massless higher spin fields,''
  Nucl.\ Phys.\ B {\bf 291}, 141 (1987);
%
  ``On the gravitational interaction of massless higher spin fields,''
  Phys.\ Lett.\ B {\bf 189}, 89 (1987).

\bibitem{Vas90} 
  M.~A.~Vasiliev,
  ``Consistent equation for interacting gauge fields of all spins in (3+1)-dimensions,''
  Phys.\ Lett.\ B {\bf 243}, 378 (1990);
  ``More on equations of motion for interacting massless fields of all spins in (3+1)-dimensions,''
  Phys.\ Lett.\ B {\bf 285}, 225 (1992);
  ``Properties of equations of motion of interacting gauge fields of all spins in (3+1)-dimensions,''
  Class.\ Quant.\ Grav.\  {\bf 8}, 1387 (1991).
  
\bibitem{Vas03} 
  M.~A.~Vasiliev,
  ``Nonlinear equations for symmetric massless higher spin fields in (A)dS(d),''
  Phys.\ Lett.\ B {\bf 567}, 139 (2003)
  [hep-th/0304049].
  

  
  
\bibitem{Klebanov} 
  I.~R.~Klebanov and A.~M.~Polyakov,
  ``AdS dual of the critical O(N) vector model,''
  Phys.\ Lett.\ B {\bf 550}, 213 (2002)
  [hep-th/0210114].
  
\bibitem{GG} 
  M.~R.~Gaberdiel and R.~Gopakumar,
  ``An $\rm AdS_3$ dual for minimal model CFTs,''
  Phys.\ Rev.\ D {\bf 83}, 066007 (2011)
  [arXiv:1011.2986 [hep-th]].
  
\bibitem{Zinoviev} 
  Y.~M.~Zinoviev,  ``On massive high spin particles in AdS,''  hep-th/0108192.
  
\bibitem{BBPT} 
  X.~Bekaert, I.~L.~Buchbinder, A.~Pashnev and M.~Tsulaia,
  ``On higher spin theory: strings, BRST, dimensional reductions,''
  Class.\ Quant.\ Grav.\  {\bf 21}, S1457 (2004)
  [hep-th/0312252].
  
 \bibitem{KO} 
  S.~M.~Kuzenko and D.~X.~Ogburn,
  ``Off-shell higher spin N=2 supermultiplets in three dimensions,''
  Phys.\ Rev.\ D {\bf 94}, no. 10, 106010 (2016)
  [arXiv:1603.04668 [hep-th]].
  
\bibitem{KT} 
  S.~M.~Kuzenko and M.~Tsulaia,
  ``Off-shell massive N=1 supermultiplets in three dimensions,''
  Nucl.\ Phys.\ B {\bf 914}, 160 (2017)
  [arXiv:1609.06910 [hep-th]].
  
\bibitem{KP1} 
  S.~M.~Kuzenko and M.~Ponds,
  ``Topologically massive higher spin gauge theories,''
  JHEP {\bf 1810}, 160 (2018)
  [arXiv:1806.06643 [hep-th]].
  
  
\bibitem{Buch2} 
  I.~L.~Buchbinder, M.~V.~Khabarov, T.~V.~Snegirev and Y.~M.~Zinoviev,
  ``Lagrangian formulation of the massive higher spin N=1 supermultiplets in $AdS_4$ space,''
  Nucl.\ Phys.\ B {\bf 942}, 1 (2019)
  [arXiv:1901.09637 [hep-th]].

  



  

\bibitem{VAsuperspace}  
 D.~V.~Volkov and V.~P.~Akulov, 
  ``Goldstone fields with spin 1/2,''
  Theor.\ Math.\ Phys.\  {\bf 18}, 28 (1974)
  [Teor.\ Mat.\ Fiz.\  {\bf 18}, 39 (1974)].
  
  \bibitem{Salam} 
  A.~Salam and J.~A.~Strathdee,
  ``Supergauge transformations,''
  Nucl.\ Phys.\ B {\bf 76}, 477 (1974).
   
  \bibitem{WB} 
  J.~Wess and J.~Bagger,
  {\it Supersymmetry and Supergravity}, 2nd Edition, Princeton University Press, Princeton, 1992. 
  
\bibitem{GGRS}
S.~J.~Gates Jr., M.~T.~Grisaru, M.~Ro\v{c}ek and W.~Siegel,
{\it Superspace, or One Thousand and One Lessons in Supersymmetry},
Benjamin/Cummings (Reading, MA),  1983, hep-th/0108200.

%
%
%

  
  
  




\bibitem{FZ}
 S.~Ferrara and B.~Zumino,
``Transformation properties of the supercurrent,''
Nucl.\ Phys.\  B {\bf 87}, 207 (1975).

\bibitem{BdeRdeW} 
  E.~Bergshoeff, M.~de Roo and B.~de Wit,
  ``Extended conformal supergravity,''
  Nucl.\ Phys.\ B {\bf 182}, 173 (1981). 

\bibitem{Curtright}
  T.~Curtright,
  ``Massless field supermultiplets with arbitrary spin,''
  Phys.\ Lett.\  B {\bf 85},   219 (1979).


\bibitem{Vas80}
  M.~A.~Vasiliev,
``Gauge form of description of massless fields with arbitrary spin,''
 Sov.\ J.\ Nucl.\ Phys.\ \ {\bf 32},  439 (1980)  [Yad.\ Fiz.\ \ {\bf 32}, 855 (1980)].
 

  \bibitem{BO}
  M.~P.~Bellon and S.~Ouvry,
  ``D = 4 supersymmetry for gauge fields of any spin,''
  Phys.\ Lett.\  B {\bf 187}, 93 (1987).

\bibitem{BO2}
  M.~P.~Bellon and S.~Ouvry,
  ``D = 4 superspace formulation for higher spin fields,''
  Phys.\ Lett.\  B {\bf 193}, 67 (1987).
  
 \bibitem{KS94}
S.~M.~Kuzenko and A.~G.~Sibiryakov,
``Free massless higher-superspin superfields on the anti-de Sitter superspace"
Phys.\ Atom.\ Nucl.\  {\bf 57}, 1257 (1994) 
   [Yad.\ Fiz.\  {\bf 57}, 1326 (1994)]
  [arXiv:1112.4612 [hep-th]].
  

\bibitem{KPS}
S.~M.~Kuzenko,  V.~V.~Postnikov and A.~G.~Sibiryakov,
``Massless gauge superfields of higher half-integer superspins,''
JETP Lett.\  {\bf 57},    534 (1993) 
[Pisma Zh.\ Eksp.\ Teor.\ Fiz.\  {\bf 57},  521 (1993)].


\bibitem{KS}
S.~M.~Kuzenko and A.~G.~Sibiryakov,
``Massless gauge superfields of higher integer superspins,''
JETP Lett.\  {\bf 57},   539 (1993)  
[Pisma Zh.\ Eksp.\ Teor.\ Fiz.\  {\bf 57}, 526 (1993)].


  
  \bibitem{BKS} 
  I.~L.~Buchbinder, S.~M.~Kuzenko and A.~G.~Sibiryakov,
  ``Quantization of higher spin superfields in the anti-De Sitter superspace,''
  Phys.\ Lett.\ B {\bf 352}, 29 (1995)
%
\bibitem{GKS1} 
  S.~J.~Gates, Jr., S.~M.~Kuzenko and A.~G.~Sibiryakov,
  ``N=2 supersymmetry of higher superspin massless theories,''
  Phys.\ Lett.\ B {\bf 412}, 59 (1997).
  
  

  
\bibitem{KMT} 
  S.~M.~Kuzenko, R.~Manvelyan and S.~Theisen,
  ``Off-shell superconformal higher spin multiplets in four dimensions,''
  JHEP {\bf 1707}, 034 (2017)
  [arXiv:1701.00682 [hep-th]]. 
  
\bibitem{HST}
P.~S.~Howe, K.~S.~Stelle and P.~K.~Townsend,
``Supercurrents,''  Nucl.\ Phys.\  B {\bf 192}, 332 (1981).

\bibitem{FT} 
E.~S.~Fradkin and A.~A.~Tseytlin,  ``Conformal supergravity,''
Phys.\ Rept.\  {\bf 119}, 233 (1985).


\bibitem{FL} 
  E.~S.~Fradkin and V.~Y.~Linetsky,
  ``Superconformal higher spin theory in the cubic approximation,''
  Nucl.\ Phys.\ B {\bf 350}, 274 (1991).
 

  
%
%
%




 
   
\bibitem{Sohnius} 
  M.~F.~Sohnius,
  ``The multiplet of currents for $N=2$ extended supersymmetry,''
  Phys.\ Lett.\  {\bf 81B}, 8 (1979).

  

\bibitem{MSW} 
  M.~Magro, I.~Sachs and S.~Wolf,
  ``Superfield Noether procedure,''
  Annals Phys.\  {\bf 298}, 123 (2002)
  [hep-th/0110131].

\bibitem{KS-Fayet} 
  Z.~Komargodski and N.~Seiberg,
  ``Comments on the Fayet-Iliopoulos term in field theory and supergravity,''
  JHEP {\bf 0906}, 007 (2009)
  [arXiv:0904.1159 [hep-th]].
  
\bibitem{SMK-Fayet} 
  S.~M.~Kuzenko,
  ``The Fayet-Iliopoulos term and nonlinear self-duality,''
  Phys.\ Rev.\ D {\bf 81}, 085036 (2010)
  [arXiv:0911.5190 [hep-th]].
  
\bibitem{KS2}
Z.~Komargodski and N.~Seiberg,
  ``Comments on supercurrent multiplets, supersymmetric field theories and
  supergravity,''
 [ arXiv:1002.2228 [hep-th]].


\bibitem{K-var}
S.~M.~Kuzenko,
``Variant supercurrent multiplets,''
JHEP {\bf 1004}, 022 (2010)  [arXiv:1002.4932 [hep-th]].
 
  
\bibitem{K-Noet} 
  S.~M.~Kuzenko,
  ``Variant supercurrents and Noether procedure,''
  Eur.\ Phys.\ J.\ C {\bf 71}, 1513 (2011)
  [arXiv:1008.1877 [hep-th]].

 
\bibitem{DS}
  T.~T.~Dumitrescu and N.~Seiberg,
  ``Supercurrents and brane currents in diverse dimensions,''
JHEP {\bf 1107}, 095 (2011)
  [arXiv:1106.0031 [hep-th]].
  

  
  
\bibitem{OS1} 
  V.~Ogievetsky and E.~Sokatchev,
  ``On vector superfield generated by supercurrent,''
  Nucl.\ Phys.\ B {\bf 124}, 309 (1977).
  
\bibitem{FZ2}
S.~Ferrara and B.~Zumino,
``Structure of conformal supergravity,''  Nucl.\ Phys.\  B {\bf 134}, 301 (1978).  

\bibitem{BK_supercurrent}
  D.~Butter and S.~M.~Kuzenko,
  ``N=2 supergravity and supercurrents,''
  JHEP {\bf 1012}, 080 (2010)
  [arXiv:1011.0339 [hep-th]].

\bibitem{KT-M11}
S.~M.~Kuzenko and G.~Tartaglino-Mazzucchelli,
  ``Three-dimensional N=2 (AdS) supergravity and associated supercurrents,''
JHEP {\bf 1112}, 052 (2011)
[arXiv:1109.0496 [hep-th]].

  

 

\bibitem{SohniusW2} 
M.~F.~Sohnius and P.~C.~West,
 ``The new minimal formulation of N=1 supergravity and its tensor calculus,''    
 in {\it Quantum Structure of Space and Time}, M. J. Duff and C. J. Isham (Eds.), 
Cambridge University Press, Cambridge, 1982, pp. 187--222. 
    
\bibitem{SohniusW3} 
 M.~Sohnius and P.~C.~West,
 ``The tensor calculus and matter coupling of the alternative minimal auxiliary field formulation of $N=1$ supergravity,''  Nucl.\ Phys.\ B {\bf 198}, 493 (1982).    


\bibitem{HL}
  P.~S.~Howe and U.~Lindstr\"om,
  ``The supercurrent in five dimensions,''
  Phys.\ Lett.\ B {\bf 103}, 422 (1981).

 \bibitem{Howe5Dsugra}
  P.~S.~Howe,
  ``Off-shell N=2 and N=4 supergravity in five-dimensions,''
in {\it Quantum Structure of Space and Time}, 
M. J. Duff and C. J. Isham, Cambridge University Press, 
1982, pp. 239--253.

\bibitem{SW} 
  M.~F.~Sohnius and P.~C.~West,
  ``An alternative minimal off-shell version of N=1 supergravity,''
  Phys.\ Lett.\  {\bf 105B}, 353 (1981).


\bibitem{GKP}
  S.~J.~Gates Jr., S.~M.~Kuzenko and J.~Phillips,
  ``The off-shell (3/2,2) supermultiplets revisited,''
  Phys.\ Lett.\  B {\bf 576}, 97 (2003)
  [arXiv:hep-th/0306288].



\bibitem{Keck} 
  B.~W.~Keck,
  ``An alternative class of supersymmetries,''
  J.\ Phys.\ A {\bf 8}, 1819 (1975).

  

\bibitem{Zumino77}
B.~Zumino, ``Nonlinear realization of supersymmetry in de Sitter space,''
Nucl.\ Phys.\  B {\bf 127}, 189 (1977).

\bibitem{IS2}
  E.~A.~Ivanov and A.~S.~Sorin,
  ``{Wess-Zumino} model as linear sigma model of spontaneously broken conformal and OSp(1,4) supersymmetries,''
  Sov.\ J.\ Nucl.\ Phys.\  {\bf 30}, 440 (1979)
  [Yad.\ Fiz.\  {\bf 30}, 853 (1979)].

\bibitem{IS1}
E.~A.~Ivanov and A.~S.~Sorin,
``Superfield formulation of OSp(1,4) supersymmetry,''
J.\ Phys.\ A  {\bf 13}, 1159 (1980).


%

  
  \bibitem{Burges:1985qq} 
  C.~J.~C.~Burges, D.~Z.~Freedman, S.~Davis and G.~W.~Gibbons,
  ``Supersymmetry in anti-de Sitter space,''
  Annals Phys.\  {\bf 167}, 285 (1986).

\bibitem{AJKL}
  A.~Adams, H.~Jockers, V.~Kumar and J.~M.~Lapan,
 ``N=1 sigma models in $AdS_4$,''
  arXiv:1104.3155 [hep-th].



\bibitem{BKsigma}
D.~Butter and S.~M.~Kuzenko,
 ``N=2 supersymmetric sigma-models in AdS,''
Phys.\ Lett.\  B {\bf 703}, 620 (2011)
  [arXiv:1105.3111 [hep-th]];
  ``The structure of N=2 supersymmetric nonlinear sigma models in $AdS_4$,''
  JHEP {\bf 1111}, 080 (2011)
  [arXiv:1108.5290 [hep-th]].
  
 
  

  
  
  
   \bibitem{FS}
 G.~Festuccia and N.~Seiberg,
``Rigid supersymmetric theories in curved superspace,''
JHEP {\bf 1106}, 114 (2011) [arXiv:1105.0689 [hep-th]].



\bibitem{Aharony:2015hix} 
  O.~Aharony, M.~Berkooz, A.~Karasik and T.~Vaknin,
  ``Supersymmetric field theories on AdS$_{p} \times$ S$^{q}$,''
  JHEP {\bf 1604}, 066 (2016)
  [arXiv:1512.04698 [hep-th]].
  
\bibitem{BK12} 
D.~Butter and S.~M.~Kuzenko,
``A dual formulation of supergravity-matter theories,''
Nucl.\ Phys.\ B {\bf 854}, 1 (2012)
  [arXiv:1106.3038 [hep-th]].
  

  

\bibitem{BK11} 
  D.~Butter and S.~M.~Kuzenko,
  ``N=2 AdS supergravity and supercurrents,''
  JHEP {\bf 1107}, 081 (2011)
    [arXiv:1104.2153 [hep-th]].
    

 

\bibitem{Kibble}
 T.~W.~B.~Kibble,
 ``Conservation laws for free fields,''
  J.\ Math.\ Phys.\, {\bf 6}, 1022 (1965).

 
  \bibitem{Migdal} 
  A.~A.~Migdal,
  ``Multicolor QCD as a dual-resonance theory,''
  Annals Phys.\  {\bf 109}, 365 (1977).

\bibitem{Makeenko} 
  Y.~M.~Makeenko,
  ``Conformal operators in quantum chromodynamics,''
  Sov.\ J.\ Nucl.\ Phys.\  {\bf 33}, 440 (1981)
  [Yad.\ Fiz.\  {\bf 33}, 842 (1981)].

\bibitem{CDT} 
N.~S.~Craigie, V.~K.~Dobrev and I.~T.~Todorov,
``Conformally covariant composite operators in quantum chromodynamics,''
  Annals Phys.\  {\bf 159}, 411 (1985).
 
\bibitem{BBvD} 
  F.~A.~Berends, G.~J.~H.~Burgers and H.~van Dam,
  ``Explicit construction of conserved currents for massless fields of arbitrary spin,''
  Nucl.\ Phys.\ B {\bf 271}, 429 (1986).  
  
 
 \bibitem{Anselmi} 
  D.~Anselmi,
  ``Theory of higher spin tensor currents and central charges,''
  Nucl.\ Phys.\ B {\bf 541}, 323 (1999)
  [hep-th/9808004]. 
 
\bibitem{Anselmi2} 
  D.~Anselmi,
  ``Higher spin current multiplets in operator product expansions,''
  Class.\ Quant.\ Grav.\  {\bf 17}, 1383 (2000)
  [hep-th/9906167].  

\bibitem{KVZ} 
  S.~E.~Konstein, M.~A.~Vasiliev and V.~N.~Zaikin,
  ``Conformal higher spin currents in any dimension and AdS / CFT correspondence,''
  JHEP {\bf 0012}, 018 (2000)
  [hep-th/0010239].

    





  
\bibitem{NSU} 
  A.~A.~Nizami, T.~Sharma and V.~Umesh,
  ``Superspace formulation and correlation functions of 3d superconformal field theories,''
  JHEP {\bf 1407}, 022 (2014)
  [arXiv:1308.4778 [hep-th]].  
  


\bibitem{Townsend} 
  P.~K.~Townsend,
  ``Cosmological  constant in supergravity,''
  Phys.\ Rev.\ D {\bf 15}, 2802 (1977);

  
\bibitem{AT} 
  A.~Achucarro and P.~K.~Townsend,
  ``A Chern-Simons action for three-dimensional anti-de Sitter supergravity theories,''
  Phys.\ Lett.\ B {\bf 180}, 89 (1986).  
  
\bibitem{BILS}
I.~A.~Bandos, E.~Ivanov, J.~Lukierski and D.~Sorokin,
``On the superconformal flatness of AdS superspaces,''
JHEP \textbf{06}, 040 (2002)
[arXiv:hep-th/0205104 [hep-th]].
  
\bibitem{KLT-M11} 
  S.~M.~Kuzenko, U.~Lindstrom and G.~Tartaglino-Mazzucchelli,
  ``Off-shell supergravity-matter couplings in three dimensions,''
  JHEP {\bf 1103}, 120 (2011)
  [arXiv:1101.4013 [hep-th]].
  
\bibitem{VanNieuwenhuizen:1981ae} 
  P.~van Nieuwenhuizen,
  ``Supergravity,''
  Phys.\ Rept.\  {\bf 68}, 189 (1981).





  \bibitem{BGK1} 
  I.~L.~Buchbinder, S.~J.~Gates and K.~Koutrolikos,
  ``Higher spin superfield interactions with the chiral supermultiplet: Conserved supercurrents and cubic vertices,''
[arXiv:1708.06262 [hep-th]]. 


\bibitem{KKvU} 
  K.~Koutrolikos, P.~Ko\v{c}i and R.~von Unge,
  ``Higher spin superfield interactions with complex linear supermultiplet: conserved supercurrents and cubic vertices,''
  JHEP {\bf 1803}, 119 (2018)
  [arXiv:1712.05150 [hep-th]].  
  
  
\bibitem{BGK-sigma} 
  I.~L.~Buchbinder, S.~J.~Gates and K.~Koutrolikos,
  ``Interaction of supersymmetric nonlinear sigma models with external higher spin superfields via higher spin supercurrents,''
  JHEP {\bf 1805}, 204 (2018)
  [arXiv:1804.08539 [hep-th]].
  
\bibitem{BGK2} 
  I.~L.~Buchbinder, S.~J.~Gates and K.~Koutrolikos,
  ``Conserved higher spin supercurrents for arbitrary spin massless supermultiplets and higher spin superfield cubic interactions,''
  JHEP {\bf 1808}, 055 (2018)
  [arXiv:1805.04413 [hep-th]].

\bibitem{BGK3} 
  I.~L.~Buchbinder, S.~J.~Gates and K.~Koutrolikos,
  ``Integer superspin supercurrents of matter supermultiplets,''
  JHEP {\bf 1905}, 031 (2019)
  [arXiv:1811.12858 [hep-th]].
  
  \bibitem{GK1} 
  S.~J.~Gates and K.~Koutrolikos,
  ``Progress on cubic interactions of arbitrary superspin supermultiplets via gauge invariant supercurrents,''
  Phys.\ Lett.\ B {\bf 797}, 134868 (2019)
  [arXiv:1904.13336 [hep-th]].


%
%
%






 
  
\bibitem{MR} 
  R.~Manvelyan and W.~Ruhl,
  ``Conformal coupling of higher spin gauge fields to a scalar field in AdS(4) and generalized Weyl invariance,''
  Phys.\ Lett.\ B {\bf 593}, 253 (2004)
  [hep-th/0403241].  
  
  \bibitem{MM} 
  R.~Manvelyan and K.~Mkrtchyan,
  ``Conformal invariant interaction of a scalar field with the higher spin field in AdS(D),''
  Mod.\ Phys.\ Lett.\ A {\bf 25}, 1333 (2010)
  [arXiv:0903.0058 [hep-th]].
  
 \bibitem{FIPT} 
  A.~Fotopoulos, N.~Irges, A.~C.~Petkou and M.~Tsulaia,
  ``Higher-spin gauge fields interacting with scalars: the Lagrangian cubic vertex,''
  JHEP {\bf 0710}, 021 (2007)
  [arXiv:0708.1399 [hep-th]]. 
  
\bibitem{FTsulaia} 
  A.~Fotopoulos and M.~Tsulaia,
``Current exchanges for reducible higher spin modes on AdS,''
 [arXiv:1007.0747 [hep-th]].
    
 \bibitem{BekaertM} 
  X.~Bekaert and E.~Meunier,
  ``Higher spin interactions with scalar matter on constant curvature spacetimes: conserved current and cubic coupling generating functions,''
  JHEP {\bf 1011}, 116 (2010)
  [arXiv:1007.4384 [hep-th]]. 

  
  

  




  


%
%
%
%
%
%
%
%
%
%
%
%
%
%
%
%
%
  
%
%
%
%

%
%
%
%

%

%
%
%
%
%
%
%
%
%
%
%
%
%
%
%
%
%
  
  

    
\bibitem{Ber} 
  F.~A.~Berezin,
  ``The method of second quantization,''
  Pure Appl.\ Phys.\  {\bf 24}, 1 (1966).
  
\bibitem{Kaku-sugra} 
  M.~Kaku and P.~K.~Townsend,
  ``Poincare supergravity as broken superconformal gravity,''
  Phys.\ Lett.\  {\bf 76B}, 54 (1978).


\bibitem{WZ78}
J.~Wess and B.~Zumino,
 ``Superfield Lagrangian for supergravity,''
 Phys.\ Lett.\  B {\bf 74}, 51 (1978).

\bibitem{old1}
K.~S.~Stelle and P.~C.~West,
``Minimal auxiliary fields for supergravity,''
Phys.\ Lett.\  B {\bf 74},  330 (1978).


\bibitem{old2}
S.~Ferrara and P.~van Nieuwenhuizen,
``The auxiliary fields of supergravity,''
Phys.\ Lett.\  B {\bf 74}, 333 (1978).



  
\bibitem{Nakayama} 
  Y.~Nakayama,
  ``Supercurrent, supervirial and superimprovement,''
  Phys.\ Rev.\ D {\bf 87}, no. 8, 085005 (2013)
  [arXiv:1208.4726 [hep-th]].
  
  \bibitem{BGLP}
I.~L.~Buchbinder, S.~J.~Gates Jr., W.~D.~Linch and J.~Phillips,
``New 4D, N = 1 superfield theory: Model of free massive superspin-3/2  multiplet,''
Phys.\ Lett.\  B {\bf 535}, 280 (2002)
[arXiv:hep-th/0201096].

\bibitem{Siegel}
  W.~Siegel,  ``Unextended superfields in extended supersymmetry,''
  Nucl.\ Phys.\  B {\bf 156}, 135 (1979).

\bibitem{Siegel78} 
  W.~Siegel,
  ``Solution to constraints in {Wess-Zumino} supergravity formalism,''
  Nucl.\ Phys.\ B {\bf 142}, 301 (1978).

  




 \bibitem{West} 
  P.~C.~West, {\it Introduction to Supersymmetry and Supergravity},
  World Scientific, Singapore, 1986 (Extended Revised Edition: 1990).
  
  
\bibitem{KU} 
  T.~Kugo and S.~Uehara,
  ``$N=1$ superconformal tensor calculus: Multiplets with external Lorentz indices and spinor derivative operators,''
  Prog.\ Theor.\ Phys.\  {\bf 73}, 235 (1985).  


  
 


\bibitem{GS80}
S.~J.~Gates  Jr. and W.~Siegel,
``(3/2, 1) superfield of O(2) supergravity,''  Nucl.\ Phys.\  {\bf B164}, 484 (1980). 

\bibitem{FV79}
  E.~S.~Fradkin and M.~A.~Vasiliev,
  ``Minimal set of auxiliary fields and S-matrix for extended supergravity,''
  Lett.\ Nuovo Cim.\  {\bf 25}, 79 (1979).

\bibitem{deWvanH}
  B.~de Wit and  J.~W.~van Holten,
  ``Multiplets of linearized SO(2) supergravity,''
  Nucl.\ Phys.\  {\bf B155}, 530 (1979).
  
\bibitem{LR2}
  U.~Lindstr\"om and M.~Ro\v{c}ek,
  ``Scalar tensor duality and $N = 1, 2$ nonlinear $\sigma$-models,''
  Nucl.\ Phys.\  B {\bf 222}, 285 (1983).

\bibitem{OS} 
  V.~I.~Ogievetsky and E.~Sokatchev,
  ``On gauge spinor superfield,''
  JETP Lett.\  {\bf 23}, 58 (1976).

 \bibitem{Fayet}
P.~Fayet, ``Fermi-Bose hypersymmetry,''
Nucl.\ Phys.\ B {\bf 113}, 135 (1976).

\bibitem{Soh-central} 
  M.~F.~Sohnius,
  ``Supersymmetry and central charges,''
  Nucl.\ Phys.\ B {\bf 138}, 109 (1978).


\bibitem{KNT} 
S.~M.~Kuzenko, J.~Novak and S.~Theisen,
  ``Non-conformal supercurrents in six dimensions,''
[arXiv:1709.09892 [hep-th]].
 
 
\bibitem{Bonora1} 
  L.~Bonora, M.~Cvitan, P.~Dominis Prester, B.~Lima de Souza and I.~Smoli\'c,
  ``Massive fermion model in 3d and higher spin currents,''
  JHEP {\bf 1605}, 072 (2016)
  [arXiv:1602.07178 [hep-th]].

\bibitem{Bonora2} 
  L.~Bonora, M.~Cvitan, P.~Dominis Prester, S.~Giaccari, B.~Lima de Souza and T.~\v{S}temberga,
  ``One-loop effective actions and higher spins,''
  JHEP {\bf 1612}, 084 (2016)
  [arXiv:1609.02088 [hep-th]].
  
\bibitem{Bonora3} 
  L.~Bonora, M.~Cvitan, P.~Dominis Prester, S.~Giaccari and T.~\v{S}temberga,
  ``One-loop effective actions and higher spins. II,''
  [arXiv:1709.01738 [hep-th]].  
  

  
  

  
\bibitem{KT-M08}
S.~M.~Kuzenko and G.~Tartaglino-Mazzucchelli,
 ``Five-dimensional superfield supergravity,''
 Phys.\ Lett.\  B {\bf 661}, 42 (2008)
  [arXiv:0710.3440 [hep-th]];
  ``5D supergravity and projective superspace,''
  JHEP {\bf 0802}, 004 (2008) [arXiv:0712.3102].


\bibitem{KLRT-M}
S.~M.~Kuzenko, U.~Lindstr\"om, M.~Ro\v cek and G.~Tartaglino-Mazzucchelli,
``4D N=2 supergravity and projective superspace,'' 
JHEP {\bf 0809}, 051 (2008) [arXiv:0805.4683].

\bibitem{KT-M-ads}
  S.~M.~Kuzenko and G.~Tartaglino-Mazzucchelli,
  ``Field theory in 4D N=2 conformally flat superspace,''
  JHEP {\bf 0810}, 001 (2008)
  [arXiv:0807.3368 [hep-th]].
  
\bibitem{Siegel79}
W.~Siegel,
``Gauge spinor superfield as a scalar multiplet,''
Phys.\ Lett.\ B {\bf 85}, 333 (1979).

\bibitem{KKvU-sc} 
  P.~Ko\v{c}i, K.~Koutrolikos and R.~von Unge,
  ``Complex linear superfields, supercurrents and supergravities,''
  JHEP {\bf 1702}, 076 (2017)
  [arXiv:1612.08706 [hep-th]].
  
\bibitem{koci-thesis}
P.~Ko\v{c}i,
{\it Aspects of Supersymmetry and Supergravity}, PhD Thesis, Masaryk University, 2019, 312p.

\bibitem{GSV} 
  O.~A.~Gelfond, E.~D.~Skvortsov and M.~A.~Vasiliev,
  ``Higher spin conformal currents in Minkowski space,''
  Theor.\ Math.\ Phys.\  {\bf 154}, 294 (2008)
  [hep-th/0601106]. 
  




\bibitem{KLT-M12} 
S.~M.~Kuzenko, U.~Lindstr\"om and G.~Tartaglino-Mazzucchelli,
``Three-dimensional (p,q) AdS superspaces and matter couplings,''
JHEP {\bf 1208}, 024 (2012)  
[arXiv:1205.4622 [hep-th]].




\bibitem{RvanN86} 
  M.~Ro\v{c}ek and P.~van Nieuwenhuizen,
  ``N $\geq$ 2 supersymmetric Chern-Simons terms as d = 3 extended conformal supergravity,''
  Class.\ Quant.\ Grav.\  {\bf 3}, 43 (1986).

\bibitem{ZupnikPak} 
  B.~M.~Zupnik and D.~G.~Pak,
  ``Superfield formulation of the simplest three-dimensional gauge theories 
  and conformal supergravities,''
  Theor.\ Math.\ Phys.\  {\bf 77}, 1070 (1988)
  [Teor.\ Mat.\ Fiz.\  {\bf 77}, 97 (1988)].


\bibitem{NG}
  H.~Nishino and S.~J.~Gates Jr.,
  ``Chern-Simons theories with supersymmetries in three dimensions,''
  Int.\ J.\ Mod.\ Phys.\  A {\bf 8}, 3371 (1993).
  


  
\bibitem{BCSS} 
  E.~Bergshoeff, S.~Cecotti, H.~Samtleben and E.~Sezgin,
  ``Superconformal sigma models in three dimensions,''
  Nucl.\ Phys.\ B {\bf 838}, 266 (2010)
  [arXiv:1002.4411 [hep-th]].  
  

%




  
 
  
  

  
  \bibitem{KLRST-M} 
  S.~M.~Kuzenko, U.~Lindstr\"om, M.~Ro\v{c}ek, I.~Sachs 
  and G.~Tartaglino-Mazzucchelli,
  ``Three-dimensional N=2 supergravity theories: From superspace to components,''
  Phys.\ Rev.\ D {\bf 89}, 085028 (2014)
  [arXiv:1312.4267 [hep-th]].


\bibitem{HIPT}
  P.~S.~Howe, J.~M.~Izquierdo, G.~Papadopoulos and P.~K.~Townsend,
  ``New supergravities with central charges and Killing spinors in 2+1 dimensions,''
  Nucl.\ Phys.\  B {\bf 467}, 183 (1996)
  [arXiv:hep-th/9505032]. 
  
  
  \bibitem{DJT1}
  S.~Deser, R.~Jackiw and S.~Templeton,
  ``Three-dimensional massive gauge theories,''
  Phys.\ Rev.\ Lett.\  {\bf 48}, 975 (1982).

\bibitem{DJT2}
  S.~Deser, R.~Jackiw and S.~Templeton,
  ``Topologically massive gauge theories,''
  Annals Phys.\  {\bf 140}, 372 (1982)
  [Erratum-ibid.\  {\bf 185}, 406 (1988)].
  
  \bibitem{DK} 
S.~Deser and J.~H.~Kay,
``Topologically massive supergravity,''
Phys.\ Lett.\ B {\bf 120}, 97 (1983).

\bibitem{Deser84}
S.~Deser,
``Cosmological topological supergravity,''
 in {\it Quantum Theory Of Gravity}, S. M. Christensen (Ed.), 
 Adam Hilger, Bristol, 1984, pp. 374-381.


  
\bibitem{BKNT-M1} 
  D.~Butter, S.~M.~Kuzenko, J.~Novak and G.~Tartaglino-Mazzucchelli,
  ``Conformal supergravity in three dimensions: New off-shell formulation,''
  JHEP {\bf 1309}, 072 (2013)
  [arXiv:1305.3132 [hep-th]]. 
  

  
\bibitem{Kuzenko12} 
S.~M.~Kuzenko,
``Prepotentials for N=2 conformal supergravity in three dimensions,''
JHEP {\bf 1212}, 021 (2012)  [arXiv:1209.3894 [hep-th]].
    
  



  





\bibitem{BKT-M} 
  D.~Butter, S.~M.~Kuzenko and G.~Tartaglino-Mazzucchelli,
  ``Nonlinear sigma models with AdS supersymmetry in three dimensions,''
  JHEP {\bf 1302}, 121 (2013)
  [arXiv:1210.5906 [hep-th]].
  
  

 \bibitem{KPT-MvU}
S.~M.~Kuzenko, J.-H.~Park, G.~Tartaglino-Mazzucchelli and R.~Unge,
``Off-shell superconformal nonlinear sigma-models in three dimensions,''
JHEP {\bf 1101}, 146  (2011)  [arXiv:1011.5727 [hep-th]].


  

  
  

  

\bibitem{Schonfeld} 
J.~F.~Schonfeld, ``A mass term for three-dimensional gauge fields,''
  Nucl.\ Phys.\ B {\bf 185}, 157 (1981).


  
\bibitem{OS2}
V.~I.~Ogievetsky and E.~Sokatchev,
``Superfield equations of motion,''
J.\ Phys.\ A {\bf 10}, 2021 (1977).


\bibitem{BL1}
N.~Berkovits and M.~M.~Leite,
``First massive state of the superstring in superspace,''
Phys.\ Lett.\ B {\bf 415}, 144 (1997) 
[hep-th/9709148].


\bibitem{BL2}
N.~Berkovits and M.~M.~Leite,
``Superspace action for the first massive
states of the superstring,''
Phys.\ Lett.\ B {\bf 454}, 38 (1999) 
[hep-th/9812153].

\bibitem{AB1}
R.~Altendorfer and J.~Bagger,
 ``Dual supersymmetry algebras from partial supersymmetry breaking,''
Phys.\ Lett.\ B {\bf 460}, 127 (1999) [hep-th/9904213].

\bibitem{AB2}
R.~Altendorfer and J.~Bagger,
``Dual anti-de Sitter superalgebras from 
partial supersymmetry breaking,''
  Phys.\ Rev.\ D {\bf 61},  104004 (2000) [hep-th/9908084].


\bibitem{BGLP1}
I.~L.~Buchbinder, S.~J.~Gates Jr., W.~D.~Linch and J.~Phillips,
``New 4D, N = 1 superfield theory: Model
of free massive superspin-3/2 multiplet,''
Phys.\ Lett.\ B {\bf 535}, 280   (2002) [hep-th/0201096].

\bibitem{BGLP2}
I.~L.~Buchbinder, S.~J.~Gates Jr., W.~D.~Linch and J.~Phillips,
``Dynamical superfield theory of free massive
superspin-1 multiplet,''
Phys.\ Lett.\ B {\bf 549}, 229 (2002) [hep-th/0207243].


\bibitem{GSS}
T.~Gregoire, M.~D.~Schwartz and Y.~Shadmi,
``Massive supergravity and deconstruction,''
JHEP {\bf 0407}, 029 (2004)
[hep-th/0403224].

\bibitem{BGKPmass} 
  I.~L.~Buchbinder, S.~James Gates, Jr., S.~M.~Kuzenko and J.~Phillips,
``Massive 4D, $N=1$ superspin 1 \& 3/2 multiplets and dualities,''
JHEP {\bf 0502}, 056 (2005)
[hep-th/0501199].



\bibitem{Gates:2005su} 
  S.~J.~Gates Jr. and S.~M.~Kuzenko,
  ``4D, N = 1 higher spin gauge superfields and quantized twistors,''
  JHEP {\bf 0510}, 008 (2005)
  [hep-th/0506255].

\bibitem{GKT-M} 
  S.~J.~Gates Jr., S.~M.~Kuzenko and G.~Tartaglino-Mazzucchelli,
  ``New massive supergravity multiplets,''
  JHEP {\bf 0702}, 052 (2007)
  [hep-th/0610333].
  
\bibitem{KNT-M}
S.~M.~Kuzenko, J.~Novak and G.~Tartaglino-Mazzucchelli,
``Higher derivative couplings and massive supergravity in three dimensions,''
JHEP {\bf 1509}, 081 (2015) 
  [arXiv:1506.09063 [hep-th]].
  
\bibitem{BSZ3} 
  I.~L.~Buchbinder, T.~V.~Snegirev and Y.~M.~Zinoviev,
  ``Lagrangian formulation of the massive higher spin supermultiplets in three dimensional space-time,''
  JHEP {\bf 1510}, 148 (2015)
  [arXiv:1508.02829 [hep-th]].

\bibitem{BSZ4} 
I.~L.~Buchbinder, T.~V.~Snegirev and Y.~M.~Zinoviev,
  ``Lagrangian description of massive higher spin supermultiplets in AdS$_{3}$ space,''
  JHEP {\bf 1708}, 021 (2017)
  [arXiv:1705.06163 [hep-th]].


\bibitem{BSZ1} 
  I.~L.~Buchbinder, T.~V.~Snegirev and Y.~M.~Zinoviev,
  ``Gauge invariant Lagrangian formulation of massive higher spin fields in $(A)dS_3$ space,''
  Phys.\ Lett.\ B {\bf 716}, 243 (2012)
  [arXiv:1207.1215 [hep-th]].

\bibitem{BSZ2} 
  I.~L.~Buchbinder, T.~V.~Snegirev and Y.~M.~Zinoviev,
  ``Frame-like gauge invariant Lagrangian formulation of massive fermionic higher spin fields in ${\rm AdS}_3$ space,''
  Phys.\ Lett.\ B {\bf 738}, 258 (2014)
  [arXiv:1407.3918 [hep-th]].



\bibitem{Metsaev} 
  R.~R.~Metsaev,
  ``Gauge invariant formulation of massive totally symmetric fermionic fields in (A)dS space,''  Phys.\ Lett.\ B {\bf 643}, 205 (2006)
  [hep-th/0609029].
  
\bibitem{Sagnotti:2003qa}
 A.~Sagnotti and M.~Tsulaia,``On higher spins and the tensionless limit of string theory,''
 Nucl.\ Phys.\ B {\bf 682}, 83 (2004)
[hep-th/0311257].


\bibitem{Sorokin:2008tf}
 D.~P.~Sorokin and M.~A.~Vasiliev, ``Reducible higher-spin multiplets in flat and AdS spaces and their geometric frame-like formulation,''
Nucl.\ Phys.\ B {\bf 809}, 110 (2009)
[arXiv:0807.0206 [hep-th]].

\bibitem{Agugliaro:2016ngl}
A.~Agugliaro, F.~Azzurli and D.~Sorokin,``Fermionic higher-spin triplets in AdS,''
Nucl.\ Phys.\ B {\bf 907}, 633 (2016)
[arXiv:1603.02251 [hep-th]].
  
\bibitem{PV1}
S.~Prokushkin and M.~A.~Vasiliev,
``Higher-spin gauge interactions for massive matter fields in 3D AdS spacetime,''
Nucl. Phys. B \textbf{545}, 385 (1999)
[arXiv:hep-th/9806236 [hep-th]].

\bibitem{PV2}
S.~F.~Prokushkin and M.~A.~Vasiliev,
``Currents of arbitrary spin in $\rm AdS_3$,''
Phys. Lett. B \textbf{464}, 53-61 (1999)
[arXiv:hep-th/9906149 [hep-th]].

\bibitem{PSV}
S.~F.~Prokushkin, A.~Y.~Segal and M.~A.~Vasiliev,
``Coordinate free action for $\rm AdS_3$ higher-spin-matter systems,''
Phys. Lett. B \textbf{478}, 333-342 (2000)
[arXiv:hep-th/9912280 [hep-th]].


\bibitem{BV}
  I.~A.~Batalin and G.~A.~Vilkovisky,
  ``Quantization of gauge theories with linearly dependent generators,''
  Phys.\ Rev.\  {\bf D28},   2567 (1983).




  
 \bibitem{FP} 
  L.~D.~Faddeev and V.~N.~Popov,
  ``Feynman diagrams for the Yang-Mills field,''
  Phys.\ Lett.\ B {\bf 25}, 29 (1967).


  
 
%
%
  
  
  


\bibitem{FH} 
  C.~Fronsdal and H.~Hata,
  ``Quantization of massless fields with arbitrary spin,''
  Nucl.\ Phys.\ B {\bf 162}, 487 (1980).

\bibitem{GGS} 
  M.~R.~Gaberdiel, R.~Gopakumar and A.~Saha,
  ``Quantum $W$-symmetry in AdS${}_3$,''
  JHEP {\bf 1102}, 004 (2011)
  [arXiv:1009.6087 [hep-th]].
  



 
%
%
\end{thebibliography}
\end{document}